\documentclass[reqno, 11pt]{gen-m-l}
\usepackage{latexsym, amssymb}
\usepackage{hyperref}
\usepackage{amsmath}
\usepackage{amsthm}
\usepackage{epsfig}
\usepackage[mathscr]{eucal}
\numberwithin{equation}{section}
\allowdisplaybreaks[1]

\renewcommand{\theequation}{\thesection.\arabic{equation}}
\renewcommand{\thesection}{\thechapter.\arabic{section}}

\allowdisplaybreaks[1]

\newcommand{\entangle}{``Entanglement and second quantization in the framework of the
fermionic projector'' (arXiv:0911.0076 [math-ph])}
\newtheorem{Def}{Def.}[section]
\newtheorem{Thm}[Def]{Theorem}

\newtheorem{Lemma}[Def]{Lemma}
\newtheorem{Remark}[Def]{Remark}
\newtheorem{Corollary}[Def]{Corollary}
\newcommand{\Proof}{{\em{Proof. }}}
\newcommand{\QED}{\ \hfill $\FBox$ \\[1em]}
\newcommand{\spc}{\;\;\;\;\;\;\;\;\;\;}
\newcommand{\bra}{\mbox{$< \!\!$ \nolinebreak}}
\newcommand{\ket}{\mbox{\nolinebreak $>$}}
\newcommand{\Bra}{\langle}
\newcommand{\Ket}{\rangle}
\newfont{\Bb}{msbm10 scaled 1095}
\newcommand\C{{\mbox{\Bb C}}}
\newcommand\Z{{\mbox{\Bb Z}}}
\newcommand\sC{{\mbox{\bf{\scriptsize{C}}}}}
\newcommand\R{{\mbox{\Bb R}}}
\newcommand\N{{\mbox{\Bb N}}}

\newcommand{\1}{\mbox{\rm 1 \hspace{-1.05 em} 1}}

\newcommand{\sR}{\mbox{\rm \scriptsize I \hspace{-.8 em} R}}

\newcommand{\Pdd}{\mbox{$\partial$ \hspace{-1.2 em} $/$}}
\newcommand{\slsh}{\mbox{ \hspace{-1.1 em} $/$}}
\newcommand{\Tr}{\mbox{\rm{Tr}\/}}
\newcommand{\tr}{\mbox{tr\/}}
\newcommand{\FBox}{\rule{2mm}{2.25mm}}
\newcommand{\OBox}{\raisebox{.6ex}{\fbox{}}\,}
\newcommand{\Sl}{\mbox{$\prec \!\!$ \nolinebreak}}
\newcommand{\Sr}{\mbox{\nolinebreak $\succ$}}
\newcommand{\Aslsh}{\mbox{ $\!\!A$ \hspace{-1.2 em} $/$}}

\newcommand{\Pexp}{\mbox{\rm{Pexp}}}
\newcommand{\eff}{{\mbox{\scriptsize{eff}}}}
\newcommand{\beq}{\begin{equation}}
\newcommand{\eeq}{\end{equation}}

\renewcommand{\H}{\mathscr{H}}

\newcounter{backupeqn}
\makeindex

\begin{document}

\frontmatter
\title{{{\vspace*{2cm}The Principle of the \\ Fermionic Projector}} \\[.5cm] \Large{Appendices}}
\author{\Large{Felix Finster} \\[2cm]
{\normalsize{\bf{Abstract}}} \\[.3cm]
\begin{quote}
\pagestyle{empty}
\footnotesize
The  ``principle of the fermionic projector'' provides a new mathematical 
framework for the formulation of physical theories and is a promising 
approach for physics beyond the standard model. The book begins with a 
brief review of relativity, relativistic quantum mechanics and classical 
gauge theories, with the emphasis on the basic physical concepts and the 
mathematical foundations. The external field problem and Klein's paradox 
are discussed and then resolved by introducing the so-called fermionic 
projector, a global object in space-time which generalizes the notion of 
the Dirac sea. The mathematical core of the book is to give a precise 
definition of the fermionic projector and to employ methods of 
hyperbolic differential equations for its detailed analysis. The 
fermionic projector makes it possible to formulate a new type of 
variational principles in space-time. The mathematical tools for the 
analysis of the corresponding Euler-Lagrange equations are developed. A 
particular variational principle is proposed which gives rise to an 
effective interaction showing many similarities to the interactions of 
the standard model.

The main chapters of the book are easily accessible for beginning 
graduate students in mathematics or physics. Several appendices provide 
supplementary material which will be useful to the experienced researcher.
\end{quote}}

\maketitle

\setcounter{page}{4}

\tableofcontents

\chapter*{Preface to the Second Online Edition}

\vspace*{-0.25em}

In the almost twelve years since this book was completed, the fermionic projector approach
evolved to what is known today as the theory of causal fermion systems.
There has been progress in several directions: the mathematical
setting was generalized, the mathematical methods were improved and enriched,
and the physical applications have been concretized and worked out in more detail.
The current status of the theory is presented in a coherent way in
the recent monograph~\cite{cfs}. An untechnical physical introduction is given in~\cite{dice2014}.

Due to these developments, parts of the present book are superseded by the
more recent research papers or the monograph~\cite{cfs}.
However, other parts of this book have not been developed further and are still up to date.
For some aspects not covered in~\cite{cfs}, the present book is still the best reference.
Furthermore, the present book is still of interest as being the first publication in which
the causal action principle was presented.
Indeed, comparing the presentation in the present book
to the later developments should give the reader a deeper understanding of why
certain constructions were modified and how they were improved.
In order to facilitate such a study, we now outline the developments which led from the present
book to the monograph~\cite{cfs}. In order not to change the original bibliography,
a list of references to more recent research papers is given at the end of this preface,
where numbers are used (whereas the original bibliography using letters is still
at the end of the book). Similar as in the first online edition, I took the opportunity
to correct a few typos. Also, I added a few footnotes 
beginning with ``{\textsf{Online version}:}''.
Apart from these changes, the online version coincides precisely
with the printed book in the AMS/IP series. In particular, all equation numbers
are the same.

Maybe the most important change in the mathematical setup was the
move from indefinite inner product spaces to Hilbert spaces, as we now explain in detail.
Clearly, the starting point of all my considerations was Dirac theory. On Dirac wave functions in
Minkowski space, one can introduce the two inner products
\begin{align}
( \Psi | \Phi) &= \int_{\sR^3} (\overline{\Psi} \gamma^0 \Phi)(t, \vec{x})\: d^3x \tag{1} \label{sprodMin} \\
\bra \Psi|\Phi \ket &= \int_{\sR^4} \overline{\Psi(x)} \Phi(x) \: d^4x \:. \tag{2} \label{stipMin}
\end{align}
The first inner product~\eqref{sprodMin} is positive definite and thus defines a scalar product.
For solutions of the Dirac equation, it is time independent due to current conservation,
making the solution space of the Dirac equation to a Hilbert space
(more generally, the scalar product can be computed by integrating the normal component
of the Dirac current over any Cauchy surface).
The inner product~\eqref{stipMin}, on the other hand,
is indefinite. It is well-defined and covariant even on wave functions which do not satisfy the Dirac equation,
giving rise to an indefinite inner product space (which can be given a Krein space structure).
It should be pointed out that the time integral in~\eqref{stipMin} in general diverges for solutions of the
Dirac equation, a problem which I always considered to be more of technical than of fundamental nature
(this technical problem can be resolved for example
by working as in~\eqref{56}--\eqref{58} with a $\delta$-normalization in the mass
parameter or by making use of the mass oscillation property as introduced in~\cite{infinite}).

The fermionic projector approach is based on the belief that on a microscopic scale (like the Planck scale),
space-time should not be modeled by Minkowski space but should have
a different, possibly discrete structure. Consequently, the Dirac equation in Minkowski space
should not be considered as being fundamental, but it should be replaced by equations of different type.
For such a more fundamental description, the scalar product~\eqref{sprodMin} is problematic, 
because it is not clear how the analog of an integral over a hypersurface should be defined,
and why this integral should be independent of the choice of the hypersurface.
The indefinite inner product~\eqref{stipMin}, however, can easily be generalized to 
for example a discrete space-time if one simply replaces the
integral in~\eqref{stipMin} by a sum over all space-time points.
Such considerations led me to regard the indefinite inner product~\eqref{stipMin} as being
more fundamental than~\eqref{sprodMin}. This is the reason why throughout this book,
we work preferably with indefinite inner product spaces. In particular, the structure of
``discrete space-time'' is introduced on an underlying indefinite inner product space
(see~\S\ref{psec13}).

My views changed gradually over the past few years. The first input which triggered this process
was obtained when developing the existence theory for the causal action principle.
While working on this problem in the simplest setting of a finite number of
space-time points~\cite{discrete}, it became clear that in
order to ensure the existence of minimizers, one needs to assume that the 
image of the fermionic projector~$P$ is a {\em{negative definite}} subspace
of the indefinite inner product space~$(H, \bra .|. \ket)$. The fact that~$P$ has
a definite image makes it possible to introduce a Hilbert space~$(\H, \langle .|. \rangle_\H)$
by setting~$\langle .|. \rangle_\H = -\bra .| P\, . \ket$ and dividing out the null space.
This construction, which was first given in~\cite[Section~1.2.2]{rrev}, gave
an underlying Hilbert space structure. However, at this time, the connection of the corresponding
scalar product to integrals over hypersurfaces as in~\eqref{sprodMin} remained obscure.

From the mathematical point of view, having an underlying Hilbert space structure has the major
benefit that functional analytic methods in Hilbert spaces become applicable.
When thinking about how to apply these methods, it became clear that also measure-theoretic
methods are useful. This led me to generalize the mathematical setting such as to allow
for the description of not only discrete, but also continuous space-times.
This setting was first introduced in~\cite{continuum} when working out the existence theory.
This analysis also clarified which constraints one must impose in order to obtain
a mathematically well-posed variational problem.

The constructions in~\cite{continuum} also inspired the notion of the {\em{universal measure}},
as we now outline. When working out the existence theory, it became clear that instead of using the
kernel of the fermionic projector, the causal action principle can be formulated equivalently
in terms of the local correlation operators~$F(x)$ which relate the Hilbert space scalar product
to the spin scalar product by
\[ \langle \psi | F(x) \phi \rangle_\H = -\Sl \psi(x) | \phi(x) \Sr_x \:. \]
In this formulation, the only a-priori structure of space-time is that of being a measure space~$(M, \mu)$.
The local correlation operators give rise to a mapping
\[ F \::\: M \rightarrow {\mathscr{F}} \:,\quad x \mapsto F(x) \:, \]
where~${\mathscr{F}}$ is the subset of finite rank operators on~$\H$ which are symmetric and 
(counting multiplicities) have at most~$2N$ positive and at most~$2N$ negative eigenvalues
(where~$N$ denotes the number of sectors).
Then, instead of working with the measure~$\mu$, the causal action can be expressed
in terms of the push-forward measure~$\rho = F_* \mu$, being a measure on~${\mathscr{F}}$
(defined by~$\rho(\Omega) = \mu(F^{-1}(\Omega))$).
As a consequence, it seems natural to leave out the measure space~$(M, \mu)$ and to work instead
directly with the measure~$\rho$ on~${\mathscr{F}}$, referred to as the universal measure.
We remark that working with~$(M, \mu)$ has the potential benefit that it is possible to
prescribe properties of the measure~$\rho$. In particular, if~$\mu$ is a discrete measure,
then so is~$\rho$ (for details see~\cite[Section~1.2]{continuum}). However, the analysis of the
causal action principle in~\cite{support} suggests
that minimizing measures are always discrete, even if one varies over all regular Borel measures
(which may have discrete and continuous components).
With this in mind, it seems unnecessary to arrange the discreteness of the measure~$\rho$
by starting with a discrete measure space~$(M, \mu)$.
Then the measure space~$(M, \mu)$ becomes obsolete.
These considerations led me to the conviction that one should work with the universal measure~$\rho$,
which should be varied within the class of all regular Borel measures. Working with general regular Borel measures
also has the advantage that it becomes possible to take convex combinations of universal measures,
which seems essential for getting the connection to second-quantized bosonic fields
(see the notions of decoherent replicas of space-time and of microscopic mixing of wave functions
in~\cite{qft} and~\cite{qftlimit}).

Combining all the above results led to the framework of {\em{causal fermion systems}}, where
a physical system is described by a Hilbert space~$(\H, \langle .|. \rangle_\H)$ and
the universal measure~$\rho$ on~${\mathscr{F}}$. This framework was first introduced in~\cite{rrev}.
Subsequently, the analytic, geometric and topological structures encoded in a causal fermion system
were worked out systematically; for an overview see~\cite[Chapter~1]{cfs}.

From the conceptual point of view, the setting of causal fermion systems
and the notion of the universal measure considerably changed both the
role of the causal action principle and the concept of what space-time is.
Namely, in the causal action principle in this book, one varies the fermionic projector~$P$ in
a given discrete space-time.
In the setting of causal fermion systems, however, one varies instead the universal measure~$\rho$,
being a measure on linear operators on an abstract Hilbert space.
In the latter formulation, there is no space-time to begin with. On the contrary, space-time is
introduced later as the support of the universal measure.
In this way, the causal action principle evolved from a variational principle for wave functions
in space-time to a variational principle for space-time itself as well as for all structures therein.

In order to complete the summary of the conceptual modifications, we remark that
the connection between the scalar product~$\langle .|. \rangle_\H$ and surface integrals as in~\eqref{sprodMin},
which had been obscure for quite a while, was finally clarified when working out Noether-like theorems
for causal variational principles~\cite{noether}. Namely, surface integrals now have a proper generalization to
causal fermion systems in terms of so-called {\em{surface layer integrals}}.
It was shown that the symmetry of the causal action under unitary transformations acting on~$\mathscr{F}$
gives rise to conserved charges which can be expressed by surface layer integrals.
For Dirac sea configurations, these conserved charges coincide with the surface
integrals~\eqref{sprodMin}.

Another major development concerns the description of {\em{neutrinos}}.
In order to explain how these developments came about, we first note that
in this book, neutrinos are modelled as left-handed massless Dirac particles (see~\S\ref{esec1}).
This has the benefit that the neutrinos drop out of the closed chain due to
chiral cancellations (see~\S\ref{esec21} and~\S\ref{esec22}).
When writing this book, I liked chiral cancellations, and I even regarded them as
a possible explanation for the fact that neutrinos appear only with one chirality.
As a side remark, I would like to mention that
I was never concerned about experimental observations which indicate that
neutrinos do have a rest mass, because I felt that these experiments are too indirect
for making a clear case. Namely, measurements only tell us that there are
fewer neutrinos on earth than expected from the number of neutrinos generated in fusion processes
in the sun. The conventional explanation for this seeming disappearance
of solar neutrinos is via neutrino oscillations, making it
necessary to consider massive neutrinos. However, it always seemed to me that
there could be other explanations for the lack of neutrinos on earth
(for example, a modification of the weak interaction or other, yet unknown fundamental forces),
in which case the neutrinos could well be massless.

My motivation for departing from massless neutrinos was not related to experimental evidence, but
had to do with problems of mathematical consistency. Namely, I noticed that left-handed neutrinos
do not give rise to stable minimizers of the causal action (see~\cite[Section~4.2]{cfs}).
This general result led me to incorporate right-handed neutrino components, and to
explain the fact that only the left-handed component is observed by the postulate that the
regularization breaks the chiral symmetry. This procedure cured the mathematical consistency
problems and had the desired side effect that neutrinos could have a rest mass, in agreement with
neutrino oscillations.

We now comment on other developments which are of more technical nature.
These developments were mainly triggered by minor errors or shortcomings in the present book.
First, Andreas Grotz noticed when working on his master thesis in 2007 that the
normalization conditions for the fermionic projector as given in~\eqref{eq:2a1} and~\eqref{eq:2a2}
are in general violated to higher order in perturbation theory. This error was corrected in~\cite{grotz}
by a rescaling procedure.
This construction showed that there are two different perturbation expansions: with and
without rescaling. The deeper meaning of these two expansions became clearer later when
working out different normalizations of the fermionic projector.
This study was initiated by the quest for a non-perturbative construction of the fermionic projector,
as was carried out in globally hyperbolic space-times in~\cite{finite, infinite}.
It turned out that in space-times of finite lifetime, one cannot work with the $\delta$-normalization in the mass
parameter as used in~\eqref{56}--\eqref{58} (the ``mass normalization''). Instead, a proper normalization
is obtained by using a scalar product~$(.|.)$ which is represented similar to~\eqref{sprodMin}
by an integral over a spacelike hypersurface (the ``spatial normalization'').
As worked out in detail in~\cite{norm} with a convenient contour integral method,
the causal perturbation expansion without rescaling realizes the spatial normalization condition,
whereas the rescaling procedure in~\cite{grotz} gives rise to the mass normalization.
The constructions in curved space-time in~\cite{finite, infinite} as well as the general connection between the
scalar product~$( .|. )$ and the surface layer integrals in~\cite{noether} showed that
the physically correct and mathematically consistent normalization condition is
the spatial normalization condition. With this in mind, the combinatorics of the causal perturbation
expansion in this book is indeed correct, but the resulting fermionic projector does not satisfy the mass
but the spatial normalization condition.

Clearly, the analysis of the continuum limit in Chapters~\ref{esec3}--\ref{esec5}
is superseded by the much more detailed analysis in~\cite[Chapters~3-5]{cfs}.
A major change concerns the treatment of the logarithmic singularities on the light cone,
as we now point out. In the present book, some of the contributions involving logarithms are
arranged to vanish by imposing that the regularization should satisfy the relation~\eqref{e:3C}.
I tried for quite a while to construct an example of a regularization which realizes this relation,
until I finally realized that there is no such regularization, for the following reason: \\[-0.5em]

\textsc{Lemma~I.}
There is no regularization which satisfies the condition~\eqref{e:3C}.

\Proof The linear combination of monomials~$M$ in~\eqref{e:3B} involves a factor~$T^{(1)}_{[2]}$,
which has a logarithmic pole on the light cone (see~\eqref{Tldef}, \eqref{Tadef}
and~\eqref{l:3.1}). Restricting attention to the corresponding
contribution~$\sim \log|\vec{\xi}|$, we have
\[ M \asymp -\frac{1}{16 \pi^3}\:
T^{(-1)}_{[0]}\: \overline{T^{(-1)}_{[0]}\: T^{(0)}_{[0]}}\:\log|\vec{\xi}| \:. \]
As a consequence,
\begin{align*}
(M & - \overline{M})\: \overline{T^{(0)}_{[0]}}^{-1} =
-\frac{\log|\vec{\xi}|}{16 \pi^3}\: \frac{\big| T^{(-1)}_{[0]} \big|^2}{\overline{T^{(0)}_{[0]}}}
\big( \overline{T^{(0)}_{[0]}} - T^{(0)}_{[0]} \big)  \\
&= -\frac{\log|\vec{\xi}|}{16 \pi^3}\: \bigg| \frac{T^{(-1)}_{[0]}}{T^{(0)}_{[0]}} \bigg|^2
\Big( \big| T^{(0)}_{[0]} \big|^2 - \big( T^{(0)}_{[0]} \big)^2 \Big)
= -\frac{\log|\vec{\xi}|}{8 \pi^3}\: \bigg| \frac{T^{(-1)}_{[0]}}{T^{(0)}_{[0]}} \bigg|^2
\, \Big(\text{Im} \,T^{(0)}_{[0]} \Big)^2 \leq 0 \:.
\end{align*}
Since this expression has a fixed sign, it vanishes in a weak evaluation on the light cone
only if it vanishes identically to the required degree.
According to~\eqref{l:3.1}, the function~$\text{Im}\, T^{(0)}_{[0]}$ is a regularization
of the distribution~$-i \pi \delta (\xi^2) \:\varepsilon (\xi^0)/(8 \pi^3)$ on the scale~$\varepsilon$.
Hence on the light cone it is of the order~$\varepsilon^{-1}$. This gives the claim.
\QED
This no-go result led me to reconsider the whole procedure of the continuum limit.
At the same time, I tried to avoid imposing relations between the regularization parameters,
which I never felt comfortable with because I wanted the continuum limit to work for at
least a generic class of regularizations. Resolving this important issue took
me a lot of time and effort. My considerations eventually led to the method of compensating the
logarithmic poles by a {\em{microlocal chiral transformation}}.
These construction as well as many preliminary considerations are given in~\cite[Section~3.7]{cfs}.

Finally, I would like to make a few comments on each chapter of the book.
Chapters~\ref{secintro}--\ref{secpfp} are still up to date, except for the generalizations
and modifications mentioned above. Compared to the presentation in~\cite{cfs},
I see the benefit that these chapters might be easier to read and might convey
a more intuitive picture of the underlying physical ideas.
Chapter~\ref{psec2} is still the best reference for the general derivation of the formalism of the
continuum limit. In~\cite[Chapter~2]{cfs} I merely explained the regularization effects in examples
and gave an overview of the methods and results in Chapter~\ref{psec2}, but without repeating
the detailed constructions. Chapter~\ref{esec2} is still the only reference where the
form of the causal action is motivated and derived step by step. Also, the
notion of state stability is introduced in detail, thus providing the basis for the
later analysis in~\cite{reg, vacstab}. As already mentioned above, the analysis in
Chapters~\ref{esec3}--\ref{secegg} is outdated. I recommend the reader to study
instead~\cite[Chapters~3--5]{cfs}. The Appendices are still valuable. I added a few
footnotes which point to later improvements and further developments.
%
\\[1.5em]
\hspace*{1cm} \hfill Felix Finster, Regensburg, August 2016 \\[0.5cm]


\centerline{\large{\bf{References}}}

\vspace*{.5em}

\newcommand\oldchapter{}
\let\oldchapter=\chapter
\renewcommand{\chapter}[2]{}

\providecommand{\bysame}{\leavevmode\hbox to3em{\hrulefill}\thinspace}
\providecommand{\MR}{\relax\ifhmode\unskip\space\fi MR }
\providecommand{\MRhref}[2]{%
  \href{http://www.ams.org/mathscinet-getitem?mr=#1}{#2}
}
\providecommand{\href}[2]{#2}

\let\chapter=\oldchapter

\include{pfp1}
\mainmatter
\setcounter{page}{123}
\include{pfp2}
\begin{appendix}
\chapter{Connection to the Fock Space Formalism}\index{Fock space}
\setcounter{equation}{0} \label{pappA}
\renewcommand{\theequation}{\thechapter.\arabic{equation}}
\renewcommand{\thesection}{\thechapter}
In this appendix it is
shown that for an observer who is making measurements only in a
subsystem of the whole physical system, the description of
a many-fermion system with the fermionic projector
is equivalent to the fermionic Fock space
formalism, provided that the number of fermions of the whole
system (including the particles of the sea) is infinite.
The following consideration applies in the same way to either
a space-time continuum or to discrete space-time.
Before beginning we point out that the action principle, from
which the fundamental physical equations can be deduced, involves
the fermions only via the Dirac action $\bra \Psi, (i \Pdd + {\mathcal{B}}
- m) \Psi \ket$. For the formulation of the Dirac action one only
needs on the fermionic Fock space the time/position operators
and the operator~$\Pdd$, which are all one-particle operators.
Therefore, we can say that many-particle operators (like for example
in the four-fermion coupling of the Fermi model)
are not essential for the formulation of the quantum field theory of
the standard model. Having this in mind,
we may here restrict attention to one-particle operators\footnote{{\textsf{Online
version}:} For the description of {\em{entanglement}},
it is indeed necessary to consider two-particle observables;
see the paper~\entangle.}.

Let $P$ be a fermionic projector acting on the vector space $H$.
The one-particle observables
correspond to operators ${\mathcal{O}}$ on $H$.
Our subsystem is described by a non-degenerate subspace $K
\subset H$; we decompose $H$ as a direct sum $H=K \oplus L$ with
$L=K^\perp$. We assume that the observables are localized in $N$;
i.e.\ they are trivial on $L$,
\begin{equation}
    {\mathcal{O}}_{|L} \;=\; 0_{|L} \: .
    \label{p:1_a6}
\end{equation}
We choose a (properly normalized) basis $\Psi_1,\ldots,\Psi_n$ of the
subspace $P(H) \subset H$, and decompose the states $\Psi_j$ in the form
\[ \Psi_j \;=\; \Psi_j^K + \Psi_j^L \spc{\mbox{with
$\Psi_j^K \in K, \Psi_j^L \in L$}} . \]
Substituting into (\ref{p:1_16}), we obtain for the
many-particle wave function the expression
\begin{equation}
\Psi \;=\; \sum_{\pi \in {\mathcal{P}}(n)}
        (-1)^{|\pi|} \left(\bigwedge_{j \in \pi} \Psi_j^K \right) \land
        \left(\bigwedge_{j \not \in \pi} \Psi_j^L \right) ,
    \label{p:1_21}
\end{equation}
where ${\mathcal{P}}(n)$ denotes the set of all subsets of $\{1,
\ldots, n\}$. For measurements in our subsystem, we must calculate
the expectation value $\bra \Psi | {\mathcal{O}} | \Psi
\ket_F$\footnote{We remark for clarity that this expectation value
does not coincide with that of a measurement in nonrelativistic
quantum mechanics. Namely, in the continuum, the scalar product
$\bra .|. \ket$ involves a time integration. But one can get a
connection to nonrelativistic measurements by considering
operators ${\mathcal{O}}$ with a special time dependence (which,
for example, act on the wave functions only in a short time
interval $[t, t +\Delta t]$).}, where the operators
${\mathcal{O}}$ act on the Fock space according to
\[ {\mathcal{O}}(\Psi_1 \land \cdots \land \Psi_n) \;=\;
({\mathcal{O}} \Psi_1) \land \cdots \land \Psi_n \:+\:
\Psi_1 \land ({\mathcal{O}} \Psi_2) \cdots \land \Psi_n \:+
\cdots +\: \Psi_1 \land \cdots \land ({\mathcal{O}} \Psi_n) \: , \]
and where $\bra .|. \ket_F$ is the scalar product on the Fock space,
induced by the scalar product $\bra .|. \ket$ on $H$.
It is useful to rewrite the expectation value with the statistical
operator $S$, i.e.
\[ \bra \Psi | {\mathcal{O}} | \Psi \ket_F \;=\; \tr_F (S \: {\mathcal{O}})
        \quad {\mbox{with}} \quad S \;=\; | \Psi \ket \bra \Psi |_F \: , \]
where $\tr_F$ denotes the trace in the Fock space.
Using~(\ref{p:1_a6}), we can take the partial trace
over $L$ and obtain, applying (\ref{p:1_21}),
\begin{eqnarray}
\label{p:1_a8} \bra \Psi | {\mathcal{O}} | \Psi \ket_F &=&
\tr_{F_K}(S^K \: {\mathcal{O}})
        \spc {\mbox{with}} \\
\label{p:1_a7}
S^K &=& \sum_{k=0}^n \sum_{ \scriptsize
        \begin{array}{cc} \scriptsize \pi, \pi^\prime \in {\mathcal{P}}(n) , \\
                \# \pi = \# \pi^\prime = k
        \end{array} }
        c_{\pi, \pi^\prime} \; | \land_{i \in \pi} \Psi_i^K \ket
                \bra \land_{j \in \pi^\prime} \Psi_j^K |_{F_K} \\
c_{\pi, \pi^\prime} &=& (-1)^{|\pi| + |\pi^\prime|} \;
\bra \land_{i \not \in \pi}
        \Psi_i^L | \land_{j \not \in \pi^\prime}
        \Psi_j^L \ket_F \: , \nonumber
\end{eqnarray}
where $\tr_{F_K}$ is the trace in the Fock space $F_K=\oplus_{k=0}^\infty
\land^k K$ generated by $K$.
Thus our subsystem is described by a statistical operator
$S^K$ on $F_K$, which is composed of mixed states consisting of
different numbers of particles.
Since the constants $c_{\pi, \pi^\prime}$ depend on the wave functions
$\Psi^L$ outside our subsystem, we can consider them as
arbitrary numbers.

In the limit when the number $n$ of particles of the
whole system tends to infinity, (\ref{p:1_a7}) goes over to a
statistical operator of the form
\begin{equation}
S^K \;=\; \sum_{k=0}^\infty \;\sum_{\alpha, \beta=0}^\infty
c^{(k)}_{\alpha \beta} \;| \Psi^{(k)}_\alpha \ket \bra \Psi^{(k)}_\beta
|_{F_K}
\label{p:A5}
\end{equation}
with arbitrary complex coefficients $c^{(k)}_{\alpha \beta}$ and
$k$-particle states $\Psi^{(k)}_\alpha \in F^k_K$. This statistical
operator differs from a general statistical operator
$S^K_{\mbox{\scriptsize{gen}}}$ in that it is diagonal on the $k$-particle
subspaces (i.e.\ that the wave functions in the ``bra''
and in the ``ket'' of (\ref{p:A5}) are both $k$-particle states);
more precisely, $S^K_{\mbox{\scriptsize{gen}}}$ has, compared to
(\ref{p:A5}), the more general form
\begin{equation}
S^K_{\mbox{\scriptsize{gen}}} \;=\; \sum_{k,l=0}^\infty \;
\sum_{\alpha, \beta=0}^\infty
c^{(k,l)}_{\alpha \beta} \;| \Psi^{(k)}_\alpha \ket \bra \Psi^{(l)}_\beta
|_{F_K}\:. \label{p:A6}
\end{equation}
We remark for clarity that a pure state of the Fock space $\Psi \in F_K$ has a
decomposition $\Psi = \sum_{k=0}^\infty \lambda_k \Psi^{(k)}$, and thus the
corresponding statistical operator is
\[ S \;=\; | \Psi \ket \bra \Psi |_{F_K} \;=\; \sum_{k,l=0}^\infty
\lambda_k \:\overline{\lambda_l} \; |\Psi^{(k)} \ket \bra \Psi^{(l)} |_{F_K}
\:. \]
This statistical operator is a special case of (\ref{p:A6}), but it is
{\em{not}} of the form (\ref{p:A5}).

The difference between (\ref{p:A5}) and (\ref{p:A6}) becomes
irrelevant if we keep in mind that all physically relevant observables
commute with the particle number operator. Namely in this case, every
expectation value reduces to the sum of the expectation values in the
$k$-particle Fock spaces,
\begin{eqnarray*}
\tr_{F_K}(S^K_{\mbox{\scriptsize{gen}}} \: {\mathcal{O}}) &=&
\sum_{k,l=0}^\infty \;\sum_{\alpha, \beta=0}^\infty
c^{(k,l)}_{\alpha \beta} \;\bra \Psi^{(l)}_\beta
\:|\:{\mathcal{O}}\:|\:
\Psi^{(k)}_\alpha \ket_{F_K} \\
&=& \sum_{k=0}^\infty \;\sum_{\alpha, \beta=0}^\infty
c^{(k,k)}_{\alpha \beta} \;\bra \Psi^{(k)}_\beta
\:|\:{\mathcal{O}}\:|\: \Psi^{(k)}_\alpha \ket_{F_K} \:.
\end{eqnarray*}
If we choose the coefficients $c^{(k)}_{\alpha \beta}$ in (\ref{p:A5})
to be $c^{(k)}_{\alpha \beta} = c^{(k,k)}_{\alpha \beta}$, these expectation
values are also obtained from the statistical operator $S^K$,
\[ \tr_{F_K}(S^K_{\mbox{\scriptsize{gen}}} \: {\mathcal{O}})
\;=\; \tr_{F_K}(S^K \: {\mathcal{O}}) \:. \] We conclude that
it is no loss of generality to describe the subsystem by the
statistical operator $S^K$.

\chapter{Some Formulas of the Light-Cone Expansion}
\label{pappLC} \setcounter{equation}{0}
This appendix is a compilation of some formulas of the
light-cone expansion\index{light-cone expansion!formulas of the}. More precisely, we list the phase-free
contribution to the light-cone expansion of the Green's functions
(cf.\ Def.\ \ref{l:def_pf}). According to Def.~\ref{l:def_pf} and
Theorem~\ref{l:thm3}, the light-cone expansion of the Green's
functions is immediately obtained by inserting ordered exponentials
into the line integrals.
Furthermore, as explained after~(\ref{fprep}), the formulas can be
applied directly to the fermionic projector; they then describe the
singularities of $\tilde{P}(x,y)$ on the light cone.
Without loss of generality, we restrict attention to the left handed
component of the Green's functions.
We compute precisely those contributions which will be of relevance
in Appendix~\ref{appC} and in Chapters~\ref{esec3}--\ref{esec5}.
The following formulas were all generated by a computer program,
see~\cite{F5} for details.

We begin with the perturbation by a chiral perturbation to first order.
The phase-free contribution (denoted by a corresponding superscript
on the equal sign) is
\begin{eqnarray}
\lefteqn{ \chi_L \:(-s \:(\chi_L \Aslsh_R + \chi_R \Aslsh_L)
\:s)(x,y) \;\stackrel{\mbox{\footnotesize{phase-free}}}{=}\;
\xi\slsh \:{\mathcal{O}}((y-x)^0) + {\mathcal{O}}((y-x)^2) } \nonumber \\
&&  +\chi_L \: S^{(0)}(x,y) \: \xi^i \int_x^y dz\:[0,1\:|\: 0]\:
(\Pdd A_{L i}) \label{l:A00} \\
&&  -\chi_L \: S^{(0)}(x,y) \: \int_x^y dz\:[0,0\:|\: 0]\: \Aslsh_L
\label{l:A01} \\
&&  +\chi_L \: S^{(0)}(x,y) \: \Aslsh_L(x) \label{l:A02} \\
&&  +\frac{1}{2} \: \chi_L \: S^{(0)}(x,y) \: \xi \slsh
\int_x^y dz\: [0,0\:|\: 0]\: (\Pdd \Aslsh_L) \\
&&  -\chi_L \: S^{(0)}(x,y) \: \xi \slsh \: \int_x^y dz\: [1,0\:|\: 0]\:
(\partial^i A_{Li}) \\
&&  +\frac{1}{2} \chi_L \: S^{(0)}(x,y) \: \xi \slsh  \: \xi^i \:
\int_x^y dz\: [0,0\:|\: 1]\: (\OBox A_{Li}) \\
&&  +\chi_L \: S^{(1)}(x,y) \: \xi^i \: \int_x^y dz\:
[0,1\:|\: 1]\: (\Pdd \OBox A_{Li}) \\
&&  +\chi_L \: S^{(1)}(x,y) \: \int_x^y dz\: [0,2\:|\: 0]\: (\OBox
\Aslsh_L) \\
&&  -2 \chi_L \: S^{(1)}(x,y) \: \int_x^y dz\: [0,0\:|\: 1]\:
(\Pdd \partial^i A_{Li})\: , \spc\spc\spc\quad \label{l:A0e}
\end{eqnarray}
where again $\xi \equiv (y-x)$. The notation~$\xi\slsh \:{\mathcal{O}}((y-x)^2)$
means that we leave out all contributions which are of the order
${\mathcal{O}}((y-x)^0)$ and have a leading factor~$\xi\slsh$.
This formula has the disadvantage that it contains partial
derivatives of the chiral potential; it would be better for physical
applications to work instead with the Yang-Mills field tensor and
the Yang-Mills current. Therefore, we introduce left and right handed
gauge-covariant derivatives $D^{L\!/\!R}$,
\[ D^L_j \;=\; \frac{\partial}{\partial x^j} \:-\: i A_{L j}
\:,\spc D^R_j \;=\; \frac{\partial}{\partial x^j} \:-\: i A_{R j}
\: , \]
and define the corresponding field tensor and current as usual by
the commutators
\beq \label{l:A1}
F^c_{jk} \;=\; i \left[ D^c_j,\:D^c_k \right] \:,\spc
j^c_l \;=\; \left[ D^{c\:k},\: F^c_{lk} \right] \spc {\mbox{($c=L$ or $R$)}}.
\eeq
In the case of an Abelian gauge field, this formula
reduces to the familiar formulas for the electromagnetic field tensor and
current,
\[ F^c_{jk} \;=\; \partial_j A_{c\:k} - \partial_k A_{c\:j} \:,\spc
j^c_l \;=\; \partial_{lk} A^k_c - \OBox A_{c\:l} \: . \]
Notice, however, that in the general case of a system of Dirac seas,
(\ref{l:A1}) involves quadratic and cubic terms in the potential.

By substituting (\ref{l:A1}) into (\ref{l:A00}--\ref{l:A0e}) and
manipulating the line integrals with integrations by parts, one can
rewrite the phase-free contribution in a way where the linear terms in the
potential are gauge invariant. For example, we can combine
(\ref{l:A00}--\ref{l:A02}) by transforming the line integrals as
\begin{eqnarray}
\lefteqn{ \xi^k \int_x^y dz \:[0,1 \:|\: 0] \: (\Pdd A_{Lk}) \;=\;
\xi^k \int_x^y dz \:[0,1 \:|\: 0] \: (\gamma^j F^L_{jk} \:+\:
\partial_k \Aslsh_L) \:+\: {\mathcal{O}}(A_L^2) } \nonumber \\
&=& \xi^k \int_x^y dz \:[0,1 \:|\: 0] \: \gamma^j F^L_{jk} \:-\: \Aslsh_L(x)
\:+\: \int_x^y dz \:[0,0 \:|\: 0] \:\Aslsh_L \:+\:
{\mathcal{O}}(A_L^2) \: .\spc
        \label{l:A2a}
\end{eqnarray}
This procedure yields (in the non-Abelian case) quadratic
and cubic terms in the potential which are {\em{not}} gauge
invariant. Fortunately, these gauge-dependent terms are all
compensated by corresponding contributions to the higher order
Feynman diagrams. We thus obtain
\begin{eqnarray*}
\lefteqn{ \chi_L \sum_{k=0}^\infty ((-s \:(\chi_L \Aslsh_R + \chi_R
\Aslsh_L))^k \:s)(x,y) \;
\stackrel{\mbox{\footnotesize{phase-free}}}{=}\;
\xi\slsh \:{\mathcal{O}}((y-x)^0) + {\mathcal{O}}((y-x)^2) } \\
&&  +\chi_L \:S^{(0)}(x,y)\: \xi^i \int_x^y dz\: [0,1\:|\: 0]\: \gamma^l F^L_{li} \\
&&  +\frac{1}{4}\:\chi_L \:S^{(0)}(x,y)\: \xi \slsh\: \int_x^y dz\: [0,0\:|\: 0]\:
\gamma^j \gamma^k\:F^L_{jk} \\
&&  -\frac{1}{2}\:\chi_L \:S^{(0)}(x,y)\: \xi \slsh\: \xi^i \int_x^y dz\:
[0,0\:|\: 1]\: j^L_i \\
&&  -i\chi_L \:S^{(0)}(x,y)\: \xi \slsh\: \xi_i \xi^j \int_x^y dz_1\:
[0,1\:|\: 1]\:F^L_{kj} \int_{z_1}^y dz_2\: [0,1\:|\: 0]\: F_L^{ki} \\
&&  +\chi_L \:S^{(1)}(x,y)\: \xi^i \int_x^y dz\: [0,1\:|\: 1]\: (\Pdd j^L_i) \\
&&  +\chi_L \:S^{(1)}(x,y)\: \int_x^y dz\: [0,2\:|\: 0]\: j^L_k \:\gamma^k \\
&&  -i\chi_L \:S^{(0)}(x,y)\: \xi \slsh\: \xi_i \xi^j \int_x^y dz_1\:
[0,1\:|\: 1]\:F^L_{kj} \int_{z_1}^y dz_2\: [0,1\:|\: 0]\: F_L^{ki} \\
&&  +i\chi_L \:S^{(1)}(x,y)\: \xi^i \xi^j \int_x^y dz_1\: [0,3\:|\: 0]\: \gamma^k
\:F^L_{kj} \int_{z_1}^y dz_2\: [0,0\:|\: 1]\: j^L_i \\
&&  +i\chi_L \:S^{(1)}(x,y)\: \xi^i \xi^j \int_x^y dz_1\: [0,2\:|\: 1]\: j^L_j
\int_{z_1}^y dz_2\: [0,1\:|\: 0]\: \gamma^l \:F^L_{li} \\
&&  -2i\chi_L \:S^{(1)}(x,y)\: \xi_i \xi^j \int_x^y dz_1\: [0,2\:|\: 1]\: F^L_{mj}
\int_{z_1}^y dz_2\: [0,2\:|\: 0]\: (\Pdd F_L^{mi}) \\
&&  -2i\chi_L \:S^{(1)}(x,y)\: \xi_i \xi^j \int_x^y dz_1\: [0,2\:|\:
1]\: (\Pdd F^L_{kj}) \int_{z_1}^y dz_2\: [0,1\:|\: 0]\: F_L^{ki} \\
&&  +i\chi_L \:S^{(1)}(x,y)\: \xi^i \xi^j \int_x^y dz_1\: [0,2\:|\: 1]\: \gamma^k F^L_{kj}
\int_{z_1}^y dz_2\: [0,2\:|\: 0]\: j^L_i \\
&&  -\frac{i}{2}\:\chi_L \:S^{(1)}(x,y)\: \xi^i \int_x^y dz_1\: [0,2\:|\: 0]\:
\gamma^j F^L_{ji} \int_{z_1}^y dz_2\: [0,0\:|\: 0]\: \gamma^k \gamma^l F^L_{kl} \\
&&  -\frac{i}{2}\:\chi_L \:S^{(1)}(x,y)\: \xi^i \int_x^y dz_1\: [0,2\:|\: 0]\:
\gamma^j \gamma^k F^L_{jk} \int_{z_1}^y dz_2\: [0,1\:|\: 0]\: \gamma^l F^L_{li} \\
&&  +2i\chi_L \:S^{(1)}(x,y)\: \xi_i \int_x^y dz_1\: [0,3\:|\: 0]\: \gamma^j F^L_{jk}
\int_{z_1}^y dz_2\: [0,1\:|\: 0]\: F_L^{ki} \\
&&  -2i\chi_L \:S^{(1)}(x,y)\: \xi^j \int_x^y dz_1\: [0,1\:|\: 1]\: F^L_{ij}
\int_{z_1}^y dz_2\: [0,1\:|\: 0]\: \gamma_k F_L^{ki} \\
&&  -2 \chi_L \:S^{(1)}(x,y)\: \xi_i \xi^j \xi^k   \\
&&\quad \times \int_x^y dz_1\: [0,4\:|\: 0]\: \gamma^l F^L_{lk}
\int_{z_1}^y dz_2\: [0,1\:|\: 1]\: F^L_{mj}
\int_{z_2}^y dz_3\:[0,1\:|\: 0]\: F_L^{mi} \\
&&  -2\chi_L \:S^{(1)}(x,y)\: \xi_i \xi^j \xi^k \\
&&\quad \times \int_x^y dz_1\:
[0,3\:|\: 1]\: \gamma^l F^L_{lk} \int_{z_1}^y dz_2\:
[0,3\:|\: 0]\: F^L_{mj} \int_{z_2}^y dz_3\:[0,1\:|\: 0]\: F_L^{mi} \\
&&  -2\chi_L \:S^{(1)}(x,y)\: \xi_i \xi^j \xi^k \\
&&\quad \times \int_x^y dz_1\:
[0,3\:|\: 1]\: F^L_{mk} \int_{z_1}^y dz_2\:
[0,3\:|\: 0]\: \gamma^l F^L_{lj} \int_{z_2}^y dz_3\:[0,1\:|\: 0]\: F_L^{mi} \\
&&  -2\chi_L \:S^{(1)}(x,y)\: \xi^i \xi_j \xi^k \\
&&\quad \times \int_x^y dz_1\:
[0,3\:|\: 1]\: F^L_{mk} \int_{z_1}^y dz_2\:
[0,3\:|\: 0]\: F_L^{mj} \int_{z_2}^y dz_3\:
[0,1\:|\: 0]\: \gamma^l F^L_{li} \:. \spc
\end{eqnarray*}
We call this formulation of the phase-free contributions purely in
terms of the Yang-Mills field tensor and the Yang-Mills current
the {\em{gauge invariant form}} \index{light-cone expansion!gauge
invariant form of the} of the
light-cone expansion\index{light-cone expansion}.

It remains to consider the scalar/pseudoscalar
perturbation; i.e.\ we must study how the dynamic mass matrices
$Y_{L\!/\!R}(x)$ show up in the light-cone expansion. We begin with
the case of a single mass matrix. To first order in the external
potential, the corresponding Feynman diagram has the light-cone
expansion
\begin{eqnarray}
\lefteqn{ \chi_L \:m\:(-s \:(-\chi_L Y_R - \chi_R
Y_L) \:s)(x,y) } \nonumber \\
&\stackrel{\mbox{\footnotesize{phase-free}}}{=}&
\frac{1}{2}\:\chi_L\: m  \:S^{(0)}(x,y)\: \xi \slsh\: \int_x^y dz\:
[0,0\:|\: 0]\: (\Pdd Y_L) \nonumber \\
&& +\chi_L\: m  \:S^{(0)}(x,y)\: Y_L(x) \:+\: {\mathcal{O}}((y-x)^0) \: . \label{l:A4a}
\end{eqnarray}
The higher orders in the chiral potentials yield no phase-free contributions.
The next orders in the mass parameter are treated similarly. The
contributions quadratic in~$m$ are
\begin{eqnarray*}
\lefteqn{ \chi_L \:m^2\: \sum_{n_1, n_2, n_3=0}^\infty
((-s \:\Aslsh_L)^{n_1} \:s\: Y_L \:s\: (-\Aslsh_R \:s)^{n_2} \:Y_R\:s\:
(-\Aslsh_R \:s)^{n_3})(x,y) } \\
&&\!\!\!\!\!\!\!\!\!\!\!\! \;
\stackrel{\mbox{\footnotesize{phase-free}}}{=}\;
\frac{i}{2}\: \chi_L\: m^2 \:S^{(0)}(x,y)\: \xi \slsh\: \int_x^y dz\:
[0,0\:|\: 0]\: Y_L \:Y_R \\
&&  +i \chi_L\: m^2 \:S^{(1)}(x,y)\: \int_x^y dz\:
[0,1\:|\: 0]\: Y_L \:\gamma^j (D_j Y_R) \\
&&  +i \chi_L\: m^2 \:S^{(1)}(x,y)\: \int_x^y dz\:
[0,1\:|\: 0]\: \gamma^j (D_j Y_L)\: Y_R \\
&&  -i \chi_L\: m^2 \:S^{(1)}(x,y)\: Y_L \int_x^y dz\:
[0,0\:|\: 0]\: \gamma^j (D_j Y_R) \\
&&  + \chi_L\: m^2 \:S^{(1)}(x,y)\: \xi^i \int_x^y dz_1\:
[0,2\:|\: 0]\: \gamma^j F^L_{ji}
\int_{z_1}^y dz_2\: [0,0\:|\: 0]\: Y_L \:Y_R \\
&&  + \chi_L\: m^2 \:S^{(1)}(x,y)\: \xi^i \int_x^y dz_1\:
[0,2\:|\: 0]\: Y_L \:Y_R \int_{z_1}^y dz_2\: [0,1\:|\: 0]\: \gamma^j
F^L_{ji} \\
&&+\xi\slsh \:{\mathcal{O}}((y-x)^0) + {\mathcal{O}}((y-x)^2) \:,
\end{eqnarray*}
whereas there is only one terms cubic in~$m$,
\begin{eqnarray*}
\lefteqn{ \chi_L \:m^3 \!\!\!\!\!\!\!\sum_{n_1, n_2, n_3, n_4=0}^\infty
\!\!\!\!\!
((-s \:\Aslsh_L)^{n_1} \:s\: Y_L \:s\: (-\Aslsh_R \:s)^{n_2} \:Y_R\:s\:
(-\Aslsh_L \:s)^{n_3} \:Y_L\:s\: (-\Aslsh_R \:s)^{n_4})(x,y) } \\
& \stackrel{\mbox{\footnotesize{phase-free}}}{=}&
\chi_L\: m^3 \:S^{(1)}(x,y)\: Y_L \int_x^y dz\:
[0,0\:|\: 0]\: Y_R \:Y_L \spc\spc\spc\spc \\
&&\;+\;\xi\slsh \:{\mathcal{O}}((y-x)^0)
+ {\mathcal{O}}((y-x)^2) \:.
\end{eqnarray*}
To the order~$\sim m^4$ and higher all contributions are on
the light cone of the order~$\xi\slsh {\mathcal{O}}((y-x)^0)
+ {\mathcal{O}}((y-x)^2)$.

The above Feynman diagrams completely characterize the Green's
functions to the order ${\mathcal{O}}((y-x)^0)$ on the light cone. Notice
that in agreement with Theorem~\ref{l:thm3} we get only a finite number of phase-free contributions.

\chapter{Normalization of Chiral Fermions}
\setcounter{equation}{0} \label{appcf}
\renewcommand{\theequation}{\thesection.\arabic{equation}}
\renewcommand{\thesection}{\thechapter.\arabic{section}}
In this appendix we describe a method for normalizing chiral
fermions. The main difficulty is that for a proper normalization one
needs to give the chiral fermions a small rest mass; this will be
discussed in Section~\ref{sec21} for a single Dirac sea in Minkowski
space. In Section~\ref{sec22} we develop a method for analyzing the
normalization of chiral fermions with a small generalized ``mass,''
whereas Section~\ref{sec23} gives the general construction including
the infrared regularization and the interaction.\index{massive
neutrino}

\section{Massive Chiral Fermions -- Preparatory Discussion}
\setcounter{equation}{0} \label{sec21}\index{massive chiral fermion}
Before introducing the infrared regularization, we need to understand how
a chiral Dirac sea can be normalized in infinite volume using some kind of
``$\delta$-normalization.'' To this end we consider a non-interacting
left-handed fermionic projector in Minkowski space,
\begin{equation}
P(x,y) \;=\; \chi_L \left. t_m(x,y) \right|_{m=0}\:, \label{eq:C1}
\end{equation}
where we set
\[ t_m \;=\; \frac{1}{2} \left(p_m - k_m \right) \]
with~$p_m$ and~$k_m$ according to~(\ref{x20}, \ref{x21}).
Naively, products of this fermionic projector
vanish due to chiral cancellations,
\begin{eqnarray}
P^2(x,y) &=& \int d^4z\: P(x,z)\: P(z,y)
\;=\; \int d^4z\: \chi_L\: t_0(x,z)\; \chi_L\: t_0(z,y) \nonumber \\
&=& \int d^4z\: \chi_L\:\chi_R \;t_0(x,z)\: t_0(z,y)
\;\stackrel{\mbox{\scriptsize{formally}}}{=}\; 0\:. \label{eq:C5}
\end{eqnarray}
However, this formal calculation has no meaning in the
formalism of causal perturbation theory~{\S}\ref{jsec2}
because in this formalism
we are not allowed to multiply Dirac seas of the same fixed mass. Instead,
we must treat the masses as variable parameters. Thus before we can give
a mathematical meaning to products of chiral Dirac seas, we must extend the
definition of a chiral Dirac sea to non-zero rest mass.

Giving chiral Dirac particles a mass is a delicate issue which often
leads to confusion and misunderstandings. Therefore, we discuss
the situation in the example~(\ref{eq:C1}) in detail.
In momentum space, the distribution $t_m$, $m \geq 0$, takes the form
\[ t_m(k) \;=\; (k \slsh+m)\: \delta(k^2-m^2)\: \Theta(-k^0)\:. \]
On the mass shell, the range of the $(4 \times 4)$-matrix $k \slsh+m$ is
two-dimensional; this corresponds to a twofold degeneracy of the eigenspaces of the Dirac
operator $(k \slsh-m)$ for any fixed $k$.  If $m=0$, the Dirac equation
splits into two separate equations for the left- and right-handed
component of the spinor, and this makes it
possible to project out half of the eigenvectors simply by multiplying by $\chi_L$,
\begin{equation}
P(k) \;=\; \chi_L\: k\slsh\: \delta(k^2)\: \Theta(-k^0)\:.
    \label{eq:C7}
\end{equation}
If $m>0$, this method cannot be applied because the left- and right-handed
subspaces are no longer invariant. In particular, the product $\chi_L t_m$ for
$m>0$ is not Hermitian and is not a solution of the Dirac equation. Nevertheless,
we can project out one of the degenerate eigenvectors as follows.
For given~$k$ on the lower mass shell we choose a vector $q$ with
\begin{equation}
k q \;=\; 0 \spc {\mbox{and}} \spc q^2 \;=\; -1\:.
    \label{eq:C9a}
\end{equation}
A short calculation shows that
\[ [t_m(k),\: \rho q \slsh] \;=\; 0 \spc {\mbox{and}}\spc (\rho q
\slsh)^2 \;=\; \1 \]
(where~$\rho$ is again the pseudoscalar matrix~(\ref{rhodef})).
This means that the matrix $\rho q \slsh$ has eigenvalues $\pm 1$, and that
the Dirac equation is invariant on the corresponding eigenspaces. Projecting
for example onto the eigenspace corresponding to the eigenvalue $-1$ gives
\begin{equation}
P_m(k) \;:=\; \frac{1}{2}\:(\1 -\rho q \slsh)\: (k \slsh+m)\:
\delta(k^2-m^2)\: \Theta(-k^0) \:. \label{eq:C8}
\end{equation}
Thus, similar to the procedure in the massless case~(\ref{eq:C7}), $P_m$ is
obtained from $t_m$ by projecting out half of the Dirac eigenstates on the lower
mass shell. But in contrast to~(\ref{eq:C7}), the construction of $P_m$ depends
on the vector field $q$, which apart from the conditions~(\ref{eq:C9a}) can be
chosen arbitrarily. A short calculation shows that $P_m$ is idempotent in the
sense that
\begin{equation}
P_m\: P_{m'} \;=\; \delta(m-m')\: P_m\:.
    \label{eq:C12}
\end{equation}

The distribution~(\ref{eq:C8}) can be regarded as a generalization of the
chiral Dirac sea~(\ref{eq:C7}) to the massive case. In order to make this
connection clearer, we now show that~(\ref{eq:C8}) reduces to~(\ref{eq:C7}) in the
limit $m \searrow 0$. Thus, for fixed $\vec{k} \neq 0$ and variable $m>0$, we let
$k$ be on the lower mass shell, $k(m) = (-\sqrt{|\vec{k}|^2 + m^2}, \vec{k})$, and
choose~$q(m)$ such that~(\ref{eq:C9a}) is satisfied. A simple example for $q$ is
\begin{equation}
q(m) \;=\; \frac{1}{m} \left(-|\vec{k}|,\: \sqrt{|\vec{k}|^2 + m^2}\:
\frac{\vec{k}}{|\vec{k}|} \right) .
    \label{eq:C9}
\end{equation}
In this example, $k$ and $mq$ coincide as $m \searrow 0$; more precisely,
\[ k-mq = {\mathcal{O}}(m^2) \:. \]
This relation holds for a large class of functions $q(m)$. Therefore, it
seems general enough to concentrate on the situation where
\begin{equation}
k-mq \;=\; m^2\:v \qquad {\mbox{with~$v(m) =
{\mathcal{O}}(m^0)$}}\:. \label{eq:C16}
\end{equation}
Solving this relation for $q$ and substituting into~(\ref{eq:C8}) gives
\begin{equation}
P_m(k) \;=\; \frac{1}{2} \left( \1 - \rho\: \frac{k \slsh}{m} + m \rho
v \slsh \right) (k \slsh+m)\: \delta(k^2-m^2)\: \Theta(-k^0)\:.
    \label{eq:C10}
\end{equation}
Using that on the mass shell $k \slsh (k \slsh+m) = m (k \slsh+m)$, we get
\begin{equation}
P_m(k) \;=\; \frac{1}{2} \left( \1 - \rho + m \rho
v \slsh \right) (k \slsh+m)\: \delta(k^2-m^2)\: \Theta(-k^0)\:.
    \label{eq:C15}
\end{equation}
If now we take the limit $m \searrow 0$, we obtain precisely~(\ref{eq:C7}), i.e.
\begin{equation}
\lim_{m \searrow 0} P_m \;=\; P \label{eq:C17}
\end{equation}
with convergence in the sense of distributions. This calculation shows
that~(\ref{eq:C8}) indeed includes~(\ref{eq:C7}) as a limiting case and that
the dependence on $q$ drops out as $m \searrow 0$.

The distribution~(\ref{eq:C8}) gives  a possible definition of a massive chiral
Dirac sea. However, it would be too restrictive to use only~(\ref{eq:C8}) as the
basis of our construction, because there are other common ways to give chiral Dirac
particles a rest mass. These alternatives are more general than~(\ref{eq:C8}) in
that the wave functions are no longer solutions of the Dirac equation. To give a
simple example, one could describe a massive left-handed Dirac sea for $m>0$ by
\begin{equation}
P_m(k) \;=\; \left( \chi_L\: k\slsh + \frac{m}{4} \right) \:\delta\!\left(
k^2 - \frac{m^2}{4} \right)\: \Theta(-k^0)\:.    \label{eq:C13}
\end{equation}
This distribution has the advantage over~(\ref{eq:C8}) that it is Lorentz
invariant, but it is clearly not a solution of the Dirac equation.  As $m
\searrow 0$, we again recover the massless chiral Dirac sea~(\ref{eq:C1}).  We
compute the operator product $P_m P_{m'}$ in momentum space,
\begin{eqnarray*}
(P_m\: P_{m'})(k) &=& \left(\chi_L\: k \slsh + \frac{m}{4}\right)
\left(\chi_L\: k \slsh + \frac{m'}{4}\right)
\delta\!\left(k^2 - \frac{m^2}{4}\right) \delta\!\left(k^2 - \frac{m'^2}{4}\right) \Theta(-k^0) \\
&=& \delta \!\left( \frac{m^2}{4} - \frac{m'^2}{4} \right)
\left( \frac{m+m'}{4}\: \chi_L\: k\slsh + \frac{m m'}{16} \right)\:
\delta\!\left(k^2 - \frac{m^2}{4}\right) \Theta(-k^0) \\
&=& \delta (m-m')
\left( \chi_L\: k\slsh + \frac{m}{8} \right)\:
\delta\!\left(k^2 - \frac{m^2}{4}\right) \Theta(-k^0) \:,
\end{eqnarray*}
where in the last step we used that $m,m'>0$. Note that in the last line the
summand $m/8$ appears (instead of the summand $m/4$ in~(\ref{eq:C13})), and
therefore $P_m$ is not idempotent in the sense~(\ref{eq:C12}). On the other
hand, one can argue that~(\ref{eq:C12}) is a too strong normalization condition,
because we are interested in the situation when the masses of the chiral
particles are arbitrarily small, and thus it seems sufficient
that~(\ref{eq:C12}) should hold in the limit $m,m' \searrow 0$. In this limit, the
problematic summands $m/4$ and $m/8$ both drop out, and thus we can state the
idempotence of $P_m$ as follows,
\begin{equation}
\lim_{m,m' \searrow 0} \left( P_m\: P_{m'} - \delta(m-m')\: P_m \right) \;=\; 0\:.
    \label{eq:C14}
\end{equation}
The above example shows that, in order to have more flexibility to give the
chiral Dirac particles a mass, it is preferable to work with the weaker
normalization condition~(\ref{eq:C14}) instead of~(\ref{eq:C12}).
Comparing with the naive calculation~(\ref{eq:C5}), one sees that
introducing the mass changes the behavior of the operator products completely,
even if the masses are arbitrarily small. Therefore, we refer to the limit
$m,m' \searrow 0$ in~(\ref{eq:C14}) as the {\em{singular mass
    limit}}\index{singular mass limit}.

For the correct understanding of the singular mass limit, it is important to
observe that, in contrast to operator products as considered in~(\ref{eq:C14}),
the formalism of the continuum limit is well-behaved as $m \searrow 0$.
Namely, in the continuum limit we consider an expansion in powers of $m$. The
different orders in $m$ have a different singular behavior on the light cone. In
particular, to every order on the light cone only a finite number of orders in
$m$ contribute. Thus to every order on the light cone, the $m$-dependence is
polynomial and therefore smooth. Expressed in terms of the kernel, the limit $m
\searrow 0$ is singular when we form the product $P(x,z)\: P(z,y)$ and
integrate over $z$ (as in~(\ref{eq:C5})). But if we take the closed chain
$P(x,y)\: P(y,x)$ and consider the singularities on the light cone, the limit
$m \searrow 0$ is regular and well-behaved. This justifies why in
Chapters~\ref{esec3}--\ref{secegg}
it was unnecessary to give the neutrinos a
mass and take the limit $m \searrow 0$ afterwards. We could treat the
neutrino sector simply as being composed of massless chiral particles. In
particular, the chiral cancellations in the formalism of the continuum
limit are consistent with the singular mass limit.

Our next goal is to develop the mathematical framework for analyzing the
singular mass limit for a fermionic projector with interaction. Clearly,
this framework should be general enough to include the examples~(\ref{eq:C8})
and~(\ref{eq:C13}). Thus we first return to~(\ref{eq:C8}). After writing $P_m$
in the form~(\ref{eq:C15}), it seems natural to interpret the leading factor as
a generalization of the chiral asymmetry matrix $X$. This is indeed convenient
in the vacuum, because introducing the operator $X_m$ by
\begin{equation}
X_m(k) \;=\; \frac{1}{2} \left( \1 - \rho - m \rho v \slsh(k) \right) ,
    \label{eq:C18}
\end{equation}
we obtain in analogy to the corresponding formulas for massless chiral
particles that
\[ P_m \;=\; X_m\: t_m \;=\; t_m\: X_m^* \:. \]
Unfortunately, the operator $X_m$ does not seem to be useful in the case
with interaction.
The reason is that $X_m$ depends on the momentum $k$, and this leads to the
following serious difficulties. First, the $k$-dependence of $X_m$ makes it very
difficult to satisfy the analogue of the causality compatibility
condition
\[ X_m^*\:(i \Pdd + {\mathcal{B}} - m) \;=\; (i \Pdd + {\mathcal{B}} - m)
\: X_m \:. \]
As a consequence, it is in general not possible to commute the chiral asymmetry
matrix through the operator products of the causal perturbation expansion; in
particular $X_m\: \tilde{t}_m$ and $\tilde{t}_m\: X_m$ do in general not
coincide (where $\tilde{t}_m$ is the interacting Dirac sea as defined via the
causal perturbation expansion). Even if we assume that there is a canonical
definition of the fermionic projector $P_m$ obtained by suitably inserting
factors of $X_m$ and $X_m^*$ into the operator product expansion for
$\tilde{t}_m$, we cannot expect that the correspondence to the massless Dirac
sea is respected; i.e.\ in the case with interaction,(\ref{eq:C17}) will in general be
violated. In order to explain how this comes about, we point out
that our argument leading to~(\ref{eq:C16}) was based on the assumption that $k$
converges to the mass cone as $m \searrow 0$. More precisely, if $\lim_{m
\searrow 0} k(m)$ is not on the mass cone, the function $v$ will diverge like
$v(m) \sim m^{-2}$, and so $X_m(k)$ will not converge to $X$ as $m \searrow 0$.
Thus $\lim_{m \searrow 0} X_m = X$ only if in this limit all the momenta are
on the mass cone. But in the causal
perturbation expansion off-shell momenta also appear (note that the Green's
functions are non-zero away from the mass cone). This means that
in the limit $m \searrow 0$, the momenta are in general not on the lower mass
cone, and so $X_m$ will not converge to $X$. Because of these problems, we
conclude that it is not admissible to first perform the perturbation expansion for
$t_m$ and to multiply by $X_m$ afterwards. Instead, the $k$-dependence of $X_m$
must be taken into account in the perturbation expansion.

At this point it is very helpful that we stated the normalization condition for
a chiral Dirac sea in the form~(\ref{eq:C14}). The key observation is that if
we substitute~(\ref{eq:C15}) into~(\ref{eq:C14}), compute the operator product
and take the limits $m,m' \searrow 0$, all contributions to~(\ref{eq:C15}) which
are at least quadratic in $m$ drop out. More precisely, if we expand $P_m$ in
the form
\begin{equation}
P_m \;=\; \left(\chi_L\: k\slsh \:+\: \frac{m}{2}\: (\1-\rho)
\:+\: \frac{m}{2}\:\rho v\slsh k\slsh + {\mathcal{O}}(m^2)
\right)\: \delta(k^2-m^2)\: \Theta(-k^0)\:, \label{eq:C19}
\end{equation}
the error term is of no relevance for the normalization
condition~(\ref{eq:C14}). Taking the inner product of~(\ref{eq:C16}) with $k$ and
using the first part of~(\ref{eq:C9a}) together with the relation $k^2=m^2$, one sees that
$vk=1$. We use this identity in~(\ref{eq:C19}) to obtain
\begin{equation}
P_m \;=\; \left(\chi_L\: k\slsh \:+\: \frac{m}{2} \:+\: \frac{m}{4}\: \rho\:
[v\slsh, k\slsh] \right) \: \delta(k^2-m^2)\: \Theta(-k^0)\:. \label{eq:C22}
\end{equation}
Writing $P_m$ in this form has the advantage that we can pull out the chiral
projectors by setting
\begin{equation}
P_m \;=\; \frac{1}{2}\:(X\:\tilde{t}_m + \tilde{t}_m\:X^*)
    \label{eq:C20}
\end{equation}
with $X=\chi_L$ and
\[ \tilde{t}_m \;=\; \; \left(k\slsh + m + \frac{m}{2}\: \rho\:
[v\slsh, k\slsh] \right) \: \delta(k^2-m^2)\: \Theta(-k^0)\:. \]
Again neglecting terms quadratic in $m$, $\tilde{t}_m$ is a solution of the
Dirac equation,
\begin{equation}
    (i \Pdd + {\mathcal{B}}_0 - m)\:\tilde{t}_m \;=\; 0\:, \label{eq:C23}
\end{equation}
where
\begin{equation}
{\mathcal{B}}_0(k) \;=\; -\frac{m}{2}\: \rho\: [v\slsh, k\slsh]\:.
\label{eq:C21}
\end{equation}
The formulation of the vacuum~(\ref{eq:C20})
and~(\ref{eq:C23}, \ref{eq:C21}) has the advantage that the
interaction can easily be introduced. Namely, in order to describe
the interaction we simply insert the external potentials into the
Dirac equation~(\ref{eq:C23}). In this way, the problems mentioned
after~(\ref{eq:C18}) get resolved. Instead of
working with a $k$-dependent chiral asymmetry matrix $X_m$, the
$k$-dependent vector field $v$ in~(\ref{eq:C15}) is now taken into
account by a perturbation ${\mathcal{B}}_0$ of the Dirac equation,
making it possible to apply perturbative methods in the spirit
of~{\S}\ref{jsec2}.

An obvious technical problem in this approach is that the
perturbation operator ${\mathcal{B}}_0$, (\ref{eq:C21}), is not of
a form previously considered in that it is nonlocal, is not causality
compatible and does not decay at infinity. This
problem will be analyzed in detail in Section~\ref{sec22}. What
makes the problem tractible is that ${\mathcal{B}}_0$ tends to
zero as $m \searrow 0$ and is {\em{homogeneous}}, meaning that its
kernel ${\mathcal{B}}_0(x,y)$ depends only on the difference
$x-y$.

Let us verify in which generality the above method~(\ref{eq:C20}, \ref{eq:C23})
applies. In the example~(\ref{eq:C13}), we can write the chiral Dirac sea in
the form~(\ref{eq:C20}) with
\begin{equation}
\tilde{t}_m \;=\; \left(k\slsh + \frac{m}{2} \right)\: \delta\!\left(k^2 -
\frac{m^2}{4} \right)\: \Theta(-k^0)\:, \label{eq:CC}
\end{equation}
and $\tilde{t}_m$ is a solution of the Dirac
equation~(\ref{eq:C23}) with ${\mathcal{B}}_0 = m/2$. Thus in this
case, ${\mathcal{B}}_0$ is a homogeneous local operator. More
generally, the method of pulling out the chiral
asymmetry~(\ref{eq:C20}) applies to any distribution $P_m$ of the
form
\[ P_m(k) \;=\; \left( \chi_L\: {\mbox{(odd)}} + {\mbox{(even)}} +
{\mathcal{O}}(m^2) \right)\: \delta(k^2- c\:m^2)\: \Theta(-k^0)\:, \]
where ``(odd)'' and ``(even)'' refer to a product of an odd and
even number of Dirac matrices, respectively (and $c$ is a
constant). Namely, the corresponding $\tilde{t}_m$ is
\[ \tilde{t}_m(k) \;=\; \left( {\mbox{(odd)}} + 2\:{\mbox{(even)}} +
{\mathcal{O}}(m^2) \right)\: \delta(k^2- c\:m^2)\: \Theta(-k^0)\:. \]
Hence the only restriction of the
method~(\ref{eq:C20}, \ref{eq:C23}) is that the {\em{right-handed
odd contribution to $P_m$}} should be {\em{of the order}}
${\mathcal{O}}(m^2)$. For example, our method does {\em{not}} apply to
\[ P_m(k) \;=\; \left(\chi_L\: k\slsh + m\: \chi_R\:f\: k \slsh + m + {\mathcal{O}}(m^2)
\right)\:\delta(k^2-m^2)\: \Theta(-k^0) \] with a scalar function
$f(k)$, although in this case the normalization
condition (\ref{eq:C14}) is satisfied. Dropping this restriction
would make it necessary to give up~(\ref{eq:C20}) and thus to
treat the trace compatibility on a level which goes far beyond
what we can accomplish here. It is our view that assuming that the
right-handed odd contribution to $P_m$ is of the order
${\mathcal{O}}(m^2)$ is a reasonable technical simplification.

We close our discussion with a comment on the
example~(\ref{eq:C13}). We saw above that $P_m$ can be written in
the form~(\ref{eq:C20}) with $\tilde{t}_m$ according
to~(\ref{eq:CC}), and that $\tilde{t}_m$ is a solution of the
Dirac equation~(\ref{eq:C23}) with the perturbation
${\mathcal{B}}_0 = m/2$. An alternative point of view is that
$\tilde{t}_m$ is a solution of the free Dirac equation of half the
mass,
\begin{equation}
(i \Pdd - M)\: \tilde{t}_m \;=\; 0 \spc {\mbox{with}} \spc M=\frac{m}{2}\:.
    \label{eq:Cmms}
\end{equation}
We refer to the method of considering a Dirac equation in which the mass
parameter is multiplied by a constant as the {\em{modified mass
    scaling}}\index{modified mass scaling}. The
modified mass scaling has the advantage that one can satisfy the normalization
conditions for chiral Dirac seas~(\ref{eq:C14}) with $P_m$
according to~(\ref{eq:C20}) and $\tilde{t}_m$ a solution of the free Dirac
equation.

\section{The Homogeneous Perturbation Expansion}
\setcounter{equation}{0} \label{sec22}
\index{perturbation expansion!homogeneous}
In the above examples we saw that there are
different methods for giving a chiral Dirac sea a rest mass, which
all correspond to inserting a suitable homogeneous operator
${\mathcal{B}}_0$ into the Dirac equation. Furthermore, we found
that the terms quadratic in the mass were irrelevant for the
normalization of the Dirac sea, and this suggests that it should
be possible to treat ${\mathcal{B}}_0$ perturbatively. This is
indeed possible, as we shall now show for a general class of
operators~${\mathcal{B}}_0$.

For simplicity, we again consider a single Dirac sea. We let
${\mathcal{B}}_0$ be a {\em{homogeneous operator}}, whose further
properties will be specified below. In order to keep track of the
different orders in perturbation theory, we multiply
${\mathcal{B}}_0$ by a small parameter $\varepsilon>0$. Exactly as
in~(\ref{eq:dir1}), we insert a parameter~$\mu$ into the
Dirac equation, which then reads
\[ (i \Pdd + \varepsilon\: {\mathcal{B}}_0 - \mu\:\1)\:\Psi \;=\; 0\:. \]
Here the Dirac operator is homogeneous and is therefore diagonal
in momentum space. Thus for given momentum $k$, the Dirac equation
reduces to the~$4 \times 4$ matrix equation
\begin{equation} \label{2Ax}
(k\slsh + \varepsilon\: {\mathcal{B}}_0(k) - \mu)\: \Psi(k) \;=\;
0\:.
\end{equation}
Our aim is to introduce and analyze the spectral projectors and
Green's functions of the Dirac operator~$i \Pdd + \varepsilon
{\mathcal{B}}_0$, where we regard $\mu$ as the eigenvalue. In
preparation, we shall now analyze the matrix equation~(\ref{2Ax})
for fixed $k$ in a perturbation expansion to first order in
$\varepsilon$. If $k^2 \neq 0$, the matrix $k\slsh$ is
diagonalizable with eigenvalues and spectral projectors
\begin{equation}
\mu_\pm \;=\; \pm \mu_k \:,\spc E_\pm \;=\; \frac{1}{2} \left( \1
\pm \frac{k\slsh}{\mu_k} \right) , \label{227}
\end{equation}
where we set $\mu_k = \sqrt{k^2}$ (if $k^2<0$, our sign convention
is such that $\mu_k $ lies in the upper complex half plane). The
eigenspaces ${\mbox{Im}}\: E_\pm$ are two-dimensional. The
spectral projectors $E_\pm$ become singular as $k^2 \to 0$. The
reason is that on the mass cone ${\mathcal{C}} := \{k \:|\:
k^2=0\}$, the matrix $k \slsh$ is not diagonalizable. We will
address this problem later and for the moment simply assume that
$k^2 \neq 0$. We next consider the Dirac operator $k\slsh +
\varepsilon {\mathcal{B}}_0$ for small $\varepsilon$. Perturbing
the eigenspaces ${\mbox{Im}}\: E_\pm$ gives rise to
two-dimensional invariant subspaces, and a standard calculation
shows that the projectors $E^\varepsilon_\pm$ onto these subspaces
are given by
\begin{equation}
E^\varepsilon_s \;=\; E_s \:+\: s\:\frac{\varepsilon}{2 \mu_k}
\left( E_s\:{\mathcal{B}}_0\:E_{\bar{s}} +
E_{\bar{s}}\:{\mathcal{B}}_0\:E_s \right) \:+\:
{\mathcal{O}}(\varepsilon^2) \label{2L}
\end{equation}
with $s=\pm$ and $\bar{s}=-s=\mp$. It remains to diagonalize the
operator $k\slsh + \varepsilon {\mathcal{B}}_0$ on the invariant
subspaces ${\mbox{Im}}\: E^\varepsilon_s$. This is carried out in
the next lemma. We choose three (possibly complex) Lorentz vectors
$(q_i)_{i=1,2,3}$ such that
\begin{equation}
\Bra q_i, k \Ket =0 \spc {\mbox{and}} \spc \Bra q_i, q_j \Ket =
-\delta_{ij}\:. \label{228}
\end{equation}
More precisely, if $k$ is time-like, we choose the $(q_i)$ as a
real orthonormal basis of the space-like hypersurface $\bra k
\ket^\perp$. If on the other hand $q$ is space-like, we choose
$q_1$ and $q_2$ real and space-like, whereas $q_3$ is
time-like and imaginary. We use the vector notation $\vec{q} =
(q_1, q_2, q_3)$ and introduce the matrices $\Sigma_{1,2,3}$ by
\begin{equation}
\vec{\Sigma} \;=\; \rho\: \vec{q} \slsh \:. \label{2sig}
\end{equation}

\begin{Lemma} \label{lemma21}
Suppose that $k^2 \neq 0$ and that for small $\varepsilon$, the
matrix $k\slsh + \varepsilon {\mathcal{B}}_0$ is diagonalizable.
Then its eigenvalues $(\mu^a_s)_{s=\pm, a=1\!/\!2}$ are given by
\begin{eqnarray}
\mu^{1\!/\!2}_+ &=& \mu_k \:+\: \varepsilon\: (\nu_+ \pm \tau_+)
\:+\: {\mathcal{O}}(\varepsilon^2) \label{2a} \\
\mu^{1\!/\!2}_- &=& -\mu_k \:+\: \varepsilon\: (\nu_- \mp \tau_-)
\:+\: {\mathcal{O}}(\varepsilon^2) \label{2b} \:,
\end{eqnarray}
where
\begin{eqnarray}
\nu_s &=& \frac{1}{2}\: \Tr \left( E_s\: {\mathcal{B}}_0 \right) \label{2ba} \\
\vec{\tau}_s &=& \frac{1}{2} \:\Tr (\vec{\Sigma}\: E_s\:
{\mathcal{B}}_0 ) \label{2bb} \\
\tau_s &=& \sqrt{(\tau^1_s)^2 + (\tau^2_s)^2 + (\tau^3_s)^2}
\label{2c} \:.
\end{eqnarray}
The corresponding spectral projectors can be written as
\begin{equation}
E^a_s \;=\; \Pi^a\:E_s \:+\: s\: \frac{\varepsilon}{2 \mu_k}
\left( \Pi^a\:E_s\: {\mathcal{B}}_0\: E_{\bar{s}} + E_{\bar{s}}
\:{\mathcal{B}}_0\: \Pi^a\:E_s \right) \:+\:
{\mathcal{O}}(\varepsilon^2) \label{2d}
\end{equation}
with
\begin{equation}
\Pi^{1\!/\!2} \;=\; \frac{1}{2} \left( \1 \pm \frac{1}{\tau_s}\:
\vec{\tau}_s \vec{\Sigma} \right) . \label{2e}
\end{equation}

If $\vec{\tau}_s=0$, the invariant subspace ${\mbox{Im}}\:
E^\varepsilon_s$ is an eigenspace to first order in $\varepsilon$; i.e.
\[ \left. (k\slsh + \varepsilon {\mathcal{B}}_0)
\right|_{\mbox{\scriptsize{Im}}\: E^\varepsilon_s} \;=\; (s\:\mu_k
+ \varepsilon\: \nu_s) \left. \1 \right|_{\mbox{\scriptsize{Im}}\:
E^\varepsilon_s} \:+\: {\mathcal{O}}(\varepsilon^2) \:. \]
\end{Lemma}
{\Proof} We restrict attention to the invariant subspace
${\mbox{Im}}\: E^\varepsilon_+$; for $E^\varepsilon_-$ the proof
is similar. A short calculation using~(\ref{2sig})
and~(\ref{227}, \ref{228}) shows that
\[ [ \Sigma_i, \:E_+] \;=\; 0\:, \spc
\Sigma_i^2 \;=\; \1 \:,\spc \Tr (\Sigma_i\: \Sigma_j\: E_+) \;=\;
2\: \delta_{ij}\:. \]
This means that the matrices $\Sigma_i$ are invariant on ${\mbox{Im}}\: E_+$,
have the eigenvalues $\pm 1$ on this subspace and are orthogonal.
Thus by choosing a suitable basis (and possibly after changing the orientation of
$\vec{\Sigma}$ by exchanging $\Sigma_1$ with $\Sigma_2$), we can
arrange that the matrices $\vec{\Sigma}|_{\mbox{\scriptsize{Im}}\:
E_+}$ coincide with the Pauli matrices $\vec{\sigma}$. To first
order in $\varepsilon$, the eigenvalues are obtained by
diagonalizing ${\mathcal{B}}_0$ on the unperturbed invariant
subspace ${\mbox{Im}}\: E_+$. A short calculation shows that
the $2 \times 2$ matrix $\nu \1 + \vec{\tau} \vec{\sigma}$ has the
eigenvalues $\nu \pm \tau$ and corresponding spectral projectors
$\Pi_{1\!/\!2}=\frac{1}{2}(\1+\frac{1} {\tau}\:\vec{\tau}
\vec{\sigma})$ with $\tau = \sqrt{(\tau_1)^2 + (\tau_2)^2 +
(\tau_3)^2}$. This gives~(\ref{2a}, \ref{2b}). Finally,
(\ref{2d}) follows from standard perturbation theory without
degeneracies. \QED

To avoid confusion, we point out that in
general $\tau_s \neq |\vec{\tau}_s|$ because~(\ref{2c}) involves
ordinary squares instead of absolute squares. In particular, it is
possible that $\tau_s=0$ although $\vec{\tau}_s \neq 0$. However,
in this case the $2 \times 2$ matrix $\varepsilon
{\mathcal{B}}_0|_{\mbox{\scriptsize{Im}} \: E_s}$ is not
diagonalizable, and thus the above lemma does not apply.

\begin{Remark} \label{remark22} \em
In the proof of the previous lemma we used that the three matrices
$\Sigma_i|_{\mbox{\scriptsize{Im}}\: E_+}$ can be represented as the
Pauli matrices $\sigma_i$. It is instructive to verify explicitly that
these matrices satisfy the correct commutation relations, for example
\[ \frac{i}{2}\: [\Sigma_1,\: \Sigma_2] |_{\mbox{\scriptsize{Im}}\: E_+}
\;=\; \Sigma_3 |_{\mbox{\scriptsize{Im}}\: E_+} \:. \]
We now give this calculation in detail. By a choice of coordinates, we
can arrange that $k=(\omega, \vec{p})$ and $q_{1\!/\!2} = (0, \vec{q}_{1\!/\!2})$.
The standard identity between the Dirac matrices $i \sigma_{jk} = \frac{\rho}{2}
\:\epsilon_{jklm}\: \sigma^{lm}$ yields that (possibly after changing the
orientation of $\vec{\Sigma}$),
\begin{equation}
i q_1\!\!\!\slsh \;\: q_2\!\!\!\slsh\; \;=\; \frac{\rho}{2 |\vec{p}|}\: [k\slsh, \gamma^0]\:.
\label{2f}
\end{equation}
From the definition of $\vec{\Sigma}$, (\ref{2sig}), one sees that
$[\Sigma_1, \Sigma_2] = -2 q_1\!\!\!\slsh \;\: q_2\!\!\!\slsh\;$, and
using~(\ref{2f}) as well as the identity $[\mu_k, \gamma^0]=0$, we conclude
that
\begin{equation}
\frac{i}{2}\: [\Sigma_1,\: \Sigma_2] \;=\; -\frac{\rho}{2 |\vec{p}|}\:
[k\slsh - \mu_k,\: \gamma^0]\:. \label{2g}
\end{equation}

In order to simplify the rhs of~(\ref{2g}) on ${\mbox{Im}}\: E_+$, we use that
$E_+$ satisfies the Dirac equation
\begin{equation}
(k\slsh - \mu_k)\:E_+ \;=\; 0\:. \label{2h}
\end{equation}
This identity allows us to replace the commutator with $k\slsh-\mu_k$ by
an anti-commu\-ta\-tor,
\begin{equation}
[k\slsh-\mu_k,\: \gamma^0]\:E_+ \;=\; \{ k\slsh-\mu_k,\: \gamma^0 \} \:E_+ \;=\;
(2\omega - 2 \mu_k \gamma^0) \:E_+\:. \label{2i}
\end{equation}
Multiplying~(\ref{2h}) by $2 \omega/\mu_k$ and adding~(\ref{2i}) gives
\[ [k\slsh-\mu_k,\: \gamma^0]\: E_+ \;=\; \frac{2\omega}{\mu_k}
\left(k\slsh - \frac{\mu^2}{\omega}\: \gamma^0 \right) E_+\:. \]
Using this identity in~(\ref{2g}) gives
\[ \frac{i}{2}\: [\Sigma_1,\: \Sigma_2]|_{\mbox{\scriptsize{Im}}\:E_+}
\;=\; \rho\: q_3 \!\!\!\slsh\;|_{\mbox{\scriptsize{Im}}\:E_+} \]
with
\[ q_3 \;=\; -\frac{\omega}{\mu_k\: |\vec{p}|} \left(k\slsh - \frac{\mu_k^2}{\omega}\:
\gamma^0 \right) , \]
and a short calculation shows that this vector $q_3$ has indeed all the
properties listed after~(\ref{228}).
\em \end{Remark}

We shall now define the spectral projectors and Green's functions
corresponding to the Dirac operator $i\Pdd + \varepsilon
{\mathcal{B}}_0$. We denote the spectrum of the matrix
in~(\ref{2Ax}) by $\sigma^\varepsilon(k)$,
\[ \sigma^\varepsilon(k) \;=\; \sigma(k\slsh + \varepsilon
{\mathcal{B}}_0(k)) \:. \]
It is natural to define the spectrum~$\sigma^\varepsilon$ of the Dirac
operator $i\Pdd + \varepsilon {\mathcal{B}}_0$ as the union of the $\sigma^\varepsilon(k)$s,
\[ \sigma^\varepsilon \;=\; \bigcup_{k \in \sR^4}
\sigma^\varepsilon(k) \:. \]
As we saw above, the matrix $k\slsh + \varepsilon {\mathcal{B}}_0(k)$
in general is not diagonalizable,
and thus we cannot introduce the spectral projectors for all $k$
pointwise. But since the diagonalizable matrices are dense in
${\mbox{Gl}}(\C^4)$, it is reasonable to assume that the matrix
$k\slsh + \varepsilon {\mathcal{B}}_0(k)$ is {\em{diagonalizable
for almost all (a.a.) $k$}}. Our formalism will involve momentum
integrals where sets of measure zero are irrelevant. Therefore, we
may in what follows restrict attention to those $k$ for which the
matrix $k\slsh + \varepsilon {\mathcal{B}}_0(k)$ is
diagonalizable. Moreover, we shall assume that ${\mathcal{B}}_0$
is {\em{smooth}} and {\em{bounded}}. According to~(\ref{227}), the
spectrum of the unperturbed Dirac operator is
$\sigma^{\varepsilon=0}=\R \cup i \R$. The next lemma shows that
the real part of the spectrum is stable under perturbations.

\begin{Lemma} \label{lemma23}
Suppose that $k^2 >0$. Then for $\varepsilon$ sufficiently small,
$\sigma^\varepsilon(k) \subset \R$.
\end{Lemma}
{\Proof} Choosing coordinates such that $k=(\omega, \vec{0})$, it
is obvious that the eigenspaces of $k\slsh$ are definite, i.e.
\[ \Sl \Psi \:|\: \Psi \Sr \neq 0 \spc {\mbox{for all eigenvectors $\Psi$}}. \]
By continuity, the eigenspaces of $k\slsh + \varepsilon
{\mathcal{B}}_0(k)$ will also be definite for sufficiently small
$\varepsilon$. As a consequence, the corresponding eigenvalues are
real, because
\[ \lambda\: \Sl \Psi \:|\: \Psi \Sr \;=\; \Sl \Psi \:|\:
(k\slsh + \varepsilon {\mathcal{B}}_0)\: \Psi \Sr \;=\; \Sl
(k\slsh + \varepsilon {\mathcal{B}}_0)\: \Psi \:|\: \Psi \Sr \;=\;
\overline{\lambda}\: \Sl \Psi \:|\: \Psi \Sr \:. \]

\vspace*{-.5cm} \QED

Unfortunately, we have a-priori no control of how the imaginary part of the
spectrum changes with $\varepsilon$. For this reason, it is most
convenient to introduce the spectral projectors for all $\mu \in
\C$, such that they vanish identically for $\mu \not \in
\sigma^\varepsilon$. For the normalization, we work with
$\delta$-distributions supported at one point in the complex
plane. More precisely, we set
\begin{eqnarray*}
\delta^2(z) &=& \delta({\mbox{Re}}\:z)\: \delta({\mbox{Im}}\: z) \\
\int_\sC d^2 z \:\cdots &=& \int_{\sR^2} d({\mbox{Re}}\:z) \:
d({\mbox{Im}}\:z) \:\cdots \:.
\end{eqnarray*}

\begin{Def} \label{def24}
For $\mu \in \C$ and $k \in \R^4$ we set
\begin{eqnarray}
p^\varepsilon_\mu(k) &=& \sum_{s=\pm, \:a=1\!/\!2} E^a_s(k)\:
\delta^2(\mu-\mu^a_s(k)) \label{2i2} \\
k^\varepsilon_\mu(k) &=& \epsilon(k^0)\: p^\varepsilon_\mu(k) \label{2j} \\
s^\varepsilon_\mu(k) &=& \int_\sC d^2\nu\; \frac{\mbox{PP}}{\mu-\nu}\:
p^\varepsilon_\nu(k) \:. \label{2k}
\end{eqnarray}
We also consider $p^\varepsilon_\mu$, $k^\varepsilon_\mu$, and $s^\varepsilon_\mu$
as multiplication operators in momentum space.
\end{Def}
In formal calculations, the operators $p^\varepsilon_\mu$ and $k^\varepsilon_\mu$
are solutions of the Dirac equation,
\[ (i \Pdd + \varepsilon {\mathcal{B}}_0 - \mu)\: p^\varepsilon_\mu \;=\;
0 \;=\; (i \Pdd + \varepsilon {\mathcal{B}}_0 - \mu)\:
k^\varepsilon_\mu \:,\]
and satisfy in analogy to~(\ref{56}--\ref{58}) the multiplication rules
\begin{eqnarray}
p^\varepsilon_\mu\: p^\varepsilon_{\mu'} &=& k^\varepsilon_\mu\: k^\varepsilon_{\mu'}
\;=\; \delta^2(\mu-\mu')\: p^\varepsilon_\mu \label{2ka} \\
p^\varepsilon_\mu\: k^\varepsilon_{\mu'} &=& k^\varepsilon_\mu\: p^\varepsilon_{\mu'}
\;=\; \delta^2(\mu-\mu')\: k^\varepsilon_\mu \label{2kb}
\end{eqnarray}
as well as the ``completeness relation''
\[ \int_\sC p^\varepsilon_\mu\: d^2\mu \;=\; \1\:. \]
Using these identities in~(\ref{2k}) yields that
\[ (i \Pdd + \varepsilon {\mathcal{B}}_0 - \mu)\: s^\varepsilon_\mu \;=\; \1\:. \]
Thus on a formal level, the operators $p_\mu^\varepsilon$,
$k^\varepsilon_\mu$ and $s^\varepsilon_\mu$ are the spectral
projectors and Green's functions of the Dirac operator,
respectively. In order to give these operators a mathematical
meaning, we can proceed as follows. Let $k$ be such that the
matrix $k\slsh + \varepsilon {\mathcal{B}}_0(k)$ can be
diagonalized. Then the functional calculus for finite matrices (as
defined e.g.\ via the approximation by polynomials) allows us to
introduce for $f \in C^1(\C)$ the matrix $f(k\slsh + \varepsilon
{\mathcal{B}}_0(k))$. Formally, we can write the functional
calculus with the spectral projectors,
\begin{equation}
\int_\sC f(\mu)\: p^\varepsilon_\mu(k)\: d^2\mu \;=\; f(k\slsh +
\varepsilon {\mathcal{B}}_0(k)) \:. \label{eq:2k}
\end{equation}
We can use this relation to give the integral in~(\ref{eq:2k}) a rigorous sense
for a.a.\ $k$. The same argument applies to $k^\varepsilon_\mu$. For
$s^\varepsilon_\mu$, we can similarly use the formal identity
\begin{equation}
\int_\sC f(\mu)\: s^\varepsilon_\mu(k)\: d^2\mu \;\stackrel{(\ref{2k})}{=}\;
\int_\sC g(\mu)\: p^\varepsilon_\nu(k)\: d^2\mu \label{2l}
\end{equation}
with
\[ g(\nu) \;=\; \int_\sC \frac{\mbox{PP}}{\mu-\nu}\: f(\mu)\: d^2\mu\:. \]
In this way, one sees that the operators $p^\varepsilon_\mu$,
$k^\varepsilon_\mu$ and $s^\varepsilon_\mu$ are well-defined when evaluated
weakly in $\mu$ and $k$.

Under additional assumptions, we can make sense of the operators in Def.~\ref{def24}
even for fixed real $\mu$. We first justify the $\delta$-distribution and the principal
part.
\begin{Lemma} \label{lemma25}
Suppose that for a given interval $I \subset \R$, the spectral projectors
$E^a_s$ in~(\ref{2i2}) are bounded uniformly in~$\mu \in I$
and~$k \in \R^4$. Then for a.a.\
$\mu \in I$, the operators $p^\varepsilon_\mu$, $k^\varepsilon_\mu$, and
$s^\varepsilon_\mu$ are well-defined distributions in momentum space.
\end{Lemma}
{\Proof} We write the Dirac equation $(k\slsh + \varepsilon
{\mathcal{B}}_0(k))\Psi =0$ in the Hamiltonian form
\[ \omega\: \Psi \;=\; H(\omega, \mu)\: \Psi \spc {\mbox{with}} \spc
H(\omega, \mu) \;=\; -\gamma^0\: (\vec{k}\slsh + \varepsilon
{\mathcal{B}}_0(\omega, \vec{k}) - \mu \1) \] and $k=(\omega,
\vec{k})$. In what follows we keep $\vec{k}$ fixed and consider
this equation for variable parameters $\omega, \mu \in \R$. The
matrix $H(\omega, \mu)$ is Hermitian with respect to the positive scalar
product $(.|.) = \Sl .|\gamma^0|. \Sr$. Thus it can be
diagonalized; we denote its eigenvalues (counting multiplicities)
by $\Omega_1 \leq \cdots \leq \Omega_4$. The min-max principle
(see~\cite{RS3}) allows us to write $\Omega_n$ as
\[ \Omega_n \;=\; \min_{U, \;\dim U=n} \;\;\max_{u \in U,\; \|u\|=1} \|H u\| \:, \]
where $\|.\|$ is the norm induced by $(.|.)$ and $U$ denotes a subspace of $\C^4$.
It follows from this representation that the $\Omega_n$ depend Lipschitz-continuously
on $\omega$ and $\mu$. Namely,
\begin{eqnarray*}
\Omega_n(\omega) &=& \min_{U, \;\dim U=n} \;\;\max_{u \in U,\; \|u\|=1} \|H(\omega) \:u\| \\
&=& \min_{U, \;\dim U=n} \;\;\max_{u \in U,\; \|u\|=1}
\|H(\omega') \:u + (H(\omega)-H(\omega')) \:u \| \\
&\leq& \min_{U, \;\dim U=n} \;\;\max_{u \in U,\; \|u\|=1} \left(\|H(\omega') \:u\|
+ \|H(\omega) - H(\omega')\|\: \|u\| \right) \\
&=& \Omega_n(\omega') + \|H(\omega) - H(\omega')\| \:.
\end{eqnarray*}
Using that ${\mathcal{B}}_0(k)$ is $C^1$ with bounded derivatives,
we obtain the estimate
\[ \|H(\omega) - H(\omega')\| \;\leq\; \| \varepsilon \gamma^0\:
({\mathcal{B}}_0(\omega)-{\mathcal{B}}_0(\omega')) \| \;\leq\;
\varepsilon c\: |\omega-\omega'| \] and thus
$\Omega_n(\omega)-\Omega_n(\omega') \leq \varepsilon c
|\omega-\omega'|$. Exchanging the roles of $\omega$ and $\omega'$
gives the bound
\begin{equation}
|\Omega_n(\omega) - \Omega_n(\omega')| \;\leq\; \varepsilon c\: |\omega-\omega'| \:.
\label{2D}
\end{equation}
A similar calculation shows that
\begin{equation}
|\Omega_n(\mu) - \Omega_n(\mu')| \;\leq\; |\mu-\mu'|\:. \label{2E}
\end{equation}

We next consider for given $n$ the equation
\begin{equation}
\omega \;=\; \Omega_n(\omega, \mu)\:. \label{2G}
\end{equation}
The following argument shows that for sufficiently small $\varepsilon$, this equation
has a unique solution $\omega_n$, which depends Lipschitz-continuously on $\mu$.
Let $\phi$ (for fixed $\mu$ and $n$) be the mapping
\[ \phi \::\: \R \to \R \;:\: \omega \mapsto \Omega_n(\omega, \mu)\:. \]
According to~(\ref{2D}),
\[ |\phi(\omega) - \phi(\omega')| \;=\; |\Omega_n(\omega) - \Omega_n(\omega')|
\;\leq\; \varepsilon c\: |\omega-\omega'| \:. \]
Thus if we choose $\varepsilon$ small enough, $\phi$ is a contraction. The Banach
fixed point theorem yields a unique fixed point $\omega_n$. The dependence on
the parameter $\mu$ is controlled by~(\ref{2D}) and~(\ref{2E}). Namely,
\begin{eqnarray*}
|\omega_n(\mu) - \omega_n(\mu')| &=& |\Omega_n(\omega_n(\mu), \mu)
- \Omega_n(\omega_n(\mu'), \mu')| \\
&\leq& \varepsilon c\: |\omega_n(\mu) - \omega_n(\mu')| \:+\: |\mu-\mu'|
\end{eqnarray*}
and thus
\begin{eqnarray}
|\omega_n(\mu) - \omega_n(\mu')| \;\leq\; (1-\varepsilon c)^{-1}\:
|\mu - \mu'|\:. \label{2F}
\end{eqnarray}

If we regard the spectral projector~(\ref{2i2}) as a distribution
in $\omega$, it is supported at those $\omega$ for which the Dirac
equation $(k\slsh + \varepsilon {\mathcal{B}}_0 - \mu) \Psi =0$
has a non-trivial solution. These are precisely the solutions
$\omega_n$ of the equation~(\ref{2G}). Thus we can write
$p^\varepsilon_\mu$ as
\begin{equation}
p^\varepsilon_\mu \;=\; \sum_{n=1}^4 E^a_s(\omega_n)\; \delta(\omega-\omega_n)
\: \delta({\mbox{Im}}\: \mu) \left| \frac{\partial \omega(\mu)}{\partial \mu}
\right| \:, \label{2H}
\end{equation}
where the parameters $a=a(n)$ and $s=s(n)$ must be chosen such that
$\mu^a_s(\omega_n) = \mu$.
Since $\omega_n(\mu)$ is Lipschitz~(\ref{2F}), the factor
$|\partial_\mu \omega_n(\mu)|$ in~(\ref{2H}) is well-defined for a.a.\ $\mu$
and is uniformly bounded. Thus $p^\varepsilon_\mu(\omega)$ is a well-defined
distribution for a.a. $\mu$. The same argument applies to $k^\varepsilon_\mu$.

It remains to justify the Green's function $s^\varepsilon_\mu$. We can write it in the
Hamiltonian framework as
\[ s^\varepsilon_\mu \;=\; \frac{\mbox{PP}}{k\slsh + \varepsilon {\mathcal{B}}_0 - \mu \: \1}
\;=\; \frac{\mbox{PP}}{\omega - H(\omega, \mu)}\: \gamma^0\:. \]
Thus denoting the spectral projectors of $H$ by $(F_n)_{n=1,\ldots,4}$, we have
\begin{equation}
s^\varepsilon_\mu(\omega) \;=\; \sum_{n=1}^4 \frac{\mbox{PP}}{\omega
- \Omega_n(\omega, \mu)} \: F_n(\omega, \mu)\: \gamma^0\:. \label{2I}
\end{equation}
According to~(\ref{2D}), $\Omega_n(\omega)$ is Lipschitz and thus differentiable
almost everywhere with $|\partial_\omega \Omega_n| \leq \varepsilon c$.
The spectral projectors
$F_n(\omega)$ can also be chosen to be Lipschitz. As a consequence, the principal
part in~(\ref{2I}) is well-defined for a.a.\ $\mu$.
\QED
This lemma involves the strong assumption that the spectral projectors $E^a_s$ must
be uniformly bounded. We shall now analyze this assumption in detail. As one sees
from~(\ref{227}) in the limit $\mu \to 0$, the spectral projectors can have poles
and thus in general are {\em{not}} uniformly bounded. Thus we need to impose an extra
condition, which we will formulate using the following
notion.\index{$\varepsilon$-definite!spectrum}
\begin{Def} \label{def27}
Let $A$ be a $4 \times 4$ matrix, which is Hermitian (with respect
to~$\Sl .|. \Sr$).
A point $\mu \in \sigma(A)$ is called {\bf{$\varepsilon$-definite}}
if there is a subset $\sigma_+ \subset \sigma(A)$ such that
\begin{enumerate}
\item[(i)] The invariant subspace $I_+$ corresponding to $\sigma_+$ is definite.
\item[(ii)] ${\mbox{dist}}(\sigma_+,\: \sigma(A) \setminus \sigma_+)>\varepsilon$.
\end{enumerate}
\end{Def}

\begin{Lemma} \label{lemma28}
If $\mu \in \sigma(A)$ is $\varepsilon$-definite, the matrix $A$ is diagonalizable
on $I_+$, and its spectral projectors $E_a$ are bounded by
\begin{equation} \label{star}
\|E_a\| \;\leq\; c \left( \frac{\|A\|}{\varepsilon} \right)^3 ,
\end{equation}
where $\|.\|$ is a matrix norm and $c$ is a constant which depends only on
the choice of the norm $\|.\|$.
\end{Lemma}
{\Proof} It clearly suffices to consider a particular matrix norm. We
introduce the positive scalar product $(.|.)=\Sl .|\gamma^0|. \Sr$, let
$\|.\|=(.|.)^\frac{1}{2}$ be the corresponding norm, and set
\[ \|A\| \;=\; \sup_{\Psi {\mbox{\scriptsize{ with }}} \|\Psi\|=1} \|A \Psi\|\:. \]
We denote the projector onto $I_+$ by $E$. $E$ can be constructed
with a functional calculus. Namely, let ${\mathcal{P}}(z)$ be a complex
polynomial satisfying the conditions
\[ {\mathcal{P}}|_{\sigma_+} \;=\; 1 \spc {\mbox{and}} \spc
{\mathcal{P}}|_{\sigma_-} \;=\; 0\:. \]
Since these are at most four conditions, ${\mathcal{P}}$ can be
chosen to be of degree three,
\[ {\mathcal{P}}(z) \;=\; \sum_{n=0}^3 c_n\: z^n \:. \]
Furthermore, the fact that $A$ is $\varepsilon$-definite can be used to
bound the coefficients $c_n$ by
\begin{equation}
|c_n| \;\leq\; \frac{C}{\varepsilon^n} \label{2n}
\end{equation}
with a suitable constant $C$ (this is easily seen from a
scaling argument). The projector $E$ is given by
$E={\mathcal{P}}(A)$, and~(\ref{2n}) gives the estimate
\begin{equation}
\|E\| \;\leq\; \sum_{n=0}^3 \frac{C}{\varepsilon^n}\: \|A\|^n
\;\leq\; C \left(\frac{\|A\|}{\varepsilon} \right)^3 ,
\label{48a}
\end{equation}
where we used $\varepsilon < \|A\|$ in the last step.

By definition, ${\mbox{Im}}\: E=I_+$ is a definite subspace. We can assume
without loss of generality that it is positive, i.e.
\[ \Sl \Psi \:|\: E\: \Psi \Sr \;\geq\; 0 \spc {\mbox{for all $\Psi$}}. \]
The matrix $A|_{I_+}$ is Hermitian with respect to the positive scalar product
$\Sl .|. \Sr|_{I_+}$. Thus it has a spectral decomposition with eigenvalues
$\mu_a$ and corresponding spectral projectors $E_a$, $a=1,\ldots,N$,
\[ A |_{I_+} \;=\; \sum_{a=1}^n \mu_a\: E_a|_{I_+} \:. \]
Extending the $E_a$ by zero to the invariant subspace corresponding to
$\sigma(A) \setminus \sigma_+$, the spectral projectors satisfy the relations
\[ E_a^* = E_a = E_a^2 \:,\spc \sum_{a=1}^N E_a = E \:,\spc
\Sl \Psi \:|\: E_a \:\Psi \Sr \geq 0 \;\;{\mbox{for all $\Psi$}}, \]
where the star denotes the adjoint with respect to $\Sl .|. \Sr$.

We introduce the operators $F$ and $F_a$ by
\[ F \;=\; \gamma^0\:E \:,\spc F_a \;=\; \gamma^0\: E_a\:. \]
It is straightforward to check that these operators have the following properties,
\begin{eqnarray}
F_a^+ &=& F_a \:,\spc (\Psi \:|\: F_a\: \Psi) \;\geq\; 0 \label{2m} \\
\sum_a F_a &=& F\:,
\end{eqnarray}
where ``$^+$'' denotes the adjoint with respect to $(.|.)$. The relations~(\ref{2m}) mean
that the $F_a$ are positive self-adjoint operators on a Hilbert space. This makes
it possible to estimate the norm of the spectral projectors as follows,
\begin{eqnarray*}
\lefteqn{ \|E_a\| \;=\; \|\gamma^0\:F_a\| \;\leq\; \|\gamma^0\|\: \|F_a\|
\;\leq\; \|F_a\| \;=\; \sup_{\Psi {\mbox{\scriptsize{ with }}} \|\Psi\|=1}
(\Psi \:|\: F_a\: \Psi) } \\
&\leq& \sup_{\Psi {\mbox{\scriptsize{ with }}} \|\Psi\|=1}
\sum_{b=1}^N (\Psi \:|\: F_b\: \Psi)
\;=\; \sup_{\Psi {\mbox{\scriptsize{ with }}} \|\Psi\|=1} (\Psi \:|\:
F\: \Psi) \spc \\
&=& \|F\| \;=\; \|\gamma^0\: E\| \: \;\leq\; \|E\|\:.
\end{eqnarray*}
We now apply~(\ref{48a}) to obtain~(\ref{star}).
\QED

\begin{Def} \label{def29}
The Dirac operator $i \Pdd + \varepsilon {\mathcal{B}}_0$ has an
{\bf{$\varepsilon$-definite kernel}}\index{$\varepsilon$-definite!kernel} if for all $\mu \in
(-\varepsilon, \varepsilon)$ and all $k$ with $\mu \in
\sigma^\varepsilon(k)$, $\mu$ is in the $\varepsilon$-definite
spectrum of the matrix $k\slsh + \varepsilon {\mathcal{B}}_0(k)$.
\end{Def}
Combining Lemma~\ref{lemma25} and Lemma~\ref{lemma28} gives the following result.
\begin{Thm} \label{thm210}
If the Dirac operator $i \Pdd + \varepsilon {\mathcal{B}}_0$ has
an $\varepsilon$-definite kernel, then its spectral projectors and
Green's functions (as given in Def.~\ref{def24}) are well-defined
distributions in momentum space for a.a.\ $\mu \in (-\varepsilon, \varepsilon)$ .
\end{Thm}

It remains to specify under which assumptions on ${\mathcal{B}}_0$
the Dirac operator has an $\varepsilon$-definite kernel. We
decompose ${\mathcal{B}}_0$ as
\begin{equation}
{\mathcal{B}}_0(k) \;=\; \alpha\:\1 + i \beta \:\rho + v\slsh +
\rho\: a \slsh + \frac{i \rho}{2}\: w_{ij}\: \sigma^{ij} \:.
\label{2o}
\end{equation}
Here $\alpha$, $\beta$, $v$, $a$, and $w$ are real potentials (namely the
scalar, pseudoscalar, vector, axial, and bilinear potentials, respectively;
clearly we assume $w$ to be anti-symmetric).
We introduce the function $\Delta(k)$ as the following combination of the
axial and bilinear potentials,
\begin{equation}
\Delta^2 \;=\; -k^2\: \Bra a, a\Ket + \Bra a, k \Ket^2 - w_{ij} k^j\:
w^{il} k_l\:. \label{2E2}
\end{equation}
The first two summands can also be written as
\begin{equation}
-k^2\: \Bra a, a\Ket + \Bra a, k \Ket^2 \;=\;
-k^2 \left(a - \frac{1}{k^2}\: \Bra a, k \Ket\: k \right)^2 \:.
\label{2D2}
\end{equation}
For timelike $k$, the vector inside the round brackets is spacelike, and
thus $(\ref{2D2}) \geq 0$. Similarly, the vector $w_{ij} k^j$ is spacelike
for $k$ timelike. We conclude that
\begin{equation}
\Delta(k) \;\geq\; 0 \spc {\mbox{if $k^2>0$.}} \label{2E3}
\end{equation}
Furthermore, $\Delta(q)$ vanishes on the mass cone
${\mathcal{C}}=\{q^2=0\}$ if and only if $q$ is collinear to the
vector $a$ and is an eigenvector of $w$,
\begin{equation}
a\;=\; \nu q \spc {\mbox{and}} \spc w_{ij} q^j \;=\; \lambda\: q_i
\spc (\nu, \lambda \in \R, \:q \in {\mathcal{C}}). \label{2G2}
\end{equation}
Expanding~(\ref{2E2}), one sees that in this case, $\Delta$ is finite
to the next order on the light cone, i.e.
\begin{equation}
\Delta(q)=0 \;\;\;\Longrightarrow\;\;\; l \equiv \lim_{k \to q}
\frac{1}{k^2}\: \Delta(k) {\mbox{ exists}}. \label{eq:2F}
\end{equation}
Qualitatively speaking, the next theorem states that the Dirac operator has an
$\varepsilon$-definite kernel if and only if the scalar potential is
non-zero and dominates the axial and bilinear potentials.
\begin{Thm} \label{thm211}
Suppose that for all $q \in {\mathcal{C}}$,
\begin{equation}
|\alpha(q)| \;>\; \frac{3}{2} + \left\{ \begin{array}{ccl}
\displaystyle \left| \frac{w_{ij}(q) a^i q^j}{\Delta(q)} \right|
&\;\;\;& {\mbox{if $\Delta(q) \neq 0$}} \\[1em]
\displaystyle \left(1+\Theta(1-2\sqrt{|l(q)|} \right)\: \sqrt{|l(q)|}
&& {\mbox{if $\Delta(q)=0$}}. \end{array} \right.
\label{2r}
\end{equation}
Then for sufficiently small $\varepsilon$, the Dirac operator $i
\Pdd + \varepsilon {\mathcal{B}}_0$ has an $\varepsilon$-definite
kernel. If conversely there is $q \in {\mathcal{C}}$ for which the
opposite inequality holds (i.e.~(\ref{2r}) with ``$>$'' replaced
by ``$<$''), then the Dirac operator has no $\varepsilon$-definite
kernel.
\end{Thm}
{\Proof} A short calculation using~(\ref{2ba}, \ref{227}, \ref{2o})
gives
\begin{equation}
\nu_\pm \;=\; \alpha \pm \frac{1}{\mu_k} \:\Bra v,k \Ket \:. \label{2p}
\end{equation}
In the special case $k\slsh = \mu_k\: \gamma^0$ and $\vec{q}\slsh = \vec{\gamma}$, we obtain
furthermore from~(\ref{2bb}) that
\[ (\tau_\pm)_r \;=\; a_r \pm w_{r 0} \spc (r=1,2,3). \]
Thus, according to~(\ref{2c}),
\[ (\tau_\pm)^2 \;=\; \sum_{r=1}^3 (a_r)^2 \pm 2\: a_r\: w_{r 0} + (w_{r 0})^2 \:, \]
and this can be written covariantly as
\begin{equation}
(\tau_\pm)^2 \;=\; -\Bra a,a \Ket + \frac{1}{\mu_k^2}\: \Bra a,k \Ket^2
-\frac{1}{\mu_k^2}\: w_{ij} k^i\: w^{il} k_l \mp \frac{2}{\mu_k}\: w_{ij} a^i k^j\:.
\label{2q}
\end{equation}
This tensor equation is valid for any time-like $k$, and it is easy to check
that it holds for spacelike $k$ as well.

Let $q \in {\mathcal{C}}$. We first consider the case
$\Delta(q)\neq0$. By continuity, $\Delta\neq0$ in a neighborhood
$U$ of $q$, and according to~(\ref{2E3}), $\Delta$ is positive in
$U$. We substitute~(\ref{2p}) and~(\ref{2q}) into~(\ref{2a})
and~(\ref{2b}). In order to remove the singularities at $\mu_k=0$,
we write the eigenvalues $\mu^a_s$ in the form
\begin{equation}
\left. \begin{array}{rcl}
\displaystyle \mu^{1\!/\!2}_+ &=&
\displaystyle \sqrt{k^2+2 \varepsilon \delta_{1\!/\!2}} + \varepsilon\:
(\alpha \pm \kappa_+) + {\mathcal{O}}(\varepsilon^2) \\[.5em]
\displaystyle \mu^{1\!/\!2}_- &=& -\sqrt{k^2+2 \varepsilon
\delta_{1\!/\!2}} + \varepsilon\: (\alpha \mp \kappa_-) +
{\mathcal{O}}(\varepsilon^2)
\end{array} \right\}, \label{2A}
\end{equation}
where we set
\[ \delta_{1\!/\!2} \;=\; \Bra v,k \Ket \pm \Delta \:,\spc
\kappa_\pm \;=\; \tau_\pm - \frac{1}{\mu_k}\: \Delta\:. \]
The functions $\kappa_\pm$ have the following expansion,
\begin{equation}
\kappa_\pm \;=\; \frac{1}{\mu_k} \left( \sqrt{\Delta^2 \mp 2 \mu_k
\:w_{ij} a^i k^j} - \Delta \right) \;=\; \mp \frac{w_{ij} a^i
k^j}{\Delta} \:+\: {\mathcal{O}}(\mu_k)\:. \label{2C}
\end{equation}
In particular, one sees that these functions are bounded locally uniformly
in $\mu_k$. Let us verify under which conditions the Dirac operator restricted
to $U$ has an $\varepsilon$-definite kernel. Suppose that $\mu^1_+ \in (-\varepsilon,
\varepsilon)$. Then, due to the square root in~(\ref{2A}),
\[ k^2 + 2 \varepsilon \delta_1 \;=\; {\mathcal{O}}(\varepsilon^2)\:. \]
It follows from~(\ref{2A}) that
\begin{eqnarray*}
\mu_\pm^2 &=& \sqrt{k^2 + 2 \varepsilon \delta_2} +
{\mathcal{O}}(\varepsilon)
\;=\;  \sqrt{{\mathcal{O}}(\varepsilon^2) + 2 \varepsilon (\delta_2-\delta_1)} + {\mathcal{O}}(\varepsilon) \\
&=& \sqrt{-4 \varepsilon\: \Delta} + {\mathcal{O}}(\varepsilon)
\;\sim\; \sqrt{\varepsilon}
\end{eqnarray*}
and therefore
\[ |\mu^1_+ - \mu^2_+| \;\sim\; \sqrt{\varepsilon} \;\gg\; \varepsilon\:. \]
Moreover, we obtain from~(\ref{2A}) and~(\ref{2C}) that
\begin{eqnarray*}
\mu^1_+ - \mu^1_- &=& 2 \mu^1_+ + 2 \varepsilon \alpha - \varepsilon (\kappa_+ - \kappa_-) + {\mathcal{O}}(\varepsilon^2) \\
&=& 2 \mu^1_+ + 2 \varepsilon \left( \alpha + \frac{w_{ij}\: a^i
k^j}{\Delta} + {\mathcal{O}}(\mu_k) \right) +
{\mathcal{O}}(\varepsilon^2)\:.
\end{eqnarray*}
Thus the condition $|\mu^1_+ - \mu^1_-|>\varepsilon$ is satisfied if
\[ \left| \alpha + \frac{w_{ij} a^i k^j}{\Delta} \right| \;>\; \frac{3}{2}\:. \]
As is proved in Lemma~\ref{lemma214} below, the eigenspace corresponding
to $\mu^1_+$ is definite. We conclude that $\mu^1_+$ is an $\varepsilon$-definite
eigenvalue of $A$. Repeating the above argument in the three other cases
$\mu^2_-, \mu^2_\pm \in (-\varepsilon, \varepsilon)$, one obtains that for
sufficiently small $\varepsilon$, the kernel of the Dirac operator is
$\varepsilon$-definite in $U$.
If conversely~(\ref{2r}) holds with ``$>$'' replaced by ``$<$'', it is
straightforward to check that the Dirac operator for small $\varepsilon$
has no $\varepsilon$-definite kernel.

It remains to consider the case $\Delta(q)=0$. We write the eigenvalues
$\mu^a_s$ as
\begin{equation} \left. \begin{array}{lcr}
\mu^{1\!/\!2}_+ &=& \sqrt{k^2 + 2 \varepsilon\: \Bra v,k \Ket}
+ \varepsilon\:(\alpha \pm \tau_+) \\[.5em]
\mu^{1\!/\!2}_- &=& -\sqrt{k^2 + 2 \varepsilon\: \Bra v,k \Ket}
+ \varepsilon\:(\alpha \mp \tau_-)\:. \end{array} \right\}
\label{62a}
\end{equation}
According to~(\ref{eq:2F}), the first three summands in~(\ref{2q}) have a finite
limit at $q$. Furthermore, (\ref{2G2}) yields that
\[ \frac{2}{\mu_k}\: w_{ij} a^i k^j \;=\; {\mathcal{O}}(\mu_k)\:. \]
We conclude that the functions $\tau_\pm$ in a neighborhood of $q$ have the expansion
\begin{equation}
\tau_\pm \;=\; \sqrt{|l|} + {\mathcal{O}}(\sqrt{|\mu_k|})\:.
\label{62b}
\end{equation}
For small $\varepsilon$, $\mu_k\sim\sqrt{\varepsilon}$, and so the
term ${\mathcal{O}}(\sqrt{\mu_k})$ is of higher order in
$\varepsilon$ and can be omitted. Furthermore, the following
continuity argument varying $l$ shows that the eigenvalues
$\mu^a_+$ and $\mu^a_-$ correspond to positive and negative
eigenvectors, respectively: If $l=0$, only the scalar and vector
potentials enter the perturbation calculation to first order in
$\varepsilon$ (see~(\ref{62a}, \ref{62b})). If only scalar
and vector potentials are present, the spectral decomposition of
the matrix $k\slsh+\varepsilon {\mathcal{B}}_0$ is easily obtained
from the identity
\[ \left[ (k\slsh + \varepsilon v\slsh + \varepsilon \alpha) - \varepsilon \alpha
\right]^2 \;=\; (k+\varepsilon v)^2\:\1\:. \]
One sees that the eigenvalues are twofold degenerate,
$\sigma^\varepsilon(k) = \{\mu_+, \mu_-\}$, and that if they
are real, the corresponding eigenspaces are definite. The parameter $l$ removes
the degeneracy of these eigenspaces, but the resulting invariant subspaces remain
definite.

Suppose that $\mu^1_+ \in (-\varepsilon, \varepsilon)$.
We consider the two subcases $2 \sqrt{|l|}>\varepsilon$ and $2 \sqrt{|l|}<\varepsilon$
separately. In the first case, $|\mu^1_s-\mu^2_s|>\varepsilon$, and thus we must arrange that
\begin{equation}
|\mu^1_+-\mu^2_\mp|>\varepsilon \spc (2 \sqrt{l}>\varepsilon). \label{2C1}
\end{equation}
In the second case, $|\mu^1_s-\mu^2_s|<\varepsilon$. Thus we must combine the eigenvalues to
pairs and consider the definite eigenspaces corresponding to the sets
$\sigma_s=\{\mu^1_s, \mu^2_s\}$, $s=\pm$, and must satisfy the condition
\begin{equation}
{\mbox{dist}}\; (\sigma_+, \sigma_-) \;>\; \varepsilon
\spc (2 \sqrt{l}<\varepsilon). \label{2C2}
\end{equation}
Evaluating~(\ref{2C1}) and~(\ref{2C2}) using~(\ref{62a}, \ref{62b}) and analyzing
similarly the three other cases $\mu^1_-, \mu^2_\pm \in (-\varepsilon, \varepsilon)$
gives the condition~(\ref{2r}).
\QED

\begin{Lemma} \label{lemma214}
Let $A$ be a Hermitian matrix (with respect to $\Sl .|. \Sr$). If $\mu \in
\sigma(A)$ is real and the corresponding invariant eigenspace $I$ is
one-dimensional, then $I$ is a definite eigenspace.
\end{Lemma}
{\Proof} Since each invariant subspace contains at least one eigenvector, $I$
is clearly an eigenspace. We must show that $I$ is definite.
Assume to the contrary that $I=\bra \Psi \ket$ is null, i.e.
\[ A \Psi = \lambda \Psi \spc{\mbox{with}}\spc
\lambda \in \R {\mbox{ and }} \Sl \Psi | \Psi \Sr =0. \]
We denote the invariant subspaces of $A$ by $(I_\mu)_{\mu \in \sigma(A)}$. Since
$I_\lambda=\bra \Psi \ket$ is one-dimensional and null, there must be an invariant
subspace $I_\mu$, $\mu \neq \lambda$, which is not orthogonal to $\Psi$,
\[ I_\mu \cap \bra \Psi \ket^\perp \;\neq\; \emptyset\:. \]
We choose on $I_\mu$ a basis $(e_1,\ldots,e_n)$ such that $A$ is in the Jordan form, i.e.
\[ A|_{I_\mu} \;=\; \left( \begin{array}{cccc} \mu & 1 & \cdots & 0 \\
0 & \mu & \cdots & 0 \\ \vdots & \vdots & \ddots & 1 \\ 0 & 0 & \cdots & \mu \end{array} \right) \:. \]
Let $k \in \{1,\ldots,n\}$ be the smallest index such that $\Sl e_k | \Psi \Sr \neq 0$. Then
\begin{eqnarray*}
\lambda\: \Sl e_k | \Psi \Sr &=& \Sl e_k \:|\: A \Psi \Sr \;=\; \Sl A e_k \:|\: \Psi \Sr \\
&=& \mu\: \Sl e_k \:|\: \Psi \Sr \:+\: \Sl e_{k-1} \:|\: \Psi \Sr \;=\;
\mu\: \Sl e_k \:|\: \Psi \Sr\:.
\end{eqnarray*}
This is a contradiction. \QED

Suppose that the homogeneous operator ${\mathcal{B}}_0$ satisfies
the condition~(\ref{2r}) in Theorem~\ref{thm211}. Then the Dirac
operator has an $\varepsilon$-definite kernel. As a consequence,
the distributions $t^\varepsilon_\mu=\frac{1}{2}(p^\varepsilon_\mu
- k^\varepsilon_\mu)$ are well-defined (see Def.~\ref{def24} and
Theorem~\ref{thm210}). Following~(\ref{eq:C20}), we introduce the
fermionic projector by
\begin{equation}\label{70a}
P^\varepsilon_\mu \;=\; \frac{1}{2} \left( X\: t^\varepsilon_\mu + t^\varepsilon_\mu\: X^* \right)
\end{equation}
with $X=\chi_L$. In order to analyze the normalization of $P^\varepsilon_\mu$,
we consider the product
\begin{equation}
P^\varepsilon_\mu \:P^\varepsilon_{\mu'} \;=\; \frac{1}{4} \left( X\: t^\varepsilon_\mu\: t^\varepsilon_{\mu'} \:X^* + X\: t^\varepsilon_\mu\: X\: t^\varepsilon_{\mu'}
+t^\varepsilon_\mu\: X^*\: t^\varepsilon_{\mu'} \:X^* \right). \label{2M}
\end{equation}
According to~(\ref{2ka}) and~(\ref{2kb}),
\begin{equation}
t^\varepsilon_\mu \:t^\varepsilon_{\mu'} \;=\; \delta^2(\mu-\mu')\: t^\varepsilon_\mu\:. \label{2K}
\end{equation}
Thus the only problem is to compute the products $t^\varepsilon_\mu X t^\varepsilon_{\mu'}$ and
$t^\varepsilon_\mu X^* t^\varepsilon_{\mu'}$. Using the relations $\chi_{L\:/\:R}=\frac{1}{2}(\1 \mp \rho)$ together
with~(\ref{2K}), this problem reduces to making mathematical sense of the
operator product
\[ t^\varepsilon_\mu \:\rho\: t^\varepsilon_{\mu'}\:. \]
It seems impossible to give this expression a meaning without
making additional assumptions on ${\mathcal{B}}_0$. For
simplicity, we shall impose a quite strong condition, which is
motivated as follows. The spectral projectors $p_\mu$
corresponding to the unperturbed Dirac operator $i \Pdd-\mu$
satisfy the relations $\rho \:p_\mu\: \rho = p_{-\mu}$ and thus
$p_\mu \:\rho\: p_\mu=0$ ($\mu>0$). It is natural to demand that
the last identity should also hold in the presence of the
homogeneous perturbation for small $\varepsilon$.
\begin{Def} \label{def212}
The kernel of the homogeneous Dirac operator $i \Pdd +
{\mathcal{B}}(\varepsilon, k)$ is {\bf{$\varepsilon$-orthogonal to
$\rho$}}\index{$\varepsilon$-orthogonal to $\rho$} if for all $\mu, \mu' \in \sigma^\varepsilon(k) \cap
(-\frac{\varepsilon}{2}, \frac{\varepsilon}{2})$, the
corresponding spectral projectors $E_\mu(k)$ and $E_{\mu'}(k)$
satisfy the condition
\begin{equation}
E_\mu \:\rho\: E_{\mu'} \;=\; 0 \:. \label{2R}
\end{equation}
\end{Def}
If the kernel of the Dirac operator is $\varepsilon$-definite and $\varepsilon$-orthogonal to
$\rho$, it follows immediately that for all $\mu, \mu' \in (-\frac{\varepsilon}{2},
\frac{\varepsilon}{2})$,
\begin{equation}
t^\varepsilon_\mu \:\rho\: t^\varepsilon_{\mu'} \;=\; 0\:. \label{2L2}
\end{equation}
Using~(\ref{2K}) and~(\ref{2L2}) in~(\ref{2M}), we see that
\[ P^\varepsilon_\mu \: P^\varepsilon_{\mu'} \;=\; \delta^2(\mu-\mu')\; \frac{1}{8} \left( X\:P^\varepsilon_\mu +
P^\varepsilon_\mu\:X^* + 2\: X\:P^\varepsilon_\mu\: X^* \right)\:. \]
Now we can take the limits $\varepsilon, \mu \searrow 0$ to obtain
\begin{equation}
\lim_{\varepsilon \searrow 0} \;\lim_{\mu, \mu' \searrow 0}
\left( P^\varepsilon_\mu \:P^\varepsilon_{\mu'} -\frac{1}{2}\:\delta^2(\mu-\mu')\:
P^\varepsilon_\mu \right) \;=\; 0\:. \label{2Q}
\end{equation}
In analogy to~(\ref{eq:C14}), this relation states that the fermionic projector
is idempotent (apart from the factor $\frac{1}{2}$ which will be treated
in Section~\ref{sec23} using the modified mass scaling).

In the remainder of this section, we analyze under which
assumptions on ${\mathcal{B}}_0$ the kernel of the Dirac operator
is $\varepsilon$-orthogonal to $\rho$. We begin with a simple
calculation in first order perturbation theory.

\begin{Lemma} \label{lemma213} Suppose that the Dirac operator
$i\Pdd + \varepsilon {\mathcal{B}}_0$ has an
$\varepsilon$-definite kernel and that the homogeneous potentials
in~(\ref{2o}) satisfy for all $k \in \R^4$ the relations
\begin{equation}
\beta(k)\;=\;0 \spc {\mbox{and}} \spc \epsilon_{ijlm}\: w^{ij}(k)\: k^l \;=\; 0\:.
\label{2O}
\end{equation}
Then for all $k$ and $\mu, \mu' \in \sigma^\varepsilon(k) \cap (-\frac{\varepsilon}{2},
\frac{\varepsilon}{2})$,
\begin{equation}
E_\mu(k) \:\rho\: E_\mu(k) \;=\; {\mathcal{O}}(\varepsilon^2)\:.
\label{2P}
\end{equation}
\end{Lemma}
{\Proof} Choose $k$ and $\mu, \mu' \in \sigma^\varepsilon(k) \cap (-\frac{\varepsilon}{2},
\frac{\varepsilon}{2})$. Since the Dirac operator has an $\varepsilon$-definite kernel,
the invariant subspace $I$ corresponding to the set $\{\mu, \mu'\} \subset
\sigma^\varepsilon(k)$ is definite (notice that $\mu \in (-\varepsilon, \varepsilon)$
and $|\mu-\mu'|<\varepsilon$). We saw in the proof of Theorem~\ref{thm211} that the
invariant subspaces ${\mbox{Im}}\: E^\varepsilon_+$ and ${\mbox{Im}}\: E^\varepsilon_+$
(with $E^\varepsilon_\pm$ according to~(\ref{2L})) are definite. Thus
$I \subset {\mbox{Im}}\: E^\varepsilon_+$ or $I \subset {\mbox{Im}}\: E^\varepsilon_-$.
Therefore, it suffices to show that for all $s=\pm$,
\begin{equation}
E^\varepsilon_s \:\rho\: E^\varepsilon_s \;=\;
{\mathcal{O}}(\varepsilon^2)\:. \label{2P2}
\end{equation}
Substituting~(\ref{2L}) and using the relations $\rho E_\pm \rho = E_\mp$, we obtain the
equivalent condition
\begin{equation}
E_s\: \{{\mathcal{B}}_0, \rho \}\: E_s \;=\; 0\:. \label{2N}
\end{equation}
This equation means that the matrix $\{{\mathcal{B}}_0, \rho \}$
must vanish on the two-dimensional subspace ${\mbox{Im}}\:E_s$.
Since on this subspace, the matrices $\vec{\Sigma}$, (\ref{2sig}),
have a representation as the Pauli matrices, we can
restate~(\ref{2N}) as the four conditions
\[ \Tr \left( E_s\: \{{\mathcal{B}}_0, \rho \} \right) \;=\; 0 \;=\;
\Tr \left( \vec{\Sigma}\: E_s\: \{{\mathcal{B}}_0, \rho \}
\right)\:. \]
Evaluating these relations using~(\ref{227}, \ref{2sig}, \ref{2o}) gives~(\ref{2O}). \QED This lemma is
not satisfactory because it gives no information on how the error
term in~(\ref{2P}) depends on $k$. More specifically, the error
term may have poles on the mass cone (and explicit calculations
show that such poles $\sim k^{-2n}$ indeed occur for $n=1$ and
$n=2$). Since in the limit $\varepsilon \searrow 0$ the kernel of
the Dirac operator is the mass cone, it is far from obvious how to
control the error term in this limit. In other words, (\ref{2P})
cannot be interpreted as ``the kernel of the Dirac operator is
$\varepsilon$-orthogonal to $\rho$ up to a small error term.''

In order to resolve this difficulty, we must proceed
non-perturbatively. In generalization of our previous ansatz $i
\Pdd + \varepsilon {\mathcal{B}}_0$, we shall consider the Dirac
operator $i \Pdd + {\mathcal{B}}^\varepsilon$, where we assume
that ${\mathcal{B}}^\varepsilon(k)$ is a homogeneous potential
which is smooth in both arguments and has the power expansion
\begin{equation}
{\mathcal{B}}^\varepsilon(k) \;=\; \varepsilon\:
{\mathcal{B}}_0(k) + \varepsilon^2\: {\mathcal{B}}_1(k) +
 \varepsilon^3\: {\mathcal{B}}_2(k) + \cdots\:. \label{2R2}
\end{equation}
The higher order potentials ${\mathcal{B}}_1,
{\mathcal{B}}_2,\ldots$ are irrelevant for Def.~\ref{def29}
because they are negligible for small $\varepsilon$. In
particular, the statement of Theorem~\ref{thm211} remains valid
without changes. Furthermore, the potentials ${\mathcal{B}}_1,
{\mathcal{B}}_2, \ldots$ should be irrelevant for the statement of
idempotence~(\ref{2Q}) because~(\ref{2Q}) involves a limit
$\varepsilon \searrow 0$. Therefore, it seems unnecessary to enter
a detailed study of these potentials. The only point of interest
is under which assumptions on ${\mathcal{B}}_0$ there exist smooth
potentials ${\mathcal{B}}_1$, ${\mathcal{B}}_2$,\ldots such that
the spectral projectors corresponding to the Dirac operator $i
\Pdd + {\mathcal{B}}^\varepsilon$ satisfy the
conditions~(\ref{2R}) exactly.

\begin{Thm} \label{thm214}
Suppose that the Dirac operator $i \Pdd + \varepsilon
{\mathcal{B}}_0$ has an $\varepsilon$-definite kernel and that the
homogeneous potentials in~(\ref{2o}) satisfy for all $k$ the
relations~(\ref{2O}). Then there is an $\varepsilon>0$ and a smooth
potential ${\mathcal{B}}^\varepsilon(k)$ having the
expansion~(\ref{2R2}) such that the kernel of the Dirac operator
$i\Pdd + {\mathcal{B}}^\varepsilon$ is $\varepsilon$-orthogonal to
$\rho$.
\end{Thm}
{\Proof} Choose $k$ and $\mu, \mu' \in \sigma^\varepsilon(k) \cap
(-\frac{\varepsilon}{2}, \frac{\varepsilon}{2})$. Similar to what was
described before~(\ref{2P2}), we know from the proof of
Theorem~\ref{thm211} that the matrix $A\equiv
k\slsh+{\mathcal{B}}^\varepsilon(k)$ has a positive and a negative
definite invariant subspace, one of which contains ${\mbox{Im}}\:
E_\mu \cup {\mbox{Im}}\: E_{\mu'}$. Again denoting the projectors
onto these subspaces by $E^\varepsilon_+$ and $E^\varepsilon_-$,
respectively, it thus suffices to show that for $s=\pm$,
\begin{equation}
E^\varepsilon_s \:\rho\: E^\varepsilon_s \;=\; 0\:. \label{2S}
\end{equation}

We first evaluate these conditions in a special spinor basis. Namely, we let
$e_1$ and $e_2$ be an orthonormal basis of ${\mbox{Im}}\: E^\varepsilon_+$ and
set $e_3=\rho e_1$, $e_4=\rho e_2$. The conditions~(\ref{2S}) imply that $e_3$ and
$e_4$ span ${\mbox{Im}}\: E^\varepsilon_-$. Using the relation $\rho^2=\1$ as well
as that the subspaces $\bra \{e_1, e_2\} \ket$ and $\bra \{e_3, e_4\} \ket$ are
invariant under $A$, we conclude that the matrices $\rho$ and $A$ are of the form
\begin{equation}
\rho \;=\; \left( \begin{array}{cc} 0 & \1 \\ \1 & 0 \end{array} \right) \:,\spc
A \;=\; \left( \begin{array}{cc} * & 0 \\ 0 & * \end{array} \right) \:,
\label{2T}
\end{equation}
where we used a block matrix notation corresponding to the splitting
\[ \C^4 \;=\; \bra \{e_1, e_2\} \ket \oplus \bra \{e_3, e_4\} \ket\:, \]
and ``$*$'' denotes an arbitrary block matrix entry. Furthermore, the
relation $\rho^*=-\1$ yields that
\[ \Sl e_3 \:|\: e_3 \Sr \;=\; -1 \;=\; \Sl e_4 \:|\: e_4 \Sr \:, \]
and thus the basis $(e_\alpha)$ is pseudo-orthonormal,
\begin{equation}
\Sl \Psi \:|\: \Phi \Sr \;=\; \sum_{\alpha=1}^4 s_\alpha \: \overline{\Psi^\alpha}
\: \Phi^\alpha \spc {\mbox{with}} \spc s_1=s_2=1,\; s_3=s_4=-1. \label{2U}
\end{equation}
We see that the matrix $\rho$ and the spin scalar product are in the usual Dirac
representation. In this representation, the fact that $A$ is block
diagonal~(\ref{2T}) can be expressed by saying that $A$ must be a real linear
combination of the 8 matrices
\begin{equation}
\1,\;\;\; \gamma^0,\;\;\; \rho \vec{\gamma},\;\;\; \rho \gamma^0 \vec{\gamma}\:. \label{2V}
\end{equation}

We next express this result in a general basis, but again in the Dirac
representation. Since the representations of the matrix $\rho$, (\ref{2T}), and
of the scalar product, (\ref{2U}), are fixed, the freedom in choosing the
basis is described by even $U(2,2)$ transformations. This group,
which we denote by $U(2,2)^{\mbox{\scriptsize{even}}}$, contains the normal
Abelian subgroup $U=\{ \exp(\vartheta \rho/2):\: \vartheta \in \R\}$. Acting by $U$
on~(\ref{2V}) gives the matrices
\begin{equation}
\1,\;\;\; ((\cosh \vartheta + \rho\: \sinh \vartheta) \:\gamma^0,\;\;\; ((\cosh \vartheta + \rho\: \sinh \vartheta)\:
\rho \vec{\gamma},\;\;\; \rho \gamma^0 \vec{\gamma}\:. \label{2V2}
\end{equation}
When the factor group $U(2,2)^{\mbox{\scriptsize{even}}} / U$ acts
on~(\ref{2V2}), the resulting transformations correspond precisely to Lorentz
transformations of the tensor indices (see Lemma~\ref{Dirtrans} for details).
Thus the conditions~(\ref{2S}) are satisfied if and only if $A$ is
of the form
\begin{equation}
A \;=\; \alpha\:\1 + ((\cosh \vartheta + \rho\: \sinh \vartheta) \:u\slsh + ((\rho\: \cosh \vartheta + \sinh \vartheta)
\: a\slsh + \rho u\slsh b\slsh \label{2Va}
\end{equation}
with a time-like vector field $u$ and two vector fields $a$ and $b$,
which are orthogonal to $u$,
\begin{equation}
\Bra u, a \Ket \;=\; 0 \;=\; \Bra u,b \Ket \:. \label{2Vb}
\end{equation}
We substitute the identity $A=k\slsh +
{\mathcal{B}}^\varepsilon(k)$ into~(\ref{2Va}) and solve for
${\mathcal{B}}^\varepsilon(k)$. Expanding in powers of
$\varepsilon$ gives the result. \QED

\section{The General Construction, Proof of Idempotence}
\setcounter{equation}{0} \label{sec23}
In this section we shall make precise what ``idempotence'' means for a fermionic projector
with chiral asymmetry in the presence of a general interaction. We proceed in several
steps. We begin with a straightforward extension of the results of Section~\ref{sec22}
to systems of Dirac seas. Then we introduce the interaction and perform the causal
perturbation expansion. After putting in an infrared regularization, we can define
the fermionic projector. Finally, idempotence is established via a singular mass limit.

We begin with a system of Dirac seas in the vacuum, described by
the mass matrix $Y$ and the chiral asymmetry matrix
$X$ (see~{\S}\ref{jsec3}). In order to give the
chiral fermions a ``small generalized mass,'' we introduce a
homogeneous operator ${\mathcal{B}}_0$ and consider for
$\varepsilon>0$ the Dirac operator $i \Pdd + \varepsilon
{\mathcal{B}}_0 - mY$. For simplicity, we assume that
${\mathcal{B}}_0$ is diagonal on the sectors and is non-trivial
only in the chiral blocks, i.e.
\[ ({\mathcal{B}}_0)^{(a \alpha)}_{(b \beta)} \;=\; \delta^a_b\: \delta^\alpha_\beta\:
{\mathcal{B}}_0^{(a \alpha)} \spc {\mbox{with}} \spc
{\mathcal{B}}_0^{(a \alpha)} = 0 {\mbox{ if }} X_a=\1. \] Then on
each sector the methods of Section~\ref{sec22} apply; let us
collect the assumptions on ${\mathcal{B}}_0$ and the main results:
For every index $(a \alpha)$ with $X^{(a \alpha)} \neq \1$ we
assume that
\begin{enumerate}
\item[(1)] ${\mathcal{B}}_0^{(a \alpha)}(k)$ depends smoothly on
$k \in \R^4$ and grows at most polynomially at infinity.
\item[(2)] The $(4 \times 4)$-matrix $k\slsh + \varepsilon
{\mathcal{B}}_0^{(a \alpha)}(k)$ is diagonalizable for a.a.\ $k$.
\item[(3)] ${\mathcal{B}}_0^{(a \alpha)}$ has the decomposition
into scalar, vector, axial, and bilinear potentials,
\[ {\mathcal{B}}_0^{(a \alpha)}(k) \;=\; \alpha\:\1 + v\slsh + \rho\: a \slsh +
\frac{i \rho}{2}\: w_{ij}\: \sigma^{ij} \:, \] such that for all
$k \in \R^4$ and $q \in {\mathcal{C}}$ the following conditions
are satisfied,
\begin{eqnarray*}
\epsilon_{ijlm}\: w^{ij}(k)\: k^l &=& 0 \\
|\alpha(q)| &>& \frac{3}{2} + \left\{ \begin{array}{ccl}
\displaystyle \left| \frac{w_{ij}(q) a^i q^j}{\Delta(q)} \right|
&\;\;\;& {\mbox{if $\Delta(q) \neq 0$}} \\[1em]
\displaystyle \left(1+\Theta(1-2\sqrt{|l(q)|} \right)\: \sqrt{|l(q)|}
&& {\mbox{if $\Delta(q)=0$}}. \end{array} \right.
\end{eqnarray*}
(with $\Delta$ and $l$ defined by~(\ref{2E2}) and~(\ref{eq:2F}),
${\mathcal{C}}=\{k \:|\: k^2=0\}$ is the mass cone).
\end{enumerate}
Then for sufficiently small $\varepsilon$, the Dirac operator $i
\Pdd + \varepsilon {\mathcal{B}}_0^{(a \alpha)}$ has an
$\varepsilon$-definite kernel (see Def.~\ref{def29} and
Theorem~\ref{thm211}). Thus for a.a.\ $\mu \in (-\varepsilon,
\varepsilon)$, the spectral projectors $p^{\varepsilon, (a
\alpha)}_\mu$, $k^{\varepsilon, (a \alpha)}_\mu$ and the Green's
functions $s^{\varepsilon, (a \alpha)}_\mu$ are well-defined
distributions in momentum space (see Def.~\ref{def24} and
Theorem~\ref{thm210}). Furthermore, the kernel of the Dirac
operator is $\varepsilon$-orthogonal to $\rho$ (see
Def.~\ref{def212} and Theorem~\ref{thm214}; for simplicity we here
omit the higher order potentials ${\mathcal{B}}_1,
{\mathcal{B}}_2,\ldots$ in~(\ref{2R2}), this is justified because
these potentials obviously drop out in the singular mass limit),
and this can be stated in the form (cf.~(\ref{2i2}))
\[ p^{\varepsilon, (a \alpha)}_\mu \:\rho\: p^{\varepsilon, (a \alpha)}_{\mu'}
\;=\; 0 \spc {\mbox{for all $\mu, \mu' \in (-\frac{\varepsilon}{2},
\frac{\varepsilon}{2})$}}\:. \]
We build up the spectral projectors $p^\varepsilon_{+\mu}$, $k^\varepsilon_{+\mu}$
and the Green's function $s^\varepsilon_{+\mu}$ of the whole system by taking
direct sums; namely,
\begin{equation}
A^\varepsilon_{+\mu} \;=\; \bigoplus_{a, \alpha} \left\{
\begin{array}{cl} A_{m_{a \alpha} + \mu} & {\mbox{if $X_a = \1$}} \\
A^{\varepsilon, (a \alpha)}_{\frac{\mu}{2}} & {\mbox{if $X_a \neq \1$}},
\end{array} \right. \label{2ds}
\end{equation}
where $A$ stands for $p$, $k$, or $s$. Note that in the chiral blocks the
mass parameter $\frac{\mu}{2}$ (and not $\mu$) is used. The purpose of this
{\em{modified mass scaling}}\index{modified mass scaling} is to get rid of the factor $\frac{1}{2}$ in the
normalization of a chiral Dirac sea~(\ref{2Q}) (also see the paragraph
after~(\ref{eq:Cmms})). The corresponding Dirac operator is
\[ i \Pdd + \varepsilon {\mathcal{B}}_0 - mY - \mu Z \:, \]
where the matrix $Z \equiv \frac{1}{2} (X+X^*)$ takes into account the modified
mass scaling. The spectral projectors satisfy the multiplication rules
\begin{equation}
\left. \begin{array}{rcl}
p^\varepsilon_{+\mu}\: p^\varepsilon_{+\mu'} &=&
k^\varepsilon_{+\mu}\: k^\varepsilon_{+\mu'} \;=\;
\delta^2(\mu-\mu')\: Z^{-1}\: p^\varepsilon_{+\mu} \\[.5em]
p^\varepsilon_{+\mu}\: k^\varepsilon_{+\mu'} &=&
k^\varepsilon_{+\mu}\: p^\varepsilon_{+\mu'} \;=\;
\delta^2(\mu-\mu')\: Z^{-1}\: k^\varepsilon_{+\mu}
\end{array} \right\} \label{lf}
\end{equation}
\begin{equation}
C^\varepsilon_{+\mu}\:\rho\:C^\varepsilon_{+\mu'} \;=\; 0
\spc {\mbox{for $\mu, \mu' \in (-\frac{\varepsilon}{2},
\frac{\varepsilon}{2})$}}\:,
\label{lg}
\end{equation}
where $C$ stands for $k$ or $p$. The Green's functions satisfy
the relations
\begin{equation}
\left. \begin{array}{rcl}
C^\varepsilon_{+\mu}\: s^\varepsilon_{+\mu'} &=&
s^\varepsilon_{+\mu}\: C^\varepsilon_{+\mu'} \;=\;
\displaystyle \frac{\mbox{PP}}{\mu-\mu'}\:
Z^{-1}\: C^\varepsilon_{+\mu} \\[.5em]
s^\varepsilon_{+\mu}\: s^\varepsilon_{+\mu'} &=&
\displaystyle \frac{\mbox{PP}}{\mu-\mu'}\:
Z^{-1}\: (s^\varepsilon_{+\mu} - s^\varepsilon_{+\mu'})\:.
\end{array} \right\} \label{lh}
\end{equation}
These multiplication rules differ from those in~{\S}\ref{jsec2} only
by the additional factor $Z^{-1}$.

To describe the interaction, we insert a potential ${\mathcal{B}}$
into the Dirac operator, which then reads
\begin{equation}
i \Pdd + {\mathcal{B}} + \varepsilon {\mathcal{B}}_0 - mY - \mu Z
\:. \label{2do}
\end{equation}
We assume that $Y$ and ${\mathcal{B}}$ have the following
properties:
\begin{enumerate}
\item[(a)] Only the chiral particles are massless, i.e.
\[ Y^{(a \alpha)}>0 \spc {\mbox{if $X_a = \1$}}. \]
\item[(b)] ${\mathcal{B}}$ is the operator of multiplication with
the Schwartz function ${\mathcal{B}}(x)$. \item[(c)] $Y$ and
${\mathcal{B}}$ are causality compatible, i.e.
\begin{equation}
X^*\: (i \Pdd + {\mathcal{B}} - mY) \;=\; (i \Pdd + {\mathcal{B}}
- mY)\:X\:. \label{2cc}
\end{equation}
\end{enumerate}
In order to introduce the spectral projectors with interaction
$\tilde{p}^\varepsilon_{+\mu}$ and $\tilde{k}^\varepsilon_{+\mu}$,
we take the operator expansion of causal perturbation
theory~{\S}\ref{jsec3} and replace the operators according to
$A \to A^\varepsilon_{+\mu}$ (with $A=p$, $k$, or $s$). All the
operator products of the resulting expansion are well-defined for
a.a.\ $\mu$ (note that $\tilde{{\mathcal{B}}}(k)$ has rapid decay
and $A^\varepsilon_{+\mu}(k)$ grows at most polynomially at
infinity).

For the infrared regularization, we proceed exactly as in~{\S}\ref{jsec6}
and replace space by the three-dimensional
torus~(\ref{7a}). Furthermore, we ``average'' the mass parameter $\mu$.
More precisely, combining~(\ref{eq:2b}) with~(\ref{70a}), the auxiliary
fermionic projector is defined by
\begin{equation}
P^{\varepsilon, \delta} \;=\; \frac{1}{2} \int_{(0,\delta) \times (-\delta, \delta)}
(X \: \tilde{t}^\varepsilon_{+\mu} + \tilde{t}^\varepsilon_{+\mu}\:X^*)\:d^2 \mu\:,
\label{2fp0}
\end{equation}
where as usual $\tilde{t}^\varepsilon_{+\mu}=\frac{1}{2}(
\tilde{p}^\varepsilon_{+\mu} - \tilde{k}^\varepsilon_{+\mu})$.
Finally, the {\em{regularized fermionic projector}} is obtained by taking
the partial trace,
\begin{equation}
(P^{\varepsilon, \delta})^a_b \;=\; \sum_{\alpha=1}^{g(a)}
\sum_{\beta=1}^{g(b)} (P^{\varepsilon, \delta})^{(a \alpha)}_{(b \beta)}\:.
\label{2fp1}
\end{equation}

Before we can prove idempotence, we need to impose the following extension
of the non-degeneracy assumption~(\ref{eq:2c}). We set
\[ \sigma^\varepsilon_{(a \alpha)} \;=\; \sigma(k\slsh + \varepsilon
{\mathcal{B}}_0^{(a \alpha)})\:. \]
\begin{Def} \label{def215}
The Dirac operator $i \Pdd + \varepsilon {\mathcal{B}}_0 - mY$ has
{\bf{$\varepsilon$-non-degenerate
    masses}}\index{$\varepsilon$-non-degenerate masses} if for all $a$ and
$\beta \neq \gamma$,
\begin{equation}\label{cndm}
\sigma^\varepsilon_{(b \beta)} \cap (-\frac{\varepsilon}{2},\frac{\varepsilon}{2})
\neq \emptyset \quad \Longrightarrow \quad \sigma^\varepsilon_{(b \gamma)} \cap
(-\varepsilon, \varepsilon) = \emptyset\:.
\end{equation}
\end{Def}

Roughly speaking, the next theorem states that the masses are
$\varepsilon$-non-degenerate if they are non-degenerate in the
massive sectors and if the homogeneous potentials in the chiral sectors
are sufficiently different from each other.
\begin{Thm} \label{thmnondeg}
Suppose that $Y$ and ${\mathcal{B}}_0$ have the following
properties:
\begin{enumerate}
\item[(i)] In the massive blocks (i.e.\ $X_a=\1$), the masses are non-degenerate,
\[ Y^{(b \beta)} \neq Y^{(b \gamma)} \spc {\mbox{if $\beta \neq \gamma$}}. \]
\item[(ii)] In the chiral blocks (i.e.\ $X_a \neq \1$), for all
$\beta \neq \gamma$ and all $q \in {\mathcal{C}}$ either
\begin{equation}
\Bra v^{(b \beta)}, q \Ket + s\: \Delta^{(b \beta)}(q) \;\neq\;
\Bra v^{(b \gamma)}, q \Ket + s'\: \Delta^{(b \gamma)}(q) \qquad
{\mbox{for all $s, s' \in \{\pm 1 \}$}} \label{cba}
\end{equation}
or else
\begin{equation}
|\alpha^{(b \beta)}(q) - \alpha^{(b \gamma)}(q)| \;>\;
2 \:+\: 2 \:|d^{(b \beta)}(q) + d^{(b \gamma)}(q)| \label{cbb}
\end{equation}
with
\[ d(q) \;=\;
\left\{ \begin{array}{ccl}
\displaystyle \left| \frac{w_{ij}(q) a^i q^j}{\Delta(q)} \right|
&\;\;\;& {\mbox{if $\Delta(q) \neq 0$}} \\[1em]
\displaystyle \sqrt{|l(q)|}
&& {\mbox{if $\Delta(q)=0$}} \end{array} \right. \]
(and $\Delta, l$ according to~(\ref{2E2}) and~(\ref{eq:2F})).
\end{enumerate}
Then for sufficiently small $\varepsilon$, the Dirac operator has
$\varepsilon$-non-degenerate masses.
\end{Thm}
{\Proof}
The condition in~{(i)} follows immediately from the fact that the
eigenvalues $\mu^a_s$ in the two sectors differ precisely by
$m (Y^{(b \beta)} - Y^{(b \gamma)})$. For part~{(ii)} we consider
the formulas for the eigenvalues~(\ref{2A}) and~(\ref{62a}). If~(\ref{cba})
holds, the eigenvalues in the two sectors all differ by contributions of
the order $\sqrt{\varepsilon}$, and so~(\ref{cndm}) is satisfied for
small~$\varepsilon$. If on the other hand~(\ref{cba}) is violated, there are
eigenvalues in two different sectors such that the square roots
in~(\ref{2A}) and/or~(\ref{62a}) coincide. Thus these eigenvalues differ by
$(\alpha + \sigma)^{(b \beta)} - (\alpha + \sigma)^{(b \gamma)}$, where
each $\sigma$ is an element of the set
$\{ \pm \kappa_+, \pm \kappa_-, \pm \tau_+, \pm \tau_-\}$. The
condition~(\ref{cbb}) guarantees that this difference is greater than
$2 \varepsilon$, and so~(\ref{cndm}) is again satisfied.
\QED

We can now state the main result of this chapter.
\begin{Thm} {\bf{(Idempotence)}} \label{thm216}
\index{fermionic projector!idempotence of the}
Consider the Dirac operator~(\ref{2do}) under the above assumptions
{(1)}--{(3)} and {(a)}--{(c)}.
Assume furthermore that the masses are $\varepsilon$-non-de\-ge\-ne\-rate
(see Def.~\ref{def215} and Theorem~\ref{thmnondeg}).
Then the corresponding fermionic projector (\ref{2fp0}, \ref{2fp1})
satisfies the identity
\begin{equation}
\lim_{\varepsilon \searrow 0} \:\lim_{\delta \searrow 0}\:
\delta \left( \int_{\sR \times T^3} \sum_{b=1}^N
P^a_b(x,z)\: P^b_c(z,y) \: d^4z \:-\: P^a_c(x,y) \right) \;=\; 0 \label{2idem}
\end{equation}
with convergence as a distribution to every order in perturbation theory.
\end{Thm}
{\Proof} Similar to~(\ref{2k}), the Green's function $s^\varepsilon_{+\mu}$
has a spectral representation in a mass parameter $\nu$. We want to
decompose $s^\varepsilon_{+\mu}$ into contributions $\dot{s}^\varepsilon_{+\mu}$
and $\breve{s}^\varepsilon_{+\mu}$ where $|\nu-\mu|$ is small and large,
respectively. To this end, we introduce in each sector the operator
\[ \left\{ \begin{array}{rcll}
\dot{s}^{\varepsilon, (a \alpha)}_\mu \!&=&\! \displaystyle
\int_{B_{\varepsilon/4}(\mu)} \frac{\mbox{PP}}{\nu - \mu}\:
p^\varepsilon_\mu\: d^2\nu & \qquad {\mbox{if $X_1 \neq \1$}} \\[1em]
\displaystyle \dot{s}_{m_{a\alpha} + \mu} \!&=&\! \displaystyle
\int_{B_{\varepsilon/2}(\mu)} \frac{\mbox{PP}}{\nu - \mu}\:
p_{m_{a \alpha}+\nu}\: d^2\nu & \qquad {\mbox{if $X_1 = \1$}}
\end{array} \right. \]
and define $\dot{s}^\varepsilon_{+\mu}$ by taking as in~(\ref{2ds}) the direct
sum. Setting $\breve{s}^\varepsilon_{+\mu} = s^\varepsilon_{+\mu} -
\dot{s}^\varepsilon_{+\mu}$, we obtain the decomposition
\begin{equation}
s^\varepsilon_{+\mu} \;=\; \dot{s}^\varepsilon_{+\mu}
+ \breve{s}^\varepsilon_{+\mu} \:. \label{le}
\end{equation}

Our first step is to show that for small $\mu$, the matrix
$\breve{s}^\varepsilon_{+\mu}(k)$ is bounded, more precisely that
\begin{equation}
\| s^\varepsilon_{+\mu}(k) \| \;\leq\; \frac{C(k)}{\varepsilon^7}
\spc {\mbox{for }} \mu \in (-\frac{\varepsilon}{2}, \frac{\varepsilon}{2})
\label{2bs}
\end{equation}
with $C(k)$ a smooth function with at most polynomial growth at
infinity (the exponent $7$ is probably not optimal,
but~(\ref{2bs}) is sufficient for our purpose). It clearly
suffices to prove~(\ref{2bs}) in a given sector $(a \alpha)$; for
simplicity the sector index will be omitted (i.e.~${\mathcal{B}}_0
\equiv {\mathcal{B}}_0^{(a \alpha)}$). Furthermore, we only
consider the case $X_a \neq \1$; the other case is analogous (and
even simpler, because in the massive sectors no homogeneous
potentials are present). We introduce the projector $E(k)$ by
\[ E \;=\; \sum_{(a, s) \in {\mathcal{S}}} E^a_s \qquad {\mbox{with}} \qquad
{\mathcal{S}} = \left\{ (a, s) {\mbox{ with }} |\mu^a_s - \mu| <
\frac{\varepsilon}{4} \right\} \] According to
Lemma~\ref{lemma28},
\begin{equation}
\|E\| \;\leq\; \frac{C_1(k)}{\varepsilon^3} \label{2grob}
\end{equation}
with $C_1(k)$ smooth with at most polynomial growth at infinity.
The matrix $\breve{s}^\varepsilon_{+\mu}$ has a simple spectral
representation,
\[ \breve{s}^\varepsilon_{+\mu} \;=\;
\sum_{(a, s) \not \in {\mathcal{S}}} \frac{1}{\mu^a_s - \mu}
\:E^a_s\:. \] Unfortunately, this representation is not suitable
for estimates, because we have no control of $\|E^a_s\|$ for
$(a,s) \not \in {\mathcal{S}}$. To avoid this problem, we rewrite
$\breve{s}^\varepsilon_{+\mu}$ as follows,
\begin{eqnarray*}
\breve{s}^\varepsilon_{+\mu} &=& \left(\sum_{(a, s) \not \in
{\mathcal{S}}} \frac{1}{\mu^a_s - \mu} \:E^a_s \:+\: \sum_{(a, s)
\in {\mathcal{S}}} \frac{1}{\mu^a_s - \mu + \varepsilon} \:E^a_s
\right)\: (\1-E) \\
&=& (k\slsh+\varepsilon {\mathcal{B}}_0 - \mu + \varepsilon
E)^{-1}\: (\1-E)\:.
\end{eqnarray*}
Introducing the ``Hamiltonian'' $H=-\gamma^0\: (\vec{k}\slsh +
\varepsilon {\mathcal{B}}_0(\omega, \vec{k}) - \mu + \varepsilon
E)$, we obtain
\begin{equation}
\breve{s}^\varepsilon_{+\mu} \;=\; (\omega-H)^{-1}\:\gamma^0\: (\1-E)\:.
\label{sform}
\end{equation}
The matrix $H(k)$ is Hermitian with respect to the positive scalar product
$(.|.) = \Sl .|\gamma^0|. \Sr$ and can thus be diagonalized, i.e.
\[ H \;=\; \sum_{n=1}^4 \Omega_n\: F_n \]
with real eigenvalues $\Omega_n$ and spectral projectors $F_n$.
Substituting into~(\ref{sform}) gives
\[ \breve{s}^\varepsilon_{+\mu} \;=\;
\sum_{n=1}^4 \frac{1}{\omega-\Omega_n}\:F_n\:\gamma^0\: (\1-E)\:, \]
and~(\ref{2grob}) yields the bound
\begin{equation}
\| \breve{s}^\varepsilon_{+\mu} \| \;\leq\; 2 \max_n
\frac{1}{|\omega - \Omega_n|}\: \|\1-E\|
\;\leq\; \frac{C_2(k)}{\varepsilon^3}\:
\max_n \frac{1}{|\omega - \Omega_n|}\:. \label{fp2}
\end{equation}
It remains to estimate the factors $|\omega-\Omega_n|$ from below.
We use that the determinant is multiplicative to obtain
\begin{eqnarray*}
\prod_{n=1}^4 (\omega-\Omega_n) &=& \det (\omega-H)
\;=\; \det (\gamma^0\:(\omega-H)) \\
&=& \det (k\slsh+\varepsilon {\mathcal{B}}_0 - \mu + \varepsilon
E) \;=\; \prod_{(a,c) \not \in {\mathcal{S}}} (\mu^a_c-\mu) \:
\prod_{(a,c) \in {\mathcal{S}}} (\mu^a_c-\mu+\varepsilon) \:.
\end{eqnarray*}
Taking the absolute value, the factors $|\mu^a_c-\mu|$ and
$|\mu^a_c-\mu+\varepsilon|$ are all greater than
$\frac{\varepsilon}{4}$, and thus
\begin{equation}
\prod_{n=1}^4 |\omega-\Omega_n| \;\geq\; \left( \frac{\varepsilon}{4} \right)^4.
\label{fp1}
\end{equation}
Since the eigenvalues of the Hermitian matrix $\omega-H$ can be estimated by
the sup-norm of the matrix,
\[ |\omega - \Omega_n| \;\leq\; \|\omega - H \|\:, \]
we can deduce from~(\ref{fp1}) that each factor $|\omega-\Omega_n|$ is
bounded by
\[ |\omega-\Omega_n| \;\geq\; \|\omega-H\|^{-3}
\left( \frac{\varepsilon}{4} \right)^4 \:. \]
Substituting this inequality into~(\ref{fp2}) gives~(\ref{2bs}).

The causal perturbation expansion expresses $\tilde{t}^\varepsilon_{+\mu}$ as a
sum of operator products of the form
\begin{equation}
\tilde{t}^\varepsilon_{+\mu} \;\asymp\;
A^\varepsilon_{+\mu}\:{\mathcal{B}}_0\:A^\varepsilon_{+\mu}
\:\cdots\:
A^\varepsilon_{+\mu}\:{\mathcal{B}}_0\:A^\varepsilon_{+\mu} \:,
\label{la}
\end{equation}
where each factor $A$ stands for $p$, $k$, or $s$. Since
$\mathcal{B}_0(k)$ has rapid decay and
$A^\varepsilon_\mu(k)$ grows at most polynomially, these operator
products are well-defined. According to~(\ref{2fp0})
and~(\ref{2fp1}), the first summand inside the brackets
in~(\ref{2idem}) can be written as
\begin{eqnarray}
\lefteqn{ \frac{1}{4} \int_0^\delta d\mu \int_0^\delta d\mu'
\sum_b \sum_{\beta, \gamma} \left(
X_a \:(t^\varepsilon_{+\mu})^{(a \alpha)}_{(b \beta)} \:
(t^\varepsilon_{+\mu'})^{(b \gamma)}_{(c \delta)} \:X_c^*
\:+\: X_a \:(t^\varepsilon_{+\mu})^{(a \alpha)}_{(b \beta)}
\:X_b\: (t^\varepsilon_{+\mu'})^{(b \gamma)}_{(c \delta)}
\right. } \nonumber \\
&& \spc \left. + (t^\varepsilon_{+\mu})^{(a \alpha)}_{(b \beta)} \:
X_b^*\:(t^\varepsilon_{+\mu'})^{(b \gamma)}_{(c \delta)} \:X_c^*
\:+\: (t^\varepsilon_{+\mu})^{(a \alpha)}_{(b \beta)} \:
X_b^* \:X_b\: (t^\varepsilon_{+\mu'})^{(b \gamma)}_{(c \delta)}
\right) . \spc\quad \label{lb}
\end{eqnarray}
When we substitute~(\ref{la}) into~(\ref{lb}), the difficult point is to
multiply the rightmost factor $A$ of the first factor $t$ to the
leftmost factor $A$ of the second factor $t$. More precisely, we must
analyze the following operator products,
\begin{eqnarray}
&&(\cdots \:A^\varepsilon_{+\mu})^{(a \alpha)}_{(b \beta)} \:
(A^\varepsilon_{+\mu'} \:\cdots)^{(b \gamma)}_{(c \delta)} \label{lc} \\
&&(\cdots \:A^\varepsilon_{+\mu})^{(a \alpha)}_{(b \beta)} \:\rho\:
(A^\varepsilon_{+\mu'} \:\cdots)^{(b \gamma)}_{(c \delta)} \label{ld}
\end{eqnarray}
with $A=p$, $k$, or $s$.

If one of the factors $A$ in~(\ref{lc}) or~(\ref{ld}) is the Green's function, we
substitute~(\ref{le}) and expand. Since $\breve{s}^\varepsilon_{+\mu}$ is
bounded~(\ref{2bs}), the products involving $\breve{s}^\varepsilon_{+\mu}$ have
a finite limit as $\delta \searrow 0$. Since the two integrals in~(\ref{lb}) give a factor
$\delta^2$, these products all drop out when the limit $\delta \searrow 0$ is taken
in~(\ref{2idem}). Thus it suffices to consider the case when the factors
$A$ in~(\ref{lc}) and~(\ref{ld}) stand for $p$, $k$, or $\dot{s}$.

Since the Dirac operator has $\varepsilon$-non-degenerate masses, the
distributions $A^\varepsilon_{+\mu}(k)$ have disjoint supports in
different sectors. More precisely, for all $\mu, \mu' \in
(-\frac{\varepsilon}{2},\frac{\varepsilon}{2})$,
\[ {\mbox{supp}}\;( A^{\varepsilon, (b \beta)}_{\mu}) \cap
{\mbox{supp}}\; (A^{\varepsilon, (b \gamma)}_{\mu'})
\;=\; \emptyset \spc {\mbox{if $\beta \neq \gamma$ and
$X_b \neq \1$}}, \]
where each factor $A$ stands for $p$, $k$, or $\dot{s}$.
A similar relation holds in the massive blocks. Therefore,
(\ref{lc}) and~(\ref{ld}) vanish if $\beta \neq \gamma$.

In the case $\beta=\gamma$, (\ref{ld}) is zero because the Dirac
operator is $\varepsilon$-orthogonal to $\rho$~(\ref{lg}). Thus,
using a matrix notation in the sectors, we only need to take into
account the operator products
\[ (\cdots\: A^\varepsilon_{+\mu})(A^\varepsilon_{+\mu'}\:\cdots) \]
with $A=p$, $k$, or $s$ (here we may again consider $s$ instead of $\dot{s}$
because, as we saw above, all factors $\breve{s}$ drop out in the
limit $\delta \searrow 0$). Now we can apply the multiplication
rules~(\ref{lf}) and~(\ref{lh}). Applying~(\ref{lf}) gives a factor
$\delta^2(\mu-\mu')$, and we can carry out the $\mu'$-integral. After
dividing by $\delta$, we can take the limits $\delta \searrow 0$ and
$\varepsilon \searrow 0$. Using that in this limit the Dirac operator
is causality compatible~(\ref{2cc}), we can ``commute $X$ through'' the
resulting operator products (see~{\S}\ref{jsec3}).
In this way, one recovers precisely the unregularized fermionic projector
$P=\lim_{\varepsilon, \delta \searrow 0} P^{\varepsilon, \delta}$.
If~(\ref{lh}) is applied, the resulting principal part is bounded after
the integrals over $\mu$ and $\mu'$ are carried out, and we can take the
limits $\delta \searrow 0$ and $\varepsilon \searrow 0$. After commuting
$X$ through the resulting operator products we find that all terms cancel.
\QED

For understanding better what the above results mean physically,
it is instructive to consider a cosmological situation where the
$4$-volume of space-time is finite. In this case, the limits
$\varepsilon, \delta \searrow 0$ in~(\ref{2idem}) are a merely
mathematical idealization corresponding to the fact that the size
of the universe is very large compared to the usual length scales
on earth. We can extrapolate from~(\ref{2fp0}, \ref{2fp1}) to get
some information on how the properly normalized physical fermionic
projector should look like: The parameter $\delta$ is to be chosen
of the order $T^{-1}$ with $T$ the lifetime of the universe (also
see~{\S}\ref{jsec6}). Then, due to the $\mu$-integral
in~(\ref{2fp0}), the Dirac seas are built up from those fermionic
states whose momenta lie                 in a thin strip around the mass cone.
Naively, the modified mass scaling implies that for neutrinos
this strip must be thinner. However, this naive picture is
misleading because the detailed form of the chiral Dirac seas
depend strongly on the homogeneous operator ${\mathcal{B}}_0$,
which is unknown. We point out that in~(\ref{2idem}) the order of
limits is essential: we must first take the infinite volume limit
and then the limit $\varepsilon \searrow 0$. This means for our
cosmology in finite $4$-volume that the homogeneous perturbation
$\varepsilon {\mathcal{B}}_0$ must be large compared to $T^{-1}$.
One possibility for realizing this is to give the neutrinos a small
rest mass. But, as shown above, the same can be achieved by more
general, possibly nonlocal potentials which do not decay at
infinity.

\chapter{The Regularized Causal Perturbation Theory}
\label{pappB} \index{perturbation expansion!regularized causal}
\renewcommand{\theequation}{\thechapter.\arabic{equation}}
\renewcommand{\thesection}{\thechapter}
\setcounter{equation}{0} \setcounter{Def}{0}
In {\S}\ref{psec26} we
gave a procedure for regularizing the formulas of the light-cone
expansion~(\ref{p:2F2}--\ref{p:2J}). We shall now derive this
regularization procedure. The basic idea is to extend the causal
perturbation expansion of~{\S}\ref{jsec2} to the case with regularization,
in such a way that the causality and gauge symmetry are preserved
for macroscopic perturbations. Using the methods of~{\S}\ref{jsec5}
one can then analyze the behavior of the so-regularized Feynman
diagrams near the light cone. For simplicity, we will restrict
attention to the first order in perturbation theory. But our
methods could be applied also to higher order Feynman diagrams,
and the required gauge symmetry suggests that our main result,
Theorem~\ref{thmpB1}, should hold to higher order in perturbation
theory as well.

We first state our assumptions on the fermionic projector of the
vacuum. As in Chapter~\ref{psec2} we describe the vacuum by a
fermionic projector~$P(x,y)$ of the form~(\ref{p:2k}) with
vector-scalar structure (\ref{p:2f2}). For small energy-momentum,
$\hat{P}$ should coincide with the unregularized fermionic
projector of the vacuum, i.e.
\begin{equation}
\hat{P}(k) \;=\; (k \slsh + m) \:\delta(k^2-m^2) \:\Theta(-k^0)
\spc {\mbox{if $|k^0| \ll E_P$ and $|\vec{k}| \ll E_P$.}}
\label{p:Bb}
\end{equation}
Furthermore, we assume that the vector component is null on the
light cone (i.e.\ that (\ref{p:2Es}) holds with
$\varepsilon_{\mbox{\scriptsize{shear}}} \ll 1$), and that $P$
satisfies all the regularity assumptions considered in {\S}\ref{psec24}
and {\S}\ref{psec25}. For simplicity, we finally
assume that $\hat{P}$ is supported inside the
lower mass cone,
\begin{equation}
{\mbox{supp }} \hat{P} \;\subset\; \overline{\mathcal{C}^\land}\label{p:Ba}
\end{equation}
(with $\mathcal{C}^\land$ according to (\ref{calcdef})). This last
condition is quite strong, but nevertheless reasonable. In
particular, it is satisfied when $P$ is composed of one-particle
states which are small perturbations of the Dirac eigenstates on the
lower mass shell.

In this appendix we shall address the question of how one can
introduce a classical external field into the system. For clarity,
we will develop our methods mainly in the example of an
electromagnetic field. As described in {\S}\ref{psec22}, we
consider the regularized fermionic projector as a model for the
fermionic projector of discrete space-time. In this sense, the
regularization specifies the microscopic structure of space-time.
Following the concept of macroscopic potentials and wave functions
introduced in {\S}\ref{psec22}, the electromagnetic field should
modify the fermionic projector only on length scales which are
large compared to the Planck length, but should leave the
microscopic structure of space-time unchanged. In order to fulfill
this requirement, we impose the following conditions. First of
all, we assume that the electromagnetic field be ``macroscopic''
in the sense that it can be described by an electromagnetic
potential $A$ which vanishes outside the low-energy region, i.e.
\begin{equation}
\hat{A}(k) \;=\; 0 \spc {\mbox{unless $|k^0| \ll E_P$ and
$|\vec{k}| \ll E_P$,}} \label{p:Bc}
\end{equation}
where $\hat{A}$ is the Fourier transform of $A$. We denote the fermionic
projector in the presence of the electromagnetic field by $P[\Aslsh]$.
In order to prevent that the electromagnetic potential might influence
the microscopic structure of space-time locally, we demand that $A$
can locally be made to zero by a gauge transformation. Thus we impose that the
usual behavior under $U(1)$ gauge transformations
\begin{equation}
P[\Aslsh + (\Pdd \Lambda)](x,y) \;=\; e^{i \Lambda(x)} \:P[\Aslsh](x,y) \:
e^{-i \Lambda(y)} \label{p:Bd}
\end{equation}
(with a real function $\Lambda$) should hold also for the regularized fermionic
projector, assuming that the involved potentials $A$ and $(A+\partial
\Lambda)$ are both macroscopic (\ref{p:Bc}). We point out that, because of the
gauge symmetry in discrete space-time (following from the freedom in choosing
the gauge (\ref{p:36})), the local phase transformations in (\ref{p:Bd}) are
irrelevant in the equations of discrete space-time, and thus the transformation
law (\ref{p:Bd}) implies the freedom to transform the electromagnetic potential
according to $\Aslsh \to \Aslsh + \Pdd \Lambda$.
Finally, we must rule out the possibility that the electromagnetic potential
might influence the microscopic structure of space-time in a nonlocal way.
For this purpose, we impose that the perturbation expansion for the
regularized fermionic projector be causal, in the sense introduced
in~{\S}\ref{jsec2}.

Let us consider how these conditions can be implemented
in the perturbation theory to first order. We first recall the
standard perturbation theory for Dirac eigenstates. For a solution $\Psi$
of the free Dirac equation $(i \Pdd - m) \Psi = 0$, the perturbation
to first order in $A$, which we denote by $\Delta \Psi[\Aslsh]$,
is given by
\begin{equation}
\Delta \Psi[\Aslsh](x) \;=\; -\int d^4y \:s_m(x,y) \:\Aslsh(y)
\:\Psi(y) \:, \label{p:Bf}
\end{equation}
where $s_m(x,y)$ is the Dirac Green's function~(\ref{51a}),
\begin{equation}
s_m(x,y) \;=\; \int \frac{d^4k}{(2 \pi)^4} \:\frac{\mbox{PP}}{k^2-m^2}
\:(k\slsh + m) \:e^{-ik(x-y)} \:. \label{p:BC}
\end{equation}
If we consider $s_m(x,y)$ as the integral kernel of an operator $s_m$ and the
potentials as multiplication operators, we can calculate $\Delta \Psi$
in the case $\Aslsh=\Pdd \Lambda$ to be
\begin{eqnarray}
\Delta \Psi[\Pdd \Lambda] &=& -s_m \:(\Pdd \Lambda) \:\Psi \;=\;
i s_m \:[i \Pdd -m,\: \Lambda] \:\Psi \nonumber \\
&=& i ((i \Pdd-m) s_m) \:\Lambda \:\Psi \:-\:
i s_m \:\Lambda\: ((i \Pdd-m) \Psi) \;=\; i \Lambda\:\Psi \:.
\label{p:Bh}
\end{eqnarray}
Thus in this case, $\Delta \Psi(x) = i \Lambda(x) \:\Psi(x)$ is simply the
contribution linear in $\Lambda$ to the phase transformed
wave function $\exp(i \Lambda(x)) \:\Psi(x)$; this shows explicitly
that the perturbation calculation is gauge invariant.

As a consequence of the regularization, the fermionic projector
$P(x,y)$ is in general not composed of Dirac eigenstates. Therefore,
we next consider a wave function $\Psi$ which is not necessarily a
solution of the free Dirac equation. But according to (\ref{p:Ba}),
we may assume that its Fourier transform $\hat{\Psi}$ has its support
in the interior of the mass cone,
\begin{equation}
{\mbox{supp }} \hat{\Psi} \;\subset\; \{ k \:|\: k^2 \geq 0 \} \:.
\label{p:Bg}
\end{equation}
In this case we can introduce $\Delta \Psi[\Aslsh]$ as follows.
The spectral projector $p_\mu$ of the free Dirac operator $i \Pdd$
corresponding to the eigenvalue $\mu \in \R$ has the form
\begin{equation}
p_\mu(x,y) \;=\; \int \frac{d^4k}{(2 \pi)^4} \:\epsilon(\mu) \:
(k \slsh+\mu) \:\delta(k^2-\mu^2) \:e^{-ik(x-y)} \label{p:BB}
\end{equation}
(see~(\ref{x20}); notice that we added the step function
$\epsilon(\mu)$ to allow for the case $\mu<0$). Since the real axis
is only part of the spectrum of the free Dirac operator (namely, the
free Dirac operator has also an imaginary spectrum), the spectral projectors
$(p_\mu)_{\mu \in \sR}$ are clearly not complete, i.e.\
$\int_{-\infty}^\infty p_\mu d\mu \neq \1$. By integrating
(\ref{p:BB}) over $\mu$,
\begin{equation}
\int_{-\infty}^\infty p_\mu(x,y)\:d\mu \;=\; \int \frac{d^4k}{(2 \pi)^4}\:
\Theta(k^2) \:e^{-ik(x-y)} \:, \label{p:BA}
\end{equation}
one sees more precisely that the operator $\int_{-\infty}^\infty p_\mu d\mu$
is the projector on all the momenta in the mass cone. But according to (\ref{p:Bg}), $\Psi$ lies
in the image of this projector, and we can thus use
the spectral projectors $p_\mu$ to decompose $\Psi$ into
eigenstates of the free Dirac operator. Each eigenstate can then be
perturbed using (\ref{p:Bf}). This leads us to introduce $\Delta \Psi[\Aslsh]$
according to
\begin{equation}
\Delta \Psi[\Aslsh] \;=\; -\int_{-\infty}^\infty d\mu\;
s_\mu \:\Aslsh\:p_\mu \:\Psi \:. \label{p:Be}
\end{equation}
This definition of $\Delta \Psi$ shows the correct behavior under gauge
transformations; namely, similar to (\ref{p:Bh}),
\begin{eqnarray}
\Delta \Psi[\Pdd \Lambda] &=& i \int_{-\infty}^\infty d\mu\;
s_\mu \:[i \Pdd -\mu,\: \Lambda] \:p_\mu\:\Psi \nonumber \\
&=& i \Lambda \left( \int_{-\infty}^\infty p_\mu\: d\mu \right) \Psi
\;\stackrel{(\ref{p:BA}, \ref{p:Bg})}{=}\; i \Lambda\:\Psi \:.
\label{p:Bi}
\end{eqnarray}

Thinking in terms of the decomposition~(\ref{Pvac}) of the fermionic
projector into the one-particle states, it seems natural
to introduce the perturbation of the fer\-mio\-nic projector $\Delta
P[\Aslsh]$ by perturbing each one-particle state according to
(\ref{p:Be}). This leads to the formula
\begin{equation}
\Delta P[\Aslsh] \;=\; -\int_{-\infty}^\infty d\mu \left(
s_\mu \:\Aslsh\:p_\mu\:P \:+\: P \:p_\mu \:\Aslsh\:s_\mu \right) \:.
\label{p:B3a}
\end{equation}
The gauge symmetry can again be verified explicitly. Namely, a calculation
similar to (\ref{p:Bi}) using (\ref{p:Ba}) yields that
\[ \Delta P[\Pdd \Lambda](x,y) \;=\; i \Lambda(x) \:P(x,y)
- i P(x,y)\: \Lambda(y) \:, \] and this is the contribution
linear in $\Lambda$ to (\ref{p:Bd}). The perturbation calculation
(\ref{p:B3a}) is immediately extended to a general perturbation
operator~${\mathcal{B}}$ by setting
\begin{equation}
\Delta P[{\mathcal{B}}] \;=\; -\int_{-\infty}^\infty d\mu \left(
s_\mu \:{\mathcal{B}}\:p_\mu\:P \:+\: P \:p_\mu
\:{\mathcal{B}}\:s_\mu \right) \:. \label{p:Bj}
\end{equation}

Let us verify if the perturbation calculation (\ref{p:Bj}) is
causal in the sense of~{\S}\ref{jsec2}. Since it seems impossible to
write (\ref{p:Bj}) in a manifestly causal form, we apply here a
different method, which allows us to analyze the causality of the perturbation
expansion in momentum space. As mentioned in~{\S}\ref{jsec5},
the causality of the perturbation expansion can be understood via
the causality of the line integrals over the external potentials
and fields which appear in the light cone expansion. More
precisely, causality means that the light-cone expansion of
$\Delta P(x,y)$ should involve only line integrals along the line
segment $\overline{xy}$, but no unbounded line integrals like for
example $\int_0^\infty d\lambda \:{\mathcal{B}}(\lambda y +
(1-\lambda) x)$. This way of understanding the causality of the
perturbation expansion yields a simple condition in momentum
space. Namely, if ${\mathcal{B}}$ has the form of a plane wave of
momentum $q$, i.e.\ ${\mathcal{B}}(x) = {\mathcal{B}}_q
\exp(-iqx)$, then the unbounded line integrals become infinite
when $q$ goes to zero (for ${\mathcal{B}}_q$ fixed), whereas
integrals along the line segment $\overline{xy}$ are clearly
bounded in this limit. Hence we can say that the perturbation
calculation (\ref{p:Bj}) is causal only if it is regular in the
limit $q \to 0$. In order to analyze this condition, we substitute
the explicit formulas (\ref{p:BC}, \ref{p:BB}) into
(\ref{p:Bj}) and obtain
\begin{eqnarray*}
\lefteqn{ \Delta P[{\mathcal{B}}](x,y) \;=\;
-\int_{-\infty}^\infty d\mu
\:\epsilon(\mu) \:\int \frac{d^4k}{(2 \pi)^4} } \nonumber \\
&&\times \left( \frac{\mbox{PP}}{(k+q)^2 - \mu^2} \:(k \slsh +q
\slsh + \mu) \:{\mathcal{B}}_q\:(k \slsh+\mu)\:\delta(k^2-\mu^2)
\:
\hat{P}(k) \;e^{-i(k+q)x + iky} \right. \nonumber \\
&&\hspace*{.7cm} \left. +\: \hat{P}(k) \:\delta(k^2-\mu^2) \:(k
\slsh +\mu) \:{\mathcal{B}}_q\: (k \slsh - q \slsh + \mu)
\:\frac{\mbox{PP}}{(k-q)^2 - \mu^2} \;e^{-ikx + i (k-q) y} \right) .
\end{eqnarray*}
We set $q = \varepsilon \check{q}$ with a fixed vector $\check{q}$ and consider
the behavior for $\epsilon \searrow 0$. Taking only the leading order in
$\varepsilon$, one can easily carry out the $\mu$-integration
and gets
\begin{eqnarray}
\lefteqn{ \Delta P[{\mathcal{B}}](x,y) \;=\;
-\frac{1}{\varepsilon}
\:\int \frac{d^4k}{(2 \pi)^4}\: e^{-ik(x-y)} } \nonumber \\
&& \times
\left( \frac{\mbox{PP}}{2 k \check{q} +
\varepsilon \check{q}^2}\: (k\slsh\: {\mathcal{B}}_q +
{\mathcal{B}}_q\: k\slsh) \: \hat{P}(k) \right. \nonumber \\
&& \hspace*{3cm} \left. \:+\:
\hat{P}(k)\:(k\slsh\: {\mathcal{B}}_q + {\mathcal{B}}_q\:
k\slsh)\: \frac{\mbox{PP}}{-2 k \check{q} + \varepsilon
\check{q}^2} \right) + {\mathcal{O}}(\varepsilon^0) \:.\qquad \label{p:Bm}
\end{eqnarray}
Since
\[ \lim_{\epsilon \searrow 0} \frac{\mbox{PP}}{2 k \check{q} +
\varepsilon \check{q}^2} \;=\; \lim_{\epsilon \searrow 0}
\frac{\mbox{PP}}{2 k \check{q} - \varepsilon \check{q}^2}
\;=\; \frac{\mbox{PP}}{2 k \check{q}} \]
in the sense of distributions in the argument $k \check{q}$ (notice that
this kind of convergence is sufficient using the regularity
of $\hat{P}$), the leading singularity of (\ref{p:Bm})
for $\varepsilon \searrow 0$ has the form
\begin{equation}
-\frac{1}{\varepsilon} \:\int \frac{d^4k}{(2 \pi)^4} \:
e^{-ik(x-y)} \:\frac{\mbox{PP}}{2 k \check{q}} \; \left[ \{
{\mathcal{B}}_q,\: k\slsh\},\: \hat{P}(k) \right] \:.
\label{p:B1a}
\end{equation}
Taking the Fourier transform in the variable $(x-y)$, it is clear
that (\ref{p:B1a}) vanishes only if the commutator/anti-commutator
combination $[ \{ {\mathcal{B}}_q, k\slsh\}, \hat{P}(k) ]$ is zero
for all $k$. Since the perturbation ${\mathcal{B}}_q$ can be
arbitrary, one sees (for example by considering a scalar
perturbation, ${\mathcal{B}}_q \sim \1$) that it is a necessary
condition for the perturbation calculation (\ref{p:Bj}) to be
regular in the limit $q \to 0$ that
\begin{equation}
    [k \slsh, \:\hat{P}(k)] \;=\; 0 \spc {\mbox{for all $k$}}.
    \label{p:Bn}
\end{equation}
This commutator vanishes only if the vector field $v(k)$ in
(\ref{p:2f2}) is a multiple of $k$, or, using the notation of {\S}\ref{psec25},
if the surface states have no shear. We conclude
that the perturbation calculation (\ref{p:Bj}) is in general not
causal.

Before resolving this causality problem, we briefly discuss how this
problem comes about. The condition (\ref{p:Bn}) can be stated
equivalently that the operator $P$ must commute with the free Dirac
operator. In other words, the perturbation calculation (\ref{p:Bj}) is causal
only if the fermionic projector of the vacuum is composed of
eigenstates of the free Dirac operator. In this formulation, our
causality problem can be understood directly. Namely, since our
perturbation method is based on the perturbation calculation (\ref{p:Bf})
for Dirac eigenstates, it is not astonishing that the method is
inappropriate for non-eigenstates, because the perturbation expansion
is then performed around the wrong unperturbed states. It is
interesting to see that this shortcoming leads to a breakdown of
causality in the perturbation expansion.

In order to comply with causality, we must modify the perturbation calculation
(\ref{p:Bj}). Our idea is to deduce the perturbation calculation for
the fermionic projector from that for a modified fermionic projector,
which satisfies the causality condition (\ref{p:Bn}). The simplest
idea for modifying the fermionic projector would be to introduce a
unitary transformation $\hat{U}(k) \in U(2,2)$ which makes the vector $v(k)$
in (\ref{p:2f2}) parallel to $k$, more precisely
\[ \hat{U}(k)^{-1} \:v_j(k) \:\gamma^j \:\hat{U}(k)
\;=\; \lambda(k) \:k \slsh \spc {\mbox{with $\lambda(k) \in \R$.}} \]
However, a unitary transformation is too restrictive because it keeps
the Lorentzian scalar product $v(k)^2$ invariant, and thus it cannot be
used for example in the case when $v(k)$ is space-like, but $k$ is
time-like. Therefore, we shall consider here a linear combination of
unitary transformations. More precisely, we introduce for $L>1$ and
$l=1,\ldots,L$ unitary operators $\hat{U}_l(k) \in U(2,2)$
and real coefficients $c_l$ such that\footnote{{\textsf{Online version}:} Taking such linear combinations
has the disadvantage that normalization and definiteness properties are not preserved.
Therefore, it is preferable to use instead the construction in the book~\cite[Appendix~F]{cfs}
(listed in the references in the preface to the second online edition).}
\begin{equation}
\sum_{l=1}^L c_l(k) = 1 \qquad{\mbox{and}}\qquad
v_j(k) \:\gamma^j \;=\; \sum_{l=1}^L c_l(k) \;
\hat{U}_l(k) \:\lambda(k)\:k \slsh\:\hat{U}_l(k)^{-1} \label{p:B8a}
\end{equation}
with $\lambda(k) \in \R$. The existence of $(\hat{U}_l, c_l)$ is guaranteed
by the fact that the $U(2,2)$ transformations comprise Lorentzian
transformations~{\S}\ref{isec5}. Clearly, the representation (\ref{p:B8a})
is not unique. According to (\ref{p:Bb}), we can choose the
transformation (\ref{p:B8a}) to be the identity in the low-energy region,
and can thus assume that
\begin{equation}
    \hat{U}_l(k) \;=\; \1 \spc {\mbox{if $|k^0| \ll E_P$ and $|\vec{k}| \ll
    E_P$.}}
    \label{p:B8b}
\end{equation}
Furthermore, the regularity assumptions and the particular properties of
the fer\-mion\-ic projector mentioned before (\ref{p:Ba}) give rise to
corresponding properties of the operators $\hat{U}_l$; this will be
specified below (see (\ref{p:Bp1}, \ref{p:Bp2})).
The operators obtained by multiplication with $\hat{U}_l(k)$ in momentum
space are denoted by $U_l$; they have in position space the kernels
\begin{equation}
U_l(x,y) \;=\; \int \frac{d^4k}{(2 \pi)^4}\: \hat{U}_l(k)\: e^{-ik(x-y)}
\:. \label{p:BU}
\end{equation}
Finally, we introduce the ``modified fermionic projector'' $Q$ by
replacing the vector field $v(k)$ in (\ref{p:2f2}) by $\lambda(k) \:k \slsh$,
i.e.
\begin{equation}
\hat{Q}(k) \;=\;  (\lambda(k) \:k\slsh \:+\: \phi(k) \:\1) \:f(k)
\:.
    \label{p:BQdef}
\end{equation}
According to (\ref{p:B8a}), the fermionic projector $P$ is obtained
from $Q$ by the transformation
\begin{equation}
    P \;=\; \sum_{l=1}^L c_l \: U_l \:Q\: U_l^{-1} \:.
    \label{p:BQtrans}
\end{equation}

The modified fermionic projector (\ref{p:BQdef}) satisfies the
condition $[\hat{Q}(k),\:k \slsh]=0$. Hence the perturbation
calculation for $Q$ does not suffer from our above causality
problem, and we can introduce $\Delta Q[{\mathcal{B}}]$ in analogy
to (\ref{p:Bj}) by
\begin{equation}
\Delta Q[{\mathcal{B}}] \;:=\; -\int_{-\infty}^\infty d\mu \left(
s_\mu \:{\mathcal{B}}\:p_\mu\:Q \:+\: Q \:p_\mu
\:{\mathcal{B}}\:s_\mu \right) \:. \label{p:Bp}
\end{equation}
We now deduce the perturbation of $P$ by applying to (\ref{p:Bp}) a
transformation analogous to that in (\ref{p:BQtrans}), namely
\begin{eqnarray}
\Delta P[{\mathcal{B}}] &:=& \sum_{l=1}^L c_l\: U_l \:\Delta
Q[{\mathcal{B}}]\:
U_l^{-1} \label{p:B50a} \\
&=& -\sum_{l=1}^L c_l \int_{-\infty}^\infty d\mu \; U_l \left(
s_\mu \:{\mathcal{B}}\:p_\mu\:Q \:+\: Q\:
p_\mu\:{\mathcal{B}}\:s_\mu \right) U_l^{-1} \:. \label{p:Bq}
\end{eqnarray}
This last transformation should not affect the causality
(in the sense of~{\S}\ref{jsec2}) because if (\ref{p:Bp}) is regular when the
momentum $q$ of the bosonic potential goes to zero, then the
transformed operator (\ref{p:B50a}) will clearly also be regular in this
limit. We call (\ref{p:Bq}) the {\em{regularized causal
perturbation}} of the fermionic projector to first order.

The perturbation calculation (\ref{p:Bq}) requires a detailed
explanation. Qualitatively speaking, the difference between
(\ref{p:Bj}) and (\ref{p:Bq}) is that the spectral projectors
$p_\mu$, the Green's functions $s_\mu$, and the perturbation
operator ${\mathcal{B}}$ have been replaced by the unitarily
transformed operators
\begin{equation}
U_l\: p_\mu\: U_l^{-1} \:,\spc U_l\: s_\mu\: U_l^{-1}
\;\;\;{\mbox{and}}\spc U_l\:{\mathcal{B}}\: U_l^{-1} \:,
    \label{p:Bq2}
\end{equation}
and that a linear combination is taken. According to
(\ref{p:B8b}), the unitary transformations in (\ref{p:Bq2}) have
no influence on the macroscopic properties of these operators,
i.e.\ on the behavior when these operators are applied to wave
functions with support in the low-energy region. But the
transformation (\ref{p:Bq2}) changes the operators on the
microscopic scale, in such a way that causality is fulfilled in
the perturbation expansion. We point out that in the case where
${\mathcal{B}}$ is the usual operator of multiplication with the
external potentials, the transformed operator $U_l {\mathcal{B}}
U_l^{-1}$ is in general no longer a multiplication operator in
position space; thus one can say that the classical potentials
have become nonlocal on the microscopic scale. In order to better
understand why the causality problem of (\ref{p:Bj}) has
disappeared in (\ref{p:Bq}), it is useful to observe that $Q$
commutes with the spectral projectors $p_\mu$. This means that $Q$
is composed of eigenstates of the Dirac operator, so that the
perturbation expansion is now performed around the correct
unperturbed states.

Let us consider a gauge transformation. In the case
${\mathcal{B}}=\Pdd \Lambda$, the perturbation (\ref{p:Bq}) is
computed to be
\begin{eqnarray}
\lefteqn{ \Delta P[\Pdd \Lambda] \;=\;
i \sum_{l=1}^L c_l \int_{-\infty}^\infty d\mu \:U_l \left(
s_\mu \:[i \Pdd - \mu,\: \Lambda] \:p_\mu\:Q \:+\:
Q\:p_\mu\: [i \Pdd-\mu,\: \Lambda]\:s_\mu \right) U_l^{-1} } \nonumber \\
&=& i \sum_{l=1}^L c_l \int_{-\infty}^\infty d\mu \:U_l \left( \Lambda
\:p_\mu\:Q \:-\: Q\:p_\mu\:\Lambda \right) U_l^{-1} \nonumber \\
&=& \sum_{l=1}^L c_l \left(
i U_l \Lambda \left(\int_{-\infty}^\infty
p_\mu\:d\mu \right) Q U_l^{-1}
\:-\: i U_l Q \left(\int_{-\infty}^\infty p_\mu\:d\mu \right)
\Lambda U_l^{-1} \right) .\qquad \label{p:B24a}
\end{eqnarray}
By construction of $\hat{Q}$, we can assume that
the distributions $\hat{P}$ and
$\hat{Q}$ have the same support, and thus
(\ref{p:Ba}) holds for $\hat{Q}$ as well,
\begin{equation}
{\mbox{supp }} \hat{Q} \;\subset\; \overline{\mathcal{C}^\land}\:. \label{p:BQsupp}
\end{equation}
Hence, according to (\ref{p:BA}), the projectors
$\int_{-\infty}^\infty p_\mu d\mu$ in (\ref{p:B24a}) can be omitted,
and we conclude that
\begin{equation}
\Delta P[\Pdd \Lambda] \;=\; \sum_{l=1}^L c_l \left(
i U_l \Lambda U_l^{-1} \:U_l Q U_l^{-1} \:-\: i
U_l Q U_l^{-1} \: U_l \Lambda U_l^{-1} \right) \:.
    \label{p:Bs}
\end{equation}
If in this formula we were allowed to replace the factors $U_l \Lambda
U_l^{-1}$ by $\Lambda$, we could substitute in (\ref{p:BQtrans}) and
would obtain the contribution linear in $\Lambda$ to the
required transformation law (\ref{p:Bd}). Indeed, the difference
between $\Lambda$ and $U_l \Lambda U_l^{-1}$ is irrelevant, as one sees in
detail as follows. We consider one summand in (\ref{p:Bs}) and set for
ease in notation $U=U_l$. According to (\ref{p:B8b}), the operators $\Lambda$
and $U \Lambda U^{-1}$ coincide macroscopically (i.e.\ when applied to
functions with support in the low-energy region), and thus
(\ref{p:Bs}) yields gauge symmetry on the macroscopic scale. However,
such a macroscopic gauge symmetry is not
sufficient for us: to ensure that the microscopic
structure of space-time is not influenced by the electromagnetic
field, it is essential that (\ref{p:Bd}) holds even on the Planck
scale. In order to show microscopic gauge invariance, we consider the
operator $U \Lambda U^{-1}$ in momentum space,
\begin{equation}
(U \Lambda U^{-1} \:f)(q) \;=\; \int \frac{d^4p}{(2 \pi)^4} \:
\hat{U}(q) \:\hat{\Lambda}(q-p) \:\hat{U}(p)^{-1} \:f(p) \:,
    \label{p:Br}
\end{equation}
where $\hat{\Lambda}$ is the Fourier transform of $\Lambda$, and $f$
is a test function in momentum space. Since we assume that the
electromagnetic potential $\Aslsh=\Pdd \Lambda$ is macroscopic (\ref{p:Bc}),
the integrand in (\ref{p:Br}) vanishes unless $q-p$ is in the
low-energy region. More precisely, we can say that
\[ |q^0 - p^0|,\: |\vec{q}-\vec{p}| \;\sim\;
l_{\mbox{\scriptsize{macro}}}^{-1} \:, \]
where $l_{\mbox{\scriptsize{macro}}}$ denotes a typical length scale
of macroscopic physics. Since the vector $q-p$ is in this sense small,
it is reasonable to expand the factor $\hat{U}(q)$ in
(\ref{p:Br}) in a Taylor series around $p$. As the operators $\hat{U}_l$
are characterized via (\ref{p:B8a}), we can assume that they have similar
regularity properties as $P$. In particular, we may assume that the
partial derivatives of $\hat{U}_l(p)$ scale in powers of
$E_P^{-1}$, in the sense that there should be a constant $c \ll
l_{\mbox{\scriptsize{macro}}} E_P$ such that
\begin{equation}
|\partial^\kappa \hat{U}_l(p)| \;\leq\; \left( \frac{c}{E_P}
\right)^{|\kappa|} \spc {\mbox{for any multi-index $\kappa$}}.
    \label{p:Bp1}
\end{equation}
From this we conclude that the Taylor expansion of $\hat{U}(q)$
around $p$ is an expansion in powers of $(l_{\mbox{\scriptsize{macro}}}
E_P)^{-1}$, and thus
\begin{eqnarray}
(U \Lambda U^{-1} \:f)(q) &=& \int \frac{d^4p}{(2 \pi)^4} \:
\hat{U}(p) \:\hat{\Lambda}(q-p) \:\hat{U}(p)^{-1} \:f(p) \nonumber \\
&& \:+\:{\mbox{(higher orders in
$(l_{\mbox{\scriptsize{macro}}} E_P)^{-1}$)}}.
\label{p:B5a}
\end{eqnarray}
Using that $\hat{\Lambda}(q-p)$ is a multiple of the identity matrix, the
factors $\hat{U}(p)$ and $\hat{U}(p)^{-1}$ in (\ref{p:B5a}) cancel each
other. We conclude that the operators $U \Lambda U^{-1}$ and $\Lambda$
coincide up to higher order in $(l_{\mbox{\scriptsize{macro}}} E_P)^{-1}$.
For the integral kernels in position space, we thus have
\begin{equation}
(U \Lambda U^{-1})(x,y) \;=\; \Lambda(x) \:\delta^4(x-y) \:+\:
{\mbox{(higher orders in
$(l_{\mbox{\scriptsize{macro}}} E_P)^{-1}$)}}.
    \label{p:Bss}
\end{equation}
We point out that this statement is much stronger than the equality
of the operators $U \Lambda U^{-1}$ and $\Lambda$
on the macroscopic scale that was mentioned at the beginning of this paragraph.
Namely, (\ref{p:Bss}) shows that these operators coincide even
microscopically,  up to a very small error term.
Notice that it was essential for the
derivation that $\Lambda$ is a scalar function (for example,
(\ref{p:Bss}) would in general be false if we replaced $\Lambda$ by
$\Aslsh$). Using (\ref{p:Bss}) in each summand of (\ref{p:Bs}) and
applying (\ref{p:BQtrans}), we conclude that
\begin{eqnarray}
\Delta P[\Pdd \Lambda](x,y) &=& i \Lambda(x) \:P(x,y) \:-\: i P(x,y)\:
\Lambda(y) \nonumber \\
&&\:+\: {\mbox{(higher orders in
$(l_{\mbox{\scriptsize{macro}}} E_P)^{-1}$)}}.
    \label{p:Bt}
\end{eqnarray}
This shows gauge symmetry of the perturbation calculation (\ref{p:Bq}).

It is interesting that, according to (\ref{p:Bt}), gauge symmetry
holds only up to an error term. This is unproblematic as long as
the length scales of macroscopic physics are large compared to the
Planck length. But (\ref{p:Bt}) indicates that the regularized
causal perturbation theory fails when energy or momentum of the
perturbation ${\mathcal{B}}$ are of the order of the Planck
energy. In this case, the distinction between the ``macroscopic''
and ``microscopic'' length scales, on which our constructions
relied from the very beginning (cf.\ (\ref{p:Bc})), can no longer
be made, and it becomes impossible to introduce a causal and gauge
invariant perturbation theory.

We conclude the discussion of the regularized causal perturbation
expansion by pointing out that our construction was based on condition
(\ref{p:Bn}), which is only a necessary condition for
causality. Hence the causality of (\ref{p:Bq}) has not yet been
proved. We shall now perform the light-cone expansion of (\ref{p:Bq}).
This will show explicitly that the light-cone expansion involves, to
leading orders in $(l_{\mbox{\scriptsize{macro}}} E_P)^{-1}$ and
$(l E_P)^{-1}$, no unbounded line integrals, thereby establishing
causality in the sense of~{\S}\ref{jsec2}.

In the remainder of this appendix, we will analyze the regularized
causal perturbation calculation (\ref{p:Bq}) near the light cone.
Our method is to first perform the light-cone expansion of $\Delta Q$,
and then to transform the resulting formulas according to (\ref{p:B50a})
 to finally obtain the light-cone expansion of $\Delta P$.
In preparation, we describe how a decomposition into
Dirac eigenstates can be used for an analysis of the operator $Q$
near the light cone. A short computation using (\ref{p:BQdef},
\ref{p:BQsupp}) yields that $\hat{Q}$ can be represented in the form
\begin{equation}
\hat{Q}(k) \;=\; \int_{-\infty}^\infty
d\mu\; w_\mu(\vec{k}) \;\epsilon(\mu) \:(k\slsh+\mu)
\:\delta(k^2-\mu^2) \:\Theta(-k^0)
    \label{p:Bx}
\end{equation}
with the real-valued distribution
\begin{equation}
w_\mu(\vec{k}) = (\phi(k) + \mu\:\lambda(k)) \:f(k) \spc{\mbox{and}}\spc
k(\vec{k})=(-\sqrt{|\vec{k}|^2+\mu^2}, \vec{k}). \label{p:BV}
\end{equation}
This representation can be understood as follows. According to
(\ref{p:BB}), the distributions $\epsilon(\mu) \:(k\slsh+\mu)
\:\delta(k^2-\mu^2)$ in the integrand of (\ref{p:Bx}) are the
spectral projectors of the free Dirac operator in momentum space.
The factor $\Theta(-k^0)$ projects out all states on the upper
mass cone, and the function $w_\mu(\vec{k})$ multiplies the states
on the lower mass shell $k=(-\sqrt{|\vec{k}|^2+\mu^2}, \vec{k})$
with a scalar weight factor. In this sense, (\ref{p:Bx}) can be
regarded as the spectral decomposition of the operator $Q$ into
Dirac eigenstates. Notice that the factor $\delta(k^2-\mu^2)
\:\Theta(-k^0)$ in (\ref{p:Bx}) is the Fourier transform of the
distribution $T_a$, (\ref{Taf}). Exactly as described for the
scalar component in {\S}\ref{psec24}, we are here interested only
in the regularization effects for large energy or momentum and may
thus disregard the logarithmic mass problem (see~{\S}\ref{jsec5} for
details). Therefore, we ``regularize'' $T_a$ according to
(\ref{Tadef}) and consider instead of (\ref{p:Bx}) the operator
\[ \hat{Q}^{\mbox{\scriptsize{reg}}}(k)
\;:=\; \int_{-\infty}^\infty d\mu\:\epsilon(\mu) \: w_\mu(\vec{k})
\:(k\slsh+\mu) \:T_{\mu^2}^{\mbox{\scriptsize{reg}}}(k) \:, \]
where $T_a^{\mbox{\scriptsize{reg}}}(k)$ is the Fourier
transform of (\ref{Tadef}). We expand the distribution
$T_{\mu^2}^{\mbox{\scriptsize{reg}}}$ in a power series in $\mu^2$,
\[ \hat{Q}^{\mbox{\scriptsize{reg}}}(k)
\;=\; \int_{-\infty}^\infty d\mu\:\epsilon(\mu)\: w_\mu(\vec{k})
\:(k\slsh+\mu) \:\sum_{n=0}^\infty \frac{1}{n!} \:
T^{(n)}(k) \:\mu^{2n} \]
with $T^{(n)}$ according to~(\ref{Tldef}).
Commuting the integral and the sum, we obtain
\begin{equation}
\hat{Q}^{\mbox{\scriptsize{reg}}}(k) \;=\;
32 \pi^3 \sum_{n=0}^\infty \frac{1}{n!} \left( g_{[n]}(\vec{k}) \:k\slsh \:+\:
h_{[n]}(\vec{k}) \right) T^{(n)}(k)
        \label{p:By1}
\end{equation}
with
\begin{eqnarray}
g_{[n]}(\vec{k}) & = & \frac{1}{32 \pi^3} \:\int_{-\infty}^\infty
d\mu \;\epsilon(\mu) \:w_\mu(\vec{k}) \:\mu^{2n} \label{p:By2} \\
h_{[n]}(\vec{k}) & = & \frac{1}{32 \pi^3} \:\int_{-\infty}^\infty
d\mu \;\epsilon(\mu) \:w_\mu(\vec{k}) \:\mu^{2n+1} \:.\label{p:By3}
\end{eqnarray}
The representation (\ref{p:By1}) is very useful because it reveals
the behavior of the operator $Q$ near the light cone. To see this,
we consider the Fourier transform of (\ref{p:By1}) in light-cone coordinates
$(s,l,x_2, x_3)$. For the Fourier transform of the factor
$T^{(n)}(k)$, we have the
representation (\ref{p:2E}). This representation can immediately be extended
to the Fourier transform of $k \slsh\: T^{(n)}(k)$
by acting on (\ref{p:2E}) with the differential operator $i \Pdd$;
more precisely in light-cone coordinates $y-x=(s,l,x_2, x_3)$,
\begin{eqnarray}
\lefteqn{ \int \frac{d^4k}{(2 \pi)^4} \:k\slsh\:
T^{(n)}(k)
\; e^{-ik(x-y)} } \nonumber \\
&=&-\frac{1}{32 \pi^3} \:(-il)^{n-2} \int_0^\infty
\left[ il \:\gamma^s \left(\frac{1}{u^{n-1}}
\right)^{\mbox{\scriptsize{reg}}} - (n-1) \:\gamma^l \left(\frac{1}{u^{n}}
\right)^{\mbox{\scriptsize{reg}}} \right] e^{-ius} \:.\spc
    \label{p:B9a}
\end{eqnarray}
In order to treat the factors $g_{[n]}$ and $h_{[n]}$ in
(\ref{p:By1}), we note that the Fourier transform of (\ref{p:By1})
can be computed similar as described in {\S}\ref{psec24} by
integrating out the transversal momenta according to (\ref{p:2q})
and analyzing the remaining two-dimensional Fourier integral
(\ref{p:2r}) with the integration-by-parts method (\ref{p:2x}). If
this is done, the functions $g_{[n]}$ and $h_{[n]}$ appear in the
integrand of (\ref{p:2r}). Our regularity assumption on the
fermionic projector of the vacuum (see {\S}\ref{psec24} and {\S}\ref{psec25})
imply that $g_{[n]}$ and $h_{[n]}$ are smooth
functions, whose partial derivatives scale in powers of
$E_P^{-1}$. Hence all derivative terms of the functions $g_{[n]}$
and $h_{[n]}$ which arise in the integration-by-parts procedure
(\ref{p:2x}) are of higher order in $(l E_P)^{-1}$. Taking into
account only the leading order in $(l E_P)^{-1}$, we thus obtain a
representation of the fermionic projector of the vacuum involving
only $g_{[n]}$ and $h_{[n]}$ at the boundary $v=\alpha_u$.
Comparing this representation with (\ref{p:2E}) and (\ref{p:B9a}),
we conclude that the Fourier transform of (\ref{p:By1}) is
obtained, to leading order in $(l E_P)^{-1}$, simply by inserting
the functions $g_{[n]}$ and $h_{[n]}$ into the integrands of
(\ref{p:2E}) and (\ref{p:B9a}), evaluated along the line
$\vec{k}=(k_x=2u, k_y=0, k_z=0)$. Thus
\begin{eqnarray}
\lefteqn{ Q^{\mbox{\scriptsize{reg}}}(s,l) \;=\;
-\sum_{n=0}^\infty \frac{1}{n!} \:(-il)^{n-1} \int_0^\infty
\left( \frac{1}{u^n} \right)^{\mbox{\scriptsize{reg}}}
\:e^{-ius} \:h_{[n]}(u) \:du } \nonumber \\
&&\!\!\!\!\!\!-\sum_{n=0}^\infty \frac{1}{n!} \:(-il)^{n-2} \int_0^\infty
\left[ il \:\gamma^s \left( \frac{1}{u^{n-1}} \right)^{\mbox{\scriptsize{reg}}}
\!\!\!\!\!- (n-1) \:\gamma^l \left( \frac{1}{u^n} \right)^{\mbox{\scriptsize{reg}}}
\right] e^{-ius} \:g_{[n]}(u) \:du \nonumber \\
&&\!\!\!\!\!\!+\: {\mbox{(higher orders in $(l E_P)^{-1}$)}}, \label{p:B3s}
\end{eqnarray}
where $h_{[n]}(u)$ and $g_{[n]}(u)$ are the functions (\ref{p:By2}, \ref{p:By3}) with $\vec{k}=(-2u,0,0)$.

The decomposition of the operator $Q$ into Dirac eigenstates
(\ref{p:Bx}) is also useful for analyzing its perturbation $\Delta Q$.
\begin{Lemma}
\label{lemmaB1} Let $B(x) \in C^2(\R^4) \cap L^1(\R^4)$ be a
matrix potential which decays so fast at infinity that the
functions $x_i {\mathcal{B}}(x)$ and $x_i x_j {\mathcal{B}}(x)$
are also $L^1$. Then the light-cone expansion of the operator
$\Delta Q[{\mathcal{B}}]$, (\ref{p:Bp}), is obtained by
regularizing the light-cone expansion of the Dirac sea to first
order in the external potential as follows. A summand of
the light-cone expansion of the Dirac sea which is proportional to
$m^p$,
\[ m^p \:{\mbox{(iterated line integrals in bosonic potentials and
fields)}} \; T^{(n)}(s,l) \:,\]
must be replaced by
\begin{eqnarray}
\lefteqn{ (-1)\:
{\mbox{(iterated line integrals in bosonic potentials and fields)}} }
\nonumber \\
&& \times (-il)^{n-1} \int_{-\infty}^\infty du
\left(\frac{1}{u^n}\right)^{\mbox{\scriptsize{reg}}} \:e^{-ius} \times
\left\{ \begin{array}{ll} h_{[\frac{p-1}{2}]} & {\mbox{for $p$ odd}} \\
g_{[\frac{p}{2}]} & {\mbox{for $p$ even}} \end{array} \right. \nonumber \\
&&+\:{\mbox{(rapid decay in $l$)}} \:+\: {\mbox{(higher orders in $(l
E_P)^{-1}, (l_{\mbox{\scriptsize{macro}}} E_P)^{-1}$)}}. \spc\label{p:Bz6}
\end{eqnarray}
A contribution $\sim m^p$ which contains a factor $(y-x)_j \gamma^j$,
\[ m^p \;{\mbox{(iterated line integrals in bosonic potentials and fields)}}
\; (y-x)_j \gamma^j \:T^{(n)}(s,l) \:, \]
is to be replaced by
\begin{eqnarray}
\lefteqn{ (-1) \:
{\mbox{(iterated line integrals in bosonic potentials and fields)}} }
\nonumber \\
&& \times (-il)^{n-1} \int_{-\infty}^\infty du
\left[ 2l \:\gamma^s \left(\frac{1}{u^n}\right)^{\mbox{\scriptsize{reg}}}
\:+\: 2i n\:\gamma^l \left(\frac{1}{u^{n+1}}\right)^{\mbox{\scriptsize{reg}}}
\right] \nonumber \\
&& \times \:e^{-ius} \:\times
\left\{ \begin{array}{ll} h_{[\frac{p-1}{2}]} & {\mbox{for
$p$ odd}} \\
g_{[\frac{p}{2}]} & {\mbox{for $p$ even}} \end{array} \right.
\:+\: {\mbox{(contributions $\sim \gamma^2, \gamma^3$)}}
\nonumber \\
&&+\:{\mbox{(rapid decay in $l$)}} \:+\: {\mbox{(higher orders in $(l
E_P)^{-1}, (l_{\mbox{\scriptsize{macro}}} E_P)^{-1}$)}} \:.\spc
\label{p:Bz7}
\end{eqnarray}
In these formulas, $g_{[n]}$ and $h_{[n]}$ are the functions
(\ref{p:By2}, \ref{p:By3}) with $\vec{k}=(-2u,0,0)$.
\end{Lemma}
{\Proof}
By substituting (\ref{p:BC}) and (\ref{p:Bx}) into
(\ref{p:Bp}), we obtain the following representation for $\Delta Q$
in momentum space,
\begin{eqnarray}
\lefteqn{ \Delta Q[{\mathcal{B}}]
\left(k+\frac{q}{2},\:k-\frac{q}{2} \right) \;=\;
-\int_{-\infty}^\infty d\mu\: \epsilon(\mu) \; (k\slsh +
\frac{q\slsh}{2} + \mu) \:{\mathcal{B}}_q \:
(k\slsh - \frac{q\slsh}{2} + \mu) } \nonumber \\
&& \times \left( \frac{\mbox{PP}}{(k+\frac{q}{2})^2-\mu^2} \:w_\mu(\vec{k}
-\frac{\vec{q}}{2}) \:T_{\mu^2}(k-\frac{q}{2}) \right. \nonumber \\
&&\hspace*{3cm} \left. \:+\:
w_\mu(\vec{k} +\frac{\vec{q}}{2}) \:T_{\mu^2}(k+\frac{q}{2}) \:
\frac{\mbox{PP}}{(k-\frac{q}{2})^2-\mu^2} \right) \nonumber \\
&=& \int_{-\infty}^\infty d\mu\: \epsilon(\mu) \; (k\slsh +
\frac{q\slsh}{2} + \mu) \:{\mathcal{B}}_q \:
(k\slsh - \frac{q\slsh}{2} + \mu) \nonumber \\
&& \hspace*{1cm} \times \;\frac{\mbox{PP}}{2 kq} \left( w_\mu(\vec{k}
+\frac{\vec{q}}{2}) \:T_{\mu^2}(k+\frac{q}{2})
\:-\: w_\mu(\vec{k}-\frac{\vec{q}}{2}) \:T_{\mu^2}(k-\frac{q}{2}) \right)
.\qquad \label{p:BX}
\end{eqnarray}
Using the methods developed in~\cite{F4}, we now perform the
light-cone expansion in momentum space and then transform back to
position space.
Since we are here interested in the regularization effects for large energy
or momentum, we may disregard the logarithmic mass problem and work on the
level of the formal light-cone expansion of \cite[Section~3]{F4} (our
constructions could be made rigorous using the resummation method of
\cite[Section~4]{F4}). As in \cite[Section~3]{F4}, we expand the
distributions $T_{\mu^2}$ in a Taylor series in $q$ and rewrite the
resulting $k$-derivatives as derivatives with respect to $\mu^2$.
This gives
\begin{equation}
T_{\mu^2}(k \pm \frac{q}{2}) \;=\; \sum_{j,r=0}^\infty c_{jr} \:
(\pm kq)^j \left(\frac{q^2}{4}\right)^r T_{\mu^2}^{(j+r)}(k) \label{p:BU2}
\end{equation}
with combinatorial factors $c_{jr}$ whose detailed form is not needed in what
follows. Next, we expand (\ref{p:BU2}) in a Taylor series in $\mu^2$
and obtain
\begin{equation}
T_{\mu^2}(k \pm \frac{q}{2}) \;=\; \sum_{n,j,r=0}^\infty c_{njr} \:
\mu^{2n} \:(\pm kq)^j \left(\frac{q^2}{4}\right)^r T^{(n+j+r)}(k)
\label{p:BT}
\end{equation}
with new combinatorial factors $c_{njr}$. We substitute the expansions
(\ref{p:BT}) into (\ref{p:BX}) and write the even and odd terms in
$kq$ together,
\begin{eqnarray}
\lefteqn{ \Delta Q[{\mathcal{B}}]
\left(k+\frac{q}{2},\:k-\frac{q}{2} \right) \;=\;
-\int_{-\infty}^\infty d\mu\: \epsilon(\mu) \; (k\slsh +
\frac{q\slsh}{2} + \mu) \:{\mathcal{B}}_q \:
(k\slsh - \frac{q\slsh}{2} + \mu) } \nonumber \\
&&\hspace*{-.6cm} \times \left( \frac{\mbox{PP}}{2kq}
\sum_{n,j,r=0,\;{\mbox{\scriptsize{$j$ even}}}}^\infty \!\!\!\!\!c_{njr} \:
\mu^{2n} \:(kq)^j \left(\frac{q^2}{4}\right)^r T^{(n+j+r)}(k)
\left( w_\mu(\vec{k} + \frac{\vec{q}}{2}) - w_\mu(\vec{k} - \frac{\vec{q}}{2})
\right) \right. \nonumber \\
&&\hspace*{-.4cm} \left. + \frac{\mbox{PP}}{2kq}
\sum_{n,j,r=0,\;{\mbox{\scriptsize{$j$ odd}}}}^\infty \!\!\!\!\!c_{njr} \:
\mu^{2n} \:(kq)^j \left(\frac{q^2}{4}\right)^r T^{(n+j+r)}(k)
\left( w_\mu(\vec{k} + \frac{\vec{q}}{2}) + w_\mu(\vec{k} - \frac{\vec{q}}{2})
\right) \right) . \nonumber\\ \label{p:BW}
\end{eqnarray}

We first consider the contributions to (\ref{p:BW}) for even $j$. These
terms contain the factor $(w_\mu(\vec{k} + \frac{\vec{q}}{2})
- w_\mu(\vec{k} - \frac{\vec{q}}{2}))$. If the distribution
$w_\mu$ were a smooth function and its
derivatives had the natural scaling behavior in powers of the Planck length,
we could immediately conclude that $|w_\mu(\vec{k} + \frac{\vec{q}}{2})
- w_\mu(\vec{k} - \frac{\vec{q}}{2})| \sim |\vec{q}| \:|\partial w_\mu|
\sim (l_{\mbox{\scriptsize{macro}}} E_P)^{-1}$, and thus all the terms for even
$j$ would be negligible. Unfortunately, the situation is more difficult
because $w_\mu$ is in general not a smooth function (cf.\
(\ref{p:BV})), and we obtain the desired regularity in $\vec{k}$
only after the $\mu$-integration has been carried out. This makes it
necessary to use the following argument. Consider one summand in (\ref{p:BW})
for even $j$. After carrying out the $\mu$-integration, this summand yields
a finite number of contributions to $\Delta Q(k+\frac{q}{2}, k-\frac{q}{2})$
of the following form,
\begin{equation}
\frac{\mbox{PP}}{kq} \:(kq)^j \left(\frac{q^2}{4}\right)^r \cdots
\:{\mathcal{B}}_q\: \cdots T^{(n+j+r)}(k) \:\left[
g(\vec{k}+\frac{\vec{q}}{2}) - g(\vec{k}+\frac{\vec{q}}{2})
\right] \:, \label{p:BT2}
\end{equation}
where each symbol ``$\cdots$'' stands for a possible factor $k \slsh$ or
$q \slsh$, and where $g$ is a scalar function, which coincides with one of the
functions $g_{[n]}$ or $h_{[n]}$ (see (\ref{p:By2}) and (\ref{p:By3})).
As already mentioned after (\ref{p:B9a}), our regularity assumptions on the
fermionic projector of the vacuum imply that the functions $g_{[n]}$
and $h_{[n]}$, and thus also $g$, are smooth, and that their derivatives scale
in powers of the Planck length. Applying the fundamental theorem of calculus,
we rewrite the square bracket in (\ref{p:BT2}) as a line integral,
\begin{equation}
(\ref{p:BT2}) \;=\; \int_{-\frac{1}{2}}^{\frac{1}{2}} d\lambda \:
\frac{\mbox{PP}}{kq} \:(kq)^j \left(\frac{q^2}{4}\right)^r \cdots
\:{\mathcal{B}}_q\: \cdots T^{(n+j+r)}(k) \:(\vec{q}
\:\vec{\nabla}) g(\vec{k}+\lambda \vec{q})  \:. \label{p:48a}
\end{equation}
We now transform (\ref{p:48a}) to position space.  Our regularity
assumptions on ${\mathcal{B}}$ mean in momentum space that
${\mathcal{B}}(q) \in C^2 \cap L^\infty$. Using furthermore the
regularity of $\vec{\nabla}g$, we can carry out the
$q$-integration in the Fourier integral.  Carrying out also the
integral over $\lambda$, we end up with a contribution to $\Delta
Q(x,y)$ of the form
\begin{equation}
\int \frac{d^4k}{(2 \pi)^4} \:T^{(n+j+r)}(k) \; F(k, x+y)
\:e^{-ik(x-y)} \label{p:BS}
\end{equation}
with a (matrix-valued) function $F$ which is differentiable in $k$ and is
of the order $(l_{\mbox{\scriptsize{macro}}} E_P)^{-1}$. In the low-energy
region, the function $g$ in (\ref{p:BT2}) is constant and thus $F$ is
homogeneous in $k$ of degree at most $j+1$. After transforming to light-cone
coordinates, this implies that (\ref{p:BS})
is close to the light cone dominated by the fermionic projector of the
vacuum, in the sense that in light-cone coordinates, $|(\ref{p:BS})| \leq
{\mbox{const}}(l) \:|P(s,l)|$.
We conclude that all summands in (\ref{p:BW}) for even $j$
are of higher order in $(l_{\mbox{\scriptsize{macro}}} E_P)^{-1}$.

It remains to consider the summands in (\ref{p:BW}) for odd $j$. In
this case, one factor $kq$ cancels the principal value, and we obtain
\begin{eqnarray}
\lefteqn{ \Delta Q[{\mathcal{B}}]
\left(k+\frac{q}{2},\:k-\frac{q}{2} \right) \;=\;
-\int_{-\infty}^\infty d\mu\: \epsilon(\mu) \; (k\slsh +
\frac{q\slsh}{2} + \mu) \:{\mathcal{B}}_q \:
(k\slsh - \frac{q\slsh}{2} + \mu) } \nonumber \\
&& \times \sum_{n,j,r=0}^\infty C_{njr} \:
\mu^{2n} \:(kq)^{2j} \left(\frac{q^2}{4}\right)^r T^{(n+2j+1+r)}(k)
\left( w_\mu(\vec{k} + \frac{\vec{q}}{2}) + w_\mu(\vec{k} - \frac{\vec{q}}{2})
\right)\nonumber \\
&& + {\mbox{(higher orders in $(l_{\mbox{\scriptsize{macro}}} E_P)^{-1}$)}}
\label{p:BR}
\end{eqnarray}
with some combinatorial factors $C_{njr}$. This formula has similarities to
the light-cone expansion of the Dirac sea in momentum space
\cite[equation (3.15)]{F4}. In \cite[Section 3]{F4}, we proceeded by
rewriting the factors $kq$ as $k$-derivatives acting on $T^{(.)}$.
When taking the Fourier transform, these $k$-derivatives were
integrated by parts onto the exponential factor $\exp(-ik(x-y))$
to yield factors $(y-x)$. After collecting and rearranging all
resulting terms, we obtained the line-integrals of the light-cone
expansion. This method can be applied also to the integrand of
(\ref{p:BR}), and we can carry out the $\mu$-integration afterwards.
We shall not go through all these constructions steps in detail here, but
merely consider what happens in principle. Whenever a $k$-derivative
$\partial_{k^i}$ acts on the factors $w_\mu$ in the integration-by-parts
procedure, we get instead of a factor $(y-x)_i \:w_\mu$
(which is obtained when the $k$-derivative acts on the exponential
$\exp(-ik(x-y))$) a factor $\partial_i w_\mu$. After carrying out the
$\mu$-integration, one sees that the resulting term is of higher order in
$(l E_P)^{-1}$. Thus we can, to leading order in $(l E_P)^{-1}$, neglect
all derivatives of the factors $w_\mu$. But then, the integration-by-parts
procedure reduces to the construction in \cite[Section~3]{F4}, and we thus
obtain precisely the line integrals of the light-cone expansion \cite{F4}.
Furthermore, we can replace the factor
$(w_\mu(\vec{k} + \frac{\vec{q}}{2}) + w_\mu(\vec{k} - \frac{\vec{q}}{2}))$
in (\ref{p:BR}) by $2 w_\mu(\vec{k})$, because a Taylor expansion of this
factor around $\vec{q}=0$ amounts, again after carrying out the
$\mu$-integration, to an expansion in powers of $(l_{\mbox{\scriptsize{macro}}}
E_P)^{-1}$, and it thus suffices to take into account the leading term
of this expansion. These considerations show that the light-cone expansion
of (\ref{p:BR}) differs from that in \cite{F4} merely by the additional
$\mu$-integration and the factor $w_\mu(\vec{k})$. Hence the light-cone
expansion of (\ref{p:BR}) is obtained from that of the Dirac sea by
the following replacements,
\begin{eqnarray*}
m^p \:T^{(n)}(x,y) &\to& \int \frac{d^4k}{(2 \pi)^4}
\int_{-\infty}^\infty d\mu \:\epsilon(\mu)\: \mu^p \:T^{(n)}(k)
\:e^{-ik(x-y)} \; w_\mu(\vec{k}) \\
\lefteqn{ \hspace*{-2cm} m^p \:(y-x)_j \gamma^j\:T^{(n)}(x,y) } \\
&\to& \int \frac{d^4k}{(2 \pi)^4}
\int_{-\infty}^\infty d\mu \:\epsilon(\mu)\: \mu^p \:(-2i k\slsh)\:T^{(n+1)}(k)
\:e^{-ik(x-y)} \; w_\mu(\vec{k})
\end{eqnarray*}
(where we used the identity $(y-x)^i T^{(n)}(x,y) = 2 \partial_{x^i}
T^{(n+1)}(x,y)$; see \cite[equation~(3.5)]{F4}).
The lemma follows by carrying out the $\mu$-integrals applying
(\ref{p:By2}, \ref{p:By3}) and by
analyzing the behavior near the light cone as explained before (\ref{p:B3s}).

\QED
From this lemma we can deduce the light-cone expansion of the regularized
fermionic projector.
\begin{Thm}
\label{thmpB1} The light-cone expansion of the regularized causal
perturbation (\ref{p:Bq}) is obtained by regularizing the
light-cone expansion of the Dirac sea to first order in the
external potential as follows. A summand of the
light-cone expansion of the Dirac sea which is proportional to
$m^p$,
\[ m^p \:{\mbox{(iterated line integrals in bosonic potentials and
fields)}} \; T^{(n)}(s,l)\:,  \]
must be replaced by (\ref{p:2F3}). A contribution $\sim m^p$ which contains
a factor $(y-x)_j \gamma^j$,
\[ m^p \;{\mbox{(iterated line integrals in bosonic potentials and fields)}}
\; (y-x)_j \gamma^j \:T^{(n)}(s,l)\:,
\] is to be replaced by (\ref{p:2N}). In these formulas, $g$, $h$,
$a$ and $b$ are the regularization functions introduced in {\S}\ref{psec24}
and {\S}\ref{psec25} (see (\ref{p:2C}, \ref{p:27x}, \ref{p:2J}, \ref{p:214})).
\end{Thm}
{\Proof}
As mentioned at the beginning of this appendix, we assume here that the
vector component is null on the light cone (\ref{p:2Es}).
Let us consider what this condition tells us about the operators $U_l$.
According to (\ref{p:B8b}), the operators $\hat{U}_l$ are trivial in the
low-energy region. Conversely,
for large energy or momentum, (\ref{p:2Es}) yields that the vector field
$v(k)$ is parallel to $k$, up to a perturbation of the order
$\varepsilon_{\mbox{\scriptsize{shear}}}$. Hence we can assume that the
transformation (\ref{p:BQtrans}) is a small perturbation of the identity,
in the sense that
\begin{equation}
c_l\: |\hat{U}_l(k) - \1| \;\sim\; \varepsilon_{\mbox{\scriptsize{shear}}}
\spc {\mbox{for all $k$.}} \label{p:Bp2}
\end{equation}

We next derive the light-cone expansion of $\Delta P$ by transforming the
result of Lemma~\ref{lemmaB1} according to (\ref{p:B50a}). Since the
transformation (\ref{p:B50a}) is small in the sense of (\ref{p:Bp2}), it
leaves the iterated line integrals in (\ref{p:Bz6}) and (\ref{p:Bz7})
unchanged to leading order in $\varepsilon_{\mbox{\scriptsize{shear}}}$.
Hence it suffices to consider the transformation of the
$u$-integrals in (\ref{p:Bz6}) and (\ref{p:Bz7}). The $u$-integral
in (\ref{p:Bz6}) is as a homogeneous scalar operator invariant under
the unitary transformations. In the $u$-integral in (\ref{p:Bz7}), on the
other hand, only the Dirac matrices $\gamma^l$ and $\gamma^s$ are modified.
More precisely, we have to leading order in
$\varepsilon_{\mbox{\scriptsize{shear}}}$,
\begin{eqnarray*}
\sum_{l=1}^L c_l \:(\hat{U}_l \gamma^s \hat{U}_l^{-1})(u, v=\alpha_u)
&=& \gamma^s + \frac{b_1(u)}{u^2} \:\gamma^l + {\mbox{(contributions $\sim
\gamma^2, \gamma^3$)}} \\
\sum_{l=1}^L c_l \:(\hat{U}_l \gamma^l \hat{U}_l^{-1})(u, v=\alpha_u)
&=& \gamma^l + \frac{b_2(u)}{u^2} \:\gamma^s + {\mbox{(contributions $\sim
\gamma^2, \gamma^3$)}}
\end{eqnarray*}
with suitable regularization functions $b_s$ and $b_l$ which are
small in the following sense,
\[ \frac{b_{1\!/\!2}(u)}{u^2} \;\sim\; \varepsilon_{\mbox{\scriptsize{shear}}}
\:. \]
Notice that in the high-energy region $u \sim E_P$, the contribution
$\sim \gamma^l$ in the integrand of (\ref{p:Bz7}) is smaller than the
contribution $\sim \gamma^s$ by a relative factor of $(l E_P)^{-1}$.
Hence we can neglect $b_2$, whereas $b_1$ must be
taken into account. We conclude that the transformation (\ref{p:B50a})
of the contributions (\ref{p:Bz6}) and (\ref{p:Bz7}) is carried out
simply by the replacement
\begin{equation}
\gamma^s \;\to\; \gamma^s \:+\: \frac{b_1(u)}{u^2} \:\gamma^l \:.
\label{p:Bz8}
\end{equation}

It remains to derive relations between the regularization functions
$g_{[n]}$, $h_{[n]}$, and $b_s$, which appear in the transformed
contributions (\ref{p:Bz6}) and (\ref{p:Bz7}), and the regularization
functions $g$, $h$, $a$, and $b$ in (\ref{p:2F3}) and (\ref{p:2N}).
For this, we apply the transformation (\ref{p:BQtrans}) to
$Q^{\mbox{\scriptsize{reg}}}$, (\ref{p:B3s}). Exactly as described above,
this transformation reduces to the replacement (\ref{p:Bz8}), and we
obtain the following expansion of the fermionic projector near the light cone,
\begin{eqnarray*}
\lefteqn{ P^{\mbox{\scriptsize{reg}}}(s,l) \;=\;
-\sum_{n=0}^\infty \frac{1}{n!} \:(-il)^{n-1} \int_0^\infty
\left( \frac{1}{u^n} \right)^{\mbox{\scriptsize{reg}}}
\:e^{-ius} \:h_{[n]}(u) \:du } \\
&&-\sum_{n=0}^\infty \frac{1}{n!} \:(-il)^{n-2} \int_0^\infty
e^{-ius} \:g_{[n]}(u) \\
&& \qquad \times
\left[ il \:\gamma^s \left( \frac{1}{u^{n-1}} \right)^{\mbox{\scriptsize{reg}}}
- (n-1) \:\gamma^l \left( \frac{1}{u^n} \right)^{\mbox{\scriptsize{reg}}}
+ il \:\gamma^s\: b(u) \left( \frac{1}{u^{n+1}}
\right)^{\mbox{\scriptsize{reg}}}
\right] \:du \\
&&+\; {\mbox{(higher orders in $\varepsilon_{\mbox{\scriptsize{shear}}}$,
$(l E_P)^{-1}$)}}.
\end{eqnarray*}
Comparing this result with the formulas for the fermionic
projector derived in {\S}\ref{psec24} and {\S}\ref{psec25} (see
(\ref{p:27b}, \ref{p:27c}) and (\ref{p:210}, \ref{p:211})), one
gets the following identities between the regularization
functions,
\[ g_{[n]}(u) \;=\; g(u) \: a(u)^n \:,\qquad
h_{[n]}(u) \;=\; h(u) \: a(u)^n \:,\qquad
b_1(u) \;=\; b(u) \:. \]

\vspace*{-.55cm} \QED

We finally explain in which sense the regularized causal
perturbation theory is unique. In order to ensure regularity of
the perturbation theory in the limit when the momentum $q$ of the
external field goes to zero, one must satisfy a causality
condition similar to (\ref{p:Bn}), and to this end one has to work
with a modified fermionic projector $Q$. Since we must modify the
direction of the vector field $v$, it is natural to describe the
transformation from $Q$ to $P$ by linear combinations of unitary
transformations (\ref{p:BQtrans}). Nevertheless, we remark that
one could just as well work with a different or more general
transformation $Q \to P$. The reason is that the particular form
of this transformation enters only in the proof of
Theorem~\ref{thmpB1}, and we use merely that this transformation
is close to the identity, in the sense similar to (\ref{p:Bp2}).
Hence the restriction to transformations of type (\ref{p:BQtrans})
is no loss in generality. Furthermore, we point out that the gauge
symmetry (\ref{p:Bt}) uniquely determines the precise form of how
the potential ${\mathcal{B}}$ enters into the perturbation
calculation (e.g.\ one may not replace ${\mathcal{B}}$ in
(\ref{p:Bq}) by $U_l^{-1} {\mathcal{B}} U_l$). We conclude that
our construction of the regularized causal perturbation theory is
canonical up to the freedom in choosing the coefficients $c_l(k)$
and the unitary transformations $\hat{U}_l(k)$. By assuming that
the unitary transformations are regular (\ref{p:Bp1}) and small
(\ref{p:Bp2}), the arbitrariness in choosing $(c_l, \hat{U}_l)$
was constrained so much that it has no influence on the
regularization of the light-cone expansion.  Indeed, the $c_l$ and
$\hat{U}_l$ do not enter the statement of Theorem~\ref{thmpB1}.
Thus we can say that the regularized causal perturbation expansion
is unique up to contributions of higher order in $(l E_P)^{-1}$,
$(l_{\mbox{\scriptsize{macro}}} E_P)^{-1}$ and
$\varepsilon_{\mbox{\scriptsize{shear}}}$.

\chapter{Linear Independence of the Basic Fractions}
\label{pappC}\index{basic fraction!linear independence of}
\renewcommand{\theequation}{\thechapter.\arabic{equation}}
\renewcommand{\thesection}{\thechapter}
\setcounter{equation}{0} \setcounter{Def}{0}
In this appendix we consider simple fractions of degree $L \geq 2$ of
the form
\beq \label{ae1}
\frac{ T^{(a_1)}_\circ \cdots T^{(a_\alpha)}_\circ \:
\overline{T^{(b_1)}_\circ \cdots T^{(b_\beta)}_\circ} }
{ T^{(c_1)}_{[0]} \cdots T^{(c_\gamma)}_{[0]} \:
\overline{T^{(d_1)}_{[0]} \cdots T^{(d_\delta)}_{[0]} } }
\eeq
with integer parameters~$\alpha, \beta \geq 1$,
$\gamma, \delta \geq 0$ which satisfy the additional conditions
\begin{eqnarray}
c_j, d_j &\in& \{-1, 0\} \label{ae3} \\
\alpha-\gamma &>& \beta-\delta \;\geq\; 1\:. \label{ae2}
\end{eqnarray}
We prove the following theorem which makes precise that the only
relations between the simple fractions are given by the
integration-by-parts rules.
\begin{Thm} \label{thmC1}
Assume that a linear combination of simple fractions~(\ref{ae1}--\ref{ae2})
vanishes when evaluated weakly on the light cone~(\ref{p:Dwe})
to leading order in~$(l E_P)^{-1}$ and~$(l_{\mbox{\scriptsize{macro}}}
E_P)^{-1}$, for any choice of~$\eta$ and the regularization functions.
Then the linear combination is trivial after suitably applying the
integration-by-parts rules.
\end{Thm}

The condition~(\ref{ae2}) ensures that the simple fractions are
asymmetric under complex conjugations. Such an asymmetry
is essential for our proof. However, (\ref{ae2}) could easily be
weakened or replaced by other asymmetry conditions. Also, (\ref{ae3}) and
the fact that the denominator involves only the square indices~$[0]$
is mainly a matter of convenience. The reason why we are content
with~(\ref{ae1}--\ref{ae2}) is that all EL equations in this book
can be expressed in terms of simple fractions of this form.

We point out that the above theorem does not imply that the
basic monomials are independent in the sense that, by choosing
suitable regularization functions, the basic regularization
parameters can be given arbitrary values. Theorem~\ref{thmC1}
states that there are no identities between the basic fractions,
but the basic regularization parameters might nevertheless be
constrained by inequalities between them (e.g.\ certain
regularization parameters might be always positive). Furthermore,
we remind that the assumptions of positivity of the scalar
component and of half occupied surface states (see the last
paragraph of {\S}\ref{psec25}) yield relations between the
regularization functions which might give additional constraints
for the regularization parameters. For these reasons, one should
in applications always verify that the values for the basic
regularization parameters obtained in the effective continuum
theory can actually be realized by suitable regularization
functions.

In the proof we will work with a class of regularization
functions for which the Fourier integrals and the weak evaluation
integral can be computed explicitly; then we will analyze in detail how
the resulting formulas depend on the regularization.
More precisely, we choose the regularization functions in~(\ref{p:D1},
\ref{p:D3}) as follows,
\begin{eqnarray}
g(u) &=& u^{\sigma-1}\: (1 + \varepsilon\: u^\nu)
\: e^{-\frac{u}{2 E_P}} \:\Theta(u)
\:,
\quad h(u) \;=\; u^{4\nu}\: g(u) \label{ae4} \\
a(u) &=& u^{8 \nu} \:,
\spc\spc\spc\spc\,\,b(u) \;=\; u^{2\nu} \label{ae6}
\end{eqnarray}
with real parameters~$\varepsilon, \sigma, \nu, E_P >0$. These regularization functions have all the properties required in~{\S}\ref{jsec5} if
$\sigma \approx 1$ and $\varepsilon \ll \nu \ll 1$; note that
the factor~$e^{-\frac{u}{2 E_P}}$
gives the desired decay on the scale of the Planck energy. Using the
decay of the integrand for large ${\mbox{Re}}\, u$, we can deform the
integration contours to obtain for any~$\rho >0$,
\beq \label{E6a}
\int_0^\infty u^{\rho-1} \:e^{-\frac{u}{2 E_P}}\: e^{-i u s}\:du
\;=\; \left(i s + \frac{1}{2 E_P} \right)^{-\rho}
\int_0^\infty v^{\rho-1} \:e^{-v}\:dv
\;=\; \frac{\Gamma(\rho)}{z^\rho} \:,
\eeq
where in the last step we set
\[ z \;=\; is + \frac{1}{2 E_P} \]
and used the definition of the gamma function.
Here the power~$z^{-\rho}$ is understood as
$\exp(-\rho \log z)$ with the logarithm defined on the complex
plane cut along the negative real axis.
By analytic continuation we we can extend~(\ref{E6a})
to~$\rho$ in the complex plane with the exception of the
poles of the gamma function,
\beq \label{aef}
\int_0^\infty u^{\rho-1} \:e^{-\frac{u}{2 E_P}}\: e^{-i u s}\:du
\;=\; \frac{\Gamma(\rho)}{z^\rho}
\:,\spc {\mbox{for $\rho \in \C \setminus \{0, -1, -2, \ldots\}$}}.
\eeq
This Fourier integral is also useful for computing
the $L^2$-scalar product of the Fourier transform via Plancherel.
Namely, under the conditions $\rho, \rho' > \frac{1}{2}$ we obtain
\[ \int_0^\infty u^{\rho + \rho' - 2} e^{-E_P u} du \;=\;
\frac{1}{2 \pi} \int_{-\infty}^\infty
\frac{\Gamma(\rho)}{z^\rho}\:
\frac{\Gamma(\rho')}{\overline{z}^{\rho'}} \:ds \:, \]
and computing the integral on the left gives
\beq \label{aeg}
\int_{-\infty}^\infty
\frac{1}{z^\rho \:\overline{z}^{\rho'}} \:ds
\;=\; 2 \pi \:E_P^{\rho + \rho' - 1}\:
\frac{\Gamma(\rho+\rho'-1)}{\Gamma(\rho)\: \Gamma(\rho')}
\spc {\mbox{for $\rho, \rho'> \frac{1}{2}$}}.
\eeq

The fact that the integral~(\ref{aef}) diverges when~$\rho$ tends to a
negative integer corresponds precisely to the logarithmic mass problem as
discussed after~(\ref{p:2D}). A short calculation shows that
the infrared regularization can be introduced here simply by subtracting
the pole, i.e. for $n \in \N_0$
\beq \label{aecntr}
\int_0^\infty \left(u^{-n-1}\right)^{\mbox{\scriptsize{reg}}}
 \:e^{-\frac{u}{2 E_P}}\: e^{-i u s}\:du
\;=\; \lim_{\rho \to -n} \left(
\frac{\Gamma(\rho)}{z^\rho} - \frac{(-1)^n}{n!} \:\frac{z^n}{\rho+n}
\right) .
\eeq
Using this formula in~(\ref{p:D1}, \ref{p:D3}), we obtain
\begin{eqnarray}
T^{(n)}_{[p]} &=& -(-il)^{n-1} \sum_{k=0,1}
\varepsilon^k\:
\frac{\Gamma(\sigma -n + (4p+k) \nu)}
{z^{\sigma -n + (4p+k) \nu}} \:-\: {\mbox{(IR-reg)}} \label{E8a} \\
T^{(n)}_{\{p\}} &=& -(-il)^{n-1} \sum_{k=0,1}
\varepsilon^k \:
\frac{\Gamma(\sigma -n + (4p+k+2) \nu)}
{z^{\sigma -n + (4p+k+2) \nu}} \:-\: {\mbox{(IR-reg)}} \:, \label{E8b}
\end{eqnarray}
where ``(IR-reg)'' means that we subtract a counter term as in~(\ref{aecntr}).

For clarity we disregard the infrared regularization for the moment and
consider only the zeroth order in~$\varepsilon$.
Substituting the obtained formulas for~$T^{(n)}_\circ$
and their complex conjugates into~(\ref{ae1}), we obtain
\begin{eqnarray}
\hspace*{-0.5cm}(\ref{ae1}) &=& \frac{(-1)^{A-C+\beta-\delta}}{(-il)^L}\;
\frac{ \Gamma(\sigma-a_1+\circ \nu) \cdots \Gamma(\sigma-a_\alpha+\circ \nu) }
{ \Gamma(\sigma-c_1) \cdots \Gamma(\sigma-c_\gamma) } \;
\frac{1}{z^{(\alpha - \gamma) \sigma - A + C + \bullet \nu}}  \nonumber \\
&& \spc\,\,\! \times \frac{ \Gamma(\sigma-b_1+\circ \nu) \cdots \Gamma(\sigma-b_\beta+\circ \nu) }
{ \Gamma(\sigma-d_1) \cdots \Gamma(\sigma-d_\delta) } \;
\frac{1}{\overline{z}^{(\beta - \delta) \sigma - B + D + \bullet \nu}} \:,\label{ag0}
\end{eqnarray}
where~$A=\sum_{j=1}^\alpha a_j$, $B=\sum_{j=1}^\beta b_j$,
$C=\sum_{j=1}^\gamma c_j$, $D=\sum_{j=1}^\delta d_j$. Here the
parameter~$\circ$ takes into account the lower indices of the
corresponding factors~$T^{(a_j)}_\circ$ or~$\overline{T^{(b_j)}_\circ}$;
more precisely for an index~$[p]$ and~$\{p\}$ it is equal to~$4 p$
and~$4p+2$, respectively. The parameter $\bullet$ stands
for the sum of the parameters~$\circ$ in the same line.
Integrating over~$s$,
\beq \label{agw}
\int_{-\infty}^\infty
\frac{ T^{(a_1)}_\circ \cdots T^{(a_\alpha)}_\circ \:
\overline{T^{(b_1)}_\circ \cdots T^{(b_\beta)}_\circ} }
{ T^{(c_1)}_{[0]} \cdots T^{(c_\gamma)}_{[0]} \:
\overline{T^{(d_1)}_{[0]} \cdots T^{(d_\delta)}_{[0]} } }
\:ds \:,
\eeq
using~(\ref{aeg}) and leaving out irrelevant prefactors, we obtain
the expression
\begin{eqnarray}
\lefteqn{ \frac{ \Gamma(\sigma-a_1+\circ \nu) \cdots \Gamma(\sigma-a_\alpha+\circ \nu)
\; \Gamma(\sigma-b_1+\circ \nu) \cdots \Gamma(\sigma-b_\beta+\circ \nu) }
{ \Gamma(\sigma-c_1) \cdots \Gamma(\sigma-c_\gamma)
\; \Gamma(\sigma-d_1) \cdots \Gamma(\sigma-d_\delta) } } \nonumber \\
&&\spc\:\times \frac{E_P^{\lambda-1}\: \Gamma(\lambda-1)}
{ \Gamma((\alpha - \gamma) \sigma - A + C + \bullet \nu)
\; \Gamma((\beta - \delta) \sigma - B + D + \bullet \nu) }\:, \label{ag1}
\end{eqnarray}
where~$\lambda$ is the sum of the arguments of the two gamma functions
in the denominator of the second line. Since the~$E_P$-dependence tells
us about~$\lambda$, we are led to a combination of gamma functions as
considered in the next lemma. Although the statement of the lemma is not
surprising, the proof is a bit delicate, and we give it in detail.
\begin{Lemma} \label{lemmaae}
Consider for given parameters~$N, M \in \N$ quotients of gamma functions
of the form
\beq \label{al1}
\frac{ \Gamma(\sigma-a_1+\nu b_1) \cdots \Gamma(\sigma-a_J+\nu b_J) }
{ \Gamma(\sigma-c_1+\nu d_1) \cdots \Gamma(\sigma-c_K+\nu d_K) } \;
\frac{1}{\Gamma(n_1 \sigma - l_1 + \nu m_1)\:
\Gamma(n_2 \sigma - l_2 + \nu m_2) }
\eeq
with integers~$J, K \geq 0$ and~$a_j, b_j, c_j, d_j, n_j, l_j, m_j
\in \Z$, which satisfy the relations
\begin{eqnarray}
n_1 + n_2 &=& N \:,\spc n_1 \;>\; n_2 \;\geq\; 1 \label{al2} \\
m_1 + m_2 &=& M\:. \label{al3}
\end{eqnarray}
If a linear combination of expressions of the form~(\ref{al1}--\ref{al3})
vanishes for all~$(\sigma, \nu)$ in an open set of~$\R^2$, then the linear
combination is trivial after suitably applying the identity
\beq \label{al4}
x \:\Gamma(x) \;=\; \Gamma(x+1)\:.
\eeq
\end{Lemma}
{\Proof} Assume that a linear combination of terms of the
form~(\ref{al1}--\ref{al3}) vanishes for all~$(\sigma, \nu)$ in an open set of~$\R^2$. By analytic continuation we can assume that the linear combination vanishes for~$\sigma$ and~$\mu$ in the whole complex plane with the exception of the poles of the gamma functions.

We first consider the asymptotics for large~$\sigma$ and~$\nu$.
If we fix~$\nu$ and choose~$\sigma$ large, we can approximate the
gamma functions with the Stirling formula
\beq \label{stirling}
\Gamma(x) \;=\; \sqrt{2 \pi} \;x^{x+\frac{1}{2}} \: e^{-x} \left( 1 +
{\mathcal{O}}(x^{-1}) \right)
\eeq
to obtain
\[ \log (\ref{al1}) \;=\; (J-K-n_1-n_2)\: \sigma \:(\log(\sigma)-1) -
\sum_{i=1,2} n_i \log(n_i)\:\sigma \:+\: {\mathcal{O}}(\log \sigma)\: . \]
Terms with a different asymptotics cannot compensate each other in the
linear combination and must therefore vanish separately. Thus we can
restrict attention to a linear combination with fixed values of the parameters
\beq
J-K-n_1-n_2 \spc {\mbox{and}} \spc
\sum_{i=1,2} n_i \log(n_i) \:. \label{aeasy1}
\eeq
More generally, we can choose~$\nu=\varepsilon \sigma$ for small
fixed~$\varepsilon \geq 0$. Then for large~$\sigma$,
\begin{eqnarray*}
\log (\ref{al1}) &=& \left( \sum_{j=1}^J (1 + \varepsilon b_i) -
\sum_{k=1}^K (1 + \varepsilon d_k) \:-\: \sum_{i=1,2} (n_i + \varepsilon m_i)
\right) \sigma \:(\log(\sigma)-1) \\
&& + \sum_{j=1}^J (1 + \varepsilon b_j)\: \log(1+\varepsilon b_j)\:\sigma
- \sum_{k=1}^K (1 + \varepsilon d_k)\: \log(1+ \varepsilon d_k)\:\sigma \\
&& - \sum_{i=1,2} (n_i+\varepsilon m_i)\: \log(n_i + \varepsilon m_i)\:\sigma
\:+\: {\mathcal{O}}(\log \sigma)\: .
\end{eqnarray*}
Expanding in powers of~$\varepsilon$, we see that the asymptotics
also determines the parameter
\beq \sum_{i=1,2} m_i\: \log(n_i) \label{aeasy2} \:.
\eeq
We can assume that the parameters~(\ref{aeasy1}, \ref{aeasy2})
are the same for all summands of our linear combination.
Combining~(\ref{al2}) with the right of~(\ref{aeasy1}), we can compute~$n_1$ and~$n_2$. Furthermore, (\ref{al3}) and~(\ref{aeasy2}) uniquely
determine~$m_1$ and~$m_2$.

By iteratively applying~(\ref{al4}) we can write each term~(\ref{al1}) as
\[ \frac{{\mathcal{P}}(\sigma, \nu)}{{\mathcal{Q}}(\sigma, \nu)} \;
\frac{ \Gamma(\sigma+\nu b_1) \cdots \Gamma(\sigma+\nu b_J) }
{ \Gamma(\sigma+\nu d_1) \cdots \Gamma(\sigma+\nu d_K) } \;
\frac{1}{\Gamma(n_1 \sigma + \nu m_1)\:
\Gamma(n_2 \sigma + \nu m_2) } \:, \]
where~${\mathcal{P}}$ and~${\mathcal{Q}}$ are polynomials in~$\sigma$
and~$\nu$ (which clearly depend also on all the integer parameters).
After bringing the summands of the linear combination on a common
denominator, the numerator must vanish identically. Thus it suffices
to consider a sum of expressions of the form
\beq \label{af3}
{\mathcal{P}}(\sigma, \nu) \;
\Gamma(n_1 \sigma + \nu m_1)\: \Gamma(n_2 \sigma + \nu m_2)
\prod_{l=1}^{L-2} \Gamma(\sigma + \nu b_l)
\eeq
with new polynomials~${\mathcal{P}}$ and parameters~$L$, $b_l$.
According to the left of~(\ref{aeasy1}), the parameter~$L$ is
the same for all summands. We denote the number of such
summands by~$P$.
For notational convenience we also write~(\ref{af3}) in the
more general form
\beq \label{af4}
{\mathcal{P}}(\sigma, \nu) \;
\prod_{l=1}^L \Gamma(n_l \sigma + \nu a_l)
\eeq
with new parameters~$n_l$ and~$a_l$.
We evaluate~(\ref{af4}) at successive points~$\sigma+p$
with~$p=0,\ldots, P-1$. Again using~(\ref{al4}), we obtain the expression
\[ {\mathcal{P}}(\sigma+p, \nu)\:
\prod_{l=1}^L \left(
\prod_{q=0}^{n_l p-1} \;(n_l \sigma + q + \nu a_l) \right)
\Gamma(n_l \sigma + \nu a_l)\:. \]
We now consider the asymptotic regime
\[ \sigma \;\gg\; \nu \;\gg\;  P \:. \]
Then~(\ref{af4}) simplifies to
\[ {\mathcal{P}}(\sigma, \nu)
\prod_{l=1}^L \;
(n_l \sigma + \nu a_l)^{n_l p}\;
\Gamma(n_l \sigma + \nu a_l) \left( 1 + {\mathcal{O}}(p \:\sigma^{-1})
+ {\mathcal{O}}(p \:\nu^{-1})
\right) . \]
It is convenient to divide by~$\prod_{l=1}^L (n_l \sigma)^{n_l p}$ (this is
possible in view of the fact that the parameters~$n_i$ and~$L$
in~(\ref{af3}) are known). This gives
\[ {\mathcal{P}}(\sigma, \nu)
\prod_{l=1}^L \;
\left[ 1 + \frac{\nu a_l}{\sigma n_l} \right]^{n_l p}\;
\Gamma(n_l \sigma + \nu a_l) \left( 1 + {\mathcal{O}}(p \:\sigma^{-1})
+ {\mathcal{O}}(p \:\nu^{-1})
\right) . \]
Since the parameters~$n_i$ and~$m_i$ in~(\ref{af3}) are also known,
we can finally divide by the factors for~$l=1,2$ to obtain the expressions
\[ F_p(\sigma, \nu) \;:=\; {\mathcal{P}}(\sigma, \nu) \prod_{l=1}^{L-2} \;
\left[ 1 + b_l\:\frac{\nu}{\sigma} \right]^{p}\;
\Gamma(n_l \sigma + \nu b_l)\:. \]
These functions satisfy the simple relations
\beq \label{af6}
F_p(\sigma, \nu) \;=\; F_0(\sigma, \nu) \:G\!\left(
\frac{\nu}{\sigma} \right)^p
\eeq
with
\beq \label{af7}
G(\lambda) \;:=\; \prod_{l=1}^{L-2}
\left[ 1 + \lambda\: b_l \right]\:.
\eeq

Let us verify that the function~$G(\lambda)$
determines all the parameters~$b_l$ in~(\ref{af3}): Assume that the
functions~$G$ and~$\tilde{G}$ corresponding to two choices of the
parameters~$b_l$ are equal. Collecting and counting common factors,
their quotient~$G(\lambda)/\tilde{G}(\lambda)$ can be written as
\[ \frac{G}{\tilde{G}}
\;=\; \prod_{i=1}^I \left[ 1 + \lambda\: b_i \right]^{q_i} \]
for some parameters~$q_i \in \Z$ satisfying the conditions
\beq \label{aecon}
\sum_{i=1}^I q_i \;=\; 0 \:.
\eeq
Here the~$b_i$ are (with a slight abuse of notation) a selection of the
parameters~$b_l$ and~$\tilde{b}_l$, which are all different from each
other, and~$I$ denotes the number of such parameters.
We must show that the powers~$q_i$ are all zero.
To this end we take the logarithm,
\[ \log G - \log {\tilde{G}} \;=\; \sum_{i=1}^I q_i
\log \left( 1 + \lambda\: b_i \right) . \]
Expanding in powers of~$\lambda$ up to the order~$I$, we obtain the conditions
\[ \sum_{i=1}^I q_i \:b_i^l \;=\; 0
\spc {\mbox{for all~$l=1,\ldots, I$.}} \]
We write these equations in the matrix form
\beq \label{aematrix}
A \Psi = 0
\eeq
with
\[ A \;=\; \left( \begin{array}{cccc}
1 & 1 & \cdots & 1 \\ b_1 & b_2 & \cdots & b_I \\
b_1^2 & b_2^2 & \cdots & b_I^2 \\
\vdots & \vdots & \ddots & \vdots \\
b_1^{I-1} & b_2^{I-1} & \cdots & b_I^{I-1}
\end{array} \right), \spc
\Psi \;=\; \left( \begin{array}{c}
q_1 b_1 \\ q_2 b_2 \\ \vdots \\ q_I b_I
\end{array} \right) . \]
An elementary consideration shows that
\beq \label{aeDet}
\det A \;=\; \prod_{1 \leq i<j \leq I} \:(b_j - b_i) \;\neq\; 0
\eeq
because the $b_j$ are all different. We conclude that the matrix~$A$
is invertible and thus~$\Psi=0$. Hence all the powers~$q_i$ vanish
whenever~$b_i \neq 0$. In the remaining case~$b_i = 0$ the corresponding
power~$q_i$ is zero because of~(\ref{aecon}).

We now return to~(\ref{af6}). By assumption the sum of the
functions~$F_p(\sigma, \nu)$ vanishes,
\[ 0 \;=\; \sum_{\alpha=1}^P F^{(\alpha)}_p(\sigma, \nu)
\spc {\mbox{for $p=0,\ldots, P-1$}}, \]
where the index~$(\alpha)$ labels the summands of the linear combination.
Using~(\ref{af6}) and keeping~$\sigma$ fixed,
we can again write these equations in matrix form~(\ref{aematrix}) with
\[ A(\nu) \;=\; \left( \begin{array}{cccc}
1 & 1 & \cdots & 1 \\ G_{(1)} & G_{(2)} & \cdots & G_{(P)} \\
G_{(1)}^2 & G_{(2)}^2 & \cdots & G_{(P)}^2 \\
\vdots & \vdots & \ddots & \vdots \\
G_{(1)}^{P-1} & G_{(2)}^{P-1} & \cdots & G_{(P)}^{P-1}
\end{array} \right), \spc
\Psi(\nu) \;=\; \left( \begin{array}{c}
F^{(1)}_0 \\ F^{(2)}_0 \\ \vdots \\ F^{(P)}_0
\end{array} \right) . \]
Suppose that~$\Psi(\nu_0) \neq 0$. Then there is~$\varepsilon>0$ such that~$\Psi(\nu) \neq 0$ for
all~$\nu \in B_\varepsilon(\nu_0)$.
Computing the determinant of~$A$
again using the formula~(\ref{aeDet}) we conclude that for each
$\nu \in B_\varepsilon(\nu_0)$, at least two of the functions
$G_{(\alpha)}$ coincide. Since there is only a finite number of
combinations to choose the indices, there must be
two indices~$(\alpha) \neq (\beta)$ such that the function~$G^{(\alpha)}-G^{(\beta)}$ has an infinite number of zeros on~$B_\varepsilon(\nu_0)B_\varepsilon(\nu_0)$. Due to analyticity, it  follows that~$G^{(\alpha)} \equiv G^{(\beta)}$,
in contradiction to our above result that the functions~$G^{(\alpha)}$
are all different. We conclude that~$\Psi(\nu_0)=0$ and thus
\[ F^{(\alpha)}_0(\sigma, \nu) \;=\; 0
\spc {\mbox{for all~$\alpha=1,\ldots P$}}\:. \]
This means that the terms~(\ref{af3}) all vanish identically.
\QED
{\em{Proof of Theorem~\ref{thmC1}. }} Consider a linear
combination~${\mathfrak{L}}$ of simple fractions which satisfies the assumptions of the theorem. We regularize according
to~(\ref{ae4}, \ref{ae6}, \ref{E8a}, \ref{E8b}), expand in powers
of~$\varepsilon$ and evaluate weakly on the light cone~(\ref{agw}).
This gives a series of terms of the general form~(\ref{ag1})
(note that due to~(\ref{ae3}) no infrared regularization is necessary in the denominator, and so the counter terms appear only in the numerator).
The scaling in~$E_P$ distinguishes between the contributions with
different values of~$\lambda$. For every basic fractions (\ref{ae1})
we introduce the parameter
\beq \label{eap2}
N = \alpha - \gamma + \beta - \delta
\eeq
and let $N_{\mbox{\scriptsize{max}}}$
be the maximum which this parameter attains
for the basic fractions in~${\mathfrak{L}}$. We restrict attention
to those contributions~(\ref{ag1}) where~$\lambda$ can be written as
\beq \label{eapar}
\lambda \;=\; N \sigma - l + m \nu
\eeq
with integer parameters~$l, m$. These contributions all come from those simple
fractions for which the parameter~(\ref{eap2}) is equal to~$N$.
Furthermore, since the counter terms
in~(\ref{E8a}, \ref{E8b}) involve no factor~$z^{-\sigma}$ (see~(\ref{aecntr})),
they contribute only for~$\lambda = n \sigma + \cdots$ with~$n<N_{\mbox{\scriptsize{max}}}$
and thus do not show up in our analysis.
By considering the contributions~(\ref{ag1}) with~$\lambda$ of the
form~(\ref{eapar}), we will show
that all simple fractions with~$N=N_{\mbox{\scriptsize{max}}}$ vanish after suitably
applying the integration-by-parts rules. Then the corresponding counter
terms also drop out, because the infrared regularization is compatible
with the integration-by-parts rules. Hence these simple fractions
completely drop out of~${\mathfrak{L}}$, and we can proceed
inductively to the analysis of the simple fractions with $N<N_{\mbox{\scriptsize{max}}}$.
This argument allows us to completely ignore the infrared
regularization in what follows.

Since the summands with different value of~$\lambda$ scale differently in~$E_P$, we can assume that the parameter~$\lambda$ is the same for all
simple fractions in~$\mathfrak{L}$. Dividing~(\ref{ag1})
through~$E_P^{\lambda-1}\: \Gamma(\lambda-1)$
and using~(\ref{ae2}), we obtain terms which are
precisely of the form as considered in Lemma~\ref{lemmaae}.
Therefore, our representation of~$\mathfrak{L}$ as a linear combination
of quotients of gamma functions~(\ref{ag1}) is unique up to
applying~(\ref{al4}). We will consider this arbitrariness later
and for the moment consider a fixed choice of summands of the
form~(\ref{ag1}).

Our goal is to get a one-to-one connection between
quotients of gamma functions of the form~(\ref{ag1}) and our original
simple fractions. Unfortunately, to zeroth order in~$\varepsilon$
one cannot reconstruct the simple fraction from the expression~(\ref{ag1}),
because~(\ref{ag1}) is symmetric in the parameters~$a_j$ and~$b_k$,
and thus it is impossible to tell which of these parameters
came from a factor~$T^{(.)}_\circ$ or~$\overline{T^{(.)}_\circ}$.
This is the reason why we need to consider the higher orders in~$\varepsilon$
as well. Out of the many terms of the general form~(\ref{agw})
we select a few terms according to the following rules,
which we apply one after the other:
\begin{enumerate}
\item[(i)] No factors~$\Gamma(\sigma + \nu)$ or~$\Gamma(\sigma+1+\nu)$
appear.
\item[(ii)] For the factor~$\Gamma((\alpha-\gamma) \sigma - n + m \nu)$ the
parameter~$m$ is maximal.
\item[(iii)] The number of factors~$\Gamma(\sigma - n + (2m+1) \nu)$ in the
numerator is minimal.
\end{enumerate}
By ``maximal'' (and similarly ``minimal'') we mean that there is no
summand for which the corresponding parameter is larger (no matter how
all other parameters look like).
Note that the factor~$\Gamma((\alpha-\gamma) \sigma - n + m \nu)$
in~(\ref{ag1}) is uniquely determined because of~(\ref{ae2}).

Assuming that~$\mathfrak{L}$ is non-trivial, the above procedure gives
us at least one term of the form~(\ref{agw}) (note that the zero-order term in~$\varepsilon$ clearly satisfies (i)). The point is that we can uniquely construct from this term a corresponding simple fraction from~${\mathfrak{L}}$,
as the following consideration shows.
According to~(i), the factors~$T^{(-1)}_{[0]}$ and~$T^{(0)}_{[0]}$
are taken into account only to lowest order in~$\varepsilon$, because
otherwise a factor of the form~$\Gamma(\sigma - n + \nu)$, $n=-1,0$,
 would appear.
In particular, one sees from~(\ref{ae3}) that we do not get
$\varepsilon$-terms of the denominator. Hence the higher orders in~$\varepsilon$
are obtained simply by expanding the numerator in~(\ref{ae1}) in powers
of~$\varepsilon$. With the rule~(ii) we arranged that {\em{all}} factors~$T^{(n)}_\circ$ with $n>0$ or $\circ \neq [0]$ are taken into account linearly in~$\varepsilon$. On the other hand, (iii)
ensures that {\em{none}} of the factors~$\overline{T^{(n)}_\circ}$
is taken into account linearly in~$\varepsilon$. Therefore, all gamma
functions in the numerator whose argument contains
an odd number times~$\nu$ belong to a factor~$T^{(n)}_\circ$.
Conversely, the gamma functions of the form~$\Gamma(\sigma-n+2 m \nu)$
belong to a factor~$\overline{T^{(n)}_\circ}$, at least when~$n > 0$
or~$\circ \neq [0]$.
In this way, the gamma functions determine the simple fraction up to
factors of~$T^{(-1)}_{[0]}$, $T^{(0)}_{[0]}$ and~$\overline{T^{(-1)}_{[0]}}$,
$\overline{T^{(0)}_{[0]}}$. But the factors~$T^{(-1)}_{[0]}$
and~$T^{(0)}_{[0]}$ can easily be determined from the argument of the factor~$\Gamma((\alpha-\gamma) \sigma - n + m \nu)$, because~$\alpha-\gamma$ gives us how many factors~$T^{(.)}_{[0]}$
we muse use, whereas~$n$ tells us about how many factors~$T^{(-1)}_{[0]}$ we
must use. Since~$\lambda$ is known, we also know the arguments of the
factor~$\Gamma((\beta - \delta) \sigma - B + D + \bullet \nu)$
in~(\ref{agw}), and this determines in turn
the factors~$\overline{T^{(-1)}_{[0]}}$ and~$\overline{T^{(0)}_{[0]}}$.

We conclude that the above construction allows us to determine
one summand of~$\mathfrak{L}$. Subtracting this summand,
we can proceed iteratively to determine all other summands of~$\mathfrak{L}$.
This construction is unique up to the transformation
of the gamma functions with~(\ref{al4}).

We conclude the proof by establishing a one-to-one correspondence
between the transformation~(\ref{al4}) of the gamma functions and the
integration-by-parts rule for the simple fraction. To every
simple fraction~(\ref{ae1}) we can associate a contribution of
the form~(\ref{ag1}) which satisfies the rules~(i)--(iii) with the following
symbolic replacements,
\begin{eqnarray*}
T^{(-1)}_{[0]}, \overline{T^{(-1)}_{[0]}} &\longrightarrow& \Gamma(\sigma+1)
\:,\spc\spc\;\:
T^{(0)}_{[0]}, \overline{T^{(0)}_{[0]}} \;\longrightarrow\; \Gamma(\sigma) \\
\lefteqn{ \hspace*{-.94cm} \left. \begin{array}{rcl}
T^{(n)}_{[p]} &\longrightarrow& \Gamma(\sigma - n + (4p+1) \nu) \\[.5em]
\overline{T^{(n)}_{[p]}} &\longrightarrow& \Gamma(\sigma - n + 4p \nu)
\end{array} \right\}
\quad {\mbox{(if~$n>0$ or~$p>0$)}} } \\
T^{(n)}_{\{p\}} &\longrightarrow& \Gamma(\sigma - n + (4p+3) \nu) \:,\spc
\overline{T^{(n)}_{\{p\}}} \;\longrightarrow\;
\Gamma(\sigma - n + (4p+2) \nu) \:.
\end{eqnarray*}
These replacement rules determine the first line in~(\ref{ag1}), whereas
the arguments of the gamma functions in the second line are obtained
as explained above by adding the arguments of the gamma functions
in the first line. If~(\ref{al4}) is applied to the gamma functions in the
first line of~(\ref{ag1}),
\[ \Gamma(\sigma - n + \circ \nu) \;\longrightarrow\;
(\sigma - (n+1) + \circ \nu)\: \Gamma(\sigma - (n+1) + \circ \nu) \:, \]
we take this into account with the following symbolic transformation
inside the simple fraction,
\[ T^{(n)}_\circ \;\longrightarrow\; \nabla T^{(n+1)}_\circ \:. \]
Here~$\nabla$ is the derivation as introduced in~(\ref{sfinte}).
Using the Leibniz rule this correspondence can be extended to composite
expressions; for example,
\begin{eqnarray*}
\lefteqn{\nabla \left( \frac{1}{T^{(n)}_\circ} \right) \;=\;
- \frac{\nabla T^{(n)}_\circ}{(T^{(n)}_\circ)^2} } \\
&\longleftrightarrow&
-\frac{(\sigma - n + \circ \sigma)\: \Gamma(\sigma - n + \circ \sigma)}
{\Gamma(\sigma - n + \circ \sigma)^2} \;=\;
-\frac{(\sigma - n + \circ \sigma)}{\Gamma(\sigma - n + \circ \sigma)} \\
\lefteqn{ \nabla(T^{(n_1)}_{\circ_1} T^{(n_2)}_{\circ_2}) \;=\;
\nabla(T^{(n_1)}_{\circ_1})\: T^{(n_2)}_{\circ_2} +
T^{(n_1)}_{\circ_1} \:\nabla(T^{(n_2)}_{\circ_2}) } \\
&\longleftrightarrow&
(2 \sigma - (n_1+n_2-2) + (\circ_1+\circ_2) \nu) \;
\Gamma(\sigma - n_1 + \circ_1 \nu) \:\Gamma(\sigma - n_1 + \circ_1 \nu)
\end{eqnarray*}
and similarly for other composite expressions.
It remains to consider the transformations of the gamma functions in the
second line of~(\ref{ag1}). Since~$\lambda$ is fixed, we can only
increment the argument of one gamma function in the denominator and at
the same time decrement the argument of the other, for example
\begin{eqnarray*}
\lefteqn{ \frac{(\alpha - \gamma) \sigma - (n+1) + \bullet_1 \nu}
{ \Gamma((\alpha - \gamma) \sigma - n + \bullet_1 \nu)
\; \Gamma((\beta - \delta) \sigma - m + \bullet_2 \nu) } } \\
&\longrightarrow&
\frac{(\beta - \delta) \sigma - m + \bullet_2 \nu)}
{ \Gamma((\alpha - \gamma) \sigma - (n+1) + \bullet_1 \nu)
\; \Gamma((\beta - \delta) \sigma - (m-1) + \bullet_2 \nu) } \:.
\end{eqnarray*}
This transformation can be related to the integration by parts rule
\[ \int_{-\infty}^\infty \nabla (\cdots) \;\overline{(\cdots)} \:ds \;\longrightarrow\;
- \int_{-\infty}^\infty (\cdots) \;\nabla \overline{(\cdots)}\:ds
\:, \] where~$(\cdots)$ and~$\overline{(\cdots)}$ stand for simple
fractions composed of~$T^{(n)}_\circ$
and~$\overline{T^{(n)}_\circ}$, respectively. As is easily verified,
these replacements rules are all compatible with each other and with
the Leibniz rule. They allow us to identify the
transformation~(\ref{al4}) with the integration-by-parts rules.
\QED

\chapter{The Commutator $[P,Q]$}
\renewcommand{\theequation}{\thechapter.\arabic{equation}}
\renewcommand{\thesection}{\thechapter}
\setcounter{equation}{0} \setcounter{Def}{0} \label{appD}
The Euler-Lagrange equations corresponding to our variational principles
involve the commutator $[P,Q]$~(\ref{e:2d}), where $Q$ is a composite expression
in the fermionic projector. In Chapter~\ref{psec2} we developed a method
with which composite expressions in the fermionic projector can be evaluated
weakly on the light cone. In this appendix we shall describe in detail
how these methods can be used to evaluate the commutator~$[P,Q]$
in the continuum limit. We begin by
collecting a few formulas from~{\S}\ref{psec26} and bring them into
a form convenient for what follows.
 The kernel $Q(x,y)$ can be written as a
linear combination of terms of the form (\ref{p:D11})
\begin{equation}
f(x,y) \; \frac{ T^{(a_1)}_\circ \cdots T^{(a_\alpha)}_\circ \:
\overline{T^{(b_1)}_\circ \cdots T^{(b_\beta)}_\circ} }
{ T^{(c_1)}_\circ \cdots T^{(c_\gamma)}_\circ \:
\overline{T^{(d_1)}_\circ \cdots T^{(d_\delta)}_\circ} } \:,
    \label{eq:Da}
\end{equation}
where $f$ is a smooth function composed of
the bosonic fields and fermionic wave functions.
Here the factors $T^{(a_j)}_\circ$ and $\overline{T^{(b_k)}_\circ}$ are
the regularized distributions of the light-cone expansion. The
quotient of monomials is called a simple fraction,
and its degree~$L$ is defined by~(\ref{p:DL}).
If $L>1$, the monomial becomes singular on the light cone when the
regularization is
removed by letting $E_P \to \infty$. In light-cone
coordinates $(s,l,x_2,x_3)$, this singular behavior on the light cone is
quantified by a weak integral over $s$ for fixed $x_2$, $x_3$, and
$l \gg E_P^{-1}$. More precisely~(\ref{p:Dwe}),
\begin{eqnarray}
\lefteqn{ \int_{-\infty}^\infty ds \; (\eta f)(s) \:
\frac{ T^{(a_1)}_\circ \cdots T^{(a_\alpha)}_\circ \:
\overline{T^{(b_1)}_\circ \cdots T^{(b_\beta)}_\circ} }
{ T^{(c_1)}_\circ \cdots T^{(c_\gamma)}_\circ \:
\overline{T^{(d_1)}_\circ \cdots T^{(d_\delta)}_\circ} }
\;=\; \frac{c_{\mbox{\scriptsize{reg}}}}{(il)^L}
\:(\eta f)(s=0) \:\log^g(E_P) \:E_P^{L-1} } \nonumber \\
&& \:+\: {\mbox{(higher orders in $(l
E_P)^{-1}$ and  $(l_{\mbox{\scriptsize{macro}}} E_P)^{-1}$)}}\:,
\spc\spc\spc\spc\quad
\label{eq:Db}
\end{eqnarray}
where $g$ is an integer, $c_{\mbox{\scriptsize{reg}}}$ is the
so-called regularization parameter, and~$\eta$ is a test function, which
must be macroscopic in the sense that its derivatives scale in powers of
$l^{-1}$ or $l_{\mbox{\scriptsize{macro}}}^{-1}$. The asymptotic
formula~(\ref{eq:Db}) applies on the upper light cone $s=0$, but
by taking the adjoint and using that $Q$ is Hermitian,
$Q(x,y)^* = Q(y,x)$, it is  immediately extended to the lower light cone.
Furthermore, we can  integrate~(\ref{eq:Db}) over $l$, $x_2$, and $x_3$,
provided that $l \gg E_P^{-1}$. In polar coordinates
$(y-x)=(t,r,\Omega=(\vartheta, \varphi))$, we thus have
\begin{eqnarray*}
\lefteqn{ \int_{-\infty}^\infty dt \int_{r_0}^\infty r^2 \:dr \int_{S^2} d\Omega
\;\eta(t,r,\Omega) \:f(x,y) \:
\frac{ T^{(a_1)}_\circ \cdots T^{(a_\alpha)}_\circ \:
\overline{T^{(b_1)}_\circ \cdots T^{(b_\beta)}_\circ} }
{ T^{(c_1)}_\circ \cdots T^{(c_\gamma)}_\circ \:
\overline{T^{(d_1)}_\circ \cdots T^{(d_\delta)}_\circ} } } \nonumber \\
&=& \log^g(E_P) \:E_P^{L-1} \:\int_{\sR \setminus [-r_0, r_0]} t^2\: dt \int_{S^2}
d\Omega \; (\eta f)_{|r=|t|} \:\frac{c_{\mbox{\scriptsize{reg}}}(\Omega)}
{(it)^L} \nonumber \\
&&\:+\: {\mbox{(higher orders in $(r E_P)^{-1}$ and
$(l_{\mbox{\scriptsize{macro}}} E_P)^{-1}$)}}\:,
\spc\spc\spc
\end{eqnarray*}
valid for every $r_0 \gg E_P^{-1}$. Next we expand the function $f$ in a
Taylor series around $(y-x) \equiv \xi=0$,
\begin{equation}
f(x,y) \;=\; \sum_J f_J(x) \:\xi^J
    \label{eq:D2a}
\end{equation}
with~$J$ a multi-index, and write the Taylor coefficients
together with the regularization parameter. Collecting all contributions,
we obtain for $Q$ the weak evaluation formula
\begin{eqnarray}
\lefteqn{ \int_{-\infty}^\infty dt \int_{r_0}^\infty r^2 \:dr
\int_{S^2} d\Omega \:\eta(t,r,\Omega) \:Q(x,y) } \nonumber \\
&=& \sum_{L=2}^{L_{\mbox{\tiny{max}}}} \:
\sum_{g=0}^{g_{\mbox{\tiny{max}}}}
\log^g(E_P) \:E_P^{L-1} \:\sum_J \:\int_{\sR \setminus [-r_0, r_0]} dt \:t^{2-L} \int_{S^2}
d\Omega \; \eta \:h_J(\Omega) \:\xi^J \nonumber \\
&&\:+\: {\mbox{(higher orders in $((r+|t|)\: E_P)^{-1}$ and
$(l_{\mbox{\scriptsize{macro}}} E_P)^{-1}$)}} \:+\: o(E_P)\qquad
\label{eq:Dz}
\end{eqnarray}
with suitable functions $h_J(\Omega)$, which depend on $L$ and $g$.
The integrand on the right side of~(\ref{eq:Dz}) is evaluated on the light
cone $r=|t|$. The maximal degree of the monomials
$L_{\mbox{\scriptsize{max}}}$ as well as $g_{\mbox{\scriptsize{max}}}$
are clearly finite parameters. Notice that the
monomials of degree $L<2$ are omitted in~(\ref{eq:Dz}); this is justified as
follows. For $L<2$, the integral (\ref{eq:Db}) diverges at most
logarithmically as $E_P \to \infty$, and furthermore has a pole
in $l$ of order at most one. Thus the corresponding contribution
to~(\ref{eq:Dz}) is at most logarithmically divergent as $E_P \to
\infty$, with bounds uniform in $r_0$. This is what we mean by $o(E_P)$.

We point out that the asymptotic expansion near the light
cone~(\ref{eq:Dz}) does not give any information on the behavior of
$Q(x,y)$ near the origin, i.e.\ when $x$ and $y$ are so close that
$r, |t| \sim E_P^{-1}$. Namely, due to the restriction $r_0 \gg
E_P^{-1}$, the region near the origin is excluded from the
integration domain. Also, near the origin the terms of higher order
in $((r+|t|) E_P)^{-1}$, which are left out in~(\ref{eq:Dz}), cannot
be neglected. As explained in detail in Appendix~\ref{pappB}, the
reason for this limitation is that near  the origin, $Q$ depends
essentially on the detailed form of the fermionic projector on the
Planck scale and thus remains undetermined within the method of
variable regularization.

Our aim is to evaluate the commutator $[P,Q]$ using the
expansion~(\ref{eq:Dz}). The main difficulty is that products of the
operators $P$ and $Q$, like for example
\begin{equation}
(Q \:P)(x,y) \;=\; \int d^4z \;Q(x,z) \:P(z,y) \:,
    \label{eq:DZ}
\end{equation}
involve $Q(x,z)$ near the origin $x=z$, where~(\ref{eq:Dz}) does not apply.  In
order to explain our strategy for dealing with this so-called {\em{problem at
the origin}}\index{problem at the origin}, we briefly discuss a simple one-dimensional example.  Assume that
we are given a function $f(x)$, $x \in \R$, and a positive integer $n$ such that
for all $x_0 \gg E_P^{-1}$ and test functions $\eta$,
\begin{eqnarray}
\lefteqn{\int_{\sR \setminus [-x_0, x_0]} f(x)\: \eta(x)\:dx } \nonumber \\
&=&\int_{\sR \setminus [-x_0, x_0]} \frac{\eta(x)}{x^n}\:dx
\;+\;{\mbox{(higher orders in $(x E_P)^{-1}$)}}. \label{eq:DA}
\end{eqnarray}
In analogy to~(\ref{eq:Dz}), this formula does not give any information on the
behavior of $f(x)$ near the origin $x=0$. Thus there are many different functions
satisfying~(\ref{eq:DA}), a typical example is
\begin{equation}
    f(x) \;=\; \frac{1}{(x-i E_P^{-1})^n}\:.
    \label{eq:DD}
\end{equation}
The question is if~(\ref{eq:DA}) is useful for analyzing the weak integral
\begin{equation}
    \int_{-\infty}^\infty f(x) \:\eta(x)\:dx \:.
    \label{eq:DB}
\end{equation}
The answer to this question depends very much on the properties of $\eta$.  If $\eta$
is an arbitrary test function with compact support, we can restrict attention to
test functions with support away from the origin, ${\mbox{supp }} \eta \subset
\R \setminus [-x_0, x_0]$.  Then~(\ref{eq:DA}) applies, and we find that $f(x)
\sim x^{-n}$.  Thus by evaluating~(\ref{eq:DB}) for suitable test functions, we
can find out that, as long as $|x| \gg E_P^{-1}$, $f(x)$ behaves like the
function $x^{-n}$, which has a pole of order $n$ at the origin.  We refer to
this statement as we can {\em{detect the pole of $f$ by testing
with $\eta$}}.
Unfortunately, the situation becomes more difficult if we assume that $\eta$
belongs to a more restricted class of functions.  Assume for example that
$\eta(x)$ is rational, goes to zero at infinity and has all its poles in the
upper half plane $\{ {\mbox{Im }} x>0\}$.  Then for $f$ as in~(\ref{eq:DD}), the
integral~(\ref{eq:DB}) can be closed to a contour integral in the lower complex
plane, and we get zero, independent of $\eta$.  This shows that when testing
only with rational functions with poles in the upper half plane, the
formula~(\ref{eq:DA}) is of no use, and we cannot detect the pole of $f$.
Indeed, the problem in~(\ref{eq:DZ}) can be understood in a similar way.
If we apply the operator product $QP$ to a test function $\eta$ and
write the result as $(QP) \eta = Q(P \eta)$, the problem of making sense out of
the integral~(\ref{eq:DZ}) can be restated by saying that $Q$ may be tested only
with the functions $P \eta$.  In other words, the test functions must lie in the
image of $P$, i.e.\ they must be negative-energy solutions of the Dirac
equation.  Thus the question is if by evaluating only with such special
functions, can we nevertheless detect the poles of $Q$, and
if yes, how can this be done?  Once these questions are settled, we can compute
the operator products $PQ, QP$ and take their difference.

For clarity we begin the analysis with the simplified situation where both
$P$ and $Q$ are homogeneous, i.e.
\begin{equation}
    P(x,y) \;=\; P(y-x) \:,\spc Q(x,y) \;=\; Q(y-x) \:.
    \label{eq:DAa}
\end{equation}
Under this assumption, the operators $P$ and $Q$ are diagonal in momentum
space,
\[ P(x,y) \;=\; \int \frac{d^4k}{(2 \pi)^4} \:\hat{P}(k) \:e^{-ik(x-y)}
\:,\;\;\;\;\; Q(x,y) \;=\; \int \frac{d^4k}{(2 \pi)^4} \:\hat{Q}(k)
\:e^{-ik(x-y)} \:, \]
and their products can be taken ``pointwise'' in $k$, i.e.\
in the example~(\ref{eq:DZ}),
\begin{equation}
    (QP)(x,y) \;=\; \int \frac{d^4k}{(2 \pi)^4} \:\hat{Q}(k)\:
    \hat{P}(k) \:e^{-ik(x-y)} \:. \label{eq:D4a}
\end{equation}
Due to this simplification, it is preferable to work in momentum space. Now
the problem at the origin becomes apparent in the Fourier integral
\begin{equation}
    \hat{Q}(k) \;=\; \int d^4 \xi \;Q(\xi) \:e^{-ik \xi} \spc
    {\mbox{($\xi \equiv y-x$)}}, \label{eq:Dc}
\end{equation}
where we must integrate over a neighborhood of $\xi =0$.  In order to handle
this problem, we must carefully keep track of how the unknown behavior of $Q$
near the origin effects the Fourier integral: If we consider the integral in
position space
\begin{equation}
\int d^4 \xi \;\eta(\xi)\: Q(\xi)
    \label{eq:De}
\end{equation}
with a smooth, macroscopic function $\eta$, we can make the unknown contribution
near the origin to the integral small by assuming that $\eta$ goes to zero
sufficiently fast near $\xi=0$.  Thus we expect that~(\ref{eq:De}) is
well-defined provided that the partial derivatives of $\eta(\xi)$ at $\xi=0$
vanish up to the order $n$,
\begin{equation}
\partial^I \eta(0) \;=\; 0 \spc {\mbox{for all $I$ with $|I| \leq n$}},
    \label{eq:Dce}
\end{equation}
where $n$ is a sufficiently large parameter (which we shall specify below).
In momentum space, the conditions~(\ref{eq:Dce}) take the form
\[ \int d^4k \; k^I \:\hat{\eta}(k) \;=\; 0 \:,\spc |I| \leq n. \]
This means that the Fourier transform of $Q$ is well-defined by
(\ref{eq:Dc}), as long as it is evaluated weakly only with test functions
$\hat{\eta}(k)$ which are orthogonal to the polynomials $k^I$.
Equivalently, we can say that $\hat{Q}(k)$ is defined only up to
the polynomials $k^I$,
\begin{equation}
    \hat{Q}(k) \;=\; \int d^4 \xi \:Q(\xi) \:e^{-ik \xi} {\mbox{ mod }}\:
    {\mathcal{P}}^n(k) \:,
    \label{eq:Dcf}
\end{equation}
where ${\mathcal{P}}^n(k)$ denotes the polynomials in $k$ of degree at most $n$.
Let us compute the Fourier integral~(\ref{eq:Dcf}) using the
expansion~(\ref{eq:Dz}), and at the same time determine the parameter $n$. Our
method is to consider~(\ref{eq:Dz}) for $\eta=\exp(-ik\xi)$ and to choose $n$ so
large that we can take the limit $r_0 \searrow 0$ to obtain the Fourier
integral~(\ref{eq:Dcf}). For each summand in~(\ref{eq:Dz}), the resulting
$t$-integral is of the form
\[ \lim_{r_0 \searrow 0} \int_{\sR \setminus [-r_0, r_0]} dt\:t^{2-L+|J|}\:e^{-i
\omega t} {\mbox{ mod }}\:{\mathcal{P}}^n(\omega) \:. \]
Here in the integrand one may distinguish between the two regions $|t| >
\omega^{-1}$, where the factor $t^{2-L+|J|}$ is regular and $e^{-i \omega t}$ is
oscillating, and $|t| < \omega^{-1}$, where the pole of $t^{2-L+|J|}$ must be
taken into account and the exponential is well-approximated by a Taylor
polynomial. Since we are interested in the scaling behavior of the integral
over the pole, it suffices to consider the region $|t| < \omega^{-1}$, and
calculating modulo ${\mathcal{P}}^n(\omega)$, the leading contribution to the
integral is
\begin{equation}
\lim_{r_0 \searrow 0} \int_{[-\omega^{-1}, \omega^{-1}] \:\setminus\:
[-r_0, r_0]} dt \:t^{2-L+|J|} \:\frac{(-i \omega t)^{n+1}}{(n+1)!}\:.
    \label{eq:Dcg}
\end{equation}
If $n \geq L-|J|-3$, the integrand is bounded near $t=0$. In the case
$n=L-|J|-4$, the limit in~(\ref{eq:Dcg}) may be defined as a principal value,
whereas for $n<L-|J|-4$, (\ref{eq:Dcg}) is ill-defined. Thus we need to assume
that $n \geq L-|J|-4$. Moreover, we must ensure that the terms of higher
order in $((r+|t|) E_P)^{-1}$, which are omitted in~(\ref{eq:Dz}), are
negligible in the Fourier integral. Since these terms are regularized on
the Planck scale, the scaling behavior of these higher order terms is in
analogy to~(\ref{eq:Dcg}) given by the integrals
\[ \int_{[-\omega^{-1}, \omega^{-1}] \:\setminus\:
[-E_P^{-1}, E_P^{-1}]} dt \:\frac{t^{3-L+|J|+n}}{(t E_P)^n} \:,\spc
n \geq 1. \]
A simple calculation shows that these integrals are negligible compared
to~(\ref{eq:Dcg}) if and only if $n \geq L-|J|-3$, and under this assumption,
they are of higher order in $\omega/E_P$. We conclude that the Fourier
transform of $Q$ has the expansion
\begin{eqnarray}
\lefteqn{ \hat{Q}(k) \;=\; \sum_{L=2}^{L_{\mbox{\tiny{max}}}}
\:\sum_{g=0}^{g_{\mbox{\tiny{max}}}} \log^g(E_P) \:E_P^{L-1} \sum_J }
\nonumber \\
&& \times \:\int_{-\infty}^\infty dt\: t^{-L+|J|+2} \int_{S^2} d\Omega\: h_J(\check{\xi})
\:\check{\xi}^J \:e^{-ik \check{\xi} t} {\mbox{ mod }}\: {\mathcal{P}}^{L-|J|-3}(k)
\nonumber \\
&& \:+\: {\mbox{(higher orders in
$k/E_P$ and $(l_{\mbox{\scriptsize{macro}}} E_P)^{-1}$)}} \:+\: o(E_P)
\:, \label{eq:Dx}
\end{eqnarray}
where $\check{\xi}$ is the ``unit null vector'' $\check{\xi} =
(1, \:\Omega \in S^2 \subset \R^3)$ and $h_J(\check{\xi}=(1,\Omega)) \equiv
h_J(\Omega)$. Carrying out the $t$-integration gives the following result.
\begin{Lemma}
\label{prpB1}
Suppose that the operator $Q$ is homogeneous. Then its Fourier transform
$\hat{Q}$ is of the form
\begin{eqnarray}
\hat{Q}(k) &=& \sum_{L=2}^{L_{\mbox{\tiny{max}}}}
 \:\sum_{g=0}^{g_{\mbox{\tiny{max}}}} \:\log^g(E_P) \:E_P^{L-1}
\:\sum_J \hat{Q}^{Lg}_J(k) {\mbox{ mod }}\: {\mathcal{P}}^{L-|J|-3}(k)
\nonumber \\
&&+\: {\mbox{(higher orders in $k/E_P$ and
$(l_{\mbox{\scriptsize{macro}}} E_P)^{-1}$)}} \:+\: o(E_P) \label{eq:Dy1}
\end{eqnarray}
with
\begin{eqnarray}
\lefteqn{ \hat{Q}^{Lg}_J(k) \;=\; -2 \pi i \:(-i \omega)^{L-|J|-3} } \nonumber \\
&&\times \int_{S^2} d\Omega \:h_J(\check{\xi})\:\check{\xi}^J \:
\times \left\{ \begin{array}{ll}
\displaystyle \frac{(\check{k} \check{\xi})^{L-|J|-3}}{(L-|J|-3)!} \:
\Theta(\omega \check{k} \check{\xi}) & {\mbox{if $|J|<L-2$}} \\[1em]
\epsilon(\omega)^{L-|J|-3} \:\delta^{(2+|J|-L)}(\check{k} \check{\xi})
& {\mbox{if $|J| \geq L-2\:,$}}
\end{array} \right.\quad \label{eq:Dy}
\end{eqnarray}
where $\omega=k^0$ is the energy and $\check{k} \equiv k/\omega$
($\epsilon$ is again the step function $\epsilon(x)=1$
for $x \geq 0$ and $\epsilon(x)=-1$ otherwise).
\end{Lemma}
{\Proof} The $t$-integral in~(\ref{eq:Dx}) is of the form
\[ \int_{-\infty}^\infty t^{-n} \:e^{-i \lambda t} \:dt
{\mbox{ mod }}\: {\mathcal{P}}^{n-1}(k) \]
with $n=L-|J|-2$ and $\lambda=k \check{\xi}$. For $n=0$, we have
\begin{equation}
\int_{-\infty}^\infty e^{-i \lambda t} \:dt \;=\; 2 \pi \:\delta(\lambda) \:.
\label{eq:Dw}
\end{equation}
The case $n<0$ follows by differentiating this equation
$(-n)$ times with respect to
$\lambda$. In order to treat the case $n>0$, we integrate~(\ref{eq:Dw}) n
times in the variable $\lambda$.
The integration constant is a polynomial in $\lambda$ of degree $n-1$ and
can thus be omitted.
\QED

Let us briefly discuss the above expansion. The
parameters $L$ and $g$ give the scaling behavior in the Planck energy.
The multi-index $J$ enters at two different points: it
determines via the factor $\omega^{-|J|}$ the dependence on the energy,
and it also influences the $S^2$-integral. This integral gives detailed
information on the behavior of $\hat{Q}(k)$ in $\check{k}$, but it is
independent of $|\omega|$. Integrating over $S^2$ takes into account
the angular dependence of the regularization
functions and of the macroscopic physical objects and tells about how the
different angles contribute to $\hat{Q}$. In the case $|J| \geq L-2$, the
integrand has a $\delta$-like singularity localized at
$\check{k} \check{\xi}=0$, and so the $S^2$-integral reduces to integrating
over the intersection of the hyperplane $\{\xi \:|\: k \xi =0\}$ with the
two-sphere $t=1=r$. This intersection is empty for time-like $k$, and is a
one-sphere for space-like $k$. As a consequence, $\hat{Q}^{Lg}_J(k)$ is zero
inside the mass cone $\mathcal{C}$ and in general non-zero outside,
without being regular on its boundary $\{ k^2=0 \}$.
If on the other hand $|J|<L-2$, the factor $\Theta(\omega \check{k} \check{\xi})$
is essential, because without this factor, we would simply have a
polynomial in $k$ of degree $L-|J|-3$, being zero modulo
${\mathcal{P}}^{L-|J|-3}(k)$. Note that for any $\check{\xi}$, the factor
$\Theta(\omega \check{k} \check{\xi})$ vanishes inside the lower mass cone
$\{\check{k}^2>0,\:\omega<0\}$, whereas it is in general non-zero otherwise.
This means that $\hat{Q}^{Lg}_J(k)$ again vanishes in the interior of
the lower mass cone and is not regular on the mass cone $\{k^2=0\}$.
The singular behavior on the lower mass cone is made more explicit in the
following lemma.
\begin{Lemma} \label{lemmaD1}
The operators $\hat{Q}^{Lg}_J(k)$, (\ref{eq:Dy}), vanish inside the lower
mass cone $\{\check{k}^2 >0,\: \omega<0\}$.
Near the lower mass cone, they have the asymptotic form
\begin{eqnarray}
\hat{Q}^{Lg}_J(k) &=& 2 \pi^2 \:i \left(-\frac{i \omega}{2} \right)^{L-|J|-3}
\: h_J(\check{k})\: \check{k}^J \: (1+O(\check{k}^2)) \nonumber \\
&&\times \left\{ \begin{array}{ll} \displaystyle \frac{\check{k}^{2
(L-|J|-2)}}{(L-|J|-2)!} \:\Theta(-\check{k}^2) & {\mbox{if $|J| \leq L-2$}} \\[1em]
(-1)^{1+|J|-L} \:\delta^{(1+|J|-L)}(\check{k}^2) & {\mbox{if $|J| > L-2\:.$}}
\end{array} \right. \label{eq:Dr}
\end{eqnarray}
\end{Lemma}
{\Proof}
Without loss of generality, we can assume that $\check{k}$ points in the
$tx$-direction of our Cartesian coordinate system, i.e.\
$\check{k}=(1,\lambda,0,0)$ with $\lambda \geq 0$. Then $\check{k}
\check{\xi}=1-\lambda \cos \vartheta$, and we can write~(\ref{eq:Dy}) in
the region $\omega<0$ as
\begin{eqnarray*}
\hat{Q}^{Lg}_J(k) &=& -2 \pi i \:(-i \omega)^{L-|J|-3} \:
\int_{-1}^1 d\cos \vartheta \int_0^{2 \pi} d\varphi\; h_J(\check{\xi})\:\check{\xi}^J \\
&&\times \left\{ \begin{array}{ll}
\displaystyle \frac{(1-\lambda \cos \vartheta)^{L-|J|-3}}{(L-|J|-3)!} \:
\Theta(\lambda \cos \vartheta -1) & {\mbox{if $|J|<L-2$}} \\[1em]
(-1)^{L-|J|-3} \:\delta^{(2+|J|-L)}(1-\lambda \cos \vartheta)
& {\mbox{if $|J| \geq L-2 \:.$}}
\end{array} \right.
\end{eqnarray*}
Inside the lower mass cone, the parameter $\lambda<1$, and the integrand is
identically equal to zero.
Outside and near the lower mass cone, $1 \leq \lambda \approx 1$, and the
integrand vanishes unless $\cos \vartheta \approx 1$. Hence to leading order
in $\check{k}^2$, we may replace $h_J(\check{\xi}) \:\check{\xi}^J$ by its value at
the coordinate pole $\vartheta=0$ and carry out the $\varphi$-integration,
\begin{eqnarray*}
\hat{Q}^{Lg}_J(k) &=& -4 \pi^2 \:i \:(-i \omega)^{L-|J|-3} \:
h_J(\check{k})\:\check{k}^J \: (1+O(\check{k}^2)) \\
&&\times \int_{-1}^1 du \:\times \left\{ \begin{array}{ll}
\displaystyle \frac{(1-\lambda u)^{L-|J|-3}}{(L-|J|-3)!} \:
\Theta(\lambda u -1) & {\mbox{if $|J|<L-2$}} \\[1em]
(-1)^{L-|J|-3} \:\delta^{(2+|J|-L)}(1-\lambda u)
& {\mbox{if $|J| \geq L-2\:.$}}
\end{array} \right.
\end{eqnarray*}
In the case $|J|<L-2$, the remaining integral is of the form
\[ \int_{-1}^1 (1-\lambda u)^n \:\Theta(\lambda u-1) \:du \;=\;
-\frac{1}{\lambda(n+1)} \:(1-\lambda)^{n+1} \:\Theta(\lambda-1) \:, \]
whereas in the case $|J| \geq L-2$,
\begin{eqnarray*}
\int_{-1}^1 \delta^{(n)}(1-\lambda u) \:du &=& \left(-\frac{d}{d\lambda}
\right)^n \int_{-1}^1 u^{-n} \:\delta(1-\lambda u) \:du \\
&=& \left(-\frac{d}{d\lambda} \right)^n (\lambda^{n-1} \:\Theta(\lambda-1)) \:.
\end{eqnarray*}
Finally, we use that $\lambda=1+O(\check{k}^2)$ and
$\check{k}^2=1-\lambda^2=2(1-\lambda)+O(\check{k}^4)$.
\QED

According to Lemma~\ref{prpB1} and Lemma~\ref{lemmaD1}, all the
information on the behavior of $Q$ on the light cone contained
in~(\ref{eq:Dz}) is encoded in momentum space in a neighborhood of the
lower mass cone.
More precisely, this information can be retrieved as follows. Due to the factor
$\omega^{-|J|}$ in~(\ref{eq:Dy}) and~(\ref{eq:Dr}),
the sum over the multi-index $J$
in~(\ref{eq:Dy1}) is an expansion in powers of $\omega^{-1}$. Thus by
considering $\hat{Q}$ for large energies, more precisely for $\omega$ in
the range
\begin{equation}
m^2 \:l_{\mbox{\scriptsize{macro}}},\: l_{\mbox{\scriptsize{macro}}}^{-1}
\;\ll\; |\omega| \;\ll\; E_P \:, \label{eq:Do}
\end{equation}
we can make the contributions for large $|J|$ small (the restriction
$|\omega| \ll E_P$ is clearly necessary because
the terms of higher order in $k/E_P$ are omitted in~(\ref{eq:Dy1})).
Hence in this energy range, the series in~(\ref{eq:Dy1}) converges fast,
and the scaling behavior in $E_P$ and $\omega$, as well as the dependence
on $\check{k}$ given explicitly in~(\ref{eq:Dr}), allow us to determine the
functions $h_J(\Omega)$ completely.

Having computed the Fourier transform of $Q$, we can now take the product with
the operator $P$ according to~(\ref{eq:D4a}).  We begin with the simplest case
where we take for $P$ one massive Dirac sea in the vacuum~(\ref{eq:2b})
with $m>0$.  In this case, $\hat{P}(k)$ is
supported inside the lower mass cone $\mathcal{C}^\land$.  According to
Lemma~\ref{lemmaD1}, the operators $\hat{Q}^{Lg}_J$ vanish identically inside
the lower mass cone.  Hence the supports of $\hat{Q}^{Lg}_J$ and $\hat{P}$
do not intersect, and it follows immediately that
\begin{equation}
\hat{Q}^{Lg}_J \:\hat{P} \;=\; 0
\:. \label{eq:Du}
\end{equation}
This means that after multiplying by $\hat{P}$, all the information
contained in the expansion of Lemma~\ref{prpB1} is lost.
We refer to this difficulty as the {\em{problem of disjoint
    supports}}\index{problem of disjoint supports}.
Using the notion introduced after~(\ref{eq:DB}), it is impossible to detect
the poles of $Q$ by testing with the negative-energy solutions of the free
Dirac equation. This situation is indeed quite similar to the example
(\ref{eq:DA}, \ref{eq:DB}) for $f$ according to~(\ref{eq:DD}) and
rational test functions with support in the
upper half plane, in particular since the Fourier transform
$\eta(k)=\int_{-\infty}^\infty \eta(x) \:\exp(-ikx)$ of such a test function is
supported in the half line $k<0$, and can thus be regarded as a one-dimensional
analogue of the negative-energy solutions of the free Dirac equation.

\begin{figure}[tb]
\begin{center}
\scalebox{0.9}
{\includegraphics{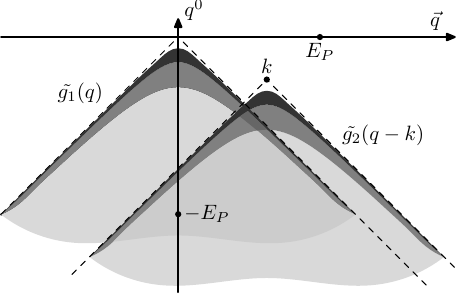}}
\caption{Example for the integrand of $\hat{M}^{Lg}$.}
\label{fig3}
\end{center}
\end{figure}

It is an instructive cross-check to see how the problem of disjoint supports
comes about if instead of analyzing the behavior of $Q$ in position
space~(\ref{eq:Dz}) and then transforming to momentum space, we work
exclusively in momentum space. For simplicity we give this qualitative argument,
which will not enter the subsequent analysis, only in the special case of a
monomial, i.e. instead of~(\ref{eq:Da}) for an expression of the form
\begin{equation}
f(x,y) \; T^{(a_1)}_\circ \cdots T^{(a_p)}_\circ \:
\overline{T^{(b_1)}_\circ \cdots T^{(b_q)}_\circ}\:.
    \label{eq:Dapoly}
\end{equation}
In this case, we can, instead of taking the product of the
factors~$T^{(a)}_\circ$ and~$\overline{T^{(b)}_\circ}$ in position
space, also compute their convolution in momentum space. As explained
in~{\S}\ref{psec23}, the singular behavior of the fermionic projector
on the light cone is determined by states near the lower mass cone. More
precisely, the main contribution to $P(x,y)$ comes from states close to the
hypersurface ${\mathcal{H}}=\{k \:|\: k \xi=0\}$, which for $\xi$ on the light cone
is tangential to the mass cone $C=\{k^2=0\}$, so that the singularity on
the light cone can be associated to the states in a neighborhood of the straight
line ${\mathcal{H}} \cap C$. For
objects derived from the fermionic projector like the regularized distributions
$T^{(a_j)}_\circ$ and $\overline{T^{(b_k)}_\circ}$, this qualitative
picture applies just as well. The Fourier transforms of the factors
$T^{(a_j)}_\circ$ and $T^{(b_k)}_\circ$
are supported in the the interior of the lower mass cone.
Thus when forming their convolution,
\begin{equation}
\hat{g}_1 \;:=\; \frac{1}{(2 \pi)^{4p}} \;\hat{T}^{(a_1)}_\circ * \cdots *
    \hat{T}^{(a_p)}_\circ \:,\spc
\hat{g}_2 \;:=\; \frac{1}{(2 \pi)^{4q}} \;\hat{T}^{(b_1)}_\circ * \cdots *
    \hat{T}^{(b_q)}_\circ \:, \label{eq:DE}
\end{equation}
the resulting convolution integrals are all finite integrals over a compact
domain (e.g., the integrand in $(\hat{T}^{(a_1)}_\circ *
\hat{T}^{(a_2)}_\circ)(k)=\int d^4q\: \hat{T}^{(a_1)}_\circ(q)\:
\hat{T}^{(a_2)}_\circ(k-q)$ vanishes unless $q$ lies in the ``diamond''
$\{q^2 \geq 0,\: q^0>0\} \cap \{(q-k)^2 \geq 0,\: (q-k)^0<0\}$).  Moreover, the
supports of $\hat{g}_1$ and $\hat{g}_2$ are again inside the lower mass
cone.  Exactly as described for the fermionic projector
in Section~\ref{sec23}, the behavior of $\hat{g}_1$ and $\hat{g}_2$ near
the lower mass cone determines the well known singularities
of $g_1$ and $g_2$ on the light cone, whereas the form of $\hat{g}_1$ and
$\hat{g}_2$ in the high-energy region away from the mass cone depends
essentially on the details of the regularization and is thus unknown.  Using
$\hat{g}_1$ and $\hat{g}_2$, we can write the Fourier transform of the
monomial as
\begin{eqnarray}
\hat{M}^{Lg}(k) &:=& \frac{1}{(2 \pi)^4} \int d^4 \xi \;
\hat{T}^{(a_1)}_\circ * \cdots * \hat{T}^{(a_p)}_\circ \;
\overline{\hat{T}^{(b_1)}_\circ * \cdots * \hat{T}^{(b_q)}_\circ}
\:e^{-i k \xi} \qquad \label{eq:D9a} \\
&=& \int d^4q \; \hat{g}_1(q) \: \hat{g}_2(q-k)\:.
    \label{eq:DC}
\end{eqnarray}
In~(\ref{eq:Da}) the monomial is multiplied by the smooth function $f$.
Thus the corresponding contribution to $\hat{Q}$ is obtained by taking the
convolution of $\hat{f}$ with $\hat{M}^{Lg}$,
\begin{equation}
\hat{Q} \;\asymp\; \hat{f} * \hat{M}^{Lg} \:.
    \label{eq:DF}
\end{equation}
Since $f$ is a macroscopic function, its Fourier transform $\hat{f}(q)$ is
localized in a neighborhood of the origin, i.e.\ in the region
$|q^0|, |\vec{q}| \sim l_{\mbox{\scriptsize{macro}}}^{-1}$.
The Taylor expansion~(\ref{eq:D2a}) corresponds
to expanding $\hat{f}$ in terms of distributions supported at the origin,
more precisely
\begin{equation}
\hat{f} \;=\; \sum_J \hat{f}_J \spc{\mbox{with}}\spc
\hat{f}_J(k) \;=\; (2 \pi)^4 \:f_J \:(i \partial_k)^J \delta^4(k) \:,
    \label{eq:DG}
\end{equation}
and substituting this expansion into~(\ref{eq:DF}) yields the expansion of
Lemma~\ref{prpB1},
\begin{equation}
    \hat{Q}^{Lg}_J \;=\; \hat{f}_J * \hat{M}^{Lg} \:.
    \label{eq:DH}
\end{equation}
Since the distributions $\hat{f}_J(q)$ are supported at $q=0$, the support
of $\hat{Q}^{Lg}_J$ coincides with that of $\hat{M}^{Lg}$. Hence to discuss
the problem of disjoint supports, we must consider $\hat{M}^{Lg}$ as given by
the integral in~(\ref{eq:DC}).  Note that this integral
differs from a convolution in that the argument of $\hat{g}_2$ has
the opposite sign; this accounts for the complex conjugation  in~(\ref{eq:D9a}).
As a consequence, the integration
domain is now not compact, and the integral is finite only due to the
regularization.  More precisely, the integration range is the intersection of
two cones, as shown in Figure~\ref{fig3} in a typical example.
We have information on the integrand only when both $q$ and $q-k$ are
close to the lower mass cone, i.e.\ when $q$ lies in the intersection
of the dark shaded regions in Figure~\ref{fig1}.  Outside of this
so-called {\em intersection region}, however, the integrand depends on
the unknown high-energy behavior of $\hat{g}_1$ or $\hat{g}_2$.  Since
the intersection region becomes large when $k$ comes close to the mass
cone and does not depend smoothly in $k$ for $k$ on the mass cone, the
contribution of the intersection region to the integral is localized
in a neighborhood of and is not regular on the mass cone $\{k^2=0\}$.
The contribution of the high-energy regions to the integral, on the
other hand, is regular in $k$ and thus well-approximated by a
polynomial in $k$.  This qualitative argument illustrates why in
Lemma~\ref{prpB1}, $\hat{Q}$ is determined only modulo a polynomial,
and why the singular behavior on the light cone (\ref{eq:Dz}) is in
momentum space encoded near the lower mass cone (see
Lemma~\ref{lemmaD1} and the discussion thereafter).  An important
conclusion from Figure~\ref{fig3} is that $\hat{Q}(k)$ is in general
not zero in the interior of the lower mass cone, and even the
intersection region gives a contribution there.  Thus for generic
regularizations or for simple regularizations like mollifying with a
smooth function, the supports of $\hat{Q}$ and $\hat{P}$ will have a
non-empty intersection, and even the singularities of $Q$ on the light
cone will contribute to the product $QP$.  On the other hand, it seems
possible that there are special regularizations for which the
contributions from the high energy regions and the intersection region
compensate each other in the integral~(\ref{eq:DC}) in such a way that
$\hat{M}^{Lg}$ indeed vanishes inside the lower mass cone.  We call
such a regularization an {\em{optimal regularization}}. According to
the method of variable regularization (see~{\S}\ref{psec22}), we want to
keep the regularization as general as possible. Therefore, we must
allow for the possibility that the regularization is optimal, and this
leads to the problem of disjoint supports.

The above consideration in momentum space gives a hint on how to resolve the
problem of disjoint supports. Namely, let us assume for a moment
that the macroscopic function $f$ has nice decay properties at infinity.
Then its Fourier transform $\hat{f}$ is a regular function. As a
consequence, the convolution~(\ref{eq:DF}) mollifies $\hat{M}^{Lg}$ on the scale
$l_{\mbox{\scriptsize{macro}}}^{-1}$, and the support of $\hat{Q}$ will be larger
than that of $\hat{M}^{Lg}$. Clearly,
$l_{\mbox{\scriptsize{macro}}}^{-1}$ is very small on the Planck scale, but
since the mass shell $\{k^2=m^2\}$ and the mass cone $\{k^2=0\}$ come asymptotically
close as the energy $|k^0|$ gets large, mollifying even on a small
scale leads to an overlap of the supports of $\hat{Q}$ and $\hat{P}$.
This is an effect which is not apparent in the expansion of
Lemma~\ref{prpB1} because by expanding $f$ in a Taylor series around
$\xi=0$, we did not use the decay properties of $f$ at infinity, and
thus we did
not see the smoothing in momentum space (cf.\ also~(\ref{eq:DG})
and~(\ref{eq:DH})). More generally, the above ``mollifying argument'' shows that
the supports of $\hat{Q}$ and $\hat{P}$ should overlap if we {\em{take into
account the macroscopic perturbations}} of $P$ and $Q$ more carefully.
Thus in order to solve the problem of disjoint supports, we shall now compute the
product $QP$ in the case of general interaction, without assuming that $P$
or $Q$ are homogeneous. Our key result will be an expansion of the operator
product around the light cone (see Theorem~\ref{thmD1} below).

Let us specify our assumptions on $P$ and $Q$. For $Q$ we merely assume that
the weak evaluation formula~(\ref{eq:Dz}) holds. For $P$, on the other hand,
we work with the formulas of the light-cone expansion, which
are of the general form
\begin{equation}
P(x,y) \;=\; \sum_{p=-1}^\infty g_p(x,y) \:
T^{(p)}(x,y) \:+\: {\mbox{(smooth contributions)}}\:.
\label{eq:bbz}
\end{equation}
Here the $g_p$ are smooth functions involving the bosonic potentials and
fields, and the unspecified smooth contributions are composed of the
fermionic wave functions as well as the non-causal contributions to the Dirac
seas (see~(\ref{1g}, \ref{fprep}) and Appendix~\ref{pappLC}).
For clarity, we shall consider the product of $Q$ with each of the summands
in~(\ref{eq:bbz}) separately, i.e.\ we will for given $p \geq -1$ compute
the product
\begin{equation}
Q\:R \spc{\mbox{with}}\spc R(x,y) \;=\; g(x,y)\:
T^{(p)}(x,y) \label{eq:bbo}
\end{equation}
and a smooth function $g$.  To avoid confusion, we recall that in the case
$p=-1$, $T^{(p)}$ is defined via a distributional
derivative; more precisely, we assume in this case that $g(x,y)$ has the form
$g=\xi_j \:f^j$ with smooth functions $f^j(x,y)$ and set similar
to~(\ref{l:7})
\begin{equation}
    g(x,y) \:T^{(-1)}(x,y) \;=\;
    -2 \:f^j(x,y) \:\frac{\partial}{\partial y^j}
    T^{(0)}(x,y) \:.
    \label{eq:bC}
\end{equation}
For technical convenience, we assume furthermore that $g$ is a Schwartz
function, $g \in {\mathcal{S}}(\R^4 \times \R^4)$, but this assumption is not
essential and could be relaxed by approximation (see the discussion
after~(\ref{eq:DA6}) below).

The contributions to $Q$ in~(\ref{eq:Dz}) are supported
on the light cone. Thus we can write them in the form
\begin{equation}
    Q(x,y) \;=\; h(x,y) \:K_{a=0}(x,y)
\end{equation}
with
\begin{equation}
    K_{a=0}(x,y) \;=\; \frac{i}{4 \pi^2}\:\delta(\xi^2) \:\epsilon(\xi^0)
    \label{eq:bbn}
\end{equation}
and a function $h(x,y)$, which in general will have a pole at the origin $x=y$. This
representation is useful because $K_a$ is a solution of the Klein-Gordon
equation, namely in momentum space \label{K_a}
\begin{equation}
    K_a(k) \;=\; \delta(k^2-a) \:\epsilon(k^0) \:,\spc a \in [0,\infty).
    \label{eq:bc}
\end{equation}
In what follows, we will also need the Green's function $S_a$ of the
Klein-Gordon equation defined by
\begin{equation}
S_a(k) \;=\; \frac{\mbox{PP}}{k^2-a} \;\equiv\; \frac{1}{2} \:\lim_{\varepsilon
\searrow 0} \sum_\pm \frac{1}{k^2-a\pm i \varepsilon}\:,\spc a \in \R.
    \label{eq:bz}
\end{equation}
As is immediately verified with contour integrals, this Green's function is for
$a>0$ causal in the sense that $S_a(x,y)$ vanishes for space-like $\xi$. On the
contrary if $a<0$, $S_a(x,y)$ vanishes for time-like $\xi$. More precisely, the
Green's function can be written as
\begin{equation}
S_a(x,y) \;=\; -\frac{1}{4 \pi}\:\delta(\xi^2) \:+\: \Theta(a \xi^2)
\:\epsilon(a) \:H_a(x,y) \:,
    \label{eq:ba}
\end{equation}
where $H_a$ is a smooth solution of the Klein-Gordon equation with power
expansion \label{H_a}
\begin{equation}
H_a(x,y) \;=\; \frac{a}{16 \pi} \:\sum_{j=0}^\infty \frac{(-1)^j}{j!\:(j+1)!}
\:\frac{a^j \:\xi^{2j}}{4^j}\:.
    \label{eq:bb}
\end{equation}
It is convenient to also introduce the Green's function \label{SJoin_a}
\begin{equation}
S^\Join_a \;=\; S_a - \Theta(a) \:H_a \:,
    \label{eq:by}
\end{equation}
which for all $a \in \R$ vanishes in time-like directions.
As one sees explicitly using~(\ref{eq:bb}) and~(\ref{eq:ba}), both $H_a(x,y)$ and
$S^\Join_a(x,y)$ are analytic in $a$ for all $a \in \R$. Similarly, a short explicit
calculation shows that $K_a(x,y)$ is
analytic for $a \in [0,\infty)$. We set \label{K^(n)} \label{S^(n)Join}
\label{H^(n)}
\[ K^{(n)} = \lim_{a \searrow 0} \left(\frac{d}{da}\right)^n K_a \:,\;\;\;\;
S^{(n)}_\Join = \left(\frac{d}{da}\right)^n S^\Join_{a \:|\:
a=0}\:,\;\;\;\; H^{(n)} = \left(\frac{d}{da}\right)^n H_{a \:|\: a=0} \:. \]
The following lemma gives the light-cone expansion for an operator product
involving two factors $K^{(.)}$. A major difference to Lemma~\ref{l:lemma1}
is that the expansion now contains unbounded line integrals; this also
requires a different method of the proof.
\begin{Lemma} \label{lemmaD2}
The operator product $K^{(l)} V K^{(r)}$ with $l,r \geq 0$ and
a scalar function $V \in {\mathcal{S}}$ has the light-cone expansion
\begin{eqnarray}
\lefteqn{ \spc \;\;\;(K^{(l)}\:V\: K^{(r)})(x,y) }  \nonumber \\
&=&\!\!\!\!\!\! -\frac{1}{2 \pi^2}\:\sum_{n=0}^\infty
\frac{1}{n!}\: \int_0^1 d\alpha\: \alpha^l\:(1-\alpha)^r\:(\alpha - \alpha^2)^n \:(\Box^n
V)_{| \alpha y + (1-\alpha) x} \; H^{(l+r+n+1)}(x,y) \nonumber \\
&&\!\!\!\!\!\!-\frac{1}{2 \pi^2}\:\sum_{n=0}^\infty
\frac{1}{n!}\: \int_{-\infty}^\infty d\alpha\: \alpha^l\:(1-\alpha)^r
\:(\alpha - \alpha^2)^n \:(\Box^n
V)_{| \alpha y + (1-\alpha) x} \; S^{(l+r+n+1)}_\Join(x,y) \nonumber \\
&&\!\!\!\!\!\!+({\mbox{non-causal contributions, smooth for $x \neq y$}}) .\label{eq:bp}
\end{eqnarray}
\end{Lemma}
We point out that, exactly as in~{\S}\ref{jsec5}, we do not study the
convergence of the infinite series in~(\ref{eq:bp}), which are merely a
convenient notation for the approximation by the partial sums. \\[.5em]

{\em{Proof of Lemma~\ref{lemmaD2}. }}
We first consider the operator product $K_a V K_b$ for $a,b>0$ in the
case when $V$ is a plane wave, $V(x) = \exp(-iqx)$. Then
in momentum space (similar to~\cite[eqn~(3.9)]{F5}), the operator product takes
the form
\begin{equation}
(K_a\:V\:K_b)\!\left(p+\frac{q}{2},\:p-\frac{q}{2}\right) \;=\;
K_a\!\left(p+\frac{q}{2}\right) \:K_b\!\left(p-\frac{q}{2}\right) \:.
    \label{eq:bd}
\end{equation}
If $q^2<0$, we get contributions when either the two upper mass shells of the
factors $K_a$ and $K_b$ intersect, or the two lower mass shells. Conversely
if $q^2>0$, we only get cross terms between the upper and lower mass shells.
Thus setting
\begin{equation}
u \;=\; \frac{a+b}{2}\:,\spc v\;=\;\frac{a-b}{2} \:,
    \label{eq:D37a}
\end{equation}
we have
\begin{eqnarray*}
\lefteqn{ (K_a\:V\:K_b)\!\left(p+\frac{q}{2},\:p-\frac{q}{2}\right)  \;=\;
-\epsilon(q^2) \:\delta \! \left( \!(p+\frac{q}{2})^2 -a \right) \:
\delta\! \left( \!(p-\frac{q}{2})^2 -b \right) } \\
&=& -\epsilon(q^2) \:\delta\!\left( \!((p-\frac{q}{2})^2 -b)
\:+\: (2 pq - 2 v) \right) \:
\delta\! \left( \!(p-\frac{q}{2})^2 -b \right) \spc \\
&=& -\epsilon(q^2) \:\delta(2pq - 2v) \:
\delta \! \left( p^2 - pq + \frac{q^2}{4}-b \right) \\
&=&-\frac{1}{2}\:\epsilon(q^2) \:\delta(pq - v) \:
\delta \! \left( p^2 + \frac{q^2}{4}-u \right) .
\end{eqnarray*}
Hence we can write our operator product as
\begin{equation}
    K_a \:V\: K_b \;=\; \frac{d}{du} A_{uv}
        \label{eq:be}
\end{equation}
where $A_{uv}$ is the operator
\begin{equation}
A_{uv}\!\left(p+\frac{q}{2},\:p-\frac{q}{2}\right) \;=\; \frac{1}{2}\;
\delta(pq-v)\; \Theta\!\left(q^2\:(p^2+\frac{q^2}{4}-u)\right) .
\label{eq:bf}
\end{equation}
Our strategy is to first derive an expansion for $A_{uv}$. Then we will
differentiate this expansion with respect to $u$ and $v$ and take the limits
$a, b \searrow 0$ to get the desired expansion for $K^{(l)} V K^{(r)}$.

The right side of~(\ref{eq:bf}) involves a product of the form $\delta(\alpha)
\:\Theta(\beta)$. This product can be transformed into a line integral as follows.
Consider for $\varepsilon>0$ the function
\begin{equation}
    f_\varepsilon(\alpha, \beta) \;=\; \frac{1}{\pi}\: \Theta(\varepsilon
    \beta - \alpha^2) \:(\varepsilon \beta - \alpha^2)^{-\frac{1}{2}}\:.
    \label{eq:bg}
\end{equation}
This function is zero unless $\beta>0$ and $\alpha \in [-\sqrt{\varepsilon
\beta},\sqrt{\varepsilon \beta}]$. As $\varepsilon \searrow 0$, the size of
this last interval tends to zero, and so $\alpha$ is confined to a smaller and
smaller neighborhood of the origin. On the other hand, the integral over
$\alpha$ stays bounded in this limit; namely,
\[ \int_{-\infty}^\infty f_\varepsilon(\alpha, \beta)\: d\alpha \;=\; \Theta(\beta)
\spc{\mbox{for all $\varepsilon>0$}}. \]
From this we conclude that
\begin{equation}
    \lim_{\varepsilon \searrow 0} f_\varepsilon(\alpha, \beta) \;=\;
    \delta(\alpha) \:\Theta(\beta)
    \label{eq:bh}
\end{equation}
with convergence as a distribution. Moreover, the formula
\begin{equation}
\int_{-\infty}^\infty \frac{\mbox{PP}}{\tau^2+\gamma}\:d\tau \;=\;
\pi\:\Theta(\gamma) \:\gamma^{-\frac{1}{2}}
    \label{eq:bi}
\end{equation}
allows one to write~(\ref{eq:bg}) as a contour integral. Putting these
relations together, we obtain for $A_{uv}$,
\begin{eqnarray}
\lefteqn{ \spc\spc A_{uv}\!\left(p+\frac{q}{2},\:p-\frac{q}{2}\right)
\;\stackrel{(\ref{eq:bf}, \ref{eq:bh})}{=}\; \frac{1}{2}\:
\lim_{\varepsilon \searrow 0} f_\varepsilon \!\left(pq-v,\:q^2\:(p^2+\frac{q^2}{4}-u)
\right) } \nonumber \\
&\stackrel{(\ref{eq:bg})}{=}&\!\!\!\!\! \frac{1}{2 \pi}\:\lim_{\varepsilon \searrow 0}\:
\frac{1}{|\varepsilon q^2|} \;\Theta
\!\left(\frac{p^2+\frac{q^2}{4}-u}{\varepsilon q^2} -
\frac{(pq-v)^2}{\varepsilon^2 q^4}\right)\!\!
\left(\frac{p^2+\frac{q^2}{4}-u}{\varepsilon q^2} -
\frac{(pq-v)^2}{\varepsilon^2 q^4}\right)^{-\frac{1}{2}} \nonumber \\
&\stackrel{(\ref{eq:bi})}{=}&\!\!\!\!\! \frac{\epsilon(q^2)}{2 \pi^2}\:\lim_{\varepsilon
\searrow 0} \int_{-\infty}^\infty \frac{\mbox{PP}}{\displaystyle
\varepsilon q^2 \tau^2 + p^2 + \frac{q^2}{4} - u -
\frac{(pq-v)^2}{\varepsilon q^2}}\;d\tau \:. \label{eq:bm}
\end{eqnarray}
After shifting the integration variable according to $\tau \to
\tau+(pq-v)/(\varepsilon q^2)$, we can identify the integrand with the Green's
function~(\ref{eq:bz}),
\begin{eqnarray}
{\lefteqn{ A_{uv}\!\left(p+\frac{q}{2},\:p-\frac{q}{2}\right)  }} \nonumber \\
&=& \frac{\epsilon(q^2)}{2 \pi^2} \;\lim_{\varepsilon \searrow 0}
\int_{-\infty}^\infty \frac{\mbox{PP}}{\displaystyle
p^2 + 2 \tau(pq-v) + \varepsilon^2 q^2 \tau^2 + \frac{q^2}{4} -
u}\;d\tau \nonumber \\
&=& \frac{\epsilon(q^2)}{2 \pi^2} \:\lim_{\varepsilon \searrow 0}
\int_{-\infty}^\infty S_{z(\varepsilon, \tau)}(p+\tau q) \:,
\label{eq:bj}
\end{eqnarray}
where $z$ is the ``mass function''
\[ z(\varepsilon, \tau) \;=\; u \:+\: 2 \tau v \:+\: (1-\varepsilon^2)\:\tau^2\:q^2 \:-\:
\frac{q^2}{4}\:. \]
If we solve~(\ref{eq:by}) for $S_a$ and substitute into~(\ref{eq:bj}), we can
take the limit $\varepsilon \searrow 0$ to obtain
\begin{equation}
A_{uv}\!\left(p+\frac{q}{2},\:p-\frac{q}{2}\right)  \;=\;
\frac{\epsilon(q^2)}{2 \pi^2} \:\int_{-\infty}^\infty (S^\Join_{z} +
\Theta(z)\: H_{z})(p+\tau q)\; d\tau
    \label{eq:bk}
\end{equation}
with
\begin{equation}
z \;\equiv\; u \:+\: 2 \tau v \:+\: (\tau^2-\frac{1}{4})\: q^2 \:.
    \label{eq:bl}
\end{equation}

The calculation so far was carried out for fixed momentum $q$ of the potential.
In order to describe the case of general $V \in {\mathcal{S}}$, we must integrate
over $q$. Furthermore, we transform to position space by integrating over $p$
(similar to~\cite[eqn~(3.10)]{F4}) and obtain
\[ A_{uv}(x,y) \;=\; \int \frac{d^4q}{(2 \pi)^4} \:\hat{V}(q) \int
\frac{d^4p}{(2 \pi)^4} \:A_{uv}(p+\frac{q}{2},p-\frac{q}{2}) \;e^{-ip(x-y)}
\:e^{-i \:\frac{q}{2}\:(x+y)} \:, \]
where $\hat{V}$ is the Fourier transform of $V$. Substituting in~(\ref{eq:bk})
and pulling out the $\tau$-integral gives
\begin{eqnarray*}
A_{uv}(x,y) &=& \frac{1}{2 \pi^2}\:\int_{-\infty}^\infty d\tau
\:\int \frac{d^4q}{(2 \pi)^4}\:\hat{V}(q)
\:\epsilon(q^2)\;e^{-i\:\frac{q}{2}\:(x+y)} \\
&&\hspace*{1cm} \times \int \frac{d^4p}{(2 \pi)^4} \:(S^\Join_{z} +
\Theta(z)\: H_{z})(p+\tau q)\;e^{-ip(x-y)}\:,
\end{eqnarray*}
and, after shifting the integration variable $p$ according to $p+\tau q
\to p$, we
can carry out the Fourier integral,
\[ \int \frac{d^4p}{(2 \pi)^4} \:(S^\Join_{z} +
\Theta(z)\: H_{z})(p+\tau q)\;e^{-ip(x-y)} \;=\;
e^{i q \tau (x-y)} \:(S^\Join_{z} + \Theta(z)\: H_{z})(x,y) \:, \]
and thus obtain
\begin{eqnarray}
A_{uv}(x,y) &=& \frac{1}{2 \pi^2}\:\int_{-\infty}^\infty d\tau
\:\int \frac{d^4q}{(2 \pi)^4}\:\hat{V}(q) \:\epsilon(q^2) \nonumber \\
&& \times \;e^{-i q((\frac{1}{2}-\tau) x +
(\frac{1}{2}+\tau) y)} \: (S^\Join_z + \Theta(z)\: H_z)(x,y)
\:. \label{eq:bn}
\end{eqnarray}
In this way, we have transformed the line integral, which
appeared in~(\ref{eq:bm}) as a contour integral in momentum space, into an
integral along the straight line $(\frac{1}{2}-\tau)x + (\frac{1}{2}+\tau)y$
through the space-time points $x$ and $y$.

The operator product $K^{(l)} V K^{(r)}$ is obtained from $A_{uv}$ by
differentiating with respect to $u, v$ and setting $u=0=v$. More precisely,
using~(\ref{eq:D37a}) and~(\ref{eq:be}),
\begin{equation}
(K^{(l)}\:V\:K^{(r)})(x,y) \;=\; \frac{1}{2^{l+r}}
\left(\frac{\partial}{\partial u} + \frac{\partial}{\partial v} \right)^l
\left(\frac{\partial}{\partial u} - \frac{\partial}{\partial v} \right)^r
\frac{\partial}{\partial u}  \left. A_{uv} \right|_{u=0=v}\:.
    \label{eq:DA1}
\end{equation}
According to~(\ref{eq:bl}), the factors $S^\Join_z$, $\Theta(z)$, and
$H_z$ in~(\ref{eq:bn}) depend implicitly on $u$ and $v$. Thus when
substituting~(\ref{eq:bn}) into~(\ref{eq:DA1}), we can carry out the partial
derivatives with the sum, product, and chain rules. Let us first
collect the terms for which all the derivatives act on the factors
$S^\Join_z$ or $H_z$. This gives the contributions to $(K^{(l)}V K^{(r)})(x,y)$
\begin{eqnarray*}
&& \frac{1}{2 \pi^2}\:\int_{-\infty}^\infty d\tau \:(\frac{1}{2}+\tau)^l \:
(\frac{1}{2}-\tau)^r
\:\int \frac{d^4q}{(2 \pi)^4}\:\hat{V}(q)
\:\epsilon(q^2)\: e^{-i\:\frac{q}{2} \:(x+y)} \\
&& \hspace*{1cm} \times\: e^{iq \tau\:(x-y)}\:(S_z^{\Join\:(l+r+1)}
+ \Theta(z)\: H_z^{(l+r+1)})(x,y)
\end{eqnarray*}
with $z=(\tau^2-\frac{1}{4}) \:q^2$. After expanding in powers of $z$ and
introducing the new integration variable $\alpha = \tau + \frac{1}{2}$, we obtain
\begin{eqnarray}
\lefteqn{ \frac{1}{2 \pi^2}\:\sum_{n=0}^\infty\:\frac{1}{n!}\:
\int_{-\infty}^\infty d\alpha\:\alpha^l\:(1-\alpha)^r \:(\alpha-\alpha^2)^n
\:\int \frac{d^4q}{(2 \pi)^4}\:\epsilon(q^2)\: (-q^2)^n \:\hat{V}(q) }
\nonumber \\
&&
\times \: e^{-iq (\alpha y + (1-\alpha)x)} \: (S_z^{\Join\:(l+r+n+1)}
+ \Theta((\alpha^2-\alpha)\: q^2)\: H_z^{(l+r+n+1)})(x,y) \:.\qquad
\label{eq:DA2}
\end{eqnarray}
If $\hat{V}$ is supported outside the mass cone, ${\mbox{supp }} \hat{V}
\subset \{q^2<0\}$, we can carry out the $q$-integral in~(\ref{eq:DA2}) and
obtain precisely the two series in~(\ref{eq:bp}).
Conversely if $\hat{V}$ is supported inside the mass cone $\{q^2>0\}$, we get
\begin{eqnarray*}
&&\hspace*{-.5cm}\frac{1}{2 \pi^2}\:\sum_{n=0}^\infty
\frac{1}{n!}\: \int_{\sR \setminus [0,1]} d\alpha\:\alpha^l\:(1-\alpha)^r\:
(\alpha - \alpha^2)^n \:(\Box^n V)_{| \alpha y + (1-\alpha) x} \; H^{(l+r+n+1)}(x,y) \\
&&\hspace*{-.3cm}+\frac{1}{2 \pi^2}\:\sum_{n=0}^\infty
\frac{1}{n!}\: \int_{-\infty}^\infty d\alpha\:\alpha^l\:(1-\alpha)^r
\:(\alpha - \alpha^2)^n \:(\Box^n V)_{| \alpha y + (1-\alpha) x} \;
S^{(l+r+n+1)}_\Join(x,y) \:.
\end{eqnarray*}
This does not coincide with the series in~(\ref{eq:bp}). Using~(\ref{eq:by})
and~(\ref{eq:ba}), the difference can be written as
\begin{equation} \begin{split}
\lefteqn{ -\frac{1}{2 \pi^2}\:\sum_{n=0}^\infty \frac{1}{n!}
\int_{-\infty}^\infty d\alpha\;\alpha^l\:(1-\alpha)^r \:(\alpha - \alpha^2)^n} \\
&  \spc \times (\Box^n V)_{| \alpha y + (1-\alpha) x}
\;\epsilon(\xi^2)\:H^{(l+r+n+1)}(x,y)\:.\; \end{split}
    \label{eq:DA3}
\end{equation}
Since unbounded line integrals are involved, this expression is clearly
non-causal. We shall now prove that~(\ref{eq:DA3}) is smooth for $x \neq y$.
Notice that the line integrals in~(\ref{eq:DA3}) are supported on the
hyperplane $\{q \:|\: q \xi = 0\}$, e.g.
\begin{equation}
    \int_{-\infty}^\infty V(\alpha y + (1-\alpha)x) \:d\alpha \;=\; \int \frac{d^4q}{(2
    \pi)^4} \:\hat{V}(q) \;2 \pi \delta(q \xi)\;
    e^{-i\:\frac{q}{2}\:(x+y)}\:.
    \label{eq:DA4}
\end{equation}
For time-like $\xi$, this hyperplane does not intersect the support of
$\hat{V}$, and thus~(\ref{eq:DA3}) vanishes identically inside the
light cone. Furthermore, (\ref{eq:DA3}) is clearly smooth in the region
$\{\xi^2<0\}$ away from the light cone. Thus it remains to show that all the
partial derivatives of~(\ref{eq:DA3}) vanish on the light cone. The
boundary values of these partial derivatives on the light cone
$\{\xi^2=0\}$ involve integrals of the form
\[ \int_{-\infty}^\infty d\alpha\: \alpha^{l+k_1} \:(1-\alpha)^{r+k_2}
\:(\alpha-\alpha^2)^n \: (\partial^K \Box^n V)_{|\alpha y + (1-\alpha) x}
\:\xi^L \]
with parameters $k_1, k_2 \geq 0$ and multi-indices $K$, $L$. Similar
to~(\ref{eq:DA4}), these integrals are supported on the hypersurface
$\{q \:|\: q \xi\}=0$, and for $\xi$ on the light cone (and $\xi \neq 0$),
this hypersurface
does not intersect the support of $\hat{V}$. We conclude that~(\ref{eq:DA2})
coincides with~(\ref{eq:bp}), both in the case when ${\mbox{supp }} \hat{V}
\subset \{q^2<0 \}$ and when ${\mbox{supp }} \hat{V} \subset \{q^2>0 \}$.
Linearity and an approximation argument near the light cone yield that~(\ref{eq:DA2})
coincides with~(\ref{eq:bp}) for general $V \in {\mathcal{S}}$.

It remains to consider the contributions when some of the derivatives
in~(\ref{eq:DA1}) act on the factor $\Theta(z)$ in~(\ref{eq:bn}).
The resulting expressions are of the form
\begin{equation}
\int \frac{d^4q}{(2 \pi)^4}\:\hat{V}(q) \:e^{-i \:\frac{q}{2}\:(x+y)}\:
\int_{-\infty}^\infty d\tau\;\epsilon(q^2)
\:\delta^{(\alpha)}(z)\:H^{(\beta)}_z(x,y) \:{\mathcal{P}}(\tau)\;e^{-iq \tau (y-x)}
    \label{eq:bq}
\end{equation}
with integers $\alpha, \beta \geq 0$ and a polynomial ${\mathcal{P}}(\tau)$.
Using the formula
\[ \delta(z) \;=\; \frac{1}{2 \pi i}\: \lim_{\varepsilon \searrow
0} \left(\frac{1}{z-i \varepsilon} - \frac{1}{z+i \varepsilon} \right) , \]
we can write the $\tau$-integral in terms of the complex integrals
\[ \lim_{\varepsilon \searrow 0} \epsilon(q^2) \:\int_{-\infty}^\infty d\tau\:
\frac{1}{((\tau^2 -\frac{1}{4})\: q^2 \pm i \varepsilon)^{\alpha+1}} \:H^{(\beta)}_z
\:{\mathcal{P}}(\tau) \:e^{-iq \tau(y-x)} \:. \]
Depending on the sign of $q(y-x)$, the integration contour can be closed either
in the upper or in the lower half plane, and the residue theorem yields
expressions of the form
\begin{equation}
    \lim_{\varepsilon \searrow 0} \:\frac{1}{(q^2 \pm i \varepsilon)^\gamma}\:
    H^{(\kappa)}(x-y) \:{\mathcal{P}}(\tau) \:e^{-iq \tau(y-x)} \spc
    {\mbox{with}}\spc \tau = \pm \frac{1}{2}
    \label{eq:br}
\end{equation}
and $\gamma \leq 2 \alpha+1$, $\beta \leq \kappa \leq \beta+\alpha$. These expressions
are well-defined distributions, and thus the $q$-integral in~(\ref{eq:bq}) is finite.
Due to the powers of $1/q^2$ in~(\ref{eq:br}), the resulting contributions to the
operator product $K^{(l)}V K^{(r)}$ are non-causal. Since the factor
$H^{(\kappa)}$ in~(\ref{eq:br}) is a polynomial in $\xi$ and $\hat{V}(q)$
in~(\ref{eq:bq}) has rapid decay, these contributions are also smooth.
\QED

The above lemma can be used to derive the light-cone expansion for the operator
product $K^{(l)} V T^{(r)}$.
\begin{Lemma} \label{lemmaD3}
For $l, r \geq 0$ and $V \in {\mathcal{S}}$,
\begin{eqnarray}
\lefteqn{\;\;\;\;
(K^{(l)}\:V\: T^{(r)})(x,y) } \nonumber \\
&=& \frac{1}{2 \pi i}\:\sum_{n=0}^\infty
\frac{1}{n!} \:\int_{-\infty}^\infty d\alpha\: \alpha^l\:(1-\alpha)^r
\:(\alpha - \alpha^2)^n \:(\Box^n
V)_{| \alpha y + (1-\alpha) x} \nonumber \\
&& \times
\; \epsilon(y^0-x^0) \: T^{(l+r+n+1)}(x,y)
\:+\:({\mbox{contributions smooth for $x \neq y$}}) \:. \qquad
\label{eq:bbv}
\end{eqnarray}
\end{Lemma}
{\Proof}
Using~(\ref{eq:by}) and the fact that
$H_a(x,y)$ is smooth in $x$ and $y$ according to~(\ref{eq:bb}), the light-cone
expansion~(\ref{eq:bp}) yields that for $p, q \geq 0$,
\begin{eqnarray}
\lefteqn{ (K^{(l)}\:V\: K^{(r)})(x,y) } \nonumber \\
&&\!\!\!\!\!\!\!\!=\, -\frac{1}{2 \pi^2}\:\sum_{n=0}^\infty
\frac{1}{n!}\: \int_{-\infty}^\infty \!\!\!d\alpha\: \alpha^l\:(1-\alpha)^r
\:(\alpha - \alpha^2)^n \:(\Box^n V)_{| \alpha y + (1-\alpha) x} \;
S^{(l+r+n+1)}(x,y) \nonumber \\
&& +({\mbox{contributions smooth for $x \neq y$}}) \:, \label{eq:bbw}
\end{eqnarray}
where $S^{(n)}=\lim_{a \searrow 0}S_a^{(n)}$. The main difference
between~(\ref{eq:bbv}) and~(\ref{eq:bbw}) is that the factors $K^{(r)}$ and
$S^{(l+r+n+1)}$ are replaced by corresponding factors
$T^{(.)}$. The method of the proof is to realize these
replacements by multiplying~(\ref{eq:bbw}) with a suitable operator from the
right.

In preparation, we rewrite the operators $S^{(.)}$ in~(\ref{eq:bbw}) in terms
of $K^{(.)}$ as follows. Using that multiplication in position space corresponds
to convolution in momentum space, we have for $a>0$,
\begin{eqnarray*}
\lefteqn{ \int K_a(x,y) \:\epsilon(y^0-x^0) \:e^{-ik \:\xi}\:d^4 \xi } \\
&=& \int_{-\infty}^\infty \frac{d\omega}{2 \pi} \; \delta(\omega^2 -
|\vec{k}|^2 - a)\:\epsilon(\omega) \; (-2i) \:\frac{\mbox{PP}}{k^0 - \omega} \\
&=& \frac{1}{i \pi} \:\frac{1}{2 \:|\omega|} \left. \frac{\mbox{PP}}{k^0 - \omega}
\right|_{\omega=-\sqrt{|\vec{k}|^2 + a}}^{\omega=\sqrt{|\vec{k}|^2 + a}}
\;=\; \frac{1}{i \pi} \:\frac{\mbox{PP}}{k^2 - a} \;=\; \frac{1}{i \pi}
\:S_a(k) \:,
\end{eqnarray*}
and thus
\[ S_a(x,y) \;=\; i \pi \:K_a(x,y) \;\epsilon(y^0-x^0) \:. \]
We differentiate with respect to $a$ and let $a \searrow 0$ to obtain
\[ S^{(n)}(x,y) \;=\; i \pi \:K^{(n)}(x,y) \; \epsilon(y^0-x^0) \:. \]
Substituting into~(\ref{eq:bbw}) gives
\begin{eqnarray}
\lefteqn{ (K^{(l)}\:V\: K^{(r)})(x,y) \;=\;
({\mbox{contributions smooth for $x \neq y$}}) } \nonumber \\
&&\:+\: \frac{1}{2 \pi i}\: \sum_{n=0}^\infty \frac{1}{n!}
\int_{-\infty}^\infty d\alpha \: \epsilon(y^0-x^0)\:
\alpha^l\:(1-\alpha)^r \:(\alpha - \alpha^2)^n  \nonumber \\
&& \hspace*{3.5cm} \times  \: (\Box^n V)_{| \alpha y + (1-\alpha) x} \;
K^{(l+r+n+1)}(x,y) \:. \label{eq:bbt}
\end{eqnarray}

The operator $T_a$, $a \geq 0$, is obtained from $K_a$ by projecting on the
negative-energy states, more precisely
\begin{equation}
    T_a \;=\; K_a \:\chi \:,
    \label{eq:bby}
\end{equation}
where $\chi$ is the multiplication operator in momentum space
\[ \chi(k) \;=\; -\Theta(-k^0) \:. \]
In position space, $\chi$ has the kernel
\begin{equation}
\chi(x,y) \;=\; \int \frac{d^4k}{(2 \pi)^4} \:\chi(k) \:e^{-ik(x-y)} \;=\;
-\frac{1}{2 \pi i} \:\lim_{\varepsilon \searrow 0} \frac{1}{y^0 - x^0 - i
\varepsilon} \: \delta^3(\vec{y}-\vec{x}) \:.
    \label{eq:bv2}
\end{equation}
If $a$ is positive, the mass shell $\{k^2=a\}$ does not intersect the
hyperplane $\{k^0=0\}$ where $\chi$ is not smooth, and thus we may
differentiate~(\ref{eq:bby}) with respect to $a$ to obtain
\begin{equation}
    T_a^{(n)} \;=\; K_a^{(n)} \: \chi \:,\spc a>0.
    \label{eq:bbx}
\end{equation}
However, difficulties arise in~(\ref{eq:bbx}) in the limit $a \searrow 0$.
Namely, the limit of the left side exist only after ``regularizing''
$T_a^{(n)}$ by subtracting a polynomial in $(y-x)$ (see~(\ref{Tadef},
\ref{Tldef}).
On the right side, the problem is that $K^{(n)}(x,y)$ behaves polynomially at
timelike infinity, whereas $\chi(x,y)$ decays for large $(y^0-x^0)$ only as
$(y^0-x^0)^{-1}$, and so the product $K^{(n)} \chi$ does not exist. To cure this
problem, we insert into~(\ref{eq:bv2}) an exponentially decaying factor by
introducing for $\kappa>0$ the kernel
\begin{equation}
\chi_\kappa(x,y) \;=\;
-\frac{1}{2 \pi i} \:\lim_{\varepsilon \searrow 0} \frac{e^{-\kappa\:|y^0-x^0|}}
{y^0 - x^0 - i \varepsilon} \: \delta^3(\vec{y}-\vec{x}) \:.
    \label{eq:bbs}
\end{equation}
The exponential factor changes the product with $K_a^{(n)}$ only by a
contribution smooth in $y-x$, and thus
\begin{eqnarray*}
\lefteqn{ T_a^{(n)}(x,y) + {\mbox{(smooth
contributions)}} } \\
&=& T_a^{(n)}(x,y) \;=\; (K_a^{(n)} \:\chi)(x,y) \;=\;
(K_a^{(n)} \:\chi_\kappa)(x,y) + {\mbox{(smooth contributions)}}\:.
\end{eqnarray*}
The very left and right of this equation converge for $a \searrow 0$, and
we conclude that
\begin{equation}
(K^{(n)} \:\chi_\kappa)(x,y) \;=\; T^{(n)}(x,y)
\:+\: {\mbox{(smooth contributions)}}. \label{eq:bbu}
\end{equation}

We multiply the operator product on the left of~(\ref{eq:bbt}) by the operator
$\chi_\kappa$. Applying~(\ref{eq:bbu}) and using that multiplying by a
smooth operator gives something smooth, we get
\begin{equation}
(K^{(l)} \:V\: K^{(r)}\: \chi_\kappa)(x,y) \;=\; (K^{(l)} \:V\:
T^{ (r)})(x,y) \:+\: {\mbox{(smooth contributions)}}.
\label{eq:bbp}
\end{equation}
It remains to show that multiplying the right side of~(\ref{eq:bbt}) by the
operator $\chi_\kappa$ gives the right side of~(\ref{eq:bbv}).
When we multiply the summands on the right side of~(\ref{eq:bbt}) by
$\chi_\kappa$, we get according to~(\ref{eq:bbs}) a convolution in
the time coordinate of the form
\begin{equation}
\lim_{\varepsilon \searrow 0} \int_{-\infty}^\infty
d\tau\:\frac{e^{-\kappa\:|\tau|}}{\tau - i \varepsilon} \:F(x;y^0-\tau,
\vec{y}) \: K^{(l+r+n+1)}(x; y^0-\tau, \vec{y}) \:.
    \label{eq:bp3}
\end{equation}
Here the function $F$ stands for the line integral in~(\ref{eq:bbt}); it is
smooth unless $y^0-\tau = x^0$. We write the convolution integral~(\ref{eq:bbp}) as
\begin{eqnarray}
&&\hspace*{-1cm} \lim_{\varepsilon \searrow 0} \int_{-\infty}^\infty
d\tau\:e^{-\kappa\:|\tau|} \: \frac{F(x;y^0-\tau, \vec{y})-F(x,y)}{\tau - i \varepsilon}
\: K^{(l+r+n+1)}(x; y^0-\tau, \vec{y}) \label{eq:bbr} \\
&&\hspace*{-1cm} +F(x,y) \;\lim_{\varepsilon \searrow 0} \int_{-\infty}^\infty
d\tau\:\frac{e^{-\kappa\:|\tau|}}{\tau - i \varepsilon} \:
K^{(l+r+n+1)}(x; y^0-\tau, \vec{y}) \:. \label{eq:bbq}
\end{eqnarray}
The term~(\ref{eq:bbq}) can be written as $F(x,y)\:(K^{(l+r+n+1)}\:
\chi_\kappa)(x,y)$, and applying (\ref{eq:bbu}) gives precisely the summands
on the right side of~(\ref{eq:bbv}). Let us prove that~(\ref{eq:bbr}) is
smooth for $x \neq y$. Thus assume that $x \neq y$. Our above construction
is Lorentz invariant in the sense that we may introduce the operator
$\chi_\kappa$ in any reference frame,
without influence on our operator products. Thus we can
choose the reference frame such that $\vec{x} \neq \vec{y}$.
The $\tau$-integral in~(\ref{eq:bbr}) can be regarded as an integral along the
straight line $\{(y^0-\tau, \vec{y})\}$. Since $\vec{x} \neq \vec{y}$, this
straight line does not intersect the point $x$. Due to causality of $K^{(.)}$,
the integrand vanishes unless $|(y^0-\tau)-x^0| \geq |\vec{y}-\vec{x}|$.
Thus if $\xi$ is spacelike, i.e.\ $|y^0-x^0| < |\vec{y}-\vec{x}|$, the integrand
vanishes in a neighborhood of $\tau=0$, and as a consequence the
integral~(\ref{eq:bbr}) is smooth in $x$ and $y$.
On the other hand if $\xi$ lies inside the cone $|y^0-x^0| > \frac{1}{2} \:
|\vec{y}-\vec{x}|$, the straight line $\{ (y^0-\tau, \vec{y})\}$ does for small
enough $\tau$ not intersect the hyperplane $\{z \:|\: z^0=x^0\}$ where $F$
is not smooth. Thus the function $F(x; y^0-\tau, \vec{y})$ is smooth in a
neighborhood of $\tau=0$, and the mean value theorem yields that the
bracket in~(\ref{eq:bbr}) is smooth. This implies that~(\ref{eq:bbr}) is
again smooth in $x$ and $y$.

It remains to show that multiplying the contributions smooth for $x \neq y$,
which are not specified in~(\ref{eq:bbt}), by the operator $\chi_\kappa$,
gives terms which are again smooth for $x \neq y$. Again using Lorentz invariance, we
can choose a reference frame such that $\vec{x} \neq \vec{y}$. Then multiplying
by $\chi_\kappa$ yields a convolution along the straight line $\{(y^0-\tau,
\vec{y})\}$, and this line integral does not intersect the point $x$. In this
way, we can avoid integrating across the origin where the contributions
in~(\ref{eq:bbt}) may be singular. We get a convolution of $\chi_\kappa$ with
a smooth function, and this is clearly finite and depends smoothly on $x$ and $y$.
\QED

After these preliminaries, we are ready to compute the operator product $QR$ in
an expansion around the light cone.  For the statement of the result, we need to
analytically extend $h_J(\Omega)$ in~(\ref{eq:Dz}) from a function on $S^2$ to a
function on Minkowski space and also regularize it at the origin: As a smooth
function on $S^2$, $h_J$ can be expanded in spherical harmonics.  Since the
spherical harmonics are the boundary values on $S^2$ of the harmonic polynomials
on $\R^3$, we have the unique expansion
\begin{equation}
h_J(\vec{x}) \;=\; \sum_{n=0}^\infty \left.
{\mathcal{P}}_n(\vec{x}) \right|_{\vec{x} \in S^2} \:,
    \label{eq:bB}
\end{equation}
where ${\mathcal{P}}_n(\vec{x})$ are suitable harmonic polynomials of degree $n$.
The smoothness of $h_J$ implies that the summands in~(\ref{eq:bB}) decay in $n$
faster than any polynomial. As a consequence, the series in~(\ref{eq:bB})
converges absolutely for any $\vec{x} \in \R^3$, and we can even extend
$h_J(\vec{x})$ to a unique function on $\C^3$.  For $\varepsilon>0$, we define the
regularization $h_J^\varepsilon$ of $h_J$ by
\begin{equation}
    h_J^\varepsilon(\xi) \;=\; h_J\!\left(\frac{\vec{\xi}}{\xi^0 - i
    \varepsilon}\right) \:.
    \label{eq:D68a}
\end{equation}

\begin{Thm} \label{thmD1}
Assume that the operator $Q$ satisfies the weak evaluation formula~(\ref{eq:Dz}).
Then the operator product $QR$ with $R$ according to~(\ref{eq:bbo}) with $p
\geq -1$ and $g \in {\mathcal{S}}(\R^4 \times \R^4)$ has the expansion
\begin{eqnarray}
\lefteqn{ \hspace*{1.1cm} (Q\:R)(x,y) \;=\; \sum_{L=2}^{L_{\mbox{\tiny{max}}}} \:
\sum_{g=0}^{g_{\mbox{\tiny{max}}}}
\log^g(E_P) \:E_P^{L-1} \:\sum_J } \label{eq:Dc1} \\
&&\hspace*{-.55cm}
\times \:(-2 \pi) \sum_{n=0}^\infty \lim_{\varepsilon \searrow 0} \int_{-\infty}^\infty
\!\!\!\!d\alpha\: (1-\alpha)^p \:(\alpha-\alpha^2)^n \:\Box_z^n \!\left(
\frac{h_J^\varepsilon(\zeta) \:\zeta^J}{(\zeta^0 - i \varepsilon)^{L-1}}\:
g(z,y)\right)_{|z=\alpha y + (1-\alpha) x} \nonumber \\
&&\times \: \epsilon(y^0-x^0)\;T^{(p+n+1)}(x,y)
\;{\mbox{ mod }} \;{\mathcal{P}}^{L-|J|-3}(\partial_x) \:R(x,y) \:+\: o(E_P) \label{eq:Dc2} \\
&&+\:{\mbox{(contributions smooth for $x \neq y$)}}
\:+\:{\mbox{(higher orders in $(l_{\mbox{\scriptsize{macro}}} E_P)^{-1}$)}}
\label{eq:Dclue}
\end{eqnarray}
with $\zeta \equiv z-x$.
\end{Thm}
{\Proof}
We first consider the case $p \geq 0$. We fix two space-time points $x_0$,
$y_0$ and set
\[ Q(\xi) \;:=\; Q(x_0, x_0+\xi) \:,\spc R(\xi) \;:=\; R(y_0-\xi, y_0)
\:. \]
We regard $Q(\xi)$ and $R(\xi)$ as the integral kernels of corresponding
homogeneous operators, which with slight abuse of notation we denote again by
$Q$ and $R$. Then in momentum space~(\ref{eq:Dc}), the operator $Q$ has the
expansion of Lemma~\ref{prpB1} and Lemma~\ref{lemmaD1}. Let us compute the
product $Q^{Lg}_J R$. According to its construction in Lemma~\ref{prpB1},
$Q^{Lg}_J(\xi)$ behaves for $\xi \neq 0$ like
\[ Q^{Lg}_J (\xi) \;=\; \delta(\xi^2) \:\epsilon(\xi^0) \:t^{-L+1}
\;h_J(\check{\xi}) \:\xi^J \:, \]
and furthermore its Fourier transform $\hat{Q}^{Lg}_J(k)$ vanishes inside the
lower mass cone (cf.\ Lemma~\ref{lemmaD1}). The distribution
\begin{equation}
    \lim_{\varepsilon \searrow 0} K^{(0)}(x,y) \left[ - 4 \pi^2 i \:(y^0 - x^0 -
    i \varepsilon)^{-L+1} \:h_J(\check{\xi})\:\xi^J \right]
    \label{eq:bbm}
\end{equation}
has these two properties, as one sees immediately from~(\ref{eq:bbn}) and
when computing the Fourier transform of~(\ref{eq:bbm}) with
contour integrals. Indeed, a short calculation shows that~(\ref{eq:bbm})
even coincides with $Q^{Lg}_J(\xi)$.

We introduce for $\varepsilon>0$ the potential
\[ V_\varepsilon(z) \;=\; -4 \pi^2 i \:(z^0 - x_0^0 - i \varepsilon)^{-L+1}\:
h_J^\varepsilon(z-x_0) \;(z-x_0)^J\: g(z,y_0) \:, \]
where $h^\varepsilon_J$ is the regularization of $h_J$~(\ref{eq:D68a}).
This potential is a Schwartz function, and thus Lemma~\ref{lemmaD3} yields that
\begin{eqnarray}
\lefteqn{\hspace*{-1cm} \;\;\;\;\; (K^{(0)}\:V^\varepsilon\: T^{(p)})(x,y) }
\nonumber \\
&& \hspace*{-1cm}=\; \frac{1}{2 \pi i}\:\sum_{n=0}^\infty \frac{1}{n!}\;
\int_{-\infty}^\infty d\alpha\: (1-\alpha)^p
\:(\alpha - \alpha^2)^n \:(\Box^n V_\varepsilon)_{| \alpha y + (1-\alpha) x}
\nonumber \\
&&\hspace*{-1cm} \;\;\;\;\;\times \:\epsilon(y^0-x^0) \; T^{(p+n+1)}(x,y)
\:+\: ({\mbox{contributions smooth for $x \neq y$}}) \:.
\label{eq:bbl}
\end{eqnarray}
We now set $x=x_0$, $y=y_0$ and take the limit $\varepsilon \searrow 0$.
On the left side of~(\ref{eq:bbl}), we can use that
$(\ref{eq:bbm})=Q^{Lg}_J(\xi)$ to obtain the operator product
$(Q^{Lg}_J R)(x_0, y_0)$. Furthermore, due
to our regularization of $h_J^\varepsilon$, the factors $(\Box^n
V_\varepsilon)(z)$ on the right side of~(\ref{eq:bbl}) are
of the form smooth function times $(z^0-x^0-i
\varepsilon)^{-l}$, and thus the limit $\varepsilon \searrow 0$ exists in each
line integral in~(\ref{eq:bbl}).
We conclude that $(Q^{Lg}_J R)(x,y)$ coincides
precisely with the series~(\ref{eq:Dc2}).
Since in~(\ref{eq:bbl}) we integrate across the origin, the higher orders in
$((r+|t|) E_P)^{-1}$ in~(\ref{eq:Dz}) (or equivalently the higher orders in
$k/E_P$ in~(\ref{eq:Dx})) yield contributions of higher order in
$(l_{\mbox{\scriptsize{macro}}} E_P)^{-1}$. Finally, calculating modulo
polynomials in~(\ref{eq:Dy1}) means in position space that $Q(x,y)$ is
determined only modulo partial derivatives of $\delta^4(x-y)$, and this gives rise
to the term ${\mbox{ mod }} {\mathcal{P}}^{L-|J|-3}(\partial_x) \:R(x,y)$
in~(\ref{eq:Dc2}). This concludes the derivation
of~(\ref{eq:Dc1}--\ref{eq:Dclue}) in the case $p \geq 0$.

In the above derivation we neglected the contributions smooth for
$x \neq y$, which are not specified in~(\ref{eq:bbl}), implicitly
assuming that they remain smooth in the limit $\varepsilon \searrow 0$.
This is justified as follows. As $\varepsilon \searrow 0$, the potential
$V_\varepsilon(z)$ becomes singular only at $z=x_0$. Thus if the
contributions smooth for $x \neq y$ in~(\ref{eq:bbl}) had a non-smooth
limit, the resulting non-smooth contributions to $(Q^{Lg}_J R)(x_0, y_0)$
would depend on $g(z,y_0)$ and its partial derivatives only at
$z=x_0$; i.e.\ they would be of the form
\begin{equation}
\partial^J_{x_0} g(x_0, y_0) \:W_J(x_0, y_0) \:, \label{eq:Dcpd}
\end{equation}
where $W_J$ are distributions independent of $g$.
Suppose that the Fourier transform $\hat{g}$ of $g(x_0+.,y_0)$ is supported inside the
upper mass shell $\{k \:|\: k^2>0, k^0>0\}$. Then due to our
$(-i \varepsilon)$-regularization, also the support of
$\hat{V}_\varepsilon$ is inside the upper mass shell. As a consequence,
the Fourier transforms of the distributions $K^{(0)}(x_0,.)$ and
$(V_\varepsilon T^{{\mbox{\scriptsize{reg}}}\:(p)})(.,y_0)$ have disjoint
supports, and the left side of~(\ref{eq:bbl}) is zero. Furthermore for
$\xi$ close to the light cone, the unbounded line integrals in~(\ref{eq:bbl})
vanish (notice that they are supported on the hypersurface $\{k \:|\:
k \xi = 0\}$). We conclude that if $\hat{g}$ is supported inside the upper
mass shell, then the contributions smooth for $x \neq y$ in~(\ref{eq:bbl})
are zero. Taking the limit $\varepsilon \searrow 0$ yields that the
contributions~(\ref{eq:Dcpd}) vanish if $\hat{g}$ is supported inside
the upper mass shell, i.e.
\begin{equation}
\int d^4k\; k^J \:\hat{g}(k) \;W_J(x_0, y_0) \;=\; 0
\label{eq:Dcq}
\end{equation}
for all $\hat{g}$ with ${\mbox{supp }} \hat{g} \subset
\{ k \:|\: k^2=0, k^0>0\}$. Since the polynomials restricted to the
upper mass cone are linearly independent, (\ref{eq:Dcq}) implies that
the contributions~(\ref{eq:Dcpd}) are all identically zero.

The extension to the case $p=-1$ follows exactly as
in~\cite[Lemma~2.2]{F5}: We pull one derivative out of the operator product,
\begin{eqnarray*}
\lefteqn{ \int d^4z \:Q(x_0,z)\: g(z,y_0)\:
T^{(-1)}(z,y) } \\
&\stackrel{(\ref{eq:bC})}{=}& -2\: \frac{\partial}{\partial y^j} \int d^4z \:
Q(x_0,z)\: f^j(z,y_0) \:T^{(0)}(z,y)_{|y=y_0} \:,
\end{eqnarray*}
substitute for the integral the expansion~(\ref{eq:Dc1}--\ref{eq:Dclue})
for $p=0$, and differentiate through.
${\mbox{ }}$ \QED

The above theorem gives the clue for understanding the operator product $QP$
as well as the commutator $[P,Q]$, as we shall now explain. Using that the
product of $Q$ with a smooth operator is smooth, we can write the operator
product $QP$ according to~(\ref{eq:bbz}) in the form
\begin{equation}
(Q\:P)(x,y) \;=\; \sum_{p=-1}^\infty (Q\:P^p)(x,y) \:+\:
{\mbox{(smooth contributions)}}
    \label{eq:DA5}
\end{equation}
with
\begin{equation}
P^p(x,y) \;=\; g_p(x,y) \:T^{(p)}(x,y) \:.
    \label{eq:DA6}
\end{equation}
The summands in~(\ref{eq:DA5}) are precisely of the form considered in
Theorem~\ref{thmD1}, with the only exception that the functions $g_p$
in~(\ref{eq:DA6}) in general have no rapid decay at infinity.
Fortunately, the behavior of the functions $g_p$ at infinity is of no relevance
to us, and we can apply Theorem~\ref{thmD1} to each summand in~(\ref{eq:DA5})
using the following approximation argument.
For fixed $x$ and $y$, we choose a Schwartz function $\eta$ which is
identically equal to one on a compact set $K \ni x,y$. Then the function $(g_p
\eta)$ has rapid decay at infinity, and Theorem~\ref{thmD1} applies to the
operator product~(\ref{eq:bbo}) with $g:=g_p \eta$. In the discussion below
leading to Corollary~\ref{corollB1}, the function $g(z)$ enters only for $z$ in
a neighborhood of $x$ and $y$. In this neighborhood, $g$ and $g_p$ coincide,
and thus the dependence on $\eta$ drops out. This shows that the behavior of
the function $g_p$ at infinity is indeed of no importance for
Corollary~\ref{corollB1} below.

Let us briefly discuss the expansion~(\ref{eq:Dc1}--\ref{eq:Dclue}).
First of all, we point out that we calculate modulo terms of the form
$\partial_x^K R(x,y)$ with $|K| \leq L-|J|-3$. This corresponds to
the fact that we have no information on the behavior of $Q$ near the origin.
For generic regularizations or simple regularizations like a cutoff in
momentum space, the terms $\partial_x^K R(x,y)$ will not be zero.
Thus in this case, the operator product $QR$ does not vanish, even when we take
for $R$ a Dirac sea of the vacuum; this is in agreement with our
consideration in momentum space after~(\ref{eq:DH}).  However, the situation is
much different if we assume that we have a regularization where the terms
$\partial^K_x R(x,y)$ all vanish.  Namely, if we then take for $R$ a Dirac sea
of the vacuum, for example
\[ R(x,y) \;=\; \frac{i}{2}\:\xi\slsh\:T^{(-1)}(x,y)
\:, \]
the integrand of the line integral in~(\ref{eq:Dc2}) is a rational function
with poles in the upper half plane, and $QR$ is zero (up to the
contributions not specified in Theorem~\ref{thmD1}). This corresponds to
our observation made after~(\ref{eq:Du}) that the poles of $Q$ cannot be
detected when testing with solutions of the free Dirac equation. The
regularizations for which the terms $\partial^K_x R(x,y)$ vanish are just
the optimal regularizations introduced after~(\ref{eq:DH}).
The main advantage compared to our earlier consideration
in momentum space after~(\ref{eq:DH}) is that the
expansion~(\ref{eq:Dc1}--\ref{eq:Dclue}) tells how the
macroscopic perturbations of $P$ and $Q$ effect the operator product. In
particular one sees that, when taking into account the macroscopic
perturbations, the operator product $QP$ does in general not vanish (even for
optimal regularizations), and thus the problem of disjoint supports disappears.

In an interacting system, the factor $g(z,y)$ in~(\ref{eq:Dc2}) is
composed of the bosonic potentials and fields. Thus in the generic situation,
the line integrals in~(\ref{eq:Dc2}) will vanish only if the operator $Q$ is
identically zero. In order to make this argument clearer, it is useful not to think
of $P$ as a fixed object, but to consider small dynamic perturbations of $P$.
More precisely, we consider perturbations of $P$ induced by perturbations of the
bosonic potentials of our physical system. In order not to disturb the
Euler-Lagrange equations, these perturbations must not be arbitrary, but should
satisfy the physical equations; a typical example are perturbations by an
electromagnetic wave. Thus we consider variations of our system by small,
physically admissible perturbations of $P$ and study the effect on the operator
product $QP$. We refer to this procedure for analyzing the operator product that
we {\em{test with physical perturbations of $P$}}\index{testing with
physical perturbations}.
Clearly, the requirement that the perturbation should
satisfy the physical equations is a strong restriction (in particular, such
perturbations are not dense in the $L^2$ topology). The reason
why it is nevertheless a reasonable concept to use physical perturbations
for testing is that these perturbations enter into~(\ref{eq:Dc2}) only along the
one-dimensional line $\{\lambda y + (1-\lambda) x\}$. In the example of the
perturbation by an electromagnetic wave, the electromagnetic field appears
in the function $g$ in~(\ref{eq:Dc2}), and by changing the location and amplitude
of the wave, we can completely determine the function $h_J(\Omega)$ as well as
the order $L-|J|-1$ of the pole of the integrand at the origin. Notice
furthermore that the summands in~(\ref{eq:Dc1}) scale in the Planck energy
exactly as the corresponding summands in the expansion~(\ref{eq:Dz}).
We conclude that by testing the operator product $QP$ with physical
perturbations of $P$, we can reconstruct the weak evaluation
formula~(\ref{eq:Dz}) completely.

Next, it is instructive to consider the behavior of the summands
in~(\ref{eq:Dc2}) as $y-x$ gets small. If $y-x$ is scaled like $(y-x)
\to \lambda \:(y-x)$ with $\lambda>0$, the variable
transformation $\alpha \to
\lambda^{-1} \alpha$ shows that as $\lambda \searrow 0$,
the line integral blows up like
$\lambda^{-(2n+p+1)}$. On the other hand, the factor
$T^{(p+n+1)}$ goes to zero like $\lambda^{2n+2p}$ in
this limit. Thus each summand in~(\ref{eq:Dc2}) scales like $\lambda^{-1+p}$.
It is remarkable that, no matter how large the order of the pole of $Q$ at
the origin is, the operator product $QP$ has at the origin a pole of at most the
order one. The reason is that in~(\ref{eq:Dc2}), we integrate over the
pole of $Q$, and this regularizes the singularity at the origin.
Since a pole of order one is integrable in three space dimensions,
we do not need to study the operator product $(QP)(x,y)$ at the
origin $x=y$. According to the light-cone expansion of
Theorem~\ref{thmD1}, the information contained in the weak evaluation
formula~(\ref{eq:Dz}) is retrieved in the operator product $QP$ by
considering the singularities on the light cone away from the origin.

The expansion of Theorem~\ref{thmD1} immediately allows us to study also the
commutator $[P,Q]$.  Namely, by taking the adjoint, $(QR)(x,y)^*=(RQ)(y,x)$, the
formula~(\ref{eq:Dc1}--\ref{eq:Dclue}) applies to the operator product $PQ$ as
well, and we can take the difference $[P,Q]=PQ-QP$.  The key observation is that
in the product $(QR)(x,y)$, the pole of $Q$ at the origin appears
in~(\ref{eq:Dc2}) together with the factor $g(z,y)$ and $z \approx x$, whereas
in the product $(RQ)(x,y)$, this pole is multiplied by $g(x,z)$ and $z \approx
y$.  Thus when testing $[P,Q](x,y)$ with perturbations of $P$, one can
distinguish between the contributions from $PQ$ and $QP$ by considering
perturbations which are localized near $y$ and $x$, respectively.

These results are summarized as follows\footnote{{\textsf{Online version}:}
A simpler and cleaner method to obtain this result is to use the so-called
method of {\em{testing on null lines}} as worked out in~\S3.5.2 in the
book~\cite{cfs} (listed in the references in the preface to the second online edition).}.
\begin{Corollary} \label{corollB1}
To every order in $E_P$, the poles of $Q(x,y)$ at the origin $x=y$ can be
detected in the commutator $[P,Q]$ by testing with physical perturbations of $P$.
\end{Corollary}

We close with a general comment on the significance of the constructions
in this appendix. Due to the problem of disjoint
supports, we could make sense out of the commutator $[P,Q]$ only after taking
into account the macroscopic perturbations of $P$ and $Q$. As a consequence,
the relevant contributions to the commutator $[P,Q]$ are
by several orders of $(l_{\mbox{\scriptsize{macro}}} E_P)^{-1}$ smaller than
expected from a simple scaling argument. This can be interpreted
that the causal structure of Minkowski space and the structure of the
Dirac seas, which are the underlying reason for the problem of
disjoint supports, have a tendency to making the commutator $[P,Q]$ small.
In this way, the causal structure and the structure of the Dirac seas seem to
correspond nicely to Euler-Lagrange equations of the form $[P,Q]=0$.

\chapter{Perturbation Calculation for the Spectral Decomposition
of $P(x,y)\: P(y,x)$} \label{appC}
\renewcommand{\theequation}{\thechapter.\arabic{equation}}
\renewcommand{\thesection}{\thechapter.\arabic{section}}
\setcounter{equation}{0} \setcounter{Def}{0}
\markboth{\ref{appC}.~PERTURBATION CALCULATION FOR THE SPECTRAL DECOMPOSITION}{}
\index{perturbation expansion!for the spectral decomposition}
In this appendix we shall develop a convenient method for analyzing
the eigenvalues and spectral projectors of the matrix $A_{xy}
\equiv P(x,y)\:P(y,x)$ and compute all contributions to the
eigenvalues needed for the derivation of the effective gauge group
in Chapter~\ref{esec4}.  Our strategy is as follows. We decompose
the fermionic projector as
\[ P \;=\; P_0 \:+\: \Delta P \]
with $P_0$ according to~(\ref{e:3d}). This gives rise to the decomposition
of $A$
\begin{equation}
A \;=\; A_0 \:+\: \Delta A \label{e:Cb1} \\
\end{equation}
with
\begin{eqnarray}
\hspace*{-0.5cm}A_0 &=& P_0(x,y)\: P_0(y,x) \label{e:Cc1} \\
\hspace*{-0.5cm}\Delta A &=& \Delta P(x,y) \: P_0(y,x) \:+\: P_0(x,y) \:\Delta P(y,x) \:+\:
\Delta P(x,y) \:\Delta P(y,x) \:. \label{e:Cd1}
\end{eqnarray}
The eigenvalues and spectral projectors of $A_0$ were computed
explicitly in Chapter~\ref{esec3}, see~(\ref{e:3spec}, \ref{e:3k}).
On the light cone, $P_0(x,y)$ has singularities
of order ${\mathcal{O}}((y-x)^{-4})$, whereas $\Delta P(x,y) =
{\mathcal{O}}((y-x)^{-2})$. Likewise, $\Delta A$ is compared to
$A_0$ of lower degree on the light cone. For this reason, $\Delta
A$ can be treated perturbatively in the sense that the eigenvalues
and spectral projectors of $A$ can be expressed to any given
degree on the light cone by a finite order perturbation
calculation. Apart from the purely computational aspects, the main
difficulty is that $A_0$ may have degenerate eigenvalues, and in
this case we need to carefully analyze whether the degeneracy is
removed by the perturbation. Our method is to first compute
projectors on invariant subspaces of $A$ ({\S}\ref{appC1}).
Analyzing the perturbation on these invariant subspaces will
then give the spectral decomposition of $A$ ({\S}\ref{appC4}).

\setcounter{backupeqn}{\value{equation}}
\section{Perturbation of Invariant Subspaces} \label{appC1}
\addtocounter{equation}{\value{backupeqn}}
We write the spectral decomposition of $A_0$ as
\[ A_0 \;=\; \sum_{k=1}^K \lambda_k\: F_k \]
with distinct eigenvalues $\lambda_k$ and corresponding spectral
projectors $F_k$. As in~{\S}\ref{esec21} we use the
convention $\lambda_1=0$. Clearly, the $F_k$ are the sum of the
spectral projectors counting multiplicities,
\begin{equation}
F_k \;=\; \sum_{n,c,s {\mbox{\scriptsize{ with }}} \lambda_{ncs}=\lambda_k}
F_{ncs} \label{e:Bc}
\end{equation}
with $\lambda_{ncs}$ and $F_{ncs}$ according to~(\ref{e:3k}). Since the
perturbation $\Delta A$ will in general split up the degenerate eigenvalues,
we cannot expect that by perturbing $F_k$ we obtain spectral projectors
of the matrix $A$. But we can form projectors $G_k$ on the space spanned
by all eigenvectors of $A$ whose eigenvalues are sufficiently close to
$\lambda_k$. The $G_k$ are most conveniently introduced using contour
integrals. We choose $\varepsilon>0$ such that
\[ |\lambda_i-\lambda_j| \;<\; 2 \varepsilon \spc {\mbox{for all
$i,j=1,\ldots,K$ and $i \neq j$}}. \]
Then we set
\begin{equation}
G_k \;=\; \frac{1}{2 \pi i} \oint_{|z-\lambda_k|=\varepsilon}
(z-A)^{-1}\: dz\:, \label{e:Ba}
\end{equation}
The Cauchy integral formula shows that $G_k$ is indeed a projector on the
desired subspace.

The integral formula~(\ref{e:Ba}) is very useful for a perturbation expansion.
To this end, we substitute~(\ref{e:Cb1}) into~(\ref{e:Ba}) and compute the
inverse with the Neumann series,
\begin{eqnarray*}
G_k &=& \frac{1}{2 \pi i} \oint_{|z-\lambda_k|=\varepsilon}
(z-A_0-\Delta A)^{-1}\: dz \\
&=& \frac{1}{2 \pi i} \oint_{|z-\lambda_k|=\varepsilon}
\left(\1 - (z-A_0)^{-1}\: \Delta A \right)^{-1}\:(z-A_0)^{-1}\: dz \\
&=& \frac{1}{2 \pi i} \oint_{|z-\lambda_k|=\varepsilon}
\sum_{n=0}^\infty \left((z-A_0)^{-1}\: \Delta A \right)^n \:(z-A_0)^{-1}\:
dz\: .
\end{eqnarray*}
Interchanging the integral with the infinite sum gives the perturbation
expansion,
\begin{equation}
G_k \;=\; \sum_{n=0}^\infty\: \frac{1}{2 \pi i}
\oint_{|z-\lambda_k|=\varepsilon}
\left((z-A_0)^{-1}\: \Delta A \right)^n \:(z-A_0)^{-1}\: dz\:,
\label{e:Bb}
\end{equation}
where $n$ is the order in perturbation theory. After substituting in the
spectral representation for $(z-A_0)^{-1}$,
\begin{equation}
(z-A_0)^{-1} \;=\; \sum_{l=1}^K \frac{F_l}{z-\lambda_l}\:, \label{e:BzA0}
\end{equation}
the contour integral in~(\ref{e:Bb}) can be carried out with residues.
For example, we obtain to second order,
\begin{eqnarray}
\lefteqn{ G_k \;=\; F_k \:+\: \sum_{l \neq k} \frac{1}{\lambda_k-\lambda_l}
\left( F_k\: \Delta A\: F_l \:+\: F_l \:\Delta A\: F_k \right)
\:+\: {\mathcal{O}}((\Delta A)^3) } \nonumber \\
&&+\sum_{l, m \neq k} \frac{1}{(\lambda_k-\lambda_{l})
(\lambda_k-\lambda_{m})}  \nonumber \\ &&\hspace*{1.5cm} \times
\left(
F_k\: \Delta A\: F_{l} \:\Delta A\: F_{m} +
F_{l}\: \Delta A\: F_k \:\Delta A\: F_{m} +
F_{l}\: \Delta A\: F_{m} \:\Delta A\: F_k \right) \nonumber \\
&&-\sum_{l \neq k} \frac{1}{(\lambda_k-\lambda_l)^2}
\nonumber \\ &&\hspace*{1.5cm} \times
\left( F_k\: \Delta A\: F_k \:\Delta A\: F_l +
F_k\: \Delta A\: F_l \:\Delta A\: F_k +
F_l\: \Delta A\: F_k \:\Delta A\: F_k \right).\spc \label{e:Bcc}
\end{eqnarray}
To order $n>2$, the corresponding formulas are clearly more complicated,
but even then they involve matrix products which are all of the form
\begin{equation}
F_{k_1} \:\Delta A\: F_{k_2} \:\Delta A\: \cdots \:F_{k_n} \:\Delta A\:
F_{k_{n+1}}\:. \label{e:Bd}
\end{equation}
Substituting in~(\ref{e:Bc}) and expanding, we can just as well consider
matrix products of the form~(\ref{e:Bd}) with the factors $F_k$ replaced
by $F_{ncs}$. Furthermore, for the computation of the eigenvalues we need
to take the expectation values of $G_k$ with certain matrix elements of
$\Delta A$. This leads us to traces of matrix products of the form
\begin{equation}
\Tr \left( F_{n_1 c_1 s_1} \:\Delta A_1\: F_{n_2 c_2 s_2} \:\Delta A_2\:
\cdots \:F_{n_l c_l s_l} \:\Delta A_l \right) \label{e:Ce1}
\end{equation}
with $l=n+1$. We refer to a trace of the form~(\ref{e:Ce1}) as a
{\em{matrix trace}}\index{matrix trace}. Our first task is to develop an efficient method
for computing matrix traces ({\S}\ref{appC2} and~{\S}\ref{appC3}); after
that we will proceed with the calculation of the eigenvalues of $A$
({\S}\ref{appC4}).

\setcounter{backupeqn}{\value{equation}}
\section{Factorization of Matrix Traces} \label{appC2}
\index{matrix trace!factorization of}
\addtocounter{equation}{\value{backupeqn}}
If one attempts to calculate a matrix trace~(\ref{e:Ce1})
directly by substituting in the formulas of the light-cone
expansion, the resulting expressions become
so complicated and involve so many Dirac matrices that they are almost
impossible to handle.  We shall now simplify the situation by giving
a procedure which allows us to factor matrix traces into a product of
so-called elementary traces, which are much easier to compute\footnote{{\textsf{Online version}:}
This factorization method is obtained in a somewhat easier way
by computing the matrix elements in a double null spinor frame as explained in~\cite[\S2.6.2]{cfs}
(see the references in the preface to the second online edition).}.
According to~(\ref{e:Cd1}), we can assume that each factor $\Delta A_j$
in~(\ref{e:Ce1}) is the product of a
contribution to $P(x,y)$ with a contribution to $P(y,x)$.  Denoting the
contributions to $P(x,y)$ by $B_j$ and using that the corresponding
contributions to $P(y,x)$ are obtained by taking the adjoint with respect to
the spin scalar product, we can write each $\Delta A_j$ in the form
\[ \Delta A_j \;=\; B_{j_1} \:B^*_{j_2}\:. \]
Inserting the completeness relation
\[ \sum_{ncs} F_{ncs} \;=\; \1 \]
and expanding gives for~(\ref{e:Ce1}) a sum of terms of the form
\begin{equation}
\Tr \left( F_{n_1 c_1 s_1} \:B_1 \: F_{n_2 c_2 s_2} \:B_2^*\:
\cdots \:F_{n_{k-1} c_{k-1} s_{k-1}}\: B_{k-1} \:F_{n_k c_k s_k} \:B_k^* \right)
\label{e:Cf1}
\end{equation}
with indices $(n_j, s_j, c_j)$ (which are in general different from those
in~(\ref{e:Ce1})) and $k=2l$.

In order to handle the sector indices in~(\ref{e:Cf1}), we introduce operators
$K_{n_1, n_2}$ which act on the sector index and map sector $n_2$ to sector
$n_1$, i.e.\ in components
\begin{equation}
    (K_{n_1 n_2})^n_{n'} \;=\; \delta^n_{n_1} \:\delta_{n' n_2} \:.
    \label{e:B1n}
\end{equation}
Then
\begin{equation}
    F_{ncs} \;=\; K_{n1} \:F_{1cs}\: K_{1n} \:.
    \label{e:Crueck}
\end{equation}
If we substitute this relation into~(\ref{e:Cf1}) and combine the operators
$K_\cdot$ and $B_j$ to ``new'' operators $B_j$, we obtain a matrix trace again
of the form~(\ref{e:Cf1}), but with all indices $n_j$ equal to one.  Therefore
in what follows we can restrict attention to the case of one sector and omit the
sector indices.  The generalization to several sectors will be straightforward
by inserting operators $K_\cdot$ into the end formulas.

We choose a space-like unit vector $u$ which is orthogonal to $\xi$
and $\overline{\xi}$. Then the imaginary vector $v=i u$ satisfies the relations
\begin{equation}
v_j \:\xi^j \;=\;0 \;=\; v_j \:\overline{\xi^j} \:,\spc v^2 \;=\; 1
\:,\spc \overline{v} \;=\; -v \:. \label{e:C61}
\end{equation}
An explicit calculation using~(\ref{e:3k}) yields that
\begin{equation}
    F_{R+} \;=\; v\slsh \:F_{L+}\:v\slsh \:,\;\;\;\;\;\;\;
    F_{L-} \;=\; \frac{1}{z} \: \xi\slsh\:v\slsh \:F_{L+}\:v\slsh\:\xi\slsh
    \:,\;\;\;\;\;\;\; F_{R-} \;=\; \frac{1}{z} \:\xi\slsh \:F_{L+} \:\xi\slsh \:.
    \label{e:C7a}
\end{equation}
Substituting these formulas into~(\ref{e:Cf1}), we obtain an expression
involving only the spectral projector $F_{L+}$, namely
\begin{equation}
(\ref{e:Cf1}) \;=\; \Tr \left( F_{L+}\:C_1 \:F_{L+}\: C_2 \cdots F_{L+}\:C_k \right)
    \label{e:Cd2}
\end{equation}
with suitable matrices $C_j$. Since the $F_{L+}$ are projectors on one-dimensional
subspaces,
\[ F_{L+} \:C\: F_{L+} \;=\; \Tr (F_{L+} \:C)\; F_{L+} \:. \]
By iteratively applying this relation in~(\ref{e:Cd2}), we get the product
of traces
\[ \Tr \left( F_{L+}\:C_1\right) \: \Tr \left(F_{L+}\: C_2 \right)
\:\cdots\: \Tr \left( F_{L+}\:C_k \right) \:. \]
If we express the matrices $C_j$ explicitly in terms of $B_j$ and $B_j^*$,
we obtain the following factorization formula,
\begin{eqnarray}
\lefteqn{ \Tr \left( F_{c_1 s_1} \:B_1 \: F_{c_2 s_2} \:F_2^*\:
\cdots \:B_{k-1} \:F_{c_k s_k} \:B_k^* \right) } \nonumber \\
&=& F^{c_1 c_2}_{s_1 s_2}(B_1)\: F^{c_2 c_3}_{s_2 s_3}(B_2^*)\:
\cdots\: F^{c_{k-1} c_k}_{s_{k-1} s_k}(B_{k-1})\:
F^{c_k c_1}_{s_k s_1}(B_k^*) \:, \label{e:Ch1}
\end{eqnarray}
where $F_{s_i s_j}^{c_i c_j}$ are the so-called {\em{elementary traces}}
defined by
\begin{equation}
\left. \begin{array}{lcl}
\!\!\!\!\!F^{LL}_{++}(B) \;=\; \Tr (F_+ \:\chi_L\: B)
&\!\!\!\!,\quad& \displaystyle F^{LR}_{++}(B) \;=\; \Tr(F_+ \:v\slsh\: \chi_L\: B)
\\[.4em]
\!\!\!\!\!F^{LL}_{+-}(B) \;=\; \Tr (\xi\slsh\: F_+ \:v\slsh\:\chi_L\: B)
&\!\!\!\!,\quad& F^{LR}_{+-}(B) \;=\; \Tr(\xi\slsh\: F_+ \:\chi_L\: B) \\[.4em]
\!\!\!\!\!F^{LL}_{-+}(B) \;=\; \displaystyle \frac{1}{z} \:\Tr (F_+ \:v\slsh\:\xi\slsh\chi_L\: B)
&\!\!\!\!,\quad& F^{LR}_{-+}(B) \;=\;
\displaystyle \frac{1}{z} \:\Tr(F_+ \:\xi\slsh\: \chi_L\: B) \\[.7em]
\!\!\!\!\!F^{LL}_{--}(B) \;=\; \displaystyle \frac{1}{z}\:\Tr (\xi\slsh\: F_+ \:\xi\slsh \:\chi_L\: B)
&\!\!\!\!,\quad& F^{LR}_{--}(B) \;=\; \displaystyle \frac{1}{z}\:\Tr(\xi\slsh\:
F_+ \:v\slsh\:\xi\slsh\: \chi_L\: B) \:.\!
\end{array} \right\} \label{e:Ce2}
\end{equation}
These formulas are also valid for the opposite chirality after the
replacements $L \leftrightarrow R$.
The elementary traces of $B^*$ are obtained by taking the complex conjugate,
\begin{equation}
\left. \begin{array}{lcl}
F^{LL}_{++}(B^*) \;=\; \overline{F^{RR}_{--}(B)} &,\;\;\;\;\;\;&
F^{LR}_{++}(B^*) \;=\; \overline{F^{LR}_{--}(B)} \\[.4em]
F^{LL}_{+-}(B^*) \;=\; \overline{F^{RR}_{+-}(B)} &,\;\;\;\;\;\;&
F^{LR}_{+-}(B^*) \;=\; \overline{F^{LR}_{+-}(B)} \\[.4em]
F^{RR}_{-+}(B^*) \;=\; \overline{F^{LL}_{-+}(B)} &,\;\;\;\;\;\;&
F^{LR}_{-+}(B^*) \;=\; \overline{F^{LR}_{-+}(B)} \\[.4em]
F^{LL}_{--}(B^*) \;=\; \overline{F^{RR}_{++}(B)} &,\;\;\;\;\;\;&
F^{LR}_{--}(B^*) \;=\; \overline{F^{LR}_{++}(B)} \:.
\end{array} \;\;\;\;\;\right\} \label{e:Cg1}
\end{equation}
The relations~(\ref{e:Ch1}--\ref{e:Cg1}) are verified by a straightforward
calculation using~(\ref{e:3k}, \ref{e:3.5}, \ref{e:C61}).

To summarize, the above procedure reduces the calculation of the matrix
trace (\ref{e:Ce1}) to the computation of the elementary traces~(\ref{e:Ce2})
for the contributions $B$ to the light-cone expansion of $P(x,y)$. Taking
the complex conjugate~(\ref{e:Cg1}), one obtains the elementary traces of
the corresponding contributions to $P(y,x)$. By applying~(\ref{e:Ch1})
and, in the case of several sectors, by suitably inserting the operators
$K_\cdot$, every matrix trace can be written as a linear combination of
products of elementary traces.

\setcounter{backupeqn}{\value{equation}}
\section{Calculation of the Matrix Traces} \label{appC3} \label{esecC2}
\addtocounter{equation}{\value{backupeqn}}
We decompose $\Delta P(x,y)$ into its odd and even
parts, denoted by $B_o$ and $B_e$,
\[ \Delta P(x,y) \;=\; B_o(x,y) \:+\: B_e(x,y)\:. \]
Explicit formulas for the fermionic projector in the presence of chiral and
scalar potentials are listed in Appendix~\ref{pappLC}.
For the purpose of this paper, only the contributions involving
the mass matrices $Y_{L\!/\!R}$ and their derivatives are of importance.  But
for completeness and for later use, we will also compute the contributions
which contain the chiral field strength and the chiral currents.
However, we will omit all contributions quadratic in the field
strength.  Namely, these contributions are related to the energy-momentum
tensor of the chiral fields, and it is therefore reasonable to postpone their
analysis until gravity is also taken into consideration.
Thus the phase-free contributions relevant here are
\begin{eqnarray*}
\chi_L \:B_e
&=& \frac{1}{2}\:\chi_L\: m  \:T^{(0)}(x,y)\: \xi \slsh\: \int_x^y dz\:
 \gamma^j \:(D_j Y_L) \\
&& +\chi_L\: m  \:T^{(0)}(x,y)\: Y_L(x) \;+\; {\mathcal{O}}(\log|\xi^2|\: \xi^0) \\
\chi_L \:B_o
&=& \frac{i}{2}\: \chi_L\: m^2 \:T^{(0)}(x,y)\: \xi \slsh\: \int_x^y dz\:
 Y_L \:Y_R \\
&&  +i \chi_L\: m^2 \:T^{(1)}(x,y)\: \int_x^y dz\:
[0,1\:|\: 0]\: Y_L \:\gamma^j (D_j Y_R) \\
&&  +i \chi_L\: m^2 \:T^{(1)}(x,y)\: \int_x^y dz\:
[0,1\:|\: 0]\: \gamma^j (D_j Y_L)\: Y_R \\
&&  -i \chi_L\: m^2 \:T^{(1)}(x,y)\: Y_L(x) \int_x^y dz\:
 \gamma^j (D_j Y_R) \\
&& +\chi_L \:T^{(0)}(x,y)\: \xi^i \int_x^y dz\: [0,1\:|\: 0]\: \gamma^l F^L_{li} \\
&&  +\frac{1}{4}\:\chi_L \:T^{(0)}(x,y)\: \xi \slsh\: \int_x^y dz\:
\gamma^j \gamma^k\:F^L_{jk} \\
&&  -\frac{1}{2}\:\chi_L \:T^{(0)}(x,y)\: \xi \slsh\: \xi^i \int_x^y dz\:
[0,0\:|\: 1]\: j^L_i \\
&&  -\chi_L \:T^{(1)}(x,y)\: \xi^i \int_x^y dz\: [0,1\:|\: 1]\: (\Pdd j^L_i) \\
&&  -\chi_L \:T^{(1)}(x,y)\: \int_x^y dz\: [0,2\:|\: 0]\: j^L_k \:\gamma^k \\
&&+\: \xi\slsh\: {\mathcal{O}}(\xi^{-2}) \:+\: \gamma^j\:F^L_{jk}
\xi^k\: {\mathcal{O}}(\xi^{-2}) \:+\:
{\mathcal{O}}(F_L^2)\:+\:{\mathcal{O}}(\log|\xi^2|\:\xi^0) \:.
\end{eqnarray*}
A straightforward calculation yields for the elementary traces
\begin{eqnarray}
\hspace*{-1cm}F^{LR}_{+-}(P_0) &=& (\deg \leq 1)
\;=\; \frac{i}{2} \:X_L \:(z\:T^{(-1)}_{[0]}) \label{e:Cet1} \\
F^{LR}_{-+}(P_0) &=& (\deg \leq 2)
\;=\; \frac{i}{2} \:X_L \:T^{(-1)}_{[0]} \\
F^{LL}_{++}(B_e) &=& (\deg \leq 1)
\;=\; Y_L(x)\: T^{(0)}_{[1]} \:+\: (\deg<1) \\
F^{LL}_{+-}(B_e) &=& (\deg \leq 0) \\
F^{LL}_{-+}(B_e) &=& (\deg \leq 1) \\
F^{LL}_{--}(B_e) &=& (\deg \leq 1) \;=\;Y_L(y) \; T^{(0)}_{[1]} \:+\: (\deg<1) \\
F^{LR}_{++}(B_o) &=& (\deg \leq 1) \label{e:B23a} \\
&=& v^j \xi^k \:\int_x^y dz\: [0,1 \:|\: 0]\:
F^L_{jk}\: T^{(0)}_{[0]} \:+\: (\deg<1) \\
&&+\frac{2i}{z-\overline{z}}\:\epsilon_{ijkl}\:\xi^i\:\overline{\xi}^j\:v^k
\int_x^y dz\:[0,1 \:|\: 0] \:F_L^{lm}\;(\xi_m\: T^{(0)}_{[0]}) \\
&&+\frac{i}{z-\overline{z}}\:\epsilon^{ijkl}\:(\xi_i \overline{\xi}_j +
\overline{\xi}_i \xi^{(0)}_j - \xi_i \xi^{(0)}_j) \:v_k\:\int_x^y dz\:
\xi^n F^L_{nl}\; T^{(0)}_{[0]} \\
F^{LR}_{+-}(B_o) &=& (\deg \leq 0) \nonumber \\
&=&\frac{i}{2} \: \int_x^y dz \:Y_L\:Y_R \;
((z\: T^{(0)}_{[2]}) \:+\: 4\:T^{(1)}_{[2]}) \:+\: (\deg<0) \\
&&-2i\: Y_L(x)\:Y_R(y) \; T^{(1)}_{[2]} \\
&&-\frac{1}{2}\: \xi_i \:\int_x^y dz\: [0,0\:|\:1] \: j_L^i\;
((z\:T^{(0)}_{[0]}) \:+\: 8\:T^{(1)}_{[0]}) \\
&&+\frac{i}{2}\:\epsilon_{ijkl}\:\frac{z\:\overline{\xi}^i -
\overline{z}\:\xi^i}{z-\overline{z}}\: \int_x^y F_L^{jk}\: (\xi^l \:T^{(0)}_{[0]}) \\
F^{LR}_{-+}(B_o) &=& (\deg \leq 1) \nonumber \\
&=&\frac{i}{2} \:\int_x^y dz \:Y_L\:Y_R \; T^{(0)}_{[2]} \:+\: (\deg<1) \\
&&-\frac{1}{2}\:\xi_i\: \int_x^y dz\: [0,0\:|\:1] \: j_L^i\; T^{(0)}_{[0]} \\
&&-\frac{i}{2}\:\epsilon_{ijkl}\:\frac{\overline{\xi}^i - \xi^i}{z-\overline{z}}\:
\int_x^y F_L^{jk}\: (\xi^l \:T^{(0)}_{[0]}) \\
F^{LR}_{--}(B_o) &=& (\deg \leq 1) \nonumber \\
&=& v^j \xi^k \:\int_x^y dz\: [1,0 \:|\: 0]\:
F^L_{jk}\: T^{(0)}_{[0]} \:+\: (\deg<1) \\
&&+\frac{i}{2}\:\epsilon^{ijkl}\: \xi_i\: v_j \int_x^y F^L_{kl}\;
T^{(0)}_{[0]} \\
&&+\frac{2i}{z-\overline{z}}\:\epsilon_{ijkl}\:\xi^i\:\overline{\xi}^j\:v^k
\int_x^y dz\:[0,1 \:|\: 0] \:F_L^{lm}\;(\xi_m\: T^{(0)}_{[0]}) \\
&&+\frac{i}{z-\overline{z}}\:\epsilon^{ijkl}\:(\xi_i \overline{\xi}_j +
\overline{\xi}_i \xi^{(0)}_j - \xi_i \xi^{(0)}_j)\:v_k \int_x^y dz\:
\xi^n F_{nl}\; T^{(0)}_{[0]}\:.\qquad \label{e:Cet2}
\end{eqnarray}
Here the totally anti-symmetric symbol $\epsilon_{ijkl}$ appears because
we applied the identity
\[ \Tr (\chi_{L\!/\!R} \:a \slsh \:b \slsh\:c \slsh\: d \slsh) \;=\;
2 \left( (ab) (cd) + (da)(bc) - (ac)(bd) \right) \:\mp\:
2i\:\epsilon_{ijkl}\: a^i b^j c^k d^l \:. \]
Therefore, the corresponding formulas for the opposite chirality are now
obtained by the replacements
\begin{equation}
L \;\longleftrightarrow\; R \:,\spc \epsilon_{ijkl}
\;\longrightarrow\; -\:\epsilon_{ijkl} \:.
\label{e:Ct}
\end{equation}
The elementary traces of the adjoints are computed via~(\ref{e:Cg1}).
All other elementary traces vanish.

Applying~(\ref{e:Ch1}) and the degree estimates for the elementary traces and
omitting all terms quadratic in the field strength, we can factor and
estimate the following matrix traces,
\begin{eqnarray}
\Tr ( F_{L+}\: \Delta A ) &=& F^{LR}_{+-}(P_0) \:F^{RL}_{-+}(B_o^*) \:+\:
F^{LR}_{+-}(B_o) \:F^{RL}_{-+}(P_0^*) \spc\nonumber \\
&&+\: F^{LL}_{++}(B_e) \:F^{LL}_{++}(B_e^*)
\:+\: (\deg <2) \label{e:CE} \\
\Tr (F_{L-}\: \Delta A) &=& F^{LR}_{-+}(P_0) \:F^{RL}_{+-}(B_o^*) \:+\:
F^{LR}_{-+}(B_o) \:F^{RL}_{+-}(P_0^*) \nonumber \\
&&+\: F^{LL}_{--}(B_e) \:F^{LL}_{--}(B_e^*)
\:+\: (\deg <2) \\
\Tr (F_{Ls} \:\Delta A\: F_{Ls} \:\Delta A) &=& \Tr (F_{Ls} \:\Delta A)
\: \Tr(F_{Ls} \:\Delta A) \;=\; (\deg <5) \label{e:C2b} \\
\Tr (F_{Ls} \:\Delta A\: F_{Rs} \:\Delta A) &=& (\deg <5) \\
\Tr (F_{L+} \:\Delta A\: F_{R-} \:\Delta A) &=&
(F^{LL}_{++}(B_e)\: F^{LR}_{+-}(P_0^*)
\:+\:F^{LR}_{+-}(P_0)\: F^{RR}_{--}(B_e^*)) \nonumber \\
&& \hspace*{-3.7cm} \times (F^{RR}_{--}(B_e)\: F^{RL}_{-+}(P_0^*)
\:+\:F^{RL}_{-+}(P_0)\: F^{LL}_{++}(B_e^*)) + (\deg <5) \;\;\;\;\;\;\;\;\;\;\; \\
\Tr (F_{L-} \:\Delta A\: F_{R+} \:\Delta A) &=&
(F^{LL}_{--}(B_e)\: F^{LR}_{-+}(P_0^*)
\:+\:F^{LR}_{-+}(P_0)\: F^{RR}_{++}(B_e^*)) \nonumber \\
&& \hspace*{-3.7cm} \times (F^{RR}_{++}(B_e)\: F^{RL}_{+-}(P_0^*)
\:+\:F^{RL}_{+-}(P_0)\: F^{LL}_{--}(B_e^*)) + (\deg <5) \label{e:CF1} \\
\Tr (F_{L+} \:\Delta A\: F_{L-} \:\Delta A) &=& 0 \;=\;
\Tr (F_{L-} \:\Delta A\: F_{L+} \:\Delta A) \:. \label{e:C2e}
\end{eqnarray}
If we consider more generally the matrix trace of order $l$, factorization gives a
linear combination of products of elementary traces as in~(\ref{e:Ch1}) (with
$k=2l$). Let us estimate the degree of each of these products.
Clearly, the number of factors $F^\cdot_{+-}$ equals the number of factors
$F^\cdot_{-+}$, we denote the number of such pairs by $p$. Furthermore, let $q$
be the number of factors $F^\cdot_\cdot(\Delta P^\cdot)$ (where $\Delta P^\cdot$
stands for either $\Delta P$ or $\Delta P^*$). According to~(\ref{e:Cd1}),
each $\Delta A$ contains at least one factor $\Delta P^\cdot$, hence $q \geq l$.
The number of factors $F^\cdot_{++}$ and $F^\cdot_{--}$ is $2(l-p)$, and we saw
above that each of these factors must involve $\Delta P^\cdot$, thus $q \geq
2(l-p)$. Adding our two upper bounds for $q$ gives the inequality $q+p \geq
3l/2$. To estimate the degrees we first note that the degree of the pair
$F_{+-}^\cdot(P_0^\cdot)\: F_{-+}^\cdot(P_0^\cdot)$ is three, and is decreased
at least by one each time a $P_0^\cdot$ is replaced by $\Delta P^\cdot$. The
total number of factors $F_{+-}^\cdot(\Delta P^\cdot)$ and
$F_{-+}^\cdot(\Delta P^\cdot)$ is $q-2(l-p)$. On the other hand, the degree of
each factor $F_{++}^\cdot$ and $F_{--}^\cdot$ is at most one. Hence the degree
of the matrix is bounded from below by $3p-(q-2(l-p))+2(l-p) = 4l-(q+p)$.
Substituting in our above lower bound for $q+p$ gives the degree estimate
\begin{equation}
\Tr ( F_{c_1 s_1} \:\Delta A\:\cdots\:  F_{c_l s_l} \:\Delta A) \;=\;
\left( \deg < \frac{5}{2}\: l \right) \;=\;
(\deg < 3l-1) \;\;\;{\mbox{ for $l \geq 3$.}} \label{e:C3ord}
\end{equation}

The above formulas are valid in the case $N=1$ of one sector. The generalization
to several sectors is done by inserting suitable operators $K_\cdot$ into the
traces. This has no effect on the degree on the light cone, and thus the
estimates of the matrix traces in~(\ref{e:CE}--\ref{e:C3ord})
hold in the general case as well.
We substitute the above results for the elementary
traces~(\ref{e:Cet1}--\ref{e:Cet2}) into (\ref{e:CE}--\ref{e:CF1}) and
insert the operators $K_\cdot$ to obtain the following explicit formulas:
\begin{eqnarray}
\lefteqn{ \Tr ( F_{nL+}\: \Delta A ) \;=\; (\deg < 2) } \nonumber \\
&&+\Tr_S \left\{I_n\:
\hat{Y}_L(x)\: \hat{Y}_L(y) \right\} T^{(0)}_{[1]}\:\overline{T^{(0)}_{[1]}}
\label{e:Ctr1} \\
&&+\frac{1}{4} \: \int_x^y dz \:\Tr_S \left\{I_n\:
\acute{Y}_L\:\grave{Y}_R \; X_R \right\} ((z \:
T^{(0)}_{[2]}) \:+\: 4\:T^{(1)}_{[2]}) \: \overline{T^{(-1)}_{[0]}} \\
&&-\Tr_S \left\{I_n\:\acute{Y}_L(x)\:\grave{Y}_R(y) \;X_R\right\}
T^{(1)}_{[2]} \: \overline{T^{(-1)}_{[0]}} \\
&&+\frac{1}{4} \: \int_y^x dz \:\Tr_S \left\{I_n\: X_L\:\acute{Y}_R\: \grave{Y}_L
\right\} (z\: T^{(-1)}_{[0]})\: \overline{T^{(0)}_{[2]}} \\
&&+\frac{i}{4}\: \xi_i \:\int_x^y dz\: [0,0\:|\:1] \: \Tr_S \left\{I_n\:j_L^i\;
X_R \right\} ((z\:T^{(0)}_{[0]}) \:+\: 8\:T^{(1)}_{[0]}) \: \overline{T^{(-1)}_{[0]}} \\
&&-\frac{i}{4}\:\xi_i \int_y^x dz\: [0,0\:|\:1] \:\Tr_S \left\{I_n\:X_L\;
j_R^i \right\} (z\:T^{(-1)}_{[0]})\: \overline{T^{(0)}_{[0]}} \\
&&+\frac{1}{4}\: \epsilon_{ijkl}\:\frac{z\:\overline{\xi}^i -
\overline{z}\:\xi^i}{z-\overline{z}}\:\xi^l\: \int_x^y \Tr_S \left \{
I_n \:F_L^{jk}\:X_R \right\} \;T^{(0)}_{[0]}\: \overline{T^{(-1)}_{[0]}} \\
&&+\frac{1}{4}\:\epsilon_{ijkl}\:\frac{\overline{\xi}^i - \xi^i}{z-\overline{z}}\:
\xi_l\: \int_y^x \Tr_x \left\{ I_n\: X_L\: F_R^{jk} \right\}\;
(z\:T^{(-1)}_{[0]})\: (\overline{T^{(0)}_{[0]}}) \\
\lefteqn{ \Tr ( F_{nL-}\: \Delta A ) \;=\; (\deg < 2) } \nonumber \\
&&+\Tr_S \left\{I_n\: \hat{Y}_L(y)\:\hat{Y}_L(x) \right\}
T^{(0)}_{[1]}\:\overline{T^{(0)}_{[1]}} \\
&&+\frac{1}{4} \:\int_x^y dz \:\Tr_S \left\{I_n\:\acute{Y}_L\:\grave{Y}_R \;X_R
\right\} T^{(0)}_{[2]} \: (\overline{z\: T^{(-1)}_{[0]}}) \\
&&+\frac{1}{4} \: \int_y^x dz \:\Tr_S \left\{I_n\:X_L\;\acute{Y}_R\:\grave{Y}_L \right\}
T^{(-1)}_{[0]}\:((\overline{z\: T^{(0)}_{[2]}}) \:+\: 4\:\overline{T^{(1)}_{[2]}}) \\
&&- \Tr_S \left\{I_n\:X_L\; \acute{Y}_R(y)\:\grave{Y}_L(x) \right\}
T^{(-1)}_{[0]}\:\overline{T^{(1)}_{[2]}} \\
&&-\frac{i}{4}\: \xi_i \int_y^x dz\: [0,0\:|\:1] \: \Tr_S \left\{I_n\:X_L\;j_R^i
\right\}  T^{(-1)}_{[0]}\:((\overline{z\:T^{(0)}_{[0]}}) \:+\: 8\:\overline{T^{(1)}_{[0]}}) \\
&&+\frac{i}{4}\:\xi_i\: \int_x^y dz\: [0,0\:|\:1] \: \Tr_S \left\{I_n\:j_L^i\;X_R\right\}
T^{(0)}_{[0]}\: (\overline{z\: T^{(-1)}_{[0]}}) \\
&&-\frac{1}{4}\:\epsilon_{ijkl}\:\frac{\overline{\xi}^i - \xi^i}{z-\overline{z}}\:
\xi^l \:\int_x^y \Tr_S \left\{ I_n\: F_L^{jk}\:X_R \right\}\: T^{(0)}_{[0]}
\:(\overline{z\: T^{(-1)}_{[0]}}) \\
&&-\frac{1}{4}\:\epsilon_{ijkl}\:\frac{z\:\overline{\xi}^i -
\overline{z}\:\xi^i}{z-\overline{z}}\: \xi^l \int_y^x
\Tr_S \left\{ I_n \:X_L\: F_R^{jk} \right\} \:T^{(-1)}_{[0]}\:\overline{T^{(0)}_{[0]}} \\
\lefteqn{ \Tr (F_{nL+} \:\Delta A\: F_{n' R-} \:\Delta A) \;=\;  (\deg < 5) } \nonumber \\
&&-\frac{1}{4}\: \Tr_S \left\{ I_n
\left(\hat{Y}_L(x) \:X_L \; T^{(0)}_{[1]} \:(\overline{z\: T^{(-1)}_{[0]}})
\:-\: X_L\: \hat{Y}_R(x) (z\: T^{(-1)}_{[0]})\: \overline{T^{(0)}_{[1]}} \right)
\right. \nonumber \\
&&\hspace*{1cm} \left.
\times \:I_{n'} \left( \hat{Y}_R(y)\: X_R \:T^{(0)}_{[1]} \:\overline{T^{(-1)}_{[0]}}
\:-\:X_R\: \hat{Y}_L(y)\: T^{(-1)}_{[0]}\: \overline{T^{(0)}_{[1]}} \right)
\right\} \label{e:Ctr2a} \\
\lefteqn{ \Tr (F_{nL-} \:\Delta A\: F_{n' R+} \:\Delta A) \;=\;  (\deg < 5) } \nonumber \\
&& -\frac{1}{4}\: \Tr_S \left\{ I_n
\left( \hat{Y}_L(y)\: X_L \:T^{(0)}_{[1]} \:\overline{T^{(-1)}_{[0]}} \:-\:
X_L\: \hat{Y}_R(y)\: T^{(-1)}_{[0]}\: \overline{T^{(0)}_{[1]}} \right) \right.
\nonumber \\
&&\hspace*{1cm} \left. \times \:I_{n'}
\left(\hat{Y}_R(x) \:X_R \; T^{(0)}_{[1]} \:(\overline{z\: T^{(-1)}_{[0]}}) \:-\:
X_R\: \hat{Y}_L(x)\; (z\: T^{(-1)}_{[0]})\: \overline{T^{(0)}_{[1]}} \right)
\right\} \;\;\qquad \label{e:Ctr2}
\end{eqnarray}

\setcounter{backupeqn}{\value{equation}}
\section{Perturbation of the Non-Zero Eigenvalues} \label{appC4}
\addtocounter{equation}{\value{backupeqn}}
In~{\S}\ref{esec4} we calculated the eigenvalues
$\lambda_{ncs}$ of $A$ in the presence of chiral and scalar
potentials to the leading degree 3, (\ref{e:3k}). Now we shall
compute the contributions to the non-zero eigenvalues of degree
two, denoted by $\Delta \lambda_{ncs}$, $n=1,\ldots,7$ (the kernel
of $A$ will be considered in~{\S}\ref{appC5}). To this end, we
need to analyze the matrix $A$ on the invariant subspaces
${\mbox{Im}} G_k$. First, we choose for fixed $k>1$ a convenient
basis of ${\mbox{Im}} G_k$ as follows. The degeneracy of the
unperturbed eigenspace ${\mbox{Im}} F_k$ can be described by the
index set $I$,
\begin{equation}
I \;=\; \left\{ (ncs) {\mbox{ with }} \lambda_{ncs} = \lambda_k \right\} .
\label{e:Bind}
\end{equation}
Note that, according to~(\ref{e:3k}), $s$ is the same for all elements $(ncs)
\in I$, provided that the eigenvalue is non-zero. The index $c$, however, may
take both values $L$ and $R$, giving rise to the partition of $I$ into $I_L$
and $I_R$,
\[ I_{L\!/\!R} \;=\; \left\{ (ncs) \in I {\mbox{ with }} c=L\!/\!R \right\} . \]
The set $I$ can be used to index a basis of $F_k$; namely we choose
\begin{equation}
(\phi_{ncs})_{(ncs) \in I} \spc {\mbox{with}}\spc 0 \neq \phi_{ncs} \;\in\;
{\mbox{Im}}\: F_{ncs}\:. \label{e:Bl}
\end{equation}
It is convenient to assume that the basis vectors are related to each other by
\begin{equation}
\phi_{n'cs} \;=\; K_{n'n}\: \phi_{ncs} \:,\spc \phi_{n' \bar{c} s} \;=\;
K_{n'n}\:v \slsh\: \phi_{ncs}\;;
    \label{e:Bh}
\end{equation}
this can clearly arranged according to~(\ref{e:Crueck}--\ref{e:C7a}).
Since $F_k$ projects onto a null space, the inner product of any two basis
vectors $\phi_{ncs}$ vanishes. Thus in order to be able to evaluate vectors
in ${\mbox{Im}} F_k$ using the scalar product, we choose a ``dual basis''
$(\phi^{ncs})_{(ncs) \in I}$ of ${\mbox{Im}} F_k^*$ given by
\begin{equation}
\phi^{ncs} \;\in\; {\mbox{Im}}\: F_{ncs}^* \:,\spc
\phi^{n'cs} \;=\; K_{n'n}\: \phi^{ncs} \:,\spc
\phi^{n' \bar{c} s} \;=\; K_{n'n}\: v\slsh\: \phi^{ncs}\:.
    \label{e:Bi}
\end{equation}
The basis vectors and their duals are orthogonal in the sense that for $(ncs)
\neq (n'c's)$,
\[ \bra \phi^{ncs} \:|\: \phi_{n'c's} \ket \;=\; \bra F_{ncs}^*\: \phi^{ncs}
\:|\: F_{n'c's}\: \phi_{n'c's} \ket \;=\;
\bra \phi^{ncs} \:|\: F_{ncs}\: F_{n'c's}\: \phi_{n'c's} \ket \;=\; 0\:. \]
We normalize the basis vectors such that
\begin{equation}
\bra \phi^{ncs} \:|\: \phi_{n'c's} \ket \;=\; \delta^n_{n'}\: \delta^c_{c'} \spc
{\mbox{for all $(ncs), (n'c's) \in I$.}} \label{e:Bj}
\end{equation}
Next we introduce a basis $(\psi_{ncs})_{(ncs) \in I}$ of the invariant
subspace ${\mbox{Im}} G_k$ by applying the projector $P_k$ to the $\phi_{ncs}$,
\begin{equation}
\psi_{ncs} \;=\; G_k\: \phi_{ncs}\:. \label{e:Br}
\end{equation}
Finally, we introduce a basis $(\psi^{ncs})_{(ncs) \in I}$ which is dual to
$(\psi_{ncs})$. We must be careful because projecting on ${\mbox{Im}} (G_k)$
and ${\mbox{Im}} (G_k)$, respectively, does not preserve the orthonormality; more
precisely,
\begin{eqnarray}
S^{ncs}_{n'c's} &\equiv& \bra\: G_k^*\: \phi^{ncs} \:|\: \psi_{n'c's} \ket \;=\;
\bra\: G_k^*\: \phi^{ncs} \:|\: G^k\: \phi_{n'c's} \ket \nonumber \\
&=& \bra\: \phi^{ncs} \:|\: G^k\:|\: \phi_{n'c's} \ket
\;\stackrel{\mbox{\tiny{in general}}}{\neq}\; \delta^n_{n'}\: \delta^c_{c'}\:.
    \label{e:Bk}
\end{eqnarray}
But $S$ is a perturbation of the identity, and thus it can be inverted within
the perturbation expansion by a Neumann series. This makes it possible to
introduce $(\psi^{ncs})_{(ncs) \in I}$ by
\begin{equation}
\psi^{ncs} \;=\; \sum_{(n'c's) \in I} (S^{-1})^{ncs}_{n'c's}\; G_k^*\:
\phi^{n'c's}\:. \label{e:Bp}
\end{equation}
A short calculation shows that this basis of ${\mbox{Im}} G_k^*$ is indeed dual
to $(\psi_{ncs})$ in the sense that
\begin{equation}
\bra \psi^{ncs} \:|\: \psi_{n'c's} \ket \;=\; \delta^n_{n'}\: \delta^c_{c'}\spc
{\mbox{for all $(ncs), (n'c's) \in I$}}. \label{e:Bg}
\end{equation}

Using the basis $(\psi_{ncs})$ and its dual $(\psi^{ncs})$, we can write down
matrix elements of $A$,
\begin{equation}
A^{ncs}_{n'c's} \;=\; \bra \psi^{ncs} \:|\: A \:|\: \psi_{n'c's} \ket \spc
{\mbox{for $(ncs),(n'c's) \in I$}}.    \label{e:Bq}
\end{equation}
From the orthonormality~(\ref{e:Bg}) one sees that $A^{ncs}_{n'c's}$ is indeed a
matrix representation for $A$ in the basis $(\psi_{ncs})$, and thus the
eigenvalues of $A$ on the invariant subspace ${\mbox{Im}} G_k$ are obtained
simply by diagonalizing this matrix. In the unperturbed case (i.e.\ if $\Delta
A=0$), the matrix $A^{ncs}_{n'c's}$ simplifies to
\begin{eqnarray*}
A^{ncs}_{n'c's} &=& \bra \phi^{ncs} \:|\: A_0 \:|\: \phi_{n'c's} \ket
\;=\; \bra \phi^{ncs} \:|\: A_0\: F_{n'c's}\: \phi_{n'c's} \ket \\
&=& \lambda_k\: \bra \phi^{ncs} \:|\: \phi_{n'c's} \ket \;=\; \lambda_k\;
\delta^n_{n'}\: \delta^c_{c'}\:,
\end{eqnarray*}
in agreement with the fact that ${\mbox{Im}} F_k$ is an eigenspace of $A_0$
corresponding to the eigenvalue $\lambda_k$. Thus we see that the matrix
elements $A^{ncs}_{n'c's}$ are to leading order on the light cone of degree 3.
In the following theorem we compute the matrix elements up to contributions of
degree $<2$.

\begin{Thm} \label{thmB1}
We consider the fermionic projector in the presence of chiral and scalar
potentials~(\ref{e:3dir}) and in composite expressions disregard all terms
quadratic in the field strength. Then for all $k=2,\ldots, K$ and
$(ncs), (n'c's) \in I$,
\begin{eqnarray}
A^{ncs}_{n'c's} &=& \lambda_k\; \delta^n_{n'}\: \delta^c_{c'} \:+\:
\delta^c_{c'}\: \Tr \left( F_{ncs}\: \Delta A\: K_{n'n} \right) \nonumber \\
&&+ \delta^c_{c'}\: \sum_{l \neq k} \frac{1}{\lambda_k - \lambda_l}\:
\Tr \left( F_{ncs}\: \Delta A\: F_l\: \Delta A\: K_{n'n} \right) \:+\:
(\deg < 2).    \label{e:Bs}
\end{eqnarray}
\end{Thm}
{\Proof} We begin by computing the matrix $S$, (\ref{e:Bk}), and its inverse.
This calculation will also illustrate how the relations~(\ref{e:Bh})
and~(\ref{e:Bi}) make it possible to rewrite expectation values as matrix
traces and thus to apply the results of~{\S}\ref{appC2} and~{\S}\ref{appC3}.
In the case $c=c'$, we obtain from~(\ref{e:Bk}) and~(\ref{e:Bh}),
\begin{eqnarray*}
S^{ncs}_{n'cs} &=& \bra \phi^{ncs} \:|\: G_k \:|\: \phi_{n'cs} \ket \;=\;
\bra \phi^{ncs} \:|\: G_k \:|\: K_{n'n}\:\phi_{ncs} \ket \\
&\stackrel{(\ref{e:Bl}, \ref{e:Bi})}{=}&
\bra F_{ncs}^*\: \phi^{ncs} \:|\: G_k\: K_{n'n}\:|\: F_{ncs}\: \phi_{ncs} \ket \\
&=& \bra \phi^{ncs} \:|\: F_{ncs}\:G_k\: K_{n'n}\: F_{ncs}\:|\: \phi_{ncs} \ket \\
&\stackrel{(*)}{=}& \Tr \left( F_{ncs}\:G_k\: K_{n'n} \right)\:
\bra \phi^{ncs} \:|\: F_{ncs}\:|\: \phi_{ncs} \ket \\
&=& \Tr \left( F_{ncs}\:G_k\: K_{n'n} \right)\:
\bra \phi^{ncs} \:|\: \phi_{ncs} \ket
\;\stackrel{(\ref{e:Bj})}{=}\; \Tr \left( F_{ncs}\:G_k\: K_{n'n} \right)\:,
\end{eqnarray*}
where in (*) we used that $F_{ncs}$ projects on a one-dimensional subspace. If
we substitute the perturbation expansion for $G_k$, (\ref{e:Bb}), into the
obtained matrix trace, the estimate~(\ref{e:C3ord}) shows that the orders $n>2$
yield contributions to $S$ of degree $<-1$. Thus it suffices to consider for $G_k$
the second order expansion~(\ref{e:Bcc}). This gives
\begin{equation}
S^{ncs}_{n'cs} \;=\; \delta^n_{n'} \:-\: \sum_{l \neq k}
\frac{1}{(\lambda_k-\lambda_l)^2}\: \Tr \left( F_{ncs}\: \Delta A\: F_l\:
\Delta A\: K_{n'n} \right) + (\deg < -1)\:.
    \label{e:Bm}
\end{equation}
Note that of the matrix trace appearing here we need to take into account only
the leading contributions of degree 5; these are easily obtained from~(\ref{e:Ctr2a})
and~(\ref{e:Ctr2}). In the case $c \neq c'$, we obtain similarly
\[ S^{ncs}_{n' \bar{c} s} \;=\; \bra \phi^{ncs} \:|\: G_k\: K_{n'n}\: v\slsh\:
\phi_{ncs} \ket \;=\; \Tr \left( F_{ncs}\: G_k\: K_{n'n}\: v\slsh \right) . \]
We again substitute in the expansion for $G_k$ (\ref{e:Bcc}). As a consequence
of the additional factor $v\slsh$, the contribution to zeroth order in $\Delta
A$ now drops out. The first order contribution to $S^{ncs}_{n' \bar{c} s}$ is
\begin{eqnarray*}
\lefteqn{ \sum_{l \neq k} \frac{1}{\lambda_k - \lambda_l}\; \left( F_{ncs}\: \Delta A\:
F_l\: K_{n'n}\: v\slsh \right) \;=\;
\frac{1}{\lambda_{ncs} - \lambda_{n'\bar{c} s}}\; \left( F_{ncs}\: \Delta A\:
F_{n' \bar{c} s}\: K_{n'n}\: v\slsh \right) } \\
&=& \frac{1}{\lambda_{ncs} - \lambda_{n'\bar{c} s}}\; \left( F_{ncs}\: \Delta A\:
K_{n'n}\: v\slsh \right) \;=\; (\deg < -1)\:, \spc \spc \spc
\end{eqnarray*}
because according to~(\ref{e:Ce2}) and~(\ref{e:B23a}) the last matrix trace
has degree $\leq 1$. Here we implicitly assumed that $\lambda_{ncs} \neq
\lambda_{n'\bar{c} s}$, because otherwise we clearly get zero.
A straightforward calculation using the factorization
formula~(\ref{e:Ch1}) as well as the estimates for the elementary traces
following~(\ref{e:Cet1}) shows that the second order contribution to
$S^{ncs}_{n' \bar{c} s}$ also is of degree $<-1$. We conclude that
\begin{equation}
S^{ncs}_{n' \bar{c} s} \;=\; (\deg < -1)\:.
    \label{e:Bn}
\end{equation}
Now we can take the inverse of the expansions~(\ref{e:Bm}) and (\ref{e:Bn}). This
gives
\begin{equation}
(S^{-1})^{ncs}_{n'c's} \;=\; \delta^n_{n'}\: \delta^c_{c'} \:+\: \delta^c_{c'}\:
\sum_{l \neq k} \frac{1}{(\lambda_k-\lambda_l)^2}\: \Tr \left( F_{ncs}\: \Delta A\: F_l\:
\Delta A\: K_{n'n} \right) + (\deg < -1)\:.    \label{e:Bo}
\end{equation}

We next compute the expectation values
\[ \bra \phi^{ncs} \:|\: A\: G_k \:|\: \phi_{n'c's} \ket \]
up to contributions of degree $<2$. The method is the same as for the above
calculation of the matrix $S$. In the case $c = c'$, we obtain the following
matrix trace,
\begin{eqnarray*}
\lefteqn{ \bra \phi^{ncs} \:|\: A\: G_k \:|\: \phi_{n' cs} \ket \;=\;
\bra \phi^{ncs} \:|\: A\: G_k\: K_{n'n} \:|\: \phi_{ncs} \ket } \\
&=& \bra \phi^{ncs} \:|\: F_{ncs}\: A\: G_k\: K_{n'n}\: F_{ncs} \:|\: \phi_{ncs} \ket
\;=\; \Tr \left( F_{ncs}\: A\: G_k\: K_{n'n} \right) .
\end{eqnarray*}
Substituting in~(\ref{e:Cb1}) and~(\ref{e:Bb}), the estimate~(\ref{e:C3ord})
shows that it suffices to take into account $G_k$ to second
order~(\ref{e:Bcc}). We get
\begin{eqnarray}
\lefteqn{ \bra \phi^{ncs} \:|\: A\: G_k \:|\: \phi_{n'cs} \ket \;=\;
\lambda_k\; \delta^n_{n'}\: \delta^c_{c'} \:+\:
\Tr \left( F_{ncs}\: \Delta A\: K_{n'n} \right) } \nonumber \\
&&+ \sum_{l \neq k} \frac{1}{\lambda_k - \lambda_l}\:
\Tr \left( F_{ncs}\: \Delta A\: F_l\: \Delta A\: K_{n'n} \right) \nonumber \\
&&- \sum_{l \neq k} \frac{\lambda_k}{(\lambda_k - \lambda_l)^2}\:
\Tr \left( F_{ncs}\: \Delta A\: F_l\: \Delta A\: K_{n'n} \right)
\:+\:(\deg < 2) . \label{e:Br1}
\end{eqnarray}
In the case $c \neq c'$, we can rewrite the expectation value as follows,
\[ \bra \phi^{ncs} \:|\: A\: G_k \:|\: \phi_{n' \bar{c} s} \ket \;=\;
\bra \phi^{ncs} \:|\: A\: G_k \: K_{n'n}\: v\slsh \:|\: \phi_{ncs} \ket
\;=\; \Tr \left( F_{ncs}\: A\: G_k\: K_{n'n}\: v\slsh \right) . \]
If we substitute in~(\ref{e:Cb1}) and~(\ref{e:Bcc}), factor the resulting
matrix traces and use the estimates of the elementary traces of~{\S}\ref{appC3},
we obtain that
\begin{equation}
    \bra \phi^{ncs} \:|\: A\: G_k \:|\: \phi_{n' \bar{c} s} \ket \;=\; (\deg < 2)\:.
    \label{e:Br2}
\end{equation}

In order to bring the matrix elements~(\ref{e:Bq}) into a suitable form, we
substitute the definitions~(\ref{e:Br}) and~(\ref{e:Bp}) into~(\ref{e:Bq}) to
obtain
\begin{eqnarray*}
A^{ncs}_{n'c's} &=& \sum_{(\tilde{n} \tilde{c} s) \in I}
(S^{-1})^{ncs}_{\tilde{n} \tilde{c} s}\; \bra G_k^*\: \phi^{\tilde{n} \tilde{c}
x} \:|\: A \:|\: G_k\: \phi_{n' c' s} \ket \\
&=& \sum_{(\tilde{n} \tilde{c} s) \in I}
(S^{-1})^{ncs}_{\tilde{n} \tilde{c} s}\; \bra \phi^{\tilde{n} \tilde{c}
x} \:|\: A \: G_k \:|\: \phi_{n' c' s} \ket \:,
\end{eqnarray*}
where in the last step we used that $G_k$ commutes with $A$ (as the projector
on an invariant subspace). Putting in the expansions~(\ref{e:Bo})
and~(\ref{e:Br1}, \ref{e:Br2}) gives the result.
\QED
If there are no degeneracies, the above theorem reduces to the well-known
formula of second order perturbation theory. The important result is that to the
considered degree on the light cone, the matrix elements $A^{ncs}_{n'c's}$ are
all zero if $c \neq c'$. In other words, the left- and right-handed components
are invariant subspaces of $A$. This fact immediately gives the
following corollary.

\begin{Corollary} \label{corC1}
Consider the fermionic projector in the presence of chiral and scalar
potentials~(\ref{e:3dir}), were in composite expressions we disregard all terms
quadratic in the field strength. Suppose that the matrix $A^{ncs}_{n'c's}$,
(\ref{e:Bs}), is diagonal in the sector indices $n,n'$ for all
$k=2,\ldots,K$. Then for $n=1,\ldots,7$, the contributions to the eigenvalues
of degree two are
\begin{eqnarray}
\Delta \lambda_{nL+} &\!\!\!=&\!\!\! \Tr (F_{nL+} \:\Delta A) \:+\: \sum_{n'=1}^8
\frac{1}{\lambda_{nL+}-\lambda_{n'R-}}\: \Tr( F_{nL+}\:\Delta A\:
F_{n'R-}\:\Delta A )  \label{e:CA1} \\
\Delta \lambda_{nL-} &\!\!\!=&\!\!\! \Tr (F_{nL-} \:\Delta A) \:+\: \sum_{n'=1}^8
\frac{1}{\lambda_{nL-}-\lambda_{n'R+}}\: \Tr( F_{nL-}\:\Delta A\:
F_{n'R+}\:\Delta A ) \:.\qquad\;\; \label{e:CB1}
\end{eqnarray}
The traces appearing here are given explicitly
by~(\ref{e:Ctr1}--\ref{e:Ctr2}), where the line integrals are in phase-free
form.  The corresponding
formulas for the opposite chirality are obtained by the
replacements~(\ref{e:Ct}).
\end{Corollary}
{\Proof} The result is an immediate consequence of Theorem~\ref{thmB1} and the
estimates~(\ref{e:CE}--\ref{e:C2e}).
\QED

\setcounter{backupeqn}{\value{equation}}
\section{Perturbation of the Kernel} \label{appC5}
\addtocounter{equation}{\value{backupeqn}}
The results of the previous section do not apply to the kernel of $A$. The
reason is that for $k=1$, the index set $I$, (\ref{e:Bind}), is
\[ I \;=\; \left\{ (ncs) {\mbox{ with }} n=8, c=L\!/\!R, s=\pm \right\} , \]
and this index set contains both elements with $s=+$ and $s=-$, giving rise
to different types of matrix elements. On the other hand, the situation for
the kernel is easier because the unperturbed spectral projector on the kernel
satisfies the relations
\begin{eqnarray}
X^*\:F_1\:X &=& 0 \label{e:BA} \\
\chi_R\: F_1\:X &=& 0 \;=\; X^*\: F_1\: \chi_L \:, \label{e:BB}
\end{eqnarray}
and furthermore we can simplify the calculations using that $\lambda_1=0$.
Using these relations, it follows that, neglecting all contributions of
degree $<2$, the dimension of the kernel is not affected by the perturbation.

\begin{Thm} \label{thmB2}
Consider the fermionic projector in the presence of chiral and scalar
potentials~(\ref{e:3dir}) and assume that the fermionic projector is
weakly causality compatible~(see Def.~\ref{defwccc}). Suppose that
in composite expressions all terms quadratic in the field strength are
discarded. Then
\[ A\: G_1 \;=\; (\deg <2)\:. \]
\end{Thm}
{\Proof} Using the definition~(\ref{e:Ba}),
\begin{eqnarray*}
A\: G_1 &=& \frac{1}{2\pi i} \oint_{|z| = \varepsilon} A\: (z-A)^{-1}\: dz
\;=\; \frac{1}{2\pi i} \oint_{|z| = \varepsilon}
\left(z\: (z-A)^{-1} \:-\: \1 \right)\: dz \\
&=& \frac{1}{2\pi i} \oint_{|z| = \varepsilon} z\: (z-A)^{-1}\: dz \:.
\end{eqnarray*}
Performing the perturbation expansion gives, similar to~(\ref{e:Bb}),
\begin{equation}
A\: G_1 \;=\; \sum_{n=0}^\infty\: \frac{1}{2 \pi i}
\oint_{|z|=\varepsilon}
z \left((z-A_0)^{-1}\: \Delta A \right)^n \:(z-A_0)^{-1}\: dz\:.
\label{e:BC}
\end{equation}
When we substitute in~(\ref{e:BzA0}) and carry out the contour integral with
residues, we get zero unless the factor $z$ is differentiated. For this to
occur, the pole at $z=0$ must be at least of order two, and thus we need to
take into account only the orders in perturbation theory $n \geq 2$. If
$n>2$, we can as in the previous section transform the matrix
products into matrix traces, and the estimate~(\ref{e:C3ord}) yields that
the resulting contributions to $A G_1$ are of degree $<2$. Thus it suffices
to consider the second order in perturbation theory,
\begin{eqnarray}
A\: G_1 &=& \frac{1}{2 \pi i} \oint_{|z|=\varepsilon} z\;
(z-A_0)^{-1}\:\Delta A\: (z-A_0)^{-1}\:\Delta A\: (z-A_0)^{-1}\: dz
\:+\: (\deg <2) \nonumber \\
&=& -\sum_{l=2}^K \frac{1}{\lambda_l} \left(
F_l \:\Delta A\: F_1 \:\Delta A\: F_1 \:+\:
F_1 \:\Delta A\: F_1 \:\Delta A\: F_l\:+\:
F_1 \:\Delta A\: F_l \:\Delta A\: F_1 \right) \nonumber \\
&&+\: (\deg <2) \label{e:BE}
\end{eqnarray}
The weak causality compatibility condition implies that
\begin{eqnarray}
X\: P(x,y) \;=\; P(x,y) \;=\; P(x,y)\: X^* \:,
\end{eqnarray}
and similarly for composite expressions in the fermionic projector.
As a consequence, the first two matrix products in~(\ref{e:BE}) vanish;
namely,
\[ \Delta A \:F_1\: \Delta A \;=\; (\Delta A\: X^*) \:F_1\: (X\: \Delta A)
\;=\; \Delta A\: (X^* \:F_1\: X)\: \Delta A \;\stackrel{(\ref{e:BA})}{=}\;
0\:. \]
In the last matrix product in~(\ref{e:BE}) we can apply~(\ref{e:BB}),
\begin{equation}
F_1 \:\Delta A\: F_l \:\Delta A\: F_1 \;=\;
F_1 \:(X\:\Delta A)\: F_l \:(\Delta A\:X^*)\: F_1 \;=\;
\chi_L\: F_1 \:\Delta A\: F_l \:\Delta A\: F_1\:\chi_R \:. \label{e:BD}
\end{equation}
Next we substitute~(\ref{e:Bc}), rewrite the resulting operator products
as matrix traces, factor these matrix traces into elementary traces, and
apply the estimates of~{\S}\ref{appC3}. This straightforward calculation
shows that the matrix product~(\ref{e:BD}) is of
degree~$<5$ on the light cone. From~(\ref{e:BE}) we conclude that $A G_1$ is
of degree~$<2$.
\QED

\end{appendix}

\backmatter
\bibliographystyle{amsalpha}

\vspace*{.3cm}%
\noindent
{\footnotesize{%
{{NWF I - Mathematik, Universit\"at Regensburg, D-93040 Regensburg, Germany}} \\
Email: {\tt{Felix.Finster@mathematik.uni-regensburg.de}}}

\printindex
\cleardoublepage
\markboth{NOTATION INDEX}{}
\addcontentsline{toc}{chapter}{Notation Index}
\twocolumn[\vspace*{2.8cm} \centerline{\bf{\Large{Notation Index}}} \vspace*{1.9cm}]

\begin{itemize}
\item[] $(M,\langle .,.\rangle)$ \,\pageref{Mrangle}
\item[] $\mathrm{E}$ \,\pageref{E}
\item[] $J_x$ \,\pageref{Jx}
\item[] $J$, $J^\lor_x$, $J^\land_x$ \,\pageref{Jlorx}
\item[] $I$, $I^\lor_x$, $I^\land_x$ \,\pageref{Ilorx}
\item[] $L$, $L^\lor_x$, $L^\land_x$ \,\pageref{Elorx}
\item[] $T_pM$ \,\pageref{Tpm}
\item[] $g_{jk}$ \,\pageref{gjk}
\item[] $\nabla$ \,\pageref{nabla}
\item[] $R^i_{jkl}$ \,\pageref{Rijkl}
\item[] $T_{jk}$ \,\pageref{Tjk}
\item[] $S$ \,\pageref{S}, \pageref{S2}, \pageref{e:2a}
\item[] $\Box$ \,\pageref{Box}
\item[] $\{.,.\}$ \,\pageref{{.,.}}
\item[] $\gamma^j$ \,\pageref{gammaj}
\item[] $\vec{\sigma}$ \,\pageref{vecsigma}
\item[] $\Pdd$ \,\pageref{Pdd}
\item[] $\Sl .|. \Sr$ \,\pageref{Sl.|.Sr}
\item[] $A^\ast$ \,\pageref{Aast}
\item[] $J^k$ \,\pageref{Jk}
\item[] $\rho$ \,\pageref{rho}
\item[] $\epsilon_{jklm}$ \,\pageref{epsilonjklm}
\item[] $\chi_L$, $\chi_R$ \,\pageref{chiL}
\item[] $(.|.)$ \,\pageref{(.|.)}, \pageref{(.|.)b}
\item[] $h$ \,\pageref{h}
\item[] $\Psi_{\vec{p} s \epsilon}$ \,\pageref{Psivecpsepsilon}
\item[] $d\mu_{\vec{p}}$ \,\pageref{dmuvecp}
\item[] $\land$ \,\pageref{land}
\item[] $\mathcal{F}$, $\mathcal{F}^n$ \,\pageref{mathcalFn}
\item[] $\hat{\Psi}_{\vec{p} s\epsilon}$, $\hat{\Psi}^\dagger_{\vec{p} s \epsilon}$ \,\pageref{hatPsidagger}
\item[] $\mathcal{G}$ \,\pageref{mathcalG}
\item[] $\mathcal{D}$ \,\pageref{mathcalD}, \pageref{mathcalD2}
\item[] $\sigma^{jk}$ \,\pageref{sigmajk}
\item[] $\Tr$ \,\pageref{Tr}, \pageref{Tr2}
\item[] $D$ \,\pageref{gcd}, \pageref{D}
\item[] $\bra .|. \ket$ \,\pageref{langle.|.rangle}
\item[] $\mathcal{B}$ \,\pageref{mathcalB}
\item[] $s_m$ \,\pageref{sm}
\item[] $s^\lor_m$, $s^\land_m$ \,\pageref{slorm}
\item[] $s^F_m$ \,\pageref{sFm}
\item[] $P^{\mbox{\scriptsize{sea}}}$ \,\pageref{Psea}, \pageref{Psea2}, \pageref{84}
\item[] $p_m$ \,\pageref{pm}
\item[] $k_m$ \,\pageref{km}
\item[] $\tilde{k}_m$ \,\pageref{tildekm}
\item[] $\tilde{p}_m$ \,\pageref{tildepm}
\item[] $\tilde{s}_m$ \,\pageref{sm2}
\item[] $p$ \,\pageref{p}
\item[] $k$ \,\pageref{k}
\item[] $X$ \,\pageref{X}
\item[] $Y$ \,\pageref{Y}
\item[] $P(x,y)$ \,\pageref{1g}, \pageref{Pfut}, \pageref{P(x,y)}
\item[] $c_{\mbox{\scriptsize{norm}}}$ \,\pageref{1g}, \pageref{Pfut}, \pageref{cnorm}
\item[] $\Psi_{\mbox{\scriptsize{in}}}$, $\Psi_{\mbox{\scriptsize{out}}}$ \,\pageref{Psiin}
\item[] $\Aslsh_L$, $\Aslsh_R$ \,\pageref{AslshR}
\item[] $\Phi$ \,\pageref{Phi}
\item[] $\Xi$ \,\pageref{Xi}
\item[] $S^\lor_{m^2}$ \,\pageref{Slorm2}
\item[] $Y_L$, $Y_R$ \,\pageref{YL}
\item[] $S^{(l)}$ \,\pageref{Sl}
\item[] $\Pexp$ \,\pageref{Pexp}
\item[] $T_{m^2}$ \,\pageref{Tm2}
\item[] $T_{m^2}^{\mbox{\scriptsize{reg}}}$ \,\pageref{Tregm2}
\item[] $T^{(l)}$ \,\pageref{T(l)}, \pageref{T^(n)}
\item[] $P^{\mbox{\scriptsize{he}}}$ \,\pageref{Phe}
\item[] $P^{\mbox{\scriptsize{le}}}$ \,\pageref{Ple}
\item[] $X^i$ \,\pageref{X^i}, \pageref{X^i2}
\item[] $\vert x\alpha \ket$ \,\pageref{vertxalpha}
\item[] $\Sl \Psi \:|\: \Phi \Sr$ \,\pageref{SlPsi|PhiSr}
\item[] $M$ \,\pageref{M}
\item[] $E_x$ \,\pageref{E_x}
\item[] $\mathcal{L}$ \,\pageref{mathcalL}
\item[] $A$ \,\pageref{A}
\item[] $|.|$ \,\pageref{|.|}
\item[] $\overline{A}$ \,\pageref{overlineA}
\item[] ${\tr}$ \,\pageref{tr}
\item[] $Q(x,y)$ \,\pageref{Q}, \pageref{e:2c}
\item[] $\mathcal{B}$ \,\pageref{mathcalB2}
\item[] $\hat{P}$ \,\pageref{hatP}
\item[] $E_P$ \,\pageref{E_P}
\item[] $s$ \,\pageref{s}
\item[] $l$ \,\pageref{l}
\item[] $u$ \,\pageref{u}
\item[] $v$ \,\pageref{v}
\item[] $l_{\mbox{\scriptsize{max}}}$ \,\pageref{lmax}
\item[] $\alpha_{\mbox{\scriptsize{max}}}$ \,\pageref{alphamax}
\item[] $\varepsilon_{\mbox{\scriptsize{shear}}}$
  \,\pageref{varepsilon_shear}
\item[] $T^{(n)}_{[p]}$ \,\pageref{T^(n)_[p]}
\item[] $T^{(n)}_{\{p\}}$ \,\pageref{T^(n)_{p}}
\item[] $\xi$ \,\pageref{xi}
\item[] $L$ \,\pageref{L}
\item[] $c_{\mbox{\scriptsize{reg}}}$ \,\pageref{c_reg}
\item[] $\mathcal{M}$ \,\pageref{mathcalM}, \pageref{mathcalM2}
\item[] $\lambda_+$, $\lambda_-$ \,\pageref{lambda_pm}
\item[] $F_+$, $F_-$ \,\pageref{F_pm}
\item[] $z^{(n)}_{[r]}$ \,\pageref{z^(n)_[r]}
\item[] $\deg$ \,\pageref{deg}
\item[] $\hat{Q}(p)$ \,\pageref{hatQ(p)}
\item[] $\mathcal{C}$ \,\pageref{mathcalC}
\item[] $\mathcal{C}^\lor$, $\mathcal{C}^\land$ \,\pageref{mathcalClor}
\item[] $A_L$, $A_R$ \,\pageref{A_L/R}
\item[] $Y_L$, $Y_R$ \,\pageref{Y_L/R}
\item[] $W_L$, $W_R$ \,\pageref{W_c}
\item[] $\nu_{nc}$ \,\pageref{nu_nc}
\item[] $I_{nc}$ \,\pageref{I_nc}
\item[] $B_p$ \,\pageref{e:3f1}
\item[] $F_p$ \,\pageref{e:3f1}
\item[] $\mathcal{B}_p$ \,\pageref{mathcalB_p}
\item[] $\mathcal{F}_p$ \,\pageref{mathcalF_p}
\item[] $\mathcal{G}_p$ \,\pageref{e:3A}
\item[] $\acute{Y}_L$, $\acute{Y}_R$ \,\pageref{acuteY_LR}
\item[] $\grave{Y}_L$, $\grave{Y}_R$ \,\pageref{graveY_LR}
\item[] $\hat{Y}_L$, $\hat{Y}_R$ \,\pageref{hatY_LR}
\item[] $I_\uparrow$, $I_\downarrow$ \,\pageref{I_arrows}
\item[] $P^q$ \,\pageref{P^q}
\item[] $P^l$ \,\pageref{P^l}
\item[] $\overline{V}$ \,\pageref{overlineV}
\item[] $Y^\eff$ \,\pageref{e:5A}
\item[] $A_L^\eff$, $A_R^\eff$ \,\pageref{e:5D}
\item[] $K_a$ \,\pageref{K_a}
\item[] $H_a$ \,\pageref{H_a}
\item[] $S^\Join_a$ \,\pageref{SJoin_a}
\item[] $K^{(n)}$ \,\pageref{K^(n)}
\item[] $S^{(n)}_\Join$ \,\pageref{S^(n)Join}
\item[] $H^{(n)}$ \,\pageref{H^(n)}

\end{itemize}


\end{document}